\documentclass[10pt,english,a4]{report}
\usepackage{times}
\usepackage[T1]{fontenc}
\usepackage[latin1]{inputenc}
\usepackage{geometry}
\usepackage{graphicx}
\usepackage{amssymb}

\makeatletter

\let\SF@@footnote\footnote
\def\footnote{\ifx\protect\@typeset@protect
    \expandafter\SF@@footnote
  \else
    \expandafter\SF@gobble@opt
  \fi
}
\expandafter\def\csname SF@gobble@opt \endcsname{\@ifnextchar[
  \SF@gobble@twobracket
  \@gobble
}
\edef\SF@gobble@opt{\noexpand\protect
  \expandafter\noexpand\csname SF@gobble@opt \endcsname}
\def\SF@gobble@twobracket[#1]#2{}

\makeatletter
\renewcommand\tableofcontents{%
    \if@twocolumn
      \@restonecoltrue\onecolumn
    \else
      \@restonecolfalse
    \fi
    \chapter*{\contentsname
      \@mkboth{\contentsname}{\contentsname}}%
    \@starttoc{toc}%
    \if@restonecol\twocolumn\fi
    }

\renewcommand\listoffigures{%
    \if@twocolumn
      \@restonecoltrue\onecolumn
    \else
      \@restonecolfalse
    \fi
    \chapter*{\listfigurename
      \@mkboth{\listfigurename}{\listfigurename}}%
    \@starttoc{lof}%
    \if@restonecol\twocolumn\fi
    }

\renewcommand\listoftables{%
    \if@twocolumn
      \@restonecoltrue\onecolumn
    \else
      \@restonecolfalse
    \fi
    \chapter*{\listtablename
      \@mkboth{\listtablename}{\listtablename}}%
    \@starttoc{lot}%
    \if@restonecol\twocolumn\fi
    }

\renewenvironment{thebibliography}[1]
     {\chapter*{\bibname
        \@mkboth{\bibname}{\bibname}}%
      \list{\@biblabel{\@arabic\c@enumiv}}%
           {\settowidth\labelwidth{\@biblabel{#1}}%
            \leftmargin\labelwidth
            \advance\leftmargin\labelsep
            \@openbib@code
            \usecounter{enumiv}%
            \let\p@enumiv\@empty
            \renewcommand\theenumiv{\@arabic\c@enumiv}}%
      \sloppy
      \clubpenalty4000
      \@clubpenalty \clubpenalty
      \widowpenalty4000%
      \sfcode`\.\@m}
     {\def\@noitemerr
       {\@latex@warning{Empty `thebibliography' environment}}%
      \endlist}

\usepackage{fancyhdr}
\usepackage{epsfig}
\usepackage{cite}
\usepackage{graphicx}
\usepackage{amsmath}
\usepackage{theorem}
\usepackage{amssymb}
\usepackage{latexsym}
\usepackage{epic}
\usepackage{setspace}
\usepackage{psfrag}

\newcounter{ind}

{\theorembodyfont{\rmfamily}%
   }
{\theorembodyfont{\rmfamily}%
   }
{\theorembodyfont{\rmfamily}%
   }
{\theorembodyfont{\rmfamily}%
   }
{\theorembodyfont{\rmfamily}%
   }
{\theorembodyfont{\rmfamily}%
   }

\newcommand {\ket} [1] {\ensuremath{\left|#1\right>}}
\newcommand {\bra} [1] {\ensuremath{\left<#1\right|}}
\newcommand {\bki} [1] {\ensuremath{\left<#1\right>}}
\newcommand {\bkii} [2] {\ensuremath{\left<#1|#2\right>}}
\newcommand {\bkiii} [3] {\ensuremath{\left<#1\left|#2\right|#3\right>}}

\usepackage{babel}
\makeatother
\begin{document}
\pagenumbering{roman} \setcounter{page}{1}

\newpage

\thispagestyle{empty}

\begin{center}\vspace*{1cm} \end{center}

\noindent \textbf{\LARGE Symmetries in QFT.}{\LARGE \par}

\begin{center}\vspace*{2cm} \end{center}

\noindent {\large K.M.Hamilton and J.F.Wheater. }{\large \par}

\begin{center}\vfill\end{center}

\noindent {\large Lecture Notes.} 

\begin{center}\vspace*{0.9cm}\end{center}

\noindent {\large Hilary term 2002}{\large \par}

\begin{center}\vspace*{0.8cm} \end{center}

\newpage
\thispagestyle{empty} \addcontentsline{toc}{chapter}{\numberline{}{Abstract}}

\noindent \begin{center}\textbf{\Large Symmetries in QFT.}\end{center}{\Large \par}

\begin{center}\vspace*{1cm} \end{center}

\begin{center}{\large $\textrm{K.M.Hamilton}^{\dagger}$ and $\textrm{J.F.Wheater}^{\ddagger}$ }\end{center}{\large \par}

\begin{center}{\large $^{\dagger}$}\emph{\large $\textrm{Denys Wilkinson Building, Keble Road, Oxford OX1 3RH, U.K.}$}\end{center}{\large \par}

\begin{center}{\large $^{\ddagger}$}\emph{\large $\textrm{Department of Physics - Theoretical Physics, University of Oxford},$}\end{center}{\large \par}

\begin{center}\emph{\large 1 Keble Road, Oxford OX1 3NP, U.K.}\end{center}{\large \par}

\begin{center}\vspace*{0.5cm} \end{center}

\noindent \begin{center}{\large January 2003}\end{center}{\large \par}

\begin{center}\vspace*{1cm} \end{center}

\begin{center}\textbf{\large Abstract}\end{center}{\large \par}

\noindent This document contains notes from the graduate lecture
course, {}``Symmetries in QFT'' given by J.F. Wheater at Oxford
University in Hilary term. The course gives an informal introduction
to QFT{\large }%
\footnote{Corrections greatly appreciated: \emph{k.hamilton1@physics.ox.ac.uk
.}%
} \textbf{\large }.

\tableofcontents{}\clearpage\pagenumbering{arabic} \setcounter{page}{1}

\chapter{Symmetries in Field Theory.}

\section{Introducing Q.F.T. and Symmetries.}

The generating functional for a quantum field theory containing a
real scalar field is given by the following\begin{equation}
Z=\int D\phi\textrm{ exp }i\int\textrm{d}^{D}x\textrm{ }{\cal {L}}\left(\phi\right)\label{1.1.1}\end{equation}
where \begin{equation}
{\cal {L}}\left(\phi\right)=\frac{1}{2}\eta^{\mu\nu}\partial_{\mu}\phi\partial_{\nu}\phi-\frac{1}{2}m^{2}\phi^{2}-V\left(\phi\right).\label{1.1.1}\end{equation}

The correlation functions in this scalar field theory are given by\begin{equation}
G\left(x_{1},...,x_{n}\right)=\frac{1}{Z}\int D\phi\textrm{ }\phi\left(x_{1}\right)...\phi\left(x_{n}\right)\textrm{ exp }i\int\textrm{d}^{D}x\textrm{ }{\cal {L}}\left(\phi\right).\label{1.1.3}\end{equation}
Suppose the potential $V\left(\phi\right)$ has the form $V\left(\phi\right)=g\phi^{4}$.
In this instance the Lagrangian has a symmetry $\phi\rightarrow-\phi$
as $L\left(\phi\right)$ is an even function of $\phi$. Before continuing
consider what is meant by the measure of the integration, $\int D\phi$.
\begin{equation}
\int D\phi=\prod_{x}\int_{-\infty}^{+\infty}\textrm{d}\phi\left(x\right)\label{1.1.4}\end{equation}
One should consider $x$ as being a label for a variable, the variable
being the value of the field at the space-time point $x$ . Hence
the product of the integrals of the values of the fields at every
point in space-time is the integral over all possible field configurations.
So what happens to the integral over all possible field configurations
under $\phi\rightarrow\phi'=-\phi$ ?\begin{equation}
\begin{array}{rcl}
\int D\phi & = & \prod_{x}\int_{-\infty}^{+\infty}\textrm{d}\phi\left(x\right)\\
 & = & \prod_{x}\int_{+\infty}^{-\infty}-\textrm{d}\phi'\left(x\right)\\
 & = & \prod_{x}\int_{-\infty}^{+\infty}\textrm{d}\phi'\left(x\right)\\
 & = & \int D\phi'\end{array}\label{1.1.5}\end{equation}
Nothing, the measure is invariant under $\phi\rightarrow\phi'=-\phi$.
Now as the measure of the integration and the action are both invariant
under this transformation of the fields the whole generating functional
is therefore invariant under $\phi\rightarrow\phi'=-\phi$, $Z\rightarrow Z$. 

The effects of this transformation on the correlation functions are
not so trivial though,\begin{equation}
\begin{array}{rcl}
G\left(x_{1},...,x_{n}\right) & = & \frac{1}{Z}\int D\phi\textrm{ }\phi\left(x_{1}\right)...\phi\left(x_{n}\right)\textrm{ exp }i\int\textrm{d}^{D}x\textrm{ }{\cal {L}}\left(\phi\right)\\
 & = & \left(-1\right)^{n}\frac{1}{Z}\int D\phi'\textrm{ }\phi'\left(x_{1}\right)...\phi'\left(x_{n}\right)\textrm{ exp }i\int\textrm{d}^{D}x\textrm{ }{\cal {L}}\left(\phi'\right)\end{array}\label{1.1.6}\end{equation}
a factor of $\left(-1\right)^{n}$ is acquired from the string of
fields before the exponent $\phi\left(x_{1}\right)...\phi\left(x_{n}\right)=\left(-1\right)^{n}\phi'\left(x_{1}\right)...\phi'\left(x_{n}\right)$.
From a purely mathematical perspective one sees that, \begin{equation}
\frac{1}{Z}\int D\phi\textrm{ }\phi\left(x_{1}\right)...\phi\left(x_{n}\right)\textrm{ exp }i\int\textrm{d}^{D}x\textrm{ }{\cal {L}}\left(\phi\right)=\frac{1}{Z}\int D\phi'\textrm{ }\phi'\left(x_{1}\right)...\phi'\left(x_{n}\right)\textrm{ exp }i\int\textrm{d}^{D}x\textrm{ }{\cal {L}}\left(\phi'\right)\label{1.1.7}\end{equation}
as the field is just a dummy integration variable which we are integrating
everywhere from $+\infty$ to $-\infty$ and because $Z\rightarrow Z$
under the transformation. Hence, \begin{equation}
\begin{array}{rcl}
G\left(x_{1},...,x_{2}\right) & = & \frac{1}{Z}\int D\phi\textrm{ }\phi\left(x_{1}\right)...\phi\left(x_{n}\right)\textrm{ exp }i\int\textrm{d}^{D}x\textrm{ }{\cal {L}}\left(\phi\right)\\
 & = & \left(-1\right)^{n}\frac{1}{Z}\int D\phi'\textrm{ }\phi'\left(x_{1}\right)...\phi'\left(x_{n}\right)\textrm{ exp }i\int\textrm{d}^{D}x\textrm{ }{\cal {L}}\left(\phi'\right)\\
 & = & \left(-1\right)^{n}\frac{1}{Z}\int D\phi\textrm{ }\phi\left(x_{1}\right)...\phi\left(x_{n}\right)\textrm{ exp }i\int\textrm{d}^{D}x\textrm{ }{\cal {L}}\left(\phi\right)\\
 & = & \left(-1\right)^{n}G\left(x_{1},...,x_{2}\right)\end{array}\label{1.1.8}\end{equation}
 For this statement to be self consistent $n$ must be even, \emph{i.e.}
no odd Green's functions. This result was seen in our perturbative
treatment of QFT in other courses where it was not so much proved
as shown. The point to take away here is that one can make important
deductions in field theory through symmetry considerations and without
resorting to perturbation theory. Also, we have derived this result
in an arbitrary number of space-time dimensions $D$. 

We can still however raise questions about the validity of this result.
One obvious point is that this result is restricted to potentials
which are even functions of $\phi$. Also, do quantum loop corrections
affect our result? Finally note that the symmetry transformation law
used to derive our result was discrete.

Consider now continuous symmetries. Let us take a similar Lagrangian
to the one above but this time with two real scalar fields $\phi_{1}$
and $\phi_{2}$. \begin{equation}
{\cal {L}}\left(\phi_{1},\phi_{2}\right)=\frac{1}{2}\eta^{\mu\nu}\partial_{\mu}\phi_{1}\partial_{\nu}\phi_{2}+\frac{1}{2}\eta^{\mu\nu}\partial_{\mu}\phi_{1}\partial_{\nu}\phi_{2}-\frac{m^{2}}{2}\left(\phi_{1}^{2}+\phi_{2}^{2}\right)-V\left(\phi_{1},\phi_{2}\right).\label{1.1.9}\end{equation}
The integration over all field configurations is obviously $\int D\phi_{1}\int D\phi_{2}$.
We will now employ a common streamlined notation for these situations.
We re-express our fields as follows\begin{equation}
\begin{array}{rcl}
\phi & = & \frac{1}{\sqrt{2}}\left(\phi_{1}+i\phi_{2}\right)\\
\phi^{\dagger} & = & \frac{1}{\sqrt{2}}\left(\phi_{1}-i\phi_{2}\right)\end{array}\textrm{i}.\textrm{e}.\textrm{ }\left(\begin{array}{c}
\phi\\
\phi^{\dagger}\end{array}\right)=\frac{1}{\sqrt{2}}\left(\begin{array}{cc}
1 & i\\
1 & -i\end{array}\right)\textrm{ }\left(\begin{array}{c}
\phi_{1}\\
\phi_{2}\end{array}\right)\label{1.1.10}\end{equation}
Note that the dagger here is \emph{not} that of canonical quantization,
$\phi^{\dagger}$ is not an annihilation operator! If that was the
case we would only have one type of creation operator and one type
of annihilation operator in decomposing $\phi$ and $\phi^{\dagger}$,
this is not the case! $\phi_{1}$ has its own creation operator and
its own annihilation operator and $\phi_{2}$ also has its own creation
operator and its own annihilation operator \emph{i.e.} there would
be two \emph{distinct} creation operators and two \emph{distinct}
annihilation operators. In the new compact notation we have,\begin{equation}
mass\textrm{ }term=-m^{2}\phi^{\dagger}\phi\label{1.1.11}\end{equation}
\begin{equation}
\begin{array}{rcl}
kinetic\textrm{ }term & = & \frac{1}{2}\eta^{\mu\nu}\left\{ \partial_{\mu}\left(\frac{\phi+\phi^{\dagger}}{\sqrt{2}}\right)\partial_{\nu}\left(\frac{\phi+\phi^{\dagger}}{\sqrt{2}}\right)\right\} +\frac{1}{2}\eta^{\mu\nu}\left\{ i\partial_{\mu}\left(\frac{\phi-\phi^{\dagger}}{\sqrt{2}}\right)i\partial_{\nu}\left(\frac{\phi-\phi^{\dagger}}{\sqrt{2}}\right)\right\} \\
 & = & \frac{1}{2}\eta^{\mu\nu}\left\{ \frac{1}{2}\left(2\partial_{\mu}\phi^{\dagger}\partial_{\nu}\phi+2\partial_{\mu}\phi\partial_{\nu}\phi^{\dagger}\right)\right\} \textrm{ }\leftarrow\textrm{ Use }\eta^{\mu\nu}\textrm{ }\mu\leftrightarrow\nu\textrm{ symmetry }\\
 & = & \eta^{\mu\nu}\partial_{\mu}\phi^{\dagger}\partial_{\nu}\phi\end{array}\label{1.1.12}\end{equation}
 The action is then,\begin{equation}
S\left[\phi,\phi^{\dagger}\right]=\int\textrm{d}^{D}x\textrm{ }\left\{ \eta^{\mu\nu}\partial_{\mu}\phi^{\dagger}\partial_{\nu}\phi-m^{2}\phi^{\dagger}\phi-V\left(\phi,\phi^{\dagger}\right)\right\} .\label{1.1.13}\end{equation}
Under a global transformation $e^{i\alpha}\in U\left(1\right)$ of
the fields: $\phi\rightarrow\phi'=e^{i\alpha}\phi$. The kinetic and
mass terms in the Lagrangian are invariant:\begin{equation}
\eta^{\mu\nu}\partial_{\mu}\phi^{\dagger}\partial_{\nu}\phi=\eta^{\mu\nu}\partial_{\mu}\left(\phi'^{\dagger}e^{+i\alpha}\right)\partial_{\nu}\left(e^{-i\alpha}\phi'\right)=\eta^{\mu\nu}\left(\partial_{\mu}\phi'^{\dagger}\partial_{\nu}\phi'\right)e^{+i\alpha}e^{-i\alpha}=\eta^{\mu\nu}\left(\partial_{\mu}\phi'^{\dagger}\partial_{\nu}\phi'\right)\label{1.1.14}\end{equation}
\begin{equation}
m^{2}\phi^{\dagger}\phi=m^{2}\phi'^{\dagger}e^{+i\alpha}e^{-i\alpha}\phi'=m^{2}\phi'^{\dagger}\phi'.\label{1.1.15}\end{equation}
 Hence to make the whole Lagrangian invariant we need to have a potential
$V\left(\phi,\phi^{\dagger}\right)$ which is invariant. This is clearly
the case if the potential is a function products of fields of the
form $\phi^{\dagger}\phi$. Hence we have that the action, \begin{equation}
S\left[\phi,\phi^{\dagger}\right]=\int\textrm{d}^{D}x\textrm{ }\left\{ \eta^{\mu\nu}\partial_{\mu}\phi^{\dagger}\partial_{\nu}\phi-m^{2}\phi^{\dagger}\phi-V\left(\phi^{\dagger}\phi\right)\right\} \label{1.1.16}\end{equation}
possesses a global $U\left(1\right)$ invariance. 

What about the measure of the integration?\begin{equation}
\int D\phi_{1}\int D\phi_{2}=\prod_{x}\int_{-\infty}^{+\infty}\textrm{d}\phi_{1}\left(x\right)\int_{-\infty}^{+\infty}\textrm{d}\phi_{2}\left(x\right).\label{1.1.17}\end{equation}
 \begin{equation}
\begin{array}{lllllllll}
\phi & \rightarrow & \phi' & = & e^{i\alpha}\phi & = & \left(\phi_{1}cos\alpha-\phi_{2}sin\alpha\right)+i\left(\phi_{1}sin\alpha+\phi_{2}cos\alpha\right) & = & \phi'_{1}+i\phi'_{2}\\
\phi^{\dagger} & \rightarrow & \phi'^{\dagger} & = & e^{-i\alpha}\phi^{\dagger} & = & \left(\phi_{1}cos\alpha-\phi_{2}sin\alpha\right)-i\left(\phi_{1}sin\alpha+\phi_{2}cos\alpha\right) & = & \phi'_{1}-i\phi'_{2}\end{array}\label{eq:1.1.18}\end{equation}
 \begin{equation}
\Rightarrow\textrm{ }\left(\begin{array}{c}
\phi'_{1}\\
\phi'_{2}\end{array}\right)=\left(\begin{array}{cc}
cos\alpha & -sin\alpha\\
sin\alpha & cos\alpha\end{array}\right)\left(\begin{array}{c}
\phi_{1}\\
\phi_{2}\end{array}\right)=\left(\begin{array}{cc}
\frac{\partial\phi'_{1}}{\partial\phi_{1}} & \frac{\partial\phi'_{1}}{\partial\phi_{2}}\\
\frac{\partial\phi'_{2}}{\partial\phi_{1}} & \frac{\partial\phi'_{2}}{\partial\phi_{2}}\end{array}\right)\left(\begin{array}{c}
\phi_{1}\\
\phi_{2}\end{array}\right)=J\left(\begin{array}{c}
\phi_{1}\\
\phi_{2}\end{array}\right)\label{1.1.19}\end{equation}

\begin{equation}
\Rightarrow\int D\phi'_{1}\int D\phi'_{2}=\prod_{x}\int_{-\infty}^{+\infty}\textrm{d}\phi'_{1}\left(x\right)\int_{-\infty}^{+\infty}\textrm{d}\phi'_{2}\left(x\right)=\prod_{x}\left(\int_{-\infty}^{+\infty}\int_{-\infty}^{+\infty}\left(\textrm{Det }J\left(x\right)\right)\textrm{d}\phi_{1}\left(x\right)\textrm{d}\phi_{2}\left(x\right)\right)\label{1.1.20}\end{equation}
Note that the Jacobian in the integral is equal to one for all $x$
($\textrm{Det }J\left(x\right)=1$)!\begin{equation}
\Rightarrow\int D\phi'_{1}\int D\phi'_{2}=\prod_{x}\int_{-\infty}^{+\infty}\textrm{d}\phi'_{1}\left(x\right)\int_{-\infty}^{+\infty}\textrm{d}\phi'_{2}\left(x\right)=\prod_{x}\int_{-\infty}^{+\infty}\textrm{d}\phi_{1}\left(x\right)\int_{-\infty}^{+\infty}\textrm{d}\phi_{2}\left(x\right)=\int D\phi_{1}\int D\phi_{2}\label{1.1.21}\end{equation}
 \emph{i.e.} the measure of the path integral is also unchanged by
the global $U\left(1\right)$ transformation of the fields. What happens
to the correlation functions?\begin{equation}
G\left(x_{1},...,x_{n},x_{n+1},...,x_{n+m}\right)=\frac{1}{Z}\int D_{\mu}\left(\phi,\phi^{\dagger}\right)\textrm{ }\phi\left(x_{1}\right)...\phi\left(x_{n}\right)\phi^{\dagger}\left(x_{n+1}\right)...\phi^{\dagger}\left(x_{n+m}\right)\textrm{ exp }iS\left[\phi,\phi^{\dagger}\right]\label{1.1.22}\end{equation}
 In the above Green's function the variables $x_{1},...,x_{n}$ apply
to the $\phi$'s and $x_{n+1},...,x_{n+m}$ apply to the $\phi^{\dagger}$'s
(note that $\int D_{\mu}\left(\phi,\phi^{\dagger}\right)$ denotes
the integration over the all possible configurations of the fields
$\phi$ and $\phi^{\dagger}$). Now consider the effect of the $U\left(1\right)$
transformation $\phi\left(x\right)\rightarrow\phi'\left(x\right)=e^{+i\alpha}\phi\left(x\right)$
and $\phi^{\dagger}\left(x\right)\rightarrow\phi'^{\dagger}\left(x\right)=e^{-i\alpha}\phi^{\dagger}\left(x\right)$.
The integration measure and the action are unchanged as we reasoned
above, the string of fields before the exponent, acquires a phase
factor $e^{i\left(m-n\right)\alpha}\phi'\left(x_{1}\right)...\phi'\left(x_{n}\right)\phi'^{\dagger}\left(x_{n+1}\right)...\phi'^{\dagger}\left(x_{n+m}\right)$.
We rewrite the Green's function in terms of $\phi'$:\begin{equation}
\begin{array}{rcl}
G\left(x_{1},...,x_{n},x_{n+1},...,x_{n+m}\right) & = & e^{i\left(m-n\right)\alpha}\frac{1}{Z}\int D_{\mu}\left(e^{-i\alpha}\phi',e^{+i\alpha}\phi'^{\dagger}\right)\textrm{ }\phi'\left(x_{1}\right)...\phi'\left(x_{n}\right)\phi'^{\dagger}\left(x_{n+1}\right)...\phi'^{\dagger}\left(x_{n+m}\right)\\
 &  & \times\textrm{exp }iS\left[e^{-i\alpha}\phi',e^{+i\alpha}\phi'^{\dagger}\right]\end{array}\label{1.1.23}\end{equation}

\noindent At the risk of being verbose the measure is:\begin{equation}
\int D_{\mu}\left(\phi,\phi^{\dagger}\right)=\int D_{\mu}\left(e^{-i\alpha}\phi',e^{+i\alpha}\phi'^{\dagger}\right)=\int D_{\mu}\left(\phi',\phi'^{\dagger}\right).\label{1.1.24}\end{equation}

\begin{equation}
\begin{array}{rcl}
G\left(x_{1},...,x_{n},x_{n+1},...,x_{n+m}\right) & = & e^{i\left(m-n\right)\alpha}\frac{1}{Z}\int D_{\mu}\left(\phi',\phi'^{\dagger}\right)\textrm{ }\phi'\left(x_{1}\right)...\phi'\left(x_{n}\right)\phi'^{\dagger}\left(x_{n+1}\right)...\phi'^{\dagger}\left(x_{n+m}\right)\\
 &  & \times\textrm{exp }iS\left[e^{-i\alpha}\phi',e^{+i\alpha}\phi'^{\dagger}\right]\end{array}\label{1.1.25}\end{equation}
Note, as before when considering the symmetry $\phi\rightarrow-\phi$,
one can step back forget any Physics or symmetries going on and note
that mathematically $\frac{1}{Z}\int D_{\mu}\left(\phi',\phi'^{\dagger}\right)\textrm{ }\phi'\left(x_{1}\right)...\phi'\left(x_{n}\right)\phi'^{\dagger}\left(x_{n+1}\right)...\phi'^{\dagger}\left(x_{n+m}\right)\times\textrm{exp }iS\left[\phi',\phi'^{\dagger}\right]$
is exactly the same as $\frac{1}{Z}\int D_{\mu}\left(\phi,\phi^{\dagger}\right)\textrm{ }\phi\left(x_{1}\right)...\phi\left(x_{n}\right)\phi^{\dagger}\left(x_{n+1}\right)...\phi^{\dagger}\left(x_{n+m}\right)\times\textrm{exp }iS\left[\phi,\phi^{\dagger}\right]$\emph{,}
that is to say both are equal to $G\left(x_{1},...,x_{n},x_{n+1},...,x_{n+m}\right)$.
This is because the $\phi$ and $\phi^{\dagger}$ fields are essentially
just {}``dummy'' variables for the integration and they are all
integrated from $-\infty$ to $+\infty$ everywhere (\emph{i.e.} the
limits of the integration are also unchanged - this was implicit in
our proof that the measure was invariant and is crucial for the last
sentence is to be correct). Analogous to the $\phi\rightarrow-\phi$
case one here has, \begin{equation}
G\left(x_{1},...,x_{n},x_{n+1},...,x_{n+m}\right)=e^{i\left(m-n\right)\alpha}G\left(x_{1},...,x_{n},x_{n+1},...,x_{n+m}\right).\label{1.1.26}\end{equation}
 For these two statements to be consistent we must have $m=n$. Therefore
only Green's functions that contain equal numbers of the different
types of fields $\phi$ and $\phi^{\dagger}$ are non vanishing!

\newpage
\section{Local Symmetries. }

Consider a local $U\left(1\right)$ transformation: $\phi\rightarrow\phi'=e^{i\alpha\left(x\right)}\phi$,
$\phi^{\dagger}\rightarrow\phi'^{\dagger}=e^{-i\alpha\left(x\right)}\phi^{\dagger}$.
Under this transformation the measure of the path integral is unchanged
as it was in the global case, the only difference is that the transformation
of variables is different at every point in space-time but is nevertheless
always of the form:\begin{equation}
J\left(x\right)=\left(\begin{array}{cc}
cos\textrm{ }\alpha\left(x\right) & -sin\textrm{ }\alpha\left(x\right)\\
sin\textrm{ }\alpha\left(x\right) & cos\textrm{ }\alpha\left(x\right)\end{array}\right)\label{1.2.1}\end{equation}
 and hence the Jacobian ($\textrm{Det }J\left(x\right)$) at every
space-time point is always one as before! \begin{equation}
\begin{array}{rcl}
\int D_{\mu}\left(\phi,\phi^{\dagger}\right) & = & \int D\phi_{1}\int D\phi_{2}\\
 & = & \prod_{x}\left(\int_{-\infty}^{+\infty}\int_{-\infty}^{+\infty}\textrm{d}\phi_{1}\left(x\right)\textrm{d}\phi_{2}\left(x\right)\right)\\
 & = & \prod_{x}\left(\int_{-\infty}^{+\infty}\int_{-\infty}^{+\infty}\left(\textrm{Det }J\left(x\right)\right)\textrm{d}\phi'_{1}\left(x\right)\textrm{d}\phi'_{2}\left(x\right)\right)\\
 & = & \prod_{x}\left(\int_{-\infty}^{+\infty}\int_{-\infty}^{+\infty}\textrm{d}\phi'_{1}\left(x\right)\textrm{d}\phi'_{2}\left(x\right)\right)\\
 & = & \int D\phi'_{1}\int D\phi'_{2}=\int D_{\mu}\left(\phi',\phi'^{\dagger}\right)\end{array}\label{1.2.2}\end{equation}
As before if we construct our potential such that it is a function
of $\phi^{\dagger}\phi$ only, then it too must be unchanged by our
transformation of the fields,\begin{equation}
\phi^{\dagger}\phi=\phi'^{\dagger}e^{+i\alpha\left(x\right)}e^{-i\alpha\left(x\right)}\phi'=\phi'^{\dagger}\phi'.\label{1.2.3}\end{equation}
The mass term is also of the form $\frac{m^{2}}{2}\phi^{\dagger}\phi$
and hence it too is invariant under these $U\left(1\right)$ transformations
of the fields. 

Once again our theory is starting to look as though it possesses $U\left(1\right)$
invariance, though this time the invariance is local. The kinetic
term in the action is however definitely not invariant. Whereas before
our fields acquired a constant phase on application of the $U\left(1\right)$
transformation now they have acquired a phase with a space-time dependence
hence we cannot merely pull the phases out in front of the derivatives
in the kinetic term and have them cancel. Instead we get the following
mess for the kinetic part of the action, \begin{equation}
\begin{array}{rcl}
S_{Kinetic}\left[\phi,\phi^{\dagger}\right] & = & \int\textrm{d}^{D}x\textrm{ }\eta^{\mu\nu}\partial_{\mu}\left(e^{-i\alpha\left(x\right)}\phi'\right)\partial_{\nu}\left(e^{+i\alpha\left(x\right)}\phi'^{\dagger}\right)\\
 & = & \int\textrm{d}^{D}x\textrm{ }\eta^{\mu\nu}\left(e^{-i\alpha\left(x\right)}\left(\partial_{\mu}\phi'-i\left(\partial_{\mu}\alpha\left(x\right)\right)\phi'\right)\times e^{+i\alpha\left(x\right)}\left(\partial_{\nu}\phi'^{\dagger}+i\left(\partial_{\nu}\alpha\left(x\right)\right)\phi'^{\dagger}\right)\right)\\
 & = & \int\textrm{d}^{D}x\textrm{ }L_{Kinetic}\left(\phi',\phi'^{\dagger}\right)+\int\textrm{d}^{D}x\textrm{ }\eta^{\mu\nu}\left(i\left(\partial_{\mu}\phi'\right)\phi'^{\dagger}\partial_{\nu}\alpha\left(x\right)-i\left(\partial_{\nu}\phi'^{\dagger}\right)\phi'\partial_{\mu}\alpha\left(x\right)\right)\\
 & + & \int\textrm{d}^{D}x\textrm{ }\eta^{\mu\nu}\left(\partial_{\mu}\alpha\left(x\right)\right)\left(\partial_{\nu}\alpha\left(x\right)\right)\phi'\phi'^{\dagger}\\
 &  & \textrm{Exploit }\eta^{\mu\nu}\textrm{ symmetry in }2\textrm{nd term }(\textrm{linear in }\alpha\left(x\right)).\\
 & = & \int\textrm{d}^{D}x\textrm{ }L_{Kinetic}\left(\phi',\phi'^{\dagger}\right)+\int\textrm{d}^{D}x\textrm{ }\eta^{\mu\nu}\left(i\left(\partial_{\mu}\phi'\right)\phi'^{\dagger}\partial_{\nu}\alpha\left(x\right)-i\left(\partial_{\mu}\phi'^{\dagger}\right)\phi'\partial_{\nu}\alpha\left(x\right)\right)\\
 & + & \int\textrm{d}^{D}x\textrm{ }\eta^{\mu\nu}\left(\partial_{\mu}\alpha\left(x\right)\right)\left(\partial_{\nu}\alpha\left(x\right)\right)\phi'\phi'^{\dagger}\\
 & = & S_{Kinetic}\left[\phi',\phi'^{\dagger}\right]+\int\textrm{d}^{D}x\textrm{ }i\eta^{\mu\nu}\left(\left(\partial_{\mu}\phi'\right)\phi'^{\dagger}-\left(\partial_{\mu}\phi'^{\dagger}\right)\phi'\right)\left(\partial_{\nu}\alpha\left(x\right)\right)\\
 & + & \int\textrm{d}^{D}x\textrm{ }\eta^{\mu\nu}\left(\partial_{\mu}\alpha\left(x\right)\right)\left(\partial_{\nu}\alpha\left(x\right)\right)\phi'\phi'^{\dagger}\end{array}\label{1.2.4}\end{equation}
 Now we have to play some games with the term linear in $\alpha\left(x\right)$.
Integrating by parts we can write it as, \begin{equation}
\begin{array}{rcl}
\int\textrm{d}^{D}x\textrm{ }i\eta^{\mu\nu}\left(\left(\partial_{\mu}\phi'\right)\phi'^{\dagger}-\left(\partial_{\mu}\phi'^{\dagger}\right)\phi'\right)\left(\partial_{\nu}\alpha\left(x\right)\right) & = & \int\textrm{d}^{D}x\textrm{ }i\eta^{\mu\nu}\partial_{\nu}\left(\alpha\left(x\right)\left(\left(\partial_{\mu}\phi'\right)\phi'^{\dagger}-\left(\partial_{\mu}\phi'^{\dagger}\right)\phi'\right)\right)\\
 & - & \int\textrm{d}^{D}x\textrm{ }i\eta^{\mu\nu}\alpha\left(x\right)\partial_{\nu}\left(\left(\partial_{\mu}\phi'\right)\phi'^{\dagger}-\left(\partial_{\mu}\phi'^{\dagger}\right)\phi'\right)\end{array}.\label{1.2.5}\end{equation}
The first term $\int\textrm{d}^{D}x\textrm{ }i\eta^{\mu\nu}\partial_{\nu}\left(\alpha\left(x\right)\left(\left(\partial_{\mu}\phi'\right)\phi'^{\dagger}-\left(\partial_{\mu}\phi'^{\dagger}\right)\phi'\right)\right)$
is a $D-divergence$ which we can write as a surface integral using
Gauss's law. The divergence is integrated over all of space-time,
so if we used Gauss's law the surface over which we would integrate
is at infinity. Assuming that the fields are vanishing at infinity
$\lim_{x\rightarrow\infty}\phi\left(x\right)=0$ the integrand will
be zero at all points on the aforementioned surface, hence the integral
is zero and we have,\begin{equation}
\int\textrm{d}^{D}x\textrm{ }i\eta^{\mu\nu}\left(\left(\partial_{\mu}\phi'\right)\phi'^{\dagger}-\left(\partial_{\mu}\phi'^{\dagger}\right)\phi'\right)\left(\partial_{\nu}\alpha\left(x\right)\right)=-\int\textrm{d}^{D}x\textrm{ }i\eta^{\mu\nu}\alpha\left(x\right)\partial_{\nu}\left(\left(\partial_{\mu}\phi'\right)\phi'^{\dagger}-\left(\partial_{\mu}\phi'^{\dagger}\right)\phi'\right).\label{1.2.6}\end{equation}
 We shall denote $i\left(\partial_{\mu}\phi'\right)\phi'^{\dagger}-i\left(\partial_{\mu}\phi'^{\dagger}\right)\phi'$
by $j_{\mu}$, so the kinetic part of the action becomes:\begin{equation}
\begin{array}{rcl}
S_{Kinetic}\left[\phi,\phi^{\dagger}\right] & = & S_{Kinetic}\left[\phi',\phi'^{\dagger}\right]-\int\textrm{d}^{D}x\textrm{ }\alpha\left(x\right)\partial_{\mu}j^{\mu}\\
 & + & \int\textrm{d}^{D}x\textrm{ }\eta^{\mu\nu}\left(\partial_{\mu}\alpha\left(x\right)\right)\left(\partial_{\nu}\alpha\left(x\right)\right)\phi'\phi'^{\dagger}.\end{array}\label{1.2.7}\end{equation}
If we take the transformation to be infinitesimal \emph{i.e.} $\alpha\left(x\right)$
very small we can neglect the term quadratic in $\alpha\left(x\right)$.
So overall we have that everything is invariant except for the kinetic
part of the action. \begin{equation}
\begin{array}{rcl}
\int D_{\mu}\left(\phi,\phi^{\dagger}\right) & = & \int D_{\mu}\left(\phi',\phi'^{\dagger}\right)\\
m^{2}\phi^{\dagger}\phi & = & m^{2}\phi'^{\dagger}\phi'\\
V\left(\phi^{\dagger}\phi\right) & = & V\left(\phi'^{\dagger}\phi'\right)\\
S_{Kinetic}\left[\phi,\phi^{\dagger}\right] & = & S_{Kinetic}\left[\phi',\phi'^{\dagger}\right]-\int\textrm{d}^{D}x\textrm{ }\alpha\left(x\right)\partial_{\mu}j^{\mu}\end{array}\label{1.2.8}\end{equation}
Now consider the Green's functions: \begin{equation}
G\left(x_{1},...,x_{n},x_{n+1},...,x_{n+m}\right)=\frac{1}{Z}\int D_{\mu}\left(\phi,\phi^{\dagger}\right)\textrm{ }\phi\left(x_{1}\right)...\phi\left(x_{n}\right)\phi^{\dagger}\left(x_{n+1}\right)...\phi^{\dagger}\left(x_{n+m}\right)\textrm{ exp }iS\left[\phi,\phi^{\dagger}\right]\label{1.2.9}\end{equation}
What about writing this in terms of the fields $\phi'$ and $\phi'$
dagger? There are two ways of doing this. The first that comes to
mind is to plug in the transformations of the fields, this gives,\begin{equation}
G\left(x_{1},...,x_{n},x_{n+1},...,x_{n+m}\right)=\frac{1}{Z}\int D_{\mu}\left(\phi',\phi'^{\dagger}\right)\textrm{ }\left(\textrm{exp }i\left(\alpha\left(x_{n+1}\right)+...+\alpha\left(x_{n+m}\right)-\alpha\left(x_{1}\right)-...-\alpha\left(x_{n}\right)\right)\right)\label{1.2.10}\end{equation}
\[
\times\textrm{ }\phi'\left(x_{1}\right)...\phi'\left(x_{n}\right)\phi'^{\dagger}\left(x_{n+1}\right)...\phi'^{\dagger}\left(x_{n+m}\right)\textrm{ exp }iS\left[\phi',\phi'^{\dagger}\right]\times\textrm{exp }-i\int\textrm{d}^{D}x\textrm{ }\alpha\left(x\right)\partial_{\mu}j^{\mu}.\]
However, as the fields are just dummy variables which get integrated
over I should be able to write the generating functional, \emph{by
its definition} , as,\begin{equation}
Z=\int D_{\mu}\left(\phi',\phi'^{\dagger}\right)\textrm{ exp }iS\left[\phi',\phi'^{\dagger}\right]\label{1.2.11}\end{equation}
 and hence the Green's function, \emph{by its definition}, as,\begin{equation}
G\left(x_{1},...,x_{n},x_{n+1},...,x_{n+m}\right)=\frac{1}{Z}\int D_{\mu}\left(\phi',\phi'^{\dagger}\right)\textrm{ }\phi'\left(x_{1}\right)...\phi'\left(x_{n}\right)\phi'^{\dagger}\left(x_{n+1}\right)...\phi'^{\dagger}\left(x_{n+m}\right)\textrm{ exp }iS\left[\phi',\phi'^{\dagger}\right]\label{1.2.12}\end{equation}
 The last two expressions for the Green's functions must be consistent.
The last equation does not depend on the function $\alpha$,\begin{equation}
\frac{\delta G}{\delta\alpha\left(y\right)}=0\label{1.2.13}\end{equation}
 so one condition for their consistency (there are undoubtedly others)
is that $G\left(x_{1},...,x_{n},x_{n+1},...,x_{n+m}\right)$ has no
dependence on $\alpha$. The previous statement is the key to understanding
the origin of Ward identities and hence this course. Generally, in
doing a transformation on some fields in a path integral one will
introduce some parameters of the transformation into the expression,
but given that the Green's function contains no such parameters by
definition, it obviously can have no dependence on any such parameters,
hence one can \emph{always} apply the above condition. 

So what is the functional derivative in this particular instance?
It can be useful when doing this to consider again the spatial coordinate
as being just a label for the {}``true'' variables that we are trying
to differentiate with respect to. We are differentiating $G$ with
respect to the function $\alpha$ \emph{at a particular point in space
$y$}! To be idiot proof this means we get a delta function every
time we differentiate an $e^{i\alpha\left(p\right)}$ and differentiating
$e^{-i\int\textrm{d}^{D}x\textrm{ }\alpha\left(x\right)\partial^{\mu}j_{\mu}\left(x\right)}$
gives $\left.\partial^{\mu}j_{\mu}\left(x\right)\right|_{x=y}e^{-i\int\textrm{d}^{D}x\textrm{ }\alpha\left(x\right)\partial^{\mu}j_{\mu}\left(x\right)}$.
These two types of term in the Green's function are the only ones
that are acted on by the differentiation as nothing else has any $\alpha's$
in it. We are essentially differentiating the following,\begin{equation}
\begin{array}{rl}
 & \frac{\delta}{\delta\alpha\left(y\right)}\left(\textrm{exp }i\left(\alpha\left(x_{n+1}\right)+...+\alpha\left(x_{n+m}\right)-\alpha\left(x_{1}\right)-...-\alpha\left(x_{n}\right)\right)\times\textrm{exp }-i\int\textrm{d}^{D}x\textrm{ }\alpha\left(x\right)\partial_{\mu}j^{\mu}\left(x\right)\right)\\
= & \frac{\delta}{\delta\alpha\left(y\right)}\left(\textrm{exp }i\left(\alpha\left(x_{n+1}\right)+...+\alpha\left(x_{n+m}\right)-\alpha\left(x_{1}\right)-...-\alpha\left(x_{n}\right)-\int\textrm{d}^{D}x\textrm{ }\alpha\left(x\right)\partial_{\mu}j^{\mu}\left(x\right)\right)\right)\\
= & i\left(\delta\left(x_{n+1}-y\right)+...+\delta\left(x_{n+m}-y\right)-\delta\left(x_{1}-y\right)-...-\delta\left(x_{n}-y\right)-\left.\partial^{\mu}j_{\mu}\left(x\right)\right|_{x=y}\right)\\
\times & \left(\textrm{exp }i\left(\alpha\left(x_{n+1}\right)+...+\alpha\left(x_{n+m}\right)-\alpha\left(x_{1}\right)-...-\alpha\left(x_{n}\right)\right)\times\textrm{exp }-i\int\textrm{d}^{D}x\textrm{ }\alpha\left(x\right)\partial_{\mu}j^{\mu}\left(x\right)\right)\end{array}\label{1.2.14}\end{equation}
\begin{equation}
\begin{array}{rcl}
\Rightarrow\frac{\delta G}{\delta\alpha\left(y\right)} & = & 0\\
 & = & \frac{1}{Z}\int D_{\mu}\left(\phi',\phi'^{\dagger}\right)\textrm{ }\left(\textrm{exp }i\left(\alpha\left(x_{n+1}\right)+...+\alpha\left(x_{n+m}\right)-\alpha\left(x_{1}\right)-...-\alpha\left(x_{n}\right)\right)\right)\\
 & \times & i\left(\delta\left(x_{n+1}-y\right)+...+\delta\left(x_{n+m}-y\right)-\delta\left(x_{1}-y\right)-...-\delta\left(x_{n}-y\right)-\left.\partial^{\mu}j_{\mu}\left(x\right)\right|_{x=y}\right)\\
 & \times & \phi'\left(x_{1}\right)...\phi'\left(x_{n}\right)\phi'^{\dagger}\left(x_{n+1}\right)...\phi'^{\dagger}\left(x_{n+m}\right)\textrm{ exp }iS\left[\phi',\phi'^{\dagger}\right]\times\textrm{exp }-i\int\textrm{d}^{D}x\textrm{ }\alpha\left(x\right)\partial_{\mu}j^{\mu}\left(x\right).\end{array}\label{1.2.15}\end{equation}
Note we haven't specified an $\alpha$, \emph{this} \emph{result}
\emph{is} \emph{true} \emph{for all} \textbf{\emph{\underbar{}}}\emph{$\alpha$}'s!
Let's see what happens with $\alpha\left(x\right)=0$, which clearly
corresponds to not transforming the fields at all:\begin{equation}
\begin{array}{rcl}
\Rightarrow\left.\frac{\delta G}{\delta\alpha\left(y\right)}\right|_{\alpha=0} & = & 0\\
 & = & \frac{1}{Z}\int D_{\mu}\left(\phi',\phi'^{\dagger}\right)\times i\left(\delta\left(x_{n+1}-y\right)+...+\delta\left(x_{n+m}-y\right)-\delta\left(x_{1}-y\right)-...\right.\\
 &  & \left.-\delta\left(x_{n}-y\right)-\left.\partial^{\mu}j_{\mu}\left(x\right)\right|_{x=y}\right)\times\phi'\left(x_{1}\right)...\phi'\left(x_{n}\right)\phi'^{\dagger}\left(x_{n+1}\right)...\phi'^{\dagger}\left(x_{n+m}\right)\textrm{ exp }iS\left[\phi',\phi'^{\dagger}\right].\end{array}\label{1.2.16}\end{equation}
 \begin{equation}
\Rightarrow\left\langle 0\left|i\left(\delta\left(x_{n+1}-y\right)+...-\delta\left(x_{n}-y\right)-\partial_{\left(y\right)}^{\mu}j_{\mu}\left(y\right)\right)\phi'\left(x_{1}\right)...\phi'\left(x_{n}\right)\phi'^{\dagger}\left(x_{n+1}\right)...\phi'^{\dagger}\left(x_{n+m}\right)\right|0\right\rangle =0\label{1.2.17}\end{equation}
This last equation is a \emph{Ward identity.} The terms $\delta\left(x-y\right)$
are known as \emph{contact} \emph{terms} they do not form an important
part of our discussion and will play no further role. Consider the
case where we have no fields before the exponent in the path integral
\emph{i.e.} $m=n=0$ in which case,\begin{equation}
\left\langle 0\left|\partial^{\mu}j_{\mu}\right|0\right\rangle =0.\label{1.2.18}\end{equation}
 Considering this at the classical level one gets $\partial^{\mu}j_{\mu}=0$
\emph{i.e.} one recovers a conserved current - Noether's theorem from
the Ward identity. 

At this point a quick summary of the process might be useful. We took
a simple scalar $\phi^{4}$ theory with fields $\phi_{1}$ and $\phi_{2}$
and rewrote it as a complex scalar field theory with fields $\phi$
and $\phi^{\dagger}$ this simplified the form of the Lagrangian considerably.
We then asked what happens to the Lagrangian when we make $U\left(1\right)$
transformations of the fields. The result was that all the terms in
the generating functional were invariant except for the kinetic part
of the action which acquired a current term $-\int\textrm{d}^{D}x\textrm{ }\alpha\left(x\right)\partial_{\mu}j^{\mu}$.
Under the transformations the Green's functions were also not invariant
and became messy under the gauge transformation $\alpha$. However
the definition of the Green's function contains no such terms and
has the same form whatever the fields are that we are integrating
over, the fields are {}``dummy'' variables, hence for consistency
we demand that the Green's function has no dependence on $\alpha$
when rewritten in terms of the transformed fields. Applying this condition
to the expression for the Green's function in terms of the transformed
fields yielded the Ward identity, which we shall see is all important
in field theory. 

Classically the symmetry (Noether) currents were exactly conserved
and the derivation of this depended on the equations of motion being
satisfied. In the path integral formalism we integrate over all possible
configurations not just the classical trajectory but we can expect
an analogous result to that of the classical case, \emph{i.e.} we
expect the expectation value of the symmetry current to be conserved
as we know that the classical equations of motion hold as quantum
averages. 

We will now attempt to generalize the formalism we just used in deriving
our first Ward identity. First we define a set of fields $\phi^{i}\left(x\right)\textrm{ }i=1,...,n$.
Next we define a local, infinitesimal unitary transformation $\Delta_{ij}\left(x\right)=\epsilon\left(x\right)^{A}t_{ij}^{A}$
belonging to some group in which the action is symmetric, where $t_{ij}^{A}$
are the group generators and $\epsilon\left(x\right)^{A}$ are the
associated infinitesimal parameters of the transformation. Finally
we consider what correlation to study, we will be general and say
that we are looking at $\left\langle 0\left|F\left(\left\{ \phi^{i}\right\} \right)\right|0\right\rangle $
where $F\left(\left\{ \phi^{i}\right\} \right)$ is some function
of the fields $\phi^{i}$. Under the transformation we have, 

\emph{\begin{equation}
\begin{array}{rcl}
\phi'^{i}\left(x\right) & \equiv & \phi^{i}\left(x\right)+\delta\phi^{i}\left(x\right)\\
 & = & \phi^{i}\left(x\right)+\Delta^{ij}\left(x\right)\phi_{j}\left(x\right)\end{array}\label{1.2.19}\end{equation}
}representing a local change of variables in the path integral. It
is evident that we have \begin{equation}
\int D\left(\phi'^{i}\right)\textrm{ }F\left(\left\{ \phi'^{i}\right\} \right)\textrm{ }e^{iS\left[\phi'\right]}=\int D\left(\phi^{i}\right)\textrm{ }F\left(\left\{ \phi^{i}\right\} \right)e^{iS\left[\phi\right]}\label{1.2.20}\end{equation}
 since the result cannot depend on the name of the integration variable. 

We now make the all important assumption that the Jacobian to to change
from $D\left(\phi'^{i}\right)$ to $D\left(\phi^{i}\right)$ is one.
This assumption is non trivial since we are dealing with complicated
functional integrals the meaning of which should really be carefully
examined. It is often possible to prove that the Jacobian of the path
integral measure is one and when possible we shall attempt to show
it. In some cases the Jacobian is not one. When the Jacobian is not
equal to one we generate \emph{anomalies} which we will talk about
in later lectures. In other words anomalies arise when a theory looks
as if it possesses a certain symmetry due to the invariance of the
action but it is not actually invariant on account of the non-invariance
of the path integral measure. Until further notice we can forget about
anomalies and take $D\left(\phi'^{i}\right)=D\left(\phi^{i}\right)$
(...at least until lecture 4 - scale invariance). Continuing from
the last equation we then have that,\begin{equation}
\begin{array}{rcl}
\int D\left(\phi^{i}\right)\left[F\left(\left\{ \phi'^{i}\right\} \right)e^{iS\left[\phi'\right]}-F\left(\left\{ \phi^{i}\right\} \right)e^{iS\left[\phi\right]}\right] & = & 0\\
\Rightarrow\int D\left(\phi^{i}\right)\delta\left\{ e^{iS\left[\phi\right]}F\left(\left\{ \phi^{i}\right\} \right)\right\}  & = & 0\end{array}\label{1.2.21}\end{equation}
The variation is due to the functions $\epsilon\left(x\right)^{A}$
so we can rewrite the variation as,\begin{equation}
\Rightarrow\int D\left(\phi^{i}\right)\int\textrm{d}^{D}x\textrm{ }\left(\frac{\delta\left\{ e^{iS\left[\phi\right]}F\left(\left\{ \phi^{i}\right\} \right)\right\} }{\delta\epsilon\left(x\right)^{A}}\epsilon\left(x\right)^{A}\right)=0\label{1.2.22}\end{equation}
note we integrate over all space-time to {}``add up'' the variation
everywhere and get the total. \begin{equation}
\begin{array}{crcl}
\Rightarrow & \int D\left(\phi^{i}\right)\int\textrm{d}^{D}x\textrm{ }\epsilon\left(x\right)^{A}\left\{ F\left(\left\{ \phi^{i}\right\} \right)\frac{\delta S\left[\phi\right]}{\delta\epsilon\left(x\right)^{A}}+\frac{\delta F\left(\left\{ \phi^{i}\right\} \right)}{\delta\epsilon\left(x\right)^{A}}\right\} e^{iS\left[\phi\right]} & = & 0\\
\Rightarrow & \int\textrm{d}^{D}x\textrm{ }\epsilon\left(x\right)^{A}\int D\left(\phi^{i}\right)\left\{ F\left(\left\{ \phi^{i}\right\} \right)\frac{\delta S\left[\phi\right]}{\delta\epsilon\left(x\right)^{A}}+\frac{\delta F\left(\left\{ \phi^{i}\right\} \right)}{\delta\epsilon\left(x\right)^{A}}\right\} e^{iS\left[\phi\right]} & = & 0\end{array}\label{1.2.23}\end{equation}
As each of the functions $\epsilon\left(x\right)^{A}$ are arbitrary
then we require that,\begin{equation}
\int D\left(\phi^{i}\right)\left\{ F\left(\left\{ \phi^{i}\right\} \right)\frac{\delta S\left[\phi\right]}{\delta\epsilon\left(x\right)^{A}}+\frac{\delta F\left(\left\{ \phi^{i}\right\} \right)}{\delta\epsilon\left(x\right)^{A}}\right\} e^{iS\left[\phi\right]}=0\textrm{ }\forall\textrm{ }A\label{1.2.24}\end{equation}
\begin{equation}
\Rightarrow\left\langle 0\left|F\left(\left\{ \phi^{i}\right\} \right)\frac{\delta S\left[\phi\right]}{\delta\epsilon\left(x\right)^{A}}\right|0\right\rangle =-\left\langle 0\left|\frac{\delta F\left(\left\{ \phi^{i}\right\} \right)}{\delta\epsilon\left(x\right)^{A}}\right|0\right\rangle \label{1.2.25}\end{equation}
 For actions which are globally invariant under some symmetry the
variation in the action produced by making the symmetry transformation
local is the divergence of the symmetry current, for unitary transformations
this is $\partial^{\nu}j_{\nu}^{A}\propto\left(\partial_{\nu}\phi'^{\dagger}\right)t^{A}\phi'-\phi'^{\dagger}t^{A}\left(\partial_{\nu}\phi'\right)$.

Now we shall consider a {}``higher'' symmetry, an $SO\left(3\right)$
global symmetry to illustrate more features of the Green's functions.
Consider the following triplet of fields and their transformation
by $R\in SO\left(3\right)$:\begin{equation}
\underline{\phi}=\left(\begin{array}{c}
\phi_{1}\\
\phi_{2}\\
\phi_{3}\end{array}\right)\textrm{ }\rightarrow\underline{\phi'}={\boldmath{R}}\underline{\phi}.\label{1.2.26}\end{equation}
 From this representation we can easily construct a globally $SO\left(3\right)$
invariant Lagrangian:\begin{equation}
\begin{array}{rcccccl}
{\cal {L}} & = & \frac{1}{2}\eta^{\mu\nu}\partial_{\mu}\underline{\phi}.\partial_{\nu}\underline{\phi} & - & m^{2}\underline{\phi}.\underline{\phi} & - & V\left(\underline{\phi}.\underline{\phi}\right)\\
 &  & \uparrow &  & \uparrow &  & \uparrow\\
 &  & Kinetic &  & Mass &  & Potential\\
 &  & term &  & term\end{array}\label{1.2.27}\end{equation}
 In our path integrals the measure of the integration is clearly invariant
in exactly the same way as it was for the fields $\phi_{1}$ and $\phi_{2}$
in the example at the end of lecture 1.\begin{equation}
\begin{array}{rcl}
\int D\underline{\phi} & = & \prod_{x}\int_{-\infty}^{+\infty}\textrm{d}\phi_{1}\left(x\right)\int_{-\infty}^{+\infty}\textrm{d}\phi_{2}\left(x\right)\int_{-\infty}^{+\infty}\textrm{d}\phi_{3}\left(x\right)\\
 & = & \prod_{x}\int_{-\infty}^{+\infty}\textrm{d}\phi'_{1}\left(x\right)\int_{-\infty}^{+\infty}\textrm{d}\phi'_{2}\left(x\right)\int_{-\infty}^{+\infty}\textrm{d}\phi'_{3}\left(x\right)\textrm{ }\times\textrm{ Det}\left(J\right)\\
 & = & \prod_{x}\int_{-\infty}^{+\infty}\textrm{d}\phi'_{1}\left(x\right)\int_{-\infty}^{+\infty}\textrm{d}\phi'_{2}\left(x\right)\int_{-\infty}^{+\infty}\textrm{d}\phi'_{3}\left(x\right)\textrm{ }\times\textrm{ Det}\left(R\right)\\
 & = & \int D\underline{\phi'}\end{array}\label{1.2.28}\end{equation}
Note that if this was a local transformation we still have no problem
as we would get the same product of ones ($\textrm{Det}\left(R\left(x\right)\right)$)'s
over all space-time points $x$. However if we were dealing with a
local $O\left(3\right)$ transformation instead then $\textrm{Det}\left(R\right)$
could be $-1$ or $+1$ and we end up with an ill-defined product
of $+1's$ and $-1's$ over all space-time points, this means that
the generating functional and Green's functions would be badly defined
in terms of the transformed fields. It is perhaps worth noting that
we stick to special $\left(\textrm{Det}=+1\right)$ groups in our
course for this reason. 

It is a general result that only combinations of the fields for the
Green's functions arguments which have a component that is a singlet
representation of the transformation group are non-vanishing! For
instance consider the following vector of Green's functions,\begin{equation}
\begin{array}{crcl}
 & G\left(x_{1}\right) & = & \frac{1}{Z}\int D\underline{\phi}\textrm{ }\phi_{1}\left(x_{1}\right)\textrm{exp }iS\left[\underline{\phi}\right]\\
\Rightarrow & G\left(x_{1}\right) & = & \frac{1}{Z}\int D\underline{\phi'}\textrm{ }R_{1i}\phi'_{i}\left(x_{1}\right)\textrm{exp }iS\left[\underline{\phi'}\right].\end{array}\label{1.2.29}\end{equation}
 As the action and integration measure are $SO\left(3\right)$ invariant
then the following integrations/Green's functions should all be equal,
we can show this if need be with a simple change of variables (no
Physics just Maths),\begin{equation}
G\left(x_{1}\right)=\frac{1}{Z}\int D\underline{\phi}\textrm{ }\phi'_{1}\left(x_{1}\right)\textrm{exp }iS\left[\underline{\phi'}\right]=\frac{1}{Z}\int D\underline{\phi'}\textrm{ }\phi'_{2}\left(x_{1}\right)\textrm{exp }iS\left[\underline{\phi'}\right]=\frac{1}{Z}\int D\underline{\phi'}\textrm{ }\phi'_{3}\left(x_{1}\right)\textrm{exp }iS\left[\underline{\phi'}\right].\label{1.2.30}\end{equation}
 This gives,\begin{equation}
G\left(x_{1}\right)=\left(R_{11}+R_{12}+R_{13}\right)\frac{1}{Z}\int D\underline{\phi'}\textrm{ }\phi'_{1}\left(x_{1}\right)\textrm{exp }iS\left[\underline{\phi'}\right]=\left(R_{11}+R_{12}+R_{13}\right)G\left(x_{1}\right).\label{1.2.31}\end{equation}
 As $R_{ij}\in SO\left(3\right)$ is arbitrary we have that $R_{11}+R_{12}+R_{13}$
is not generally equal to one thus for consistency in the definition
of the Green's functions it must be the case that $G\left(x_{1}\right)=0$.
This result can also be obtained by imposing the (usual) constraint
that the Green's function (by definition) cannot depend on any \emph{parameter(s)}
of the transformation,

\begin{equation}
R=\left(\begin{array}{ccc}
sin\phi sin\psi & -cos\phi sin\psi & -cos\psi\\
cos\phi sin\theta-cos\theta cos\psi sin\phi & cos\theta cos\phi cos\psi+sin\theta sin\phi & -cos\theta sin\psi\\
cos\theta cos\phi+cos\psi sin\theta sin\phi & -cos\phi cos\psi sin\theta+cos\theta sin\phi & sin\theta sin\psi\end{array}\right)\label{1.2.32}\end{equation}
\begin{equation}
\frac{\delta G\left(x_{1}\right)}{\delta\theta}=\frac{\delta G\left(x_{1}\right)}{\delta\psi}=\frac{\delta G\left(x_{1}\right)}{\delta\phi}=0.\label{1.2.33}\end{equation}
Note $SO\left(3\right)$ transformations only have three \emph{parameters}
( = number of generators) not as many as the number of elements in
${\boldmath{R}}$ so they can only give at most three conditions by
this method. 

The property of non-singlet combinations of fields having vanishing
Green's functions was also seen in the $U\left(1\right)$ case just
considered where we saw that only strings of fields (before the exponent
in the path integral) containing an equal number of $\phi$'s and
$\phi^{\dagger}$'s were non-zero. So in the case of $SO\left(3\right)$
this means we are interested in things like,\begin{equation}
\begin{array}{crcl}
 & G\left(x_{1},x_{2}\right) & = & \int D\underline{\phi}\textrm{ }\underline{\phi^{T}}\left(x_{1}\right)\underline{\phi}\left(x_{2}\right)\textrm{ exp }iS\left[\underline{\phi}\right]\\
 &  & = & \int D\underline{\phi'}\textrm{ }\underline{\phi'^{T}}\left(x_{1}\right)R^{T}R\underline{\phi'}\left(x_{2}\right)\textrm{ exp }iS\left[\underline{\phi'}\right]\\
 &  & = & \int D\underline{\phi'}\textrm{ }\underline{\phi'^{T}}\left(x_{1}\right).\underline{\phi'}\left(x_{2}\right)\textrm{ exp }iS\left[\underline{\phi'}\right].\end{array}\label{1.2.34}\end{equation}

When dealing with Green's functions in this way it is important to
realize that just because the string of fields isn't invariant under
the symmetry transformation does not rule out the fact that it may
have a singlet component (\emph{i.e.} it does not mean it is equal
to zero) it could have a singlet component inside it somewhere. One
must project out the irreducible representations of the symmetry group
from the string of fields preceding the exponent in the path integral
and then look for the singlet component.

\newpage
\section{Gauge Symmetries.}

In this lecture we generalize our knowledge of local $U\left(1\right)$
symmetries to more complicated symmetry groups. Consider the following
general unitary symmetry transformation ,\begin{equation}
\phi_{a}\left(x\right)\rightarrow\phi'_{a}\left(x\right)=U_{ab}\phi_{b}\left(x\right)\label{1.3.1}\end{equation}
 \begin{equation}
U_{ik}^{\dagger}U_{kj}=\delta_{ij}\label{1.3.2}\end{equation}
where a repeated roman index implies a sum over it. We shall restrict
ourselves to special unitary transformation \emph{i.e.} $\textrm{Det}\left(U\right)=+1$,
this will avoid the nasty problem of getting ill-defined products
of $-1$'s and $+1$'s in transforming the integration measure. 

The kinetic term in the action for our system of fields transforms
as follows,\begin{equation}
\begin{array}{rcl}
S_{k} & = & \int\textrm{d}^{D}x\textrm{ }\eta^{\mu\nu}\partial_{\mu}\phi_{a}^{\dagger}\partial_{\nu}\phi_{a}\\
 & \rightarrow & \int\textrm{d}^{D}x\textrm{ }\eta^{\mu\nu}\partial_{\mu}\left(\phi_{b}^{\dagger}U_{ba}^{\dagger}\right)\partial_{\nu}\left(U_{ac}\phi_{c}\right).\end{array}\label{1.3.3}\end{equation}
 If the transformation is a global one then the $U_{ba}^{\dagger}$
and $U_{ac}$ terms can be brought in front of the derivatives to
give $\delta_{bc}$. If we take the potential to be once again a function
of $\phi_{a}^{\dagger}\phi_{a}$ as before then both it and the mass
term $m^{2}\phi_{a}^{\dagger}\phi_{a}$ will be invariant whether
the transformation is global or not: \begin{equation}
\begin{array}{rcl}
\phi_{a}^{\dagger}\phi_{a} & = & \phi_{b}'^{\dagger}U_{ba}^{\dagger}\left(x\right)U_{ac}\left(x\right)\phi'_{c}\\
 & = & \phi_{b}'^{\dagger}\delta_{bc}\phi'_{c}\\
 & = & \phi_{b}'^{\dagger}\phi'_{b}.\end{array}\label{1.3.4}\end{equation}

Next we will study the effect of making this general (special) unitary
symmetry a local symmetry. To begin with we define the symmetry transformation
in more detail. Writing a general transformation in terms of its parameters
$\alpha_{A}\left(x\right)$ and its generators $t_{A}$ we have, \begin{equation}
U\left(x\right)=\textrm{exp }i\sum_{A}\alpha_{A}\left(x\right)t^{A}.\label{1.3.5}\end{equation}
 The generators are defined by the following relations,\begin{equation}
\begin{array}{rcl}
\left\{ t^{A},t^{B}\right\}  & = & i\sum_{c}f^{ABC}t^{C}\\
Tr\left(t^{A}t^{B}\right) & = & \delta^{AB}\end{array}\label{1.3.6}\end{equation}
 $f^{ABC}$ are numbers known as the structure constants of the group.
Now we rewrite the kinetic part of our action in terms of the new
variables $\phi'_{a}=U_{ab}^{-1}\left(x\right)\phi_{b}$. For brevity
I will drop the indices on the fields and transformation matrix so
from now on $\phi_{a}^{\dagger}U_{ab}^{-1}\left(x\right)$ is $\phi^{\dagger}U^{-1}\left(x\right)$,
$U_{ab}\left(x\right)\phi_{b}$ is $U\left(x\right)\phi$ \emph{etc}...
\begin{equation}
\begin{array}{rcl}
S_{k}\left[U^{\dagger}\phi',\phi'^{\dagger}U\right] & = & \int\textrm{d}^{D}x\textrm{ }\eta^{\mu\nu}\partial_{\mu}\left(\phi'^{\dagger}U\right)\partial_{\nu}\left(U^{\dagger}\phi'\right)\\
 & = & \int\textrm{d}^{D}x\textrm{ }\eta^{\mu\nu}\left(\left(\partial_{\mu}\phi'^{\dagger}\right)U+\phi'^{\dagger}\partial_{\mu}U\right)\left(U^{\dagger}\partial_{\nu}\phi'+\left(\partial_{\nu}U^{\dagger}\right)\phi'\right)\\
 & = & \int\textrm{d}^{D}x\textrm{ }\eta^{\mu\nu}\partial_{\mu}\phi'^{\dagger}\partial_{\nu}\phi'\\
 & + & \int\textrm{d}^{D}x\textrm{ }\eta^{\mu\nu}\left\{ \left(\partial_{\mu}\phi'^{\dagger}\right)\left(U\partial_{\nu}U^{\dagger}\right)\phi'+\phi'^{\dagger}\left(\partial_{\mu}U\right)U^{\dagger}\partial_{\nu}\phi'\right\} \\
 & + & \int\textrm{d}^{D}x\textrm{ }\eta^{\mu\nu}\phi'^{\dagger}\left(\partial_{\mu}U\right)\left(\partial_{\nu}U^{\dagger}\right)\phi'\end{array}\label{1.3.7}\end{equation}
Now we need to have a think about terms like $\partial_{\mu}U\left(x\right)$.
We omit the $\sum_{A}$, take the repeated indices to represent sums.
\begin{equation}
\begin{array}{rcl}
\partial_{\mu}U\left(x\right) & = & \partial_{\mu}\textrm{exp }i\alpha_{A}\left(x\right)t^{A}\\
 & = & \partial_{\mu}\left(1+i\alpha_{A}\left(x\right)t^{A}-\frac{1}{2}\left(\alpha_{A}\left(x\right)t^{A}\right)\left(\alpha_{B}\left(x\right)t^{B}\right)+...\right)\\
 & = & i\left(\partial_{\mu}\alpha_{A}\left(x\right)\right)t^{A}+O\left(\alpha^{2}\right)\end{array}\label{1.3.8}\end{equation}
 So for infinitesimal transformations we can safely drop terms of
order $\alpha_{A}\left(x\right)^{2}$ and above. Back to the action,
the second term,\begin{equation}
\int\textrm{d}^{D}x\textrm{ }\eta^{\mu\nu}\left\{ \left(\partial_{\mu}\phi'^{\dagger}\right)\left(U\partial_{\nu}U^{\dagger}\right)\phi'+\phi'^{\dagger}\left(\partial_{\mu}U\right)U^{\dagger}\partial_{\nu}\phi'\right\} \label{1.3.9}\end{equation}
 neglecting terms of order $\alpha_{A}\left(x\right)^{2}$ and above
becomes,\begin{equation}
-i\int\textrm{d}^{D}x\textrm{ }\eta^{\mu\nu}\left\{ \left(\partial_{\mu}\phi'^{\dagger}\right)\left(\partial_{\nu}\alpha_{A}\left(x\right)\right)t^{A}\phi'-\phi'^{\dagger}\left(\partial_{\mu}\alpha_{A}\left(x\right)\right)t^{A}\left(\partial_{\nu}\phi'\right)\right\} .\label{1.3.10}\end{equation}
 If we exploit the symmetry of $\eta_{\mu\nu}$ we get,\begin{equation}
-\int d^{D}x\textrm{ }\eta^{\mu\nu}\left(\partial_{\mu}\alpha_{A}\left(x\right)\right)i\left\{ \left(\partial_{\nu}\phi'^{\dagger}\right)t^{A}\phi'-\phi'^{\dagger}t^{A}\left(\partial_{\nu}\phi'\right)\right\} .\label{1.3.11}\end{equation}
We now define our current,\begin{equation}
j_{\nu}^{A}=i\left(\partial_{\nu}\phi'^{\dagger}\right)t^{A}\phi'-i\phi'^{\dagger}t^{A}\left(\partial_{\nu}\phi'\right),\label{1.3.12}\end{equation}
 which makes the middle term in the action equal to,\begin{equation}
-\int\textrm{d}^{D}x\textrm{ }\left(\partial^{\nu}\alpha_{A}\left(x\right)\right)j_{\nu}^{A}=-\int\textrm{d}^{D}x\textrm{ }\partial^{\nu}\left(\alpha_{A}\left(x\right)j_{\nu}^{A}\right)+\int\textrm{d}^{D}x\textrm{ }\alpha_{A}\left(x\right)\partial^{\nu}j_{\nu}^{A}.\label{1.3.13}\end{equation}
 On the right hand side of equation \ref{1.3.13} is a 4-divergence
which we can express as a surface integral. If all the fields $\phi'$
and $\phi'^{\dagger}$ inside $j_{\nu}^{A}$ vanish at infinity (the
surface would be integrating over) then so does $j_{\nu}^{A}$ and
the surface integral is zero and,\begin{equation}
-\int\textrm{d}^{D}x\textrm{ }\left(\partial^{\nu}\alpha_{A}\left(x\right)\right)j_{\nu}^{A}=\int\textrm{d}^{D}x\textrm{ }\alpha_{A}\left(x\right)\partial^{\nu}j_{\nu}^{A}.\label{1.3.14}\end{equation}
 The kinetic part of the action is then, \begin{equation}
\begin{array}{rcl}
S_{k}\left[\phi,\phi^{\dagger}\right] & = & S_{k}\left[U^{\dagger}\phi',\phi'^{\dagger}U\right]\\
 & = & S_{k}\left[\phi',\phi'^{\dagger}\right]+\int\textrm{d}^{D}x\textrm{ }\alpha_{A}\left(x\right)\partial^{\nu}j_{\nu}^{A}+\int\textrm{d}^{D}x\textrm{ }\eta^{\mu\nu}\phi'^{\dagger}\left(\partial_{\mu}U\right)\left(\partial_{\nu}U^{\dagger}\right)\phi'.\end{array}\label{1.3.15}\end{equation}
 As we are neglecting terms of order $\alpha_{A}\left(x\right)^{2}$
and above the last term in \ref{1.3.15} is effectively zero since
its first non-zero contribution is of order $\alpha_{A}\left(x\right)^{2}$.
This gives us that the \emph{entire} action is invariant up to a current
term which arises from the variance of the kinetic term.\begin{equation}
\begin{array}{rcl}
S_{k}\left[\phi,\phi^{\dagger}\right] & = & S_{k}\left[U^{\dagger}\phi',\phi'^{\dagger}U\right]\\
 & = & S_{k}\left[\phi',\phi'^{\dagger}\right]+\int\textrm{d}^{D}x\textrm{ }\alpha_{A}\left(x\right)\partial^{\nu}j_{\nu}^{A}\end{array}\label{1.3.16}\end{equation}

From this one obtains the Ward identities in the usual way by functionally
differentiating with respect to the parameters of the transformation.
Recall that we must have some singlet representation of the fields
in the Green's function for it to be non-vanishing. This requires
that we have an equal number of $\phi$ and $\phi^{\dagger}$ fields
(this is essentially stating global charge conservation). \begin{equation}
\begin{array}{rcl}
G\left(x_{1},..,x_{n},x_{n+1},...,x_{n+m}\right) & = & \frac{1}{Z}\int D_{\mu}\left(\phi,\phi^{\dagger}\right)\phi^{\dagger}\left(x_{1}\right)\phi\left(x_{1}\right)...\phi^{\dagger}\left(x_{n}\right)\phi\left(x_{n}\right)\textrm{exp }iS\left[\phi,\phi^{\dagger}\right]\\
 & = & \frac{1}{Z}\int D_{\mu}\left(\phi',\phi'^{\dagger}\right)\phi'^{\dagger}\left(x_{n+1}\right)e^{-i\alpha_{A}\left(x_{n+1}\right)t^{A}}e^{i\alpha_{A}\left(x_{1}\right)t^{A}}\phi'\left(x_{1}\right)...\\
 & \times & \phi'^{\dagger}\left(x_{n+m}\right)e^{-i\alpha_{A}\left(x_{n+m}\right)t^{A}}e^{i\alpha_{A}\left(x_{n}\right)t^{A}}\phi'\left(x_{n}\right)\textrm{ exp }iS\left[\phi,\phi^{\dagger}\right]\\
 & \times & \textrm{exp }\int\textrm{d}^{D}x\textrm{ }\alpha_{A}\left(x\right)\partial^{\nu}j_{\nu}^{A}\end{array}\label{1.3.17}\end{equation}
As usual, by definition of the Green's function we have also,\begin{equation}
G\left(x_{1},..,x_{n},x_{n+1},...,x_{n+m}\right)=\frac{1}{Z}\int D_{\mu}\left(\phi,\phi^{\dagger}\right)\phi^{\dagger}\left(x_{n+1}\right)\phi\left(x_{1}\right)...\phi^{\dagger}\left(x_{n+m}\right)\phi\left(x_{n}\right)\textrm{exp }iS\left[\phi,\phi^{\dagger}\right]\label{1.3.18}\end{equation}
which has no dependence on any of the $\alpha_{A}\left(x\right)$'s.
Functional differentiation with respect to the parameters $\alpha_{A}\left(x\right)$
and setting $\alpha_{A}\left(x\right)=0\textrm{ }\forall\textrm{ }A$
gives the familiar looking form for the Ward identities:\begin{equation}
\begin{array}{rcl}
\left<0\left|\left(\delta\left(x_{n+1}-y\right)t^{A}+...+\delta\left(x_{n+m}-y\right)t^{A}-\delta\left(x_{1}-y\right)t^{A}-...-\delta\left(x_{n}-y\right)t^{A}-\partial_{\left(y\right)}^{\nu}j_{\nu}^{A}\left(y\right)\right)\right.\right.\\
\times\left.\left.\phi^{\dagger}\left(x_{n+1}\right)\phi\left(x_{1}\right)...\phi^{\dagger}\left(x_{n+m}\right)\phi\left(x_{n}\right)\right|0\right> & = & 0\end{array}.\label{1.3.19}\end{equation}

An obvious question to ask now is what do these currents couple to?
To answer this we invoke the gauge principle and introduce new \emph{gauge}
fields to make the Lagrangian completely invariant. Given that our
fields $\phi$ transform as $\phi\rightarrow\phi'=U\left(x\right)\phi$
and $\phi^{\dagger}\rightarrow\phi'^{\dagger}=\phi^{\dagger}U^{\dagger}\left(x\right)$
this means, \begin{equation}
\begin{array}{lcl}
\partial_{\nu}\phi' & = & \left(\partial_{\nu}U\left(x\right)\right)\phi+U\left(x\right)\left(\partial_{\nu}\phi\right)\\
\partial_{\nu}\phi'^{\dagger} & = & \left(\partial_{\nu}\phi^{\dagger}\right)U^{\dagger}\left(x\right)+\phi^{\dagger}\left(\partial_{\nu}U^{\dagger}\left(x\right)\right)\end{array}\label{1.3.20}\end{equation}
 which we can rewrite as,\begin{equation}
\begin{array}{lcl}
\partial_{\nu}\phi' & = & U\left(x\right)\left(U^{\dagger}\left(x\right)\left(\partial_{\nu}U\left(x\right)\right)\phi+\partial_{\nu}\phi\right)\\
\partial_{\nu}\phi'^{\dagger} & = & \left(\left(\partial_{\nu}\phi^{\dagger}\right)+\phi^{\dagger}\left(\partial_{\nu}U^{\dagger}\left(x\right)\right)U\left(x\right)\right)U^{\dagger}\left(x\right)\end{array}\label{1.3.21}\end{equation}
 Sandwiching these two terms together gives a kinetic term which isn't
gauge invariant.\begin{equation}
\partial_{\nu}\phi'^{\dagger}\partial^{\nu}\phi'=\left(\left(\partial_{\nu}\phi^{\dagger}\right)+\phi^{\dagger}\left(\partial_{\nu}U^{\dagger}\left(x\right)\right)U\left(x\right)\right)\left(U^{\dagger}\left(x\right)\left(\partial^{\nu}U\left(x\right)\right)\phi+\partial^{\nu}\phi\right)\label{1.3.22}\end{equation}
To make this gauge invariant we need to introduce a gauge field $A_{\mu}$
which transforms as $A'_{\mu}=U\left(x\right)A_{\mu}U^{\dagger}\left(x\right)-\left(\partial_{\mu}U\left(x\right)\right)U^{\dagger}\left(x\right)$.
This gauge field is added to the normal derivative $\partial_{\mu}$
to form the covariant derivative $D_{\mu}=\partial_{\mu}+A_{\mu}$
so-called because it transforms like the fields do:

\begin{equation}
\begin{array}{rcl}
D'_{\mu}\phi'\left(x\right) & = & \partial_{\mu}\phi'\left(x\right)+A'_{\mu}\phi'\left(x\right)\\
 & = & \left(\partial_{\mu}U\left(x\right)\right)\phi\left(x\right)+U\left(x\right)\partial_{\mu}\phi\left(x\right)+U\left(x\right)A_{\mu}\phi\left(x\right)-\left(\partial_{\mu}U\left(x\right)\right)\phi\left(x\right)\\
 & = & U\left(x\right)\left(\partial_{\mu}+A_{\mu}\right)\phi\left(x\right)\\
 & = & U\left(x\right)D_{\mu}\phi\left(x\right)\end{array}\label{1.3.23}\end{equation}
Note that in future we will not write down the explicit space-time
dependence of $U$, $U=U\left(x\right)$ unless otherwise stated.
If we now replace the derivative in the kinetic part of our action
$\partial_{\mu}$ by the covariant derivative $D_{\mu}$ we will find
that we have something which is now gauge invariant \emph{i.e.} replace\begin{equation}
\eta^{\mu\nu}\partial_{\mu}\phi^{\dagger}\partial_{\nu}\phi\rightarrow\eta^{\mu\nu}\left(D_{\mu}\phi\right)^{\dagger}D_{\nu}\phi.\label{1.3.24}\end{equation}
 It is worth making a few points about the field $A_{\mu}$ that we
have introduced. Firstly $A_{\mu}$ is actually an anti-Hermitian
matrix. The anti-Hermitian property is easy to prove, imagine $A_{\mu}=0$,
this is called a \emph{pure} \emph{gauge} configuration as gauge transformed
versions of the field consist only of the gauge transformation's inhomogeneous
term, then \begin{equation}
\begin{array}{crcl}
 & A'_{\mu} & = & UA_{\mu}U^{\dagger}-\left(\partial_{\mu}U\right)U^{\dagger}\\
 &  & = & -\left(\partial_{\mu}U\right)U^{\dagger}\\
\Rightarrow & A_{\mu}'^{\dagger} & = & -U\left(\partial_{\mu}U^{\dagger}\right)\\
\\\Rightarrow & A'_{\mu} & = & U\left(\partial_{\mu}U^{\dagger}\right)-\partial_{\mu}\left(UU^{\dagger}\right)\\
 &  & = & U\left(\partial_{\mu}U^{\dagger}\right)-\partial_{\mu}\left(I\right)\\
 &  & = & U\left(\partial_{\mu}U^{\dagger}\right)\\
\\\Rightarrow & A_{\mu}'^{\dagger} & = & -A'_{\mu}\left(x\right)\end{array}\label{1.3.25}\end{equation}
Also for global transformations ($\partial_{\mu}U=0$) $A_{\mu}$
is a matrix living in the adjoint representation of the transformation
group \emph{i.e.} in this case we have $A'_{\mu}=UA_{\mu}U^{\dagger}$.
In light of this fact it is possible to write the matrix $A_{\mu}$
in terms of the generators of the group $A_{\mu}\left(x\right)=\sum_{A}A_{\mu A}\left(x\right)t^{A}$.
Note that in the case of the Abelian gauge field the covariant derivative
appears to contradict the definition above, in massless QED we have
${\cal {L}}=\bar{\psi}i\not D\psi=\bar{\psi}\left(i\not\partial+\not A\right)\psi$
\emph{i.e.} $D_{\mu}=\partial_{\mu}-iA_{\mu}$ which implies that
the gauge field $A_{\mu}$ transforms as $A_{\mu}\rightarrow UA_{\mu}U^{\dagger}-i\left(\partial_{\mu}U\right)U^{\dagger}$,
this is just the usual $A_{\mu}\rightarrow A_{\mu}+\partial_{\mu}\alpha\left(x\right)$
for $U=e^{-i\alpha\left(x\right)}$. This peculiarity is an annoying
relic in the general literature and arises because as the Lagrangian
there the kinetic term has an $i$ attached to it but the gauge field
does not. With this definition the $A_{\mu}$ field in QED is Hermitian,
this is easy to see as one just has an extra factor of $i$ floating
around in the above calculation where we proved it was anti-Hermitian.
Henceforth we use the $\bar{\psi}\left(i\not\partial+\not A\right)\psi$
form of the QED Lagrangian. A better definition of the QED Lagrangian
would be $\bar{\psi}i\left(\not\partial+\not A\right)\psi$ in which
case $A_{\mu}$ does transform as $A'_{\mu}=UA_{\mu}U^{\dagger}-\left(\partial_{\mu}U\right)U^{\dagger}$
and we have a better analogy with the Lagrangians and gauge transformations
of non-Abelian gauge theories. 

$A_{\mu}$ is sometimes referred to as a \emph{connection} which means
it does for gauge theory what the Christoffel symbols, $\Gamma_{\beta\gamma}^{\alpha}$,
do for General Relativity. In General Relativity the covariant derivative
of a vector $V^{\mu}$ is $D_{\nu}V^{\mu}=\partial_{\nu}V^{\mu}+\Gamma_{\nu\lambda}^{\mu}V^{\lambda}$
and in our gauge theory we have $D_{\nu}\phi^{a}=\partial_{\nu}\phi^{a}+A_{\mu}^{ab}\phi^{b}$.
This is perhaps interesting from the point of view that General Relativity
is a gauge theory locally invariant under transformations in the Poincare
group. It is also worth noting that $\Gamma_{\beta\gamma}^{\alpha}$
doesn't transform as a tensor except for global Poincare transformations
(it has an inhomogeneous term in its transformation law),\begin{equation}
\Gamma_{\lambda\mu}^{'\kappa}=\frac{\partial x'^{\kappa}}{\partial x^{\alpha}}\frac{\partial x'^{\beta}}{\partial x'^{\lambda}}\frac{\partial x^{\gamma}}{\partial x'^{\mu}}\Gamma_{\beta\lambda}^{\alpha}+\frac{\partial^{2}x^{\alpha}}{\partial x'^{\lambda}\partial x'^{\mu}}\frac{\partial x'^{\kappa}}{\partial x^{\alpha}}.\label{1.3.26}\end{equation}
For global Poincare symmetry we have $\frac{\partial x^{\alpha}}{\partial x'^{\beta}}$
is the Lorentz transformation matrix plus some matrix of translations
\emph{i.e.} $\frac{\partial^{2}x^{\alpha}}{\partial x'^{\beta}\partial x'^{\gamma}}=0\textrm{ }\forall\textrm{ }\alpha,\beta,\gamma$
and hence $\Gamma_{\beta\lambda}^{\alpha}$ would transform as a tensor,
the inhomogeneous term is identically zero. Now if $\Gamma_{\beta\lambda}^{\alpha}$
transforms as a tensor by the equivalence principal it is possible
to transform to a frame where it would be zero at a given point and
hence it would be zero everywhere \emph{i.e.} $\Gamma_{\beta\lambda}^{\alpha}$
wouldn't exist - in special relativity $\Gamma_{\beta\lambda}^{\alpha}$
is effectively transforming as a tensor, it doesn't exist. This is
analogous to our gauge field $A_{\mu}$ it also has an inhomogeneous
term in its transformation law $-\left(\partial_{\mu}U\right)U^{\dagger}$.

The only thing left to do now is turn our $A_{\mu}\left(x\right)$
field into something physical. To do that we have to give it a kinetic
term in the action. The kinetic term should be gauge invariant. The
following relation is essential to constructing these terms.\begin{equation}
\begin{array}{rcl}
D'_{\mu}D'_{\nu}\phi' & = & D'_{\mu}\left(UD_{\nu}\phi\right)\\
 & = & \left(\partial_{\mu}+A'_{\mu}\right)\left(UD_{\nu}\phi\right)\\
 & = & \left(\partial_{\mu}+UA_{\mu}U^{\dagger}-\left(\partial_{\mu}U\right)U^{\dagger}\right)\left(UD_{\nu}\phi\right)\\
 & = & \left(\partial_{\mu}U\right)D_{\nu}\phi+U\partial_{\mu}D_{\nu}\phi+UA_{\mu}D_{\nu}\phi-\left(\partial_{\mu}U\right)D_{\nu}\phi\\
 & = & UD_{\mu}D_{\nu}\phi\end{array}\label{1.3.27}\end{equation}
\begin{equation}
\Rightarrow\left[D'_{\mu}D'_{\nu}\right]\phi'=U\left[D_{\mu},D_{\nu}\right]\phi\label{1.3.28}\end{equation}
We denote $F_{\mu\nu}=\left[D_{\mu},D_{\nu}\right]=\partial_{\nu}A_{\mu}-\partial_{\mu}A_{\nu}+\left[A_{\mu},A_{\nu}\right]$.
From the above we see that, \begin{equation}
F_{\mu\nu}\phi\rightarrow F'_{\mu\nu}\phi=UF_{\mu\nu}\phi=UF_{\mu\nu}U^{\dagger}U\phi\label{1.3.29}\end{equation}
from which it is clear that $F_{\mu\nu}$ transforms as, \begin{equation}
F_{\mu\nu}\rightarrow F'_{\mu\nu}=UF_{\mu\nu}U^{\dagger}.\label{1.3.30}\end{equation}
To make the $A_{\mu}$ field physical we then take its gauge invariant,
Lorentz invariant kinetic term to be \begin{equation}
\begin{array}{rcl}
Tr\left[F_{\mu\nu}F^{\mu\nu}\right]\rightarrow Tr\left[F'_{\mu\nu}F'^{\mu\nu}\right] & = & Tr\left[UF_{\mu\nu}U^{\dagger}UF^{\mu\nu}U^{\dagger}\right]\\
 & = & Tr\left[UF_{\mu\nu}F^{\mu\nu}U^{\dagger}\right].\end{array}\label{1.3.31}\end{equation}
 It is a result that the trace of a product of matrices is independent
of the ordering of the product hence,\begin{equation}
\begin{array}{rcl}
Tr\left[UF_{\mu\nu}F^{\mu\nu}U^{\dagger}\right] & = & Tr\left[F_{\mu\nu}F^{\mu\nu}U^{\dagger}U\right]\\
 & = & Tr\left[F_{\mu\nu}F^{\mu\nu}\right]\end{array}.\label{1.3.32}\end{equation}
This means $Tr\left[F_{\mu\nu}F^{\mu\nu}\right]$ is indeed gauge
invariant, \begin{equation}
\begin{array}{rcl}
Tr\left[F'_{\mu\nu}F'^{\mu\nu}\right] & = & Tr\left[UF_{\mu\nu}F^{\mu\nu}U^{\dagger}\right]\\
 & = & Tr\left[F_{\mu\nu}F^{\mu\nu}\right]\end{array}.\label{1.3.33}\end{equation}

\chapter{Conformal Symmetry.}

\section{The Conformal Group.}

This lecture is about space-time symmetries. We start off with the
usual scalar field action,\begin{equation}
S\left[\phi\right]=\int\textrm{d}^{D}x\textrm{ }\frac{1}{2}\partial^{\mu}\phi\partial_{\mu}\phi-m^{2}\phi^{2}-V\left(\phi\right).\label{2.1.1}\end{equation}
 In natural units $\left(\hbar=1\right)$ we have that $S\left[\phi\right]$
is dimensionless. For the dimensions of the terms in the action to
be consistent we then require that $m$ have dimensions of $length^{-1}$
as $\partial_{\mu}$ has dimensions of $length^{-1}$. This action
is invariant for transformations of the Poincare group \emph{i.e.}
it is invariant under spatial translations, rotations and Lorentz
boosts (we are integrating over all of space-time). As a result of
these symmetries we already have some important consequences. Consider
the effect of translational invariance on the Green's functions for
instance:\begin{equation}
\begin{array}{rcl}
G\left(x_{1},...,x_{n}\right) & = & \frac{1}{Z}\int D\phi\textrm{ }\phi\left(x_{1}\right)...\phi\left(x_{n}\right)\textrm{ exp }iS\left[\phi\right]\\
\phi\left(x_{\mu}\right) & = & \phi'\left(x_{\mu}+a_{\mu}\right)\\
S\left[\phi\right] & = & S\left[\phi'\right]\end{array}\label{2.1.2}\end{equation}
The measure is also invariant, consider again the spatial coordinate
$x$ as the index labeling the true variable in the path integral
\emph{i.e.} field. The translation amounts to re-labeling this index
(by a constant shift), the path integral measure is the product of
the integrals over all possible values of the field at each (and every)
space-time point so we do not expect the measure to change, roughly
speaking, \begin{equation}
\begin{array}{rcl}
\prod_{x}\textrm{d}\phi\left(x\right) & = & \prod_{x-a}\textrm{d}\phi\left(x-a\right)\\
 & = & \prod_{x-a}\textrm{d}\phi'\left(x\right)\\
 & = & \prod_{x}\textrm{d}\phi'\left(x\right)\end{array}\label{2.1.3}\end{equation}

\begin{equation}
\begin{array}{rcl}
\Rightarrow\textrm{ }G\left(x_{1},...,x_{n}\right) & = & \frac{1}{Z}\int D\phi'\textrm{ }\phi'\left(x_{1}+a\right)...\phi\left(x_{n}+a\right)\textrm{ exp }iS\left[\phi'\right]\\
 & = & G\left(x_{1}+a,...,x_{n}+a\right)\end{array}.\label{2.1.4}\end{equation}
 So for an action and an integration measure which are invariant under
spatial translations one has that the Green's functions are also translation
invariant and hence they can only depend on where their arguments
are relative to each other \emph{i.e.} Differences in the positions
of their external points. Similarly one can deduce from Lorentz/rotational
invariance that the Green's functions can only be functions of things
like $\left(x_{1}-x_{2}\right)^{\mu}\left(x_{1}-x_{2}\right)_{\mu}$,
$\left(x_{1}-x_{2}\right)^{\mu}\left(x_{1}-x_{3}\right)_{\mu}$ \emph{etc}...

In actions with $m=0$ we have an additional {}``special'' symmetry
to consider for particular choices of $V\left(\phi\right)$. We start
this study by considering the dimensions of the various pieces of
the action. Let the fields $\phi$ have dimensions of $length^{A}$
this means that for the kinetic part of the action we have overall
dimensions $length^{2A-2+D}$, we have a product of two fields giving
$length^{2A}$, two partial derivatives, $length^{-2}$, and an $\int\textrm{d}^{D}x$
giving $length^{D}$. At the start we said the action was dimensionless
so this means $A=\frac{2-D}{2}$. The potential term should also be
dimensionless. Say $V\left(\phi\right)=g\phi^{P}$ with $g$ a dimensionless
coupling constant, this has dimensions $length^{\frac{P}{2}\left(2-D\right)}$
but when we add in the effect of the integral $\int\textrm{d}^{D}x$
the potential part of the action actually has dimensions $length^{\frac{P}{2}\left(2-D\right)+D}$
which of course must actually be dimensionless \emph{i.e.} $P=2D/\left(D-2\right)$.
So in $D=4$ space-time dimensions we have $V\left(\phi\right)=g\phi^{4}$,
in $D=3$ space-time dimensions we have $V\left(\phi\right)=g\phi^{6}$. 

The {}``special symmetry'' which we alluded to earlier is a symmetry
of the action under \emph{scale transformations} $x'^{\mu}=\lambda x^{\mu}$.
Under such transformations the dimensionless action that we deduced
above transforms as follows,\begin{equation}
\begin{array}{rcl}
S\left[\phi\right] & = & \int\textrm{d}^{D}x\textrm{ }\frac{1}{2}\partial^{\mu}\phi\left(x\right)\partial_{\mu}\phi\left(x\right)-g\phi\left(x\right)^{\frac{2D}{D-2}}\\
 & = & \lambda^{-D}\int\textrm{d}^{D}x'\textrm{ }\lambda^{2}\frac{1}{2}\partial'^{\mu}\phi\left(\frac{x'}{\lambda}\right)\partial'_{\mu}\phi\left(\frac{x'}{\lambda}\right)-g\phi\left(\frac{x'}{\lambda}\right)^{\frac{2D}{D-2}}\end{array}\label{2.1.5}\end{equation}
where $\partial'_{\mu}=\frac{\partial}{\lambda\partial x^{\mu}}$.
\begin{equation}
\Rightarrow S\left[\phi\right]=\int\textrm{d}^{D}x'\frac{1}{2}\partial'^{\mu}\left(\lambda^{-\frac{\left(D-2\right)}{2}}\phi\left(\frac{x'}{\lambda}\right)\right)\partial'_{\mu}\left(\lambda^{-\frac{\left(D-2\right)}{2}}\phi\left(\frac{x'}{\lambda}\right)\right)-g\left(\lambda^{-\frac{\left(D-2\right)}{2}}\phi\left(\frac{x'}{\lambda}\right)\right)^{\frac{2D}{D-2}}.\label{2.1.6}\end{equation}
If we now define that the fields transform under scale transformations
as $\phi'\left(x'\right)=\lambda^{-\frac{\left(D-2\right)}{2}}\phi\left(\frac{x'}{\lambda}\right)$
then we have the action is invariant under scale transformations:\begin{equation}
\begin{array}{rcl}
S\left[\phi\right] & = & \int\textrm{d}^{D}x\textrm{ }\frac{1}{2}\partial^{\mu}\phi\left(x\right)\partial_{\mu}\phi\left(x\right)-g\phi\left(x\right)^{\frac{2D}{D-2}}\\
 & = & S\left[\phi'\right]\\
 & = & \int\textrm{d}^{D}x\textrm{ }\frac{1}{2}\partial^{\mu}\phi'\left(x\right)\partial_{\mu}\phi'\left(x\right)-g\phi'\left(x\right)^{\frac{2D}{D-2}}\end{array}\label{2.1.7}\end{equation}
 For these scale transformations the measure is not invariant! \emph{}\begin{equation}
\begin{array}{rcl}
\int D\phi & = & \prod_{x_{i}}\int\textrm{d}\phi\left(x_{i}\right)\\
 & = & \prod_{x_{i}}\lambda^{\frac{D-2}{2}}\int\textrm{d}\phi'\left(x'_{i}\right)\\
 & = & \left(\prod_{x_{i}}\lambda^{\frac{D-2}{2}}\right)\int D\phi'.\end{array}\label{2.1.8}\end{equation}
However at the front of the definition of the Green's functions we
have a factor $\frac{1}{Z}$. The generating functional $Z$ acquires
exactly the same factor when the measure is transformed $\left(\prod_{x_{i}}\lambda^{\frac{D-2}{2}}\right)$
in exactly the same way and so the two factors in the numerator and
denominator cancel. As a result of this scale invariance the form
of the 2-point function is fixed! \begin{equation}
\begin{array}{rcl}
G\left(x_{1}-x_{2}\right) & = & \frac{1}{Z}\int D\phi\textrm{ }\lambda^{\frac{D-2}{2}}\phi'\left(\lambda x_{1}\right)\lambda^{\frac{D-2}{2}}\phi'\left(\lambda x_{2}\right)e^{iS\left[\phi'\left(x'\right)\right]}\\
 & = & \lambda^{D-2}\frac{1}{Z}\int D\phi\textrm{ }\phi'\left(\lambda x_{1}\right)\phi'\left(\lambda x_{2}\right)e^{iS\left[\phi'\left(x'\right)\right]}\\
 & = & \lambda^{D-2}G\left(\lambda\left(x_{1}-x_{2}\right)\right)\end{array}\label{2.1.9}\end{equation}
This last equation in fact means that the 2-point function must be
of the form,\begin{equation}
G\left(x_{1}-x_{2}\right)\sim\frac{1}{\left|x_{1}-x_{2}\right|^{D-2}}.\label{2.1.10}\end{equation}
 This result is true in Minkowski and Euclidean space. 

So far we have considered scale transformations and Poincare transformations
separately. The combination of the Poincare transformations and scale
transformations are known as the \emph{conformal} \emph{transformations}.
Note that Poincare and scale transformations preserve the angles between
things. Here when we talk of angles we are talking about angles in
whatever $D$ dimensional space we are in \emph{i.e.} if we are in
Minkowski space we mean the angle between things in $3+1$ space-time
dimensions not $3$ spatial dimensions. Perhaps a better, more formal
definition of a conformal (angle preserving) transformation is that
it is a transformation where the ratio,\begin{equation}
\frac{\textrm{d}x^{\alpha}}{\left|\textrm{d}x\right|}\frac{\textrm{d}y_{\alpha}}{\left|\textrm{d}y\right|}\label{2.1.11}\end{equation}
is unchanged. Clearly rotations, translations, Lorentz boosts and
scalings all satisfy this requirement.

Under the general (infinitesimal) coordinate transformation $x\rightarrow x'=x+\epsilon\left(x\right)$
the metric tensor undergoes the following transformation to first
order in $\epsilon\left(x\right)$, \begin{equation}
\begin{array}{rcl}
g'_{\mu\nu} & = & g_{\alpha\beta}\frac{\partial x'^{\alpha}}{\partial x^{\mu}}\frac{\partial x'^{\beta}}{\partial x^{\nu}}\\
 & = & g_{\alpha\beta}\left(\delta_{\mu}^{\alpha}+\partial_{\mu}\epsilon^{\alpha}\left(x\right)\right)\left(\delta_{\nu}^{\beta}+\partial_{\nu}\epsilon^{\beta}\left(x\right)\right)\\
 & = & g_{\mu\nu}+g_{\mu\beta}\partial_{\nu}\epsilon^{\beta}\left(x\right)+g_{\alpha\nu}\partial_{\mu}\epsilon^{\alpha}\left(x\right)\end{array}\label{2.1.12}\end{equation}
 If it is the case that the transformation in the line above is of
the form $g_{\mu\nu}\rightarrow g'_{\mu\nu}=g_{\mu\nu}\times\left(1+\Lambda\left(x\right)\right)$
then the metric is just being multiplied by some number (at each space-time
point $x$) and the transformation is a local conformal transformation
\emph{i.e.} it is (locally) angle preserving. We would like to find
functions $\epsilon\left(x\right)$ that have this property. Let us
take the case that our initial metric is flat $g_{\mu\nu}=\eta_{\mu\nu}$
in which case we want $\partial_{\nu}\epsilon_{\mu}\left(x\right)+\partial_{\mu}\epsilon_{\nu}\left(x\right)=\Lambda\left(x\right)\eta_{\mu\nu}$.
Contracting the last equation gives $2\partial^{\mu}\epsilon_{\mu}\left(x\right)=D\Lambda\left(x\right)$
where $D$ is the number of space-time dimensions we are considering.
Substituting in for $\Lambda\left(x\right)$ we have, \begin{equation}
\Rightarrow\partial_{\nu}\epsilon_{\mu}\left(x\right)+\partial_{\mu}\epsilon_{\nu}\left(x\right)=\frac{2}{D}\partial^{\kappa}\epsilon_{\kappa}\left(x\right)\eta_{\mu\nu}.\label{2.1.13}\end{equation}
differentiate this with respect to $\partial^{\mu}$ to get, \begin{equation}
\Rightarrow\partial^{\mu}\partial_{\mu}\epsilon_{\nu}\left(x\right)=\left(\frac{2}{D}-1\right)\partial_{\nu}\left(\partial^{\mu}\epsilon_{\mu}\left(x\right)\right),\label{2.1.14}\end{equation}
and differentiate again with respect to $\partial^{\nu}$,\begin{equation}
\Rightarrow\left(\frac{2}{D}-2\right)\partial^{\mu}\partial_{\mu}\left(\partial^{\nu}\epsilon_{\nu}\left(x\right)\right)=0.\label{2.1.15}\end{equation}
 This equation has two solutions. Either $D=1$ or $\partial^{\mu}\partial_{\mu}\left(\partial^{\nu}\epsilon_{\nu}\left(x\right)\right)=0.$
To solve for $\epsilon_{\nu}\left(x\right)$ consider \ref{2.1.13}
and \ref{2.1.14}. Acting on \ref{2.1.13} with $\partial^{\lambda}\partial_{\lambda}$
we get,\begin{equation}
\partial^{\lambda}\partial_{\lambda}\partial_{\nu}\epsilon_{\mu}\left(x\right)+\partial^{\lambda}\partial_{\lambda}\partial_{\mu}\epsilon_{\nu}\left(x\right)=\frac{2}{D}\partial^{\lambda}\partial_{\lambda}\partial^{\kappa}\epsilon_{\kappa}\left(x\right)\eta_{\mu\nu}\label{2.1.16}\end{equation}
 but the equation which we are trying to solve tells us that $\partial^{\lambda}\partial_{\lambda}\partial^{\kappa}\epsilon_{\kappa}\left(x\right)=0$
so\begin{equation}
\begin{array}{rccccl}
 & \partial^{\lambda}\partial_{\lambda}\partial_{\nu}\epsilon_{\mu}\left(x\right) & + & \partial^{\lambda}\partial_{\lambda}\partial_{\mu}\epsilon_{\nu}\left(x\right) & = & 0\\
\Rightarrow &  &  & \partial_{\nu}\partial^{\lambda}\partial_{\lambda}\epsilon_{\mu}\left(x\right) & = & -\partial_{\mu}\partial^{\lambda}\partial_{\lambda}\epsilon_{\nu}\left(x\right)\end{array}\label{2.1.17}\end{equation}
 If we now act on \ref{2.1.14} with $\partial_{\mu}$ we find,\begin{equation}
\partial_{\mu}\partial^{\lambda}\partial_{\lambda}\epsilon_{\nu}\left(x\right)=\left(\frac{2}{D}-1\right)\partial_{\mu}\partial_{\nu}\left(\partial^{\lambda}\epsilon_{\lambda}\left(x\right)\right).\label{2.1.18}\end{equation}
 Looking at the above equation we see that the right hand side is
symmetric in $\mu$ and $\nu$ so the left hand side must be also
however we have just shown that $\partial_{\nu}\partial^{\lambda}\partial_{\lambda}\epsilon_{\mu}\left(x\right)=-\partial_{\mu}\partial^{\lambda}\partial_{\lambda}\epsilon_{\nu}\left(x\right)$
so the only way that we can have these two equations consistent is
if $D=2$ or $\partial_{\mu}\partial^{\lambda}\partial_{\lambda}\epsilon_{\nu}\left(x\right)=0$
and hence $\partial_{\mu}\partial_{\nu}\left(\partial^{\lambda}\epsilon_{\lambda}\left(x\right)\right)=0$. 

We shall not discuss the case where $D=2$ but try and solve $\partial_{\mu}\partial_{\nu}\left(\partial^{\lambda}\epsilon_{\lambda}\left(x\right)\right)=0$.
With a little thought one sees that the general solution to this equation
is $\partial^{\lambda}\epsilon_{\lambda}\left(x\right)=A+B^{\lambda}x_{\lambda}$
which means $\epsilon_{\mu}\left(x\right)$ must have the general
solution, \begin{equation}
\epsilon_{\mu}\left(x\right)=a_{\mu}+b_{\mu\nu}x^{\nu}+c_{\mu\nu\lambda}x^{\nu}x^{\lambda}.\label{2.1.19}\end{equation}
 For a non vanishing $x^{\nu}x^{\lambda}$ term we must have $c_{\mu\nu\lambda}$
symmetric in $\nu$ and $\lambda$. This is trivial to show, first
use the fact that $\nu$ and $\lambda$ are dummy variables, summed
over, so we can call them whatever we like, we can call $\nu$ $\lambda$
and call $\lambda$ $\nu$ \begin{equation}
\begin{array}{rrcl}
\Rightarrow & c_{\mu\nu\lambda}x^{\nu}x^{\lambda} & = & c_{\mu\lambda\nu}x^{\lambda}x^{\nu}\\
\Rightarrow & c_{\mu\nu\lambda}x^{\nu}x^{\lambda} & = & c_{\mu\lambda\nu}x^{\nu}x^{\lambda}.\end{array}\label{2.1.20}\end{equation}

Now we have a form for the $\epsilon_{\mu}\left(x\right)$'s where
the $\epsilon_{\mu}\left(x\right)$'s are performing a rescaling of
the metric. If we plug our general solution for $\epsilon_{\mu}\left(x\right)$
back into our original equation for $\epsilon_{\mu}\left(x\right)$
\emph{i.e.} \begin{equation}
\partial_{\nu}\epsilon_{\mu}\left(x\right)+\partial_{\mu}\epsilon_{\nu}\left(x\right)=\frac{2}{D}\partial^{\kappa}\epsilon_{\kappa}\left(x\right)\eta_{\mu\nu}\label{2.1.21}\end{equation}
we can get constraints on the coefficients $b_{\mu\nu}$ and $c_{\mu\nu\lambda}$.
First consider $\partial_{\nu}\epsilon_{\mu}\left(x\right)$,\begin{equation}
\begin{array}{rcl}
\partial_{\nu}\epsilon_{\mu}\left(x\right) & = & \partial_{\nu}\left(a_{\mu}+b_{\mu\rho}x^{\rho}+c_{\mu\rho\beta}x^{\rho}x^{\beta}\right)\\
 & = & \partial_{\nu}\left(b_{\mu\rho}x^{\rho}+c_{\mu\rho\beta}x^{\rho}x^{\beta}\right)\\
 & = & b_{\mu\rho}\delta_{\nu}^{\rho}+c_{\mu\rho\beta}\delta_{\nu}^{\rho}x^{\beta}+c_{\mu\rho\beta}x^{\rho}\delta_{\nu}^{\beta}\\
 & = & b_{\mu\nu}+\left(c_{\mu\nu\rho}+c_{\mu\rho\nu}\right)x^{\rho}\\
 & = & b_{\mu\nu}+2c_{\mu\nu\rho}x^{\rho}.\end{array}\label{2.1.22}\end{equation}
Substituting this into \ref{2.1.18} gives,\begin{equation}
b_{\mu\nu}+2c_{\mu\nu\rho}x^{\rho}+b_{\nu\mu}+2c_{\nu\mu\rho}x^{\rho}=\frac{2}{D}b_{\lambda}^{\lambda}\eta_{\mu\nu}+\frac{4}{D}c_{\,\,\lambda\rho}^{\lambda}x^{\rho}\eta_{\mu\nu}.\label{2.1.23}\end{equation}
Comparing the coefficients of $x^{\rho}$ on either side of the equation
one gets the following constraints on the forms of $b_{\mu\nu}$ and
$c_{\mu\nu\lambda}$, \begin{equation}
\begin{array}{rcl}
b_{\mu\nu}+b_{\nu\mu} & = & \frac{2}{D}\eta_{\mu\nu}b_{\lambda}^{\lambda}\\
c_{\mu\nu\rho}+c_{\nu\mu\rho} & = & \frac{2}{D}\eta_{\mu\nu}c_{\,\,\lambda\rho}^{\lambda}.\end{array}\label{2.1.23}\end{equation}
 Let us write $b_{\mu\nu}$ in the form, \begin{equation}
\begin{array}{rcl}
b_{\mu\nu} & = & \left(b_{\mu\nu}-\frac{b_{\lambda}^{\lambda}}{D}\eta_{\mu\nu}\right)+\frac{b_{\lambda}^{\lambda}}{D}\eta_{\mu\nu}\\
 & = & M_{\mu\nu}+\frac{b_{\lambda}^{\lambda}}{D}\eta_{\mu\nu}.\end{array}\label{2.1.24}\end{equation}
 Substituting this into $b_{\mu\nu}+b_{\nu\mu}=\frac{2}{D}\eta_{\mu\nu}b_{\lambda}^{\lambda}$
we have,\begin{equation}
\begin{array}{crcl}
 & M_{\mu\nu}+\frac{b_{\lambda}^{\lambda}}{D}\eta_{\mu\nu}+M_{\nu\mu}+\frac{b_{\lambda}^{\lambda}}{D}\eta_{\nu\mu} & = & \frac{2}{D}\eta_{\mu\nu}b_{\lambda}^{\lambda}\\
\Rightarrow & M_{\mu\nu} & = & -M_{\nu\mu}\end{array}\label{2.1.25}\end{equation}
 \emph{i.e.} we can write $b_{\mu\nu}$ as the sum of an antisymmetric
matrix $M_{\mu\nu}$ and a symmetric matrix $\eta_{\mu\nu}\Lambda$,\begin{equation}
\begin{array}{crcl}
\Rightarrow & b_{\mu\nu}+b_{\nu\mu} & = & M_{\mu\nu}+\eta_{\mu\nu}\Lambda+M_{\nu\mu}+\eta_{\nu\mu}\Lambda\\
 &  & = & \left(M+M^{T}\right)_{\mu\nu}+2\eta_{\mu\nu}\Lambda\\
 &  & = & 2\eta_{\mu\nu}\Lambda\\
 &  & = & 2\eta_{\mu\nu}\frac{1}{D}b_{\lambda}^{\lambda}.\end{array}\label{2.1.26}\end{equation}
\begin{equation}
\begin{array}{ccccc}
b_{\mu\nu} & = & M_{\mu\nu} & + & \eta_{\mu\nu}\Lambda\\
 &  & \uparrow &  & \uparrow\\
 &  & Lorentz &  & Scale\\
 &  & Transformations &  & Transformations.\\
 &  & (incl\textrm{ }rotations)\end{array}\label{2.1.27}\end{equation}

Turning to the constraint on $c_{\mu\nu\lambda}$ we have,\begin{equation}
c_{\mu\nu\rho}+c_{\nu\mu\rho}=\frac{2}{D}\eta_{\mu\nu}c_{\,\,\lambda\rho}^{\lambda}.\label{2.1.28}\end{equation}
We now wish to constrain $c_{\mu\nu\lambda}$ with an analysis similar
to $b_{\mu\nu}$. We can do this by remembering that a non-vanishing
$x^{\nu}x^{\lambda}$ term in $\epsilon_{\mu}\left(x\right)$ requires
that $c_{\mu\nu\lambda}$ is symmetric in $\nu$ and $\lambda$. \begin{equation}
\begin{array}{rcl}
c_{\mu\nu\rho} & = & \frac{2}{D}\eta_{\mu\nu}c_{\,\,\lambda\rho}^{\lambda}-c_{\nu\mu\rho}\\
 & = & \frac{2}{D}\eta_{\mu\nu}c_{\,\,\lambda\rho}^{\lambda}-c_{\nu\rho\mu}\\
 & = & \frac{2}{D}\eta_{\mu\nu}c_{\,\,\lambda\rho}^{\lambda}-\frac{2}{D}\eta_{\nu\rho}c_{\,\,\lambda\mu}^{\lambda}+\frac{2}{D}\eta_{\rho\mu}c_{\,\,\lambda\nu}^{\lambda}-c_{\mu\rho\nu}\\
 & = & \frac{2}{D}\eta_{\mu\nu}c_{\,\,\lambda\rho}^{\lambda}-\frac{2}{D}\eta_{\nu\rho}c_{\,\,\lambda\mu}^{\lambda}+\frac{2}{D}\eta_{\rho\mu}c_{\,\,\lambda\nu}^{\lambda}-c_{\mu\nu\rho}\end{array}\label{2.1.29}\end{equation}
\[
\Rightarrow c_{\mu\nu\rho}=\frac{4}{D}\eta_{\mu\nu}c_{\,\,\lambda\rho}^{\lambda}-\frac{4}{D}\eta_{\nu\rho}c_{\,\,\lambda\mu}^{\lambda}+\frac{4}{D}\eta_{\rho\mu}c_{\,\,\lambda\nu}^{\lambda}\]

\newpage
\section{Conformal Symmetry and the Dilatation Current.}

In this lecture we round up our study of scale transformations. From
last lecture we had that for have a general conformal transformation,\begin{equation}
\begin{array}{ccccccccccccc}
x_{\mu} & \rightarrow & x'_{\mu} & = & x_{\mu} & + & a_{\mu} & + & \Lambda x_{\mu} & + & M_{\mu\nu}x^{\nu} & + & \frac{4}{D}\left(\eta_{\mu\nu}c_{\rho}+\eta_{\mu\rho}c_{\nu}-\eta_{\nu\rho}c_{\mu}\right)x^{\rho}x^{\nu}\\
 &  &  &  &  &  & \uparrow &  & \uparrow &  & \uparrow &  & \uparrow\\
 &  &  &  &  &  & Translations &  & Scaling &  & Rotations/ &  & Special\textrm{ }Conformal\\
 &  &  &  &  &  &  &  &  &  & Boosts &  & Transformations\end{array}\label{2.2.1}\end{equation}
If our theory is symmetric under such transformations then associated
with the symmetry should be Ward identities and conserved currents
as in the previously studied theories. We expect to find Ward identities
for both rigid rescaling $\Lambda\ne0$ as well as the special conformal
transformations. 

In the last lecture we saw that the following action was invariant
under rescaling $x\rightarrow x'=\lambda x$ ($\lambda$ is a constant),\begin{equation}
S=\int\textrm{d}^{D}x\textrm{ }\frac{1}{2}\partial^{\mu}\phi\partial_{\mu}\phi-g\phi^{\frac{2D}{D-2}}\label{2.2.2}\end{equation}
 with the fields transforming as $\phi'\left(x'\right)=\lambda^{-\frac{2D}{D-2}}\phi\left(x\right)$.
We shall now make the following infinitesimal change in coordinates
(we neglect terms less than of order $\epsilon\left(x\right)^{2}$),\begin{equation}
x'_{\mu}=\left(1+\epsilon\left(x\right)\right)x_{\mu}\label{2.2.3}\end{equation}
\begin{equation}
\phi'\left(x'\right)=\left(1+\epsilon\left(x\right)\right)^{-\frac{\left(D-2\right)}{2}}\phi\left(x\right).\label{2.2.4}\end{equation}
The Jacobian of this transformation is needed for the integral measure
$d^{D}x$ in the action, it is given by,\begin{equation}
\begin{array}{rcl}
J\left(x,x'\right) & = & \left|\frac{\partial x'^{\mu}}{\partial x^{\nu}}\right|\\
 & = & \left|\delta_{\nu}^{\mu}\left(1+\epsilon\left(x\right)\right)+x^{\mu}\partial_{\nu}\epsilon\left(x\right)\right|\\
 & = & \left|\begin{array}{cccc}
1+\epsilon\left(x\right)+x^{1}\partial_{1}\epsilon\left(x\right) & x^{1}\partial_{2}\epsilon\left(x\right) & x^{1}\partial_{3}\epsilon\left(x\right) & .\\
x^{2}\partial_{1}\epsilon\left(x\right) & 1+\epsilon\left(x\right)+x^{2}\partial_{2}\epsilon\left(x\right) & . & .\\
x^{3}\partial_{1}\epsilon\left(x\right) & . & . & .\\
. & . & . & 1+\epsilon\left(x\right)+x^{D}\partial_{D}\epsilon\left(x\right)\end{array}\right|\end{array}\label{2.2.5}\end{equation}
With a little thought one can see that all of the off diagonal terms
in the determinant above will only contribute terms of order $\epsilon\left(x\right)^{2}$
and above so we can forget about them at all points in evaluating
the determinant making it the product of the diagonal terms. To first
order in $\epsilon\left(x\right)$ the determinant is,\begin{equation}
J\left(x,x'\right)=1+D\epsilon\left(x\right)+x^{\mu}\partial_{\mu}\epsilon\left(x\right)+O\left(\epsilon\left(x\right)^{2}\right).\label{2.2.6}\end{equation}
\begin{equation}
\Rightarrow\int\textrm{d}^{D}x\textrm{ }=\int\textrm{d}^{D}x'\textrm{ }\frac{1}{J\left(x,x'\right)}\sim\int\textrm{d}^{D}x'\textrm{ }\left(1-D\epsilon\left(x\right)-x^{\mu}\partial_{\mu}\epsilon\left(x\right)\right)\label{2.2.7}\end{equation}
 We attempt now to evaluate $\partial_{\mu}\phi$ in the new variables
so as to ultimately work out the transformed action. \begin{equation}
\begin{array}{rcl}
\frac{\partial\phi\left(x\right)}{\partial x^{\mu}} & = & \frac{\partial x'^{\rho}}{\partial x^{\mu}}\frac{\partial}{\partial x'^{\rho}}\left(\left(1+\epsilon\left(x\right)\right)^{\frac{D-2}{2}}\phi'\left(x'\right)\right)\\
 & \approx & \left(\delta_{\mu}^{\rho}\left(1+\epsilon\left(x\right)\right)+x^{\rho}\frac{\partial\epsilon\left(x\right)}{\partial x^{\mu}}\right)\frac{\partial}{\partial x'^{\rho}}\left(\left(1+\frac{D-2}{2}\epsilon\left(x\right)\right)\phi'\left(x'\right)\right)\\
 & = & \left(\delta_{\mu}^{\rho}\left(1+\epsilon\left(x\right)\right)+x^{\rho}\frac{\partial\epsilon\left(x\right)}{\partial x^{\mu}}\right)\left(\frac{\partial\phi'\left(x'\right)}{\partial x'^{\rho}}+\frac{D-2}{2}\epsilon\left(x\right)\frac{\partial\phi'\left(x'\right)}{\partial x'^{\rho}}\right)\\
 &  & +\left(\delta_{\mu}^{\rho}\left(1+\epsilon\left(x\right)\right)+x^{\rho}\frac{\partial\epsilon\left(x\right)}{\partial x^{\mu}}\right)\frac{D-2}{2}\phi'\left(x'\right)\frac{\partial\epsilon\left(x\right)}{\partial x'^{\rho}}\\
 & = & \delta_{\mu}^{\rho}\left(1+\epsilon\left(x\right)\right)\left(\frac{\partial\phi'\left(x'\right)}{\partial x'^{\rho}}+\frac{D-2}{2}\epsilon\left(x\right)\frac{\partial\phi'\left(x'\right)}{\partial x'^{\rho}}\right)+x^{\rho}\frac{\partial\phi'\left(x'\right)}{\partial x'^{\rho}}\frac{\partial\epsilon\left(x\right)}{\partial x^{\mu}}\\
 &  & +\frac{D-2}{2}\delta_{\mu}^{\rho}\phi'\left(x'\right)\frac{\partial\epsilon\left(x\right)}{\partial x'^{\rho}}+O\left(\epsilon^{2}\right)\\
 & = & \frac{\partial\phi'\left(x'\right)}{\partial x'^{\mu}}+\frac{D-2}{2}\epsilon\left(x\right)\frac{\partial\phi'\left(x'\right)}{\partial x'^{\mu}}+\epsilon\left(x\right)\frac{\partial\phi'\left(x'\right)}{\partial x'^{\mu}}+x^{\rho}\frac{\partial\phi'\left(x'\right)}{\partial x'^{\rho}}\frac{\partial\epsilon\left(x\right)}{\partial x^{\mu}}\\
 &  & +\frac{D-2}{2}\phi'\left(x'\right)\frac{\partial\epsilon\left(x\right)}{\partial x'^{\mu}}+O\left(\epsilon^{2}\right)\\
 & = & \frac{\partial\phi'\left(x'\right)}{\partial x'^{\mu}}+\left(1+\frac{D-2}{2}\right)\epsilon\left(x\right)\frac{\partial\phi'\left(x'\right)}{\partial x'^{\mu}}+x^{\rho}\frac{\partial\phi'\left(x'\right)}{\partial x'^{\rho}}\frac{\partial\epsilon\left(x\right)}{\partial x^{\mu}}+\frac{D-2}{2}\phi'\left(x'\right)\frac{\partial\epsilon\left(x\right)}{\partial x'^{\mu}}+O\left(\epsilon^{2}\right)\\
 & = & \left(1+\frac{D}{2}\epsilon\left(x\right)\right)\frac{\partial\phi'\left(x'\right)}{\partial x'^{\mu}}+x^{\rho}\frac{\partial\phi'\left(x'\right)}{\partial x'^{\rho}}\frac{\partial\epsilon\left(x\right)}{\partial x^{\mu}}+\frac{D-2}{2}\phi'\left(x'\right)\frac{\partial\epsilon\left(x\right)}{\partial x'^{\mu}}+O\left(\epsilon^{2}\right)\end{array}\label{2.2.8}\end{equation}
For ease of notation we denote differentiation with respect to the
transformed coordinates $x'$ $\frac{\partial}{\partial x'^{\mu}}$
as $\partial'_{\mu}$ \emph{i.e.}\begin{equation}
\partial_{\mu}\phi\left(x\right)=\left(1+\frac{D}{2}\epsilon\left(x\right)\right)\partial'_{\mu}\phi'\left(x'\right)+x^{\rho}\partial'_{\rho}\phi'\left(x'\right)\partial_{\mu}\epsilon\left(x\right)+\frac{D-2}{2}\phi'\left(x'\right)\partial'_{\mu}\epsilon\left(x\right)+O\left(\epsilon^{2}\right).\label{2.2.9}\end{equation}
 This means the kinetic term in the Lagrangian is,\begin{equation}
\begin{array}{rcl}
\frac{1}{2}\partial^{\mu}\phi\left(x\right)\partial_{\mu}\phi\left(x\right) & = & \frac{1}{2}\left(\partial'_{\mu}\phi'\left(x'\right)+\frac{D}{2}\epsilon\left(x\right)\partial'_{\mu}\phi'\left(x'\right)+x^{\rho}\partial'_{\rho}\phi'\left(x'\right)\partial_{\mu}\epsilon\left(x\right)+\frac{D-2}{2}\phi'\left(x'\right)\partial'_{\mu}\epsilon\left(x\right)\right)\\
 &  & \times\left(\partial'^{\mu}\phi'\left(x'\right)+\frac{D}{2}\epsilon\left(x\right)\partial'^{\mu}\phi'\left(x'\right)+x^{\rho}\partial'_{\rho}\phi'\left(x'\right)\partial^{\mu}\epsilon\left(x\right)+\frac{D-2}{2}\phi'\left(x'\right)\partial'^{\mu}\epsilon\left(x\right)\right)\\
 & = & \frac{1}{2}\partial'^{\mu}\phi'\left(x'\right)\partial'_{\mu}\phi'\left(x'\right)+\frac{1}{2}D\epsilon\left(x\right)\partial'^{\mu}\phi'\left(x'\right)\partial'_{\mu}\phi'\left(x'\right)+x^{\rho}\partial'_{\rho}\phi'\left(x'\right)\partial'_{\mu}\phi'\left(x'\right)\partial^{\mu}\epsilon\left(x\right)\\
 &  & +\frac{D-2}{2}\phi'\left(x'\right)\partial'^{\mu}\epsilon\left(x\right)\partial'_{\mu}\phi'\left(x'\right)+O\left(\epsilon^{2}\right)\\
 & = & \frac{1}{2}\left(\partial'^{\mu}\phi'\left(x'\right)\partial'_{\mu}\phi'\left(x'\right)\right)\left(1+D\epsilon\left(x\right)\right)+\left(\partial'_{\mu}\phi'\left(x'\right)\right)\left(x^{\rho}\partial'_{\rho}\phi'\left(x'\right)\partial^{\mu}\epsilon\left(x\right)\right)\\
 &  & +\left(\partial'_{\mu}\phi'\left(x'\right)\right)\left(\frac{D-2}{2}\phi'\left(x'\right)\partial'^{\mu}\epsilon\left(x\right)\right).\end{array}\label{2.2.10}\end{equation}
In the action all terms in the Lagrangian will be multiplied by the
Jacobian $\frac{1}{J\left(x,x'\right)}=1-D\epsilon\left(x\right)-x^{\mu}\partial_{\mu}\epsilon\left(x\right)+O\left(\epsilon^{2}\right).$
So we have even more terms of order $\epsilon^{2}$ and above to find
and then drop from the kinetic part.\begin{equation}
\begin{array}{rcl}
\int\textrm{d}^{D}x\frac{1}{2}\partial^{\mu}\phi\left(x\right)\partial_{\mu}\phi\left(x\right) & = & \int\textrm{d}^{D}x'\textrm{ }\left(1-D\epsilon\left(x\right)-x^{\lambda}\partial_{\lambda}\epsilon\left(x\right)\right)\left(\frac{1}{2}\left(\partial'^{\mu}\phi'\left(x'\right)\partial'_{\mu}\phi'\left(x'\right)\right)\left(1+D\epsilon\left(x\right)\right)\right.\\
 &  & +\left.\left(\partial'_{\mu}\phi'\left(x'\right)\right)\left(x^{\rho}\partial'_{\rho}\phi'\left(x'\right)\partial^{\mu}\epsilon\left(x\right)+\frac{D-2}{2}\phi'\left(x'\right)\partial'^{\mu}\epsilon\left(x\right)\right)\right)\\
 & = & \int\textrm{d}^{D}x'\textrm{ }\left(\frac{1}{2}\left(\partial'^{\mu}\phi'\left(x'\right)\partial'_{\mu}\phi'\left(x'\right)\right)\left(1+D\epsilon\left(x\right)\right)+\left(\partial'_{\mu}\phi'\left(x'\right)\right)\left(x^{\rho}\partial'_{\rho}\phi'\left(x'\right)\partial^{\mu}\epsilon\left(x\right)\right.\right.\\
 &  & \left.\left.+\frac{D-2}{2}\phi'\left(x'\right)\partial'^{\mu}\epsilon\left(x\right)\right)\right)-\frac{1}{2}\left(\partial'^{\mu}\phi'\left(x'\right)\partial'_{\mu}\phi'\left(x'\right)\right)\left(D\epsilon\left(x\right)+x^{\lambda}\partial_{\lambda}\epsilon\left(x\right)\right)\\
 & = & \int\textrm{d}^{D}x'\textrm{ }\frac{1}{2}\left(\partial'^{\mu}\phi'\left(x'\right)\partial'_{\mu}\phi'\left(x'\right)\right)+\left(\partial'_{\mu}\phi'\left(x'\right)\right)\left(x^{\rho}\partial'_{\rho}\phi'\left(x'\right)\partial^{\mu}\epsilon\left(x\right)\right)\\
 &  & +\left(\partial'_{\mu}\phi'\left(x'\right)\right)\left(\frac{D-2}{2}\phi'\left(x'\right)\partial'^{\mu}\epsilon\left(x\right)\right)-\frac{1}{2}x^{\lambda}\partial_{\lambda}\epsilon\left(x\right)\left(\partial'^{\mu}\phi'\left(x'\right)\partial'_{\mu}\phi'\left(x'\right)\right)\\
 & = & \int\textrm{d}^{D}x'\textrm{ }\frac{1}{2}\left(\partial'^{\mu}\phi'\left(x'\right)\partial'_{\mu}\phi'\left(x'\right)\right)\left(1-x^{\lambda}\partial_{\lambda}\epsilon\left(x\right)\right)\\
 &  & +\left(\partial'_{\mu}\phi'\left(x'\right)\right)\left(x^{\rho}\partial'_{\rho}\phi'\left(x'\right)\partial^{\mu}\epsilon\left(x\right)+\frac{D-2}{2}\phi'\left(x'\right)\partial'^{\mu}\epsilon\left(x\right)\right)\end{array}\label{2.2.11}\end{equation}
 For the potential part of the action we have the following (dropping
terms $O\left(\epsilon^{2}\right)$ and above):\begin{equation}
\begin{array}{rcl}
-g\int\textrm{d}^{D}x\textrm{ }\phi\left(x\right)^{\frac{2D}{D-2}} & = & -g\int\textrm{d}^{D}x'\textrm{ }\left(1-D\epsilon\left(x\right)-x^{\lambda}\partial_{\lambda}\epsilon\left(x\right)\right)\left(\left(1+\epsilon\left(x\right)\right)^{\frac{\left(D-2\right)}{2}}\phi'\left(x'\right)\right)^{\frac{2D}{D-2}}\\
 & = & -g\int\textrm{d}^{D}x'\textrm{ }\left(1-D\epsilon\left(x\right)-x^{\lambda}\partial_{\lambda}\epsilon\left(x\right)\right)\left(1+\epsilon\left(x\right)\right)^{D}\phi'\left(x'\right)^{\frac{2D}{D-2}}\\
 & = & -g\int\textrm{d}^{D}x'\textrm{ }\left(1-D\epsilon\left(x\right)-x^{\lambda}\partial_{\lambda}\epsilon\left(x\right)\right)\left(1+D\epsilon\left(x\right)\right)\phi'\left(x'\right)^{\frac{2D}{D-2}}\\
 & = & -g\int\textrm{d}^{D}x'\textrm{ }\left(1-x^{\lambda}\partial_{\lambda}\epsilon\left(x\right)\right)\phi'\left(x'\right)^{\frac{2D}{D-2}}\end{array}.\label{2.2.12}\end{equation}
 So the effect of the transformation / change of variables on the
action as a whole is,\begin{equation}
\int\textrm{d}^{D}x\textrm{ }{\cal {L}}\left(\phi\right)=\int\textrm{d}^{D}x'\textrm{ }{\cal {L}}\left(\phi'\right)\left(1-x^{\lambda}\partial_{\lambda}\epsilon\left(x\right)\right)+\left(\partial'_{\mu}\phi'\left(x'\right)\right)\left(x^{\rho}\partial'_{\rho}\phi'\left(x'\right)\partial^{\mu}\epsilon\left(x\right)+\frac{D-2}{2}\phi'\left(x'\right)\partial'^{\mu}\epsilon\left(x\right)\right)\label{2.2.13}\end{equation}
\begin{equation}
\Rightarrow S\left[\phi\right]=S\left[\phi'\right]+\int\textrm{d}^{D}x'\textrm{ }-L\left(\phi'\right)x^{\mu}\partial_{\mu}\epsilon\left(x\right)+\left(\partial'_{\mu}\phi'\left(x'\right)\right)\left(x^{\rho}\partial'_{\rho}\phi'\left(x'\right)\partial^{\mu}\epsilon\left(x\right)+\frac{D-2}{2}\phi'\left(x'\right)\partial'^{\mu}\epsilon\left(x\right)\right).\label{2.2.14}\end{equation}
To first order in $\epsilon\left(x\right)$ we have $\epsilon\left(x\right)=\epsilon\left(x'\right)$.
This is shown by Taylor expanding $\epsilon\left(x'\right)$:\begin{equation}
\begin{array}{rcl}
\epsilon\left(x'^{\mu}\right) & = & \epsilon\left(x^{\mu}+\epsilon\left(x^{\mu}\right)x^{\mu}\right)=\epsilon\left(x^{\mu}\right)+\left.\frac{\partial\epsilon\left(y^{\mu}\right)}{\partial y^{\nu}}\right|_{y^{\mu}=x^{\mu}}\epsilon\left(x^{\mu}\right)x^{\nu}+...\\
 & = & \epsilon\left(x^{\mu}\right)+O\left(\epsilon\left(x^{\mu}\right)^{2}\right).\end{array}\label{2.2.15}\end{equation}
Also,\begin{equation}
\begin{array}{rcl}
\partial'_{\nu}\epsilon\left(x'\right) & = & \frac{\partial x^{\lambda}}{\partial x'^{\nu}}\partial_{\lambda}\epsilon\left(x\right)\\
 & = & \left(\frac{\partial}{\partial x'^{\nu}}\left(1+\epsilon\left(x\right)\right)^{-1}x'^{\lambda}\right)\partial_{\lambda}\epsilon\left(x\right)\\
 & = & \left(\frac{\partial}{\partial x'^{\nu}}\left(1-\epsilon\left(x\right)\right)x'^{\lambda}\right)\partial_{\lambda}\epsilon\left(x\right)\\
 & = & \left(\frac{\partial x'^{\lambda}}{\partial x'^{\nu}}-\frac{\partial\left(\epsilon\left(x\right)x'^{\lambda}\right)}{\partial x'^{\nu}}\right)\partial_{\lambda}\epsilon\left(x\right)\\
 & = & \partial_{\nu}\epsilon\left(x\right)+O\left(\epsilon^{2}\right)\end{array}.\label{2.2.16}\end{equation}
If we now use these identities on the second term in the transformed
version of the action, equation \ref{2.2.14},\begin{equation}
\int\textrm{d}^{D}x'\textrm{ }-{\cal {L}}\left(\phi'\right)x^{\mu}\partial_{\mu}\epsilon\left(x\right)+\left(\partial'_{\mu}\phi'\left(x'\right)\right)\left(x^{\rho}\partial'_{\rho}\phi'\left(x'\right)\partial^{\mu}\epsilon\left(x\right)+\frac{D-2}{2}\phi'\left(x'\right)\partial'^{\mu}\epsilon\left(x\right)\right),\label{2.2.17}\end{equation}
 we get,\begin{equation}
\begin{array}{rl}
 & \int\textrm{d}^{D}x'\textrm{ }-{\cal {L}}\left(\phi'\right)x^{\mu}\partial'_{\mu}\epsilon\left(x'\right)+\left(\partial'_{\mu}\phi'\left(x'\right)\right)\left(x^{\rho}\partial'_{\rho}\phi'\left(x'\right)\partial'^{\mu}\epsilon\left(x'\right)+\frac{D-2}{2}\phi'\left(x'\right)\partial'^{\mu}\epsilon\left(x'\right)\right)\\
= & \int\textrm{d}^{D}x'\textrm{ }-{\cal {L}}\left(\phi'\right)x_{\mu}\partial'^{\mu}\epsilon\left(x'\right)+\left(\partial'_{\mu}\phi'\left(x'\right)\right)\left(x^{\rho}\partial'_{\rho}\phi'\left(x'\right)+\frac{D-2}{2}\phi'\left(x'\right)\right)\partial'^{\mu}\epsilon\left(x'\right)\\
= & \int\textrm{d}^{D}x'\textrm{ }\left(-{\cal {L}}\left(\phi'\right)x_{\mu}+\left(\partial'_{\mu}\phi'\left(x'\right)\right)\left(x^{\rho}\partial'_{\rho}\phi'\left(x'\right)+\frac{D-2}{2}\phi'\left(x'\right)\right)\right)\partial'^{\mu}\epsilon\left(x'\right).\end{array}\label{2.2.18}\end{equation}
If we use the product rule for differentiating / integration by parts
we can write this as,\begin{equation}
\begin{array}{rl}
= & \int\textrm{d}^{D}x'\textrm{ }\partial'^{\mu}\left(\textrm{ }\left(-{\cal {L}}\left(\phi'\right)x_{\mu}+\left(\partial'_{\mu}\phi'\left(x'\right)\right)\left(x^{\rho}\partial'_{\rho}\phi'\left(x'\right)+\frac{D-2}{2}\phi'\left(x'\right)\right)\right)\epsilon\left(x'\right)\right)\\
- & \int\textrm{d}^{D}x'\textrm{ }\epsilon\left(x'\right)\partial'^{\mu}\left(-{\cal {L}}\left(\phi'\right)x_{\mu}+\left(\partial'_{\mu}\phi'\left(x'\right)\right)\left(x^{\rho}\partial'_{\rho}\phi'\left(x'\right)+\frac{D-2}{2}\phi'\left(x'\right)\right)\right).\end{array}\label{2.2.19}\end{equation}
As the first term is a D-divergence of some function of the field
then if we make the usual assumption that all fields and their first
derivatives \emph{etc} vanish at infinity then the first term goes
to zero (we write the volume integral of the D-divergence over all
space-time as a surface integral, the surface being at infinity).
So the transformed version of the action is,\begin{equation}
S\left[\phi\right]=S\left[\phi'\right]-\int\textrm{d}^{D}x'\textrm{ }\epsilon\left(x'\right)\partial'^{\mu}\left(-L\left(\phi'\right)x_{\mu}+\left(\partial'_{\mu}\phi'\left(x'\right)\right)\left(x^{\rho}\partial'_{\rho}\phi'\left(x'\right)+\frac{D-2}{2}\phi'\left(x'\right)\right)\right).\label{2.2.20}\end{equation}
 As $x'_{\mu}=\left(1+\epsilon\left(x\right)\right)x_{\mu}$ and we
have an $\epsilon\left(x'\right)$ at the beginning of the whole integrand,
neglecting terms of order $\epsilon^{2}$ and above, we can just swap
$x_{\mu}$ and $x^{\rho}$ for $x'_{\mu}$ and $x'^{\rho}$: \begin{equation}
S\left[\phi\right]=S\left[\phi'\right]+\int\textrm{d}^{D}x'\textrm{ }\epsilon\left(x'\right)\partial'^{\mu}\left(L\left(\phi'\right)x'_{\mu}-\left(\partial'_{\mu}\phi'\left(x'\right)\right)\left(x'^{\rho}\partial'_{\rho}\phi'\left(x'\right)+\frac{D-2}{2}\phi'\left(x'\right)\right)\right)\label{2.2.21}\end{equation}
\begin{equation}
\Rightarrow S\left[\phi\right]=S\left[\phi'\right]+\int\textrm{d}^{D}x'\textrm{ }\epsilon\left(x'\right)\partial'^{\mu}j_{\mu}\left(x\right)\label{2.2.22}\end{equation}
 where we have defined the \emph{dilatation current,\begin{equation}
j_{\mu}=L\left(\phi'\right)x'_{\mu}-\left(\partial'_{\mu}\phi'\left(x'\right)\right)\left(x'^{\rho}\partial'_{\rho}\phi'\left(x'\right)+\frac{D-2}{2}\phi'\left(x'\right)\right).\label{2.2.23}\end{equation}
} The dilatation current is quite novel on account of it having $x'_{\mu}$'s
in it. The dilatation current is a common feature of field theories
incorporating gravitational effects, it couples to a scalar field,
the \emph{dilaton}. 

The Ward identities are obtained in the exact same way as before,
we consider a general Green's function, make a change of variables
(do the transformation) and then differentiate with respect to the
parameter of the transformation. The measure of the path integral
is invariant for exactly the same reason that it was in the last lecture
(we pick up factors of $\left(1-\frac{D-2}{2}\epsilon\left(x\right)\right)$
in the measure which cancel with those picked up in the measure of
$\frac{1}{Z}$). 

Up to now our discussion of our scale invariant scalar field theory
has been at a pseudo classical level. If we are to think about the
theory as a quantum field theory we will have to think about UV divergences.
Consider for instance the first order correction to the two-point
function in $\phi^{4}$ theory, this is a loop on top of the free
particle propagator. To make our theory meaningful we have to be able
to deal with the infinities produced by integrating over loop momenta,
we have to renormalize it, which means introducing a dimensionful
parameter, a cut off, into the otherwise dimensionless theory. Introducing
dimensionful parameters into our theory breaks scale invariance. An
example of this can be seen by trying to introducing mass terms into
the theory. Consider the action of lecture 4 with $x'^{\mu}=\lambda x^{\mu}$
for constant $\lambda$, \begin{equation}
\begin{array}{rcl}
S\left[\phi\right] & = & \int\textrm{d}^{D}x\textrm{ }\frac{1}{2}\partial^{\mu}\phi\left(x\right)\partial_{\mu}\phi\left(x\right)-m^{2}\phi^{2}\left(x\right)\\
 & = & \lambda^{-D}\int\textrm{d}^{D}x'\textrm{ }\left(\lambda^{2}\frac{1}{2}\partial'^{\mu}\phi\left(\frac{x'}{\lambda}\right)\partial'_{\mu}\phi\left(\frac{x'}{\lambda}\right)-m^{2}\phi\left(\frac{x'}{\lambda}\right)^{2}\right)\end{array}\label{2.2.24}\end{equation}
\begin{equation}
\phi'\left(x'\right)=\lambda^{-\left(\frac{D-2}{2}\right)}\phi\left(\frac{x'}{\lambda}\right)\label{2.2.25}\end{equation}
\begin{equation}
\begin{array}{crcl}
\Rightarrow & S\left[\phi\right] & = & \int\textrm{d}^{D}x'\textrm{ }\frac{1}{2}\partial'^{\mu}\phi'\left(x'\right)\partial'_{\mu}\phi'\left(x'\right)-\frac{m^{2}}{\lambda^{2}}\phi'\left(x'\right)^{2}\\
 &  & = & S\left[\phi'\right]+\int\textrm{d}^{D}x\textrm{ }m^{2}\left(1-\frac{1}{\lambda^{2}}\right)\phi^{2}.\end{array}\label{2.2.26}\end{equation}

\chapter{Principles of Gauge Field Theory Quantization\label{cha:Principles-of-Gauge}.}

\section{Faddeev-Popov Gauge Fixing and Ghosts. }

This lecture deals with how to quantize gauge theories, specifically
we describe how to get around the problem of integrating over too
many/artificial (gauge) degrees of freedom in the path integral by
the Faddeev Popov gauge fixing procedure. 

We will illustrate Faddeev Popov gauge fixing on QED though the method
generalizes to the other gauge theories. The problem is that in the
path integral formalism transition probabilities are obtained by integrating
over all possible physical paths, but if we naively write the generating
functional of QED as an integral over the gauge field $A_{\mu}$ we
are integrating over more than just the number of \emph{physical}
paths because all fields $A_{\mu}$ related by a gauge transformation
represent the same physical configuration (they lie on the same \emph{gauge
orbit i.e.} they are related by a gauge transformation). They give
rise to the same $\vec{E}$ and $\vec{B}$ \emph{etc} fields that
we observe. There are actually an infinite number of gauges to choose
from which sure enough means that we over count the number of paths
by a factor of infinity, ideally we want to isolate and divide out
this {}``infinite group volume factor''. This means we must somehow
constrain our integral over the gauge field $A_{\mu}$ such that we
are only integrating over all possible physically inequivalent paths. 

To solve this problem it is tempting to insert a condition into the
path integral like say $\delta\left(\partial_{\mu}A^{\mu}\right)$
(Lorentz gauge condition) or $\delta\left(\vec{\nabla}.\vec{A}\right)$
(Coulomb gauge condition) which would set the contributions from all
other gauge transformed versions of $A^{\mu}$ to zero \emph{i.e.}
we want something to fix the gauge (we only want to cut across a gauge
orbit once when integrating through the function space of $A_{\mu}$).
Ideally we should do general derivations and so we should consider
inserting a general gauge condition $F\left(A_{\mu}\right)=0$ with
$F$ an arbitrary function of the fields (and their derivatives \emph{etc}).
However we cannot trivially insert such delta functions as this changes
the measure of the integration $DA_{\mu}$. Say we had an ordinary
integral and we wanted to fix some variable in it $x$ to be some
value, we could fix it with $\delta\left(x-a\right)$ or equivalently
$\delta\left(f\left(x\right)\right)$ where $f\left(x\right)$ is
some function which is zero at $x=a$. Both delta functions will set
$x=a$ but the latter does something else as well, \begin{equation}
\delta\left(f\left(x\right)\right)=\frac{\delta\left(x-a\right)}{\left|f'\left(x\right)\right|}\label{3.1.1}\end{equation}
 The difference between the two choices of delta function is a factor
$\frac{1}{\left|f'\left(x\right)\right|}$, thus there will be some
ambiguity in our choice of constraint which we will have to sort out.

Before plunging into the full field theory calculation with it's infinite
number of degrees of freedom (functional integrals) it is useful to
consider the problem (and solution) in the context of something with
a finite number of degrees of freedom (function integrals). 

\begin{equation}
I=\int\textrm{d}^{2}r\textrm{ }e^{-iS\left(r\right)}\label{3.1.2}\end{equation}
Consider the above integral, it is invariant under the \emph{gauge}
\emph{transformation} $\theta\rightarrow\theta'=\theta+\alpha$. Now
we would like to \emph{choose} \emph{a} \emph{gauge} so that we only
integrate over \emph{gauge} \emph{inequivalent} situations . This
means we have to restrict the integral so that we only integrate $e^{-iS\left(r\right)}$
along a line starting at the origin and moving out in the $xy$ plane.
Say the function $\theta=\phi\left(r\right)$ defines this line. $\phi\left(r\right)$
should be a many to one function in $r$ and $\theta$ \emph{i.e.}
for any given value of $r$ there is only one value of $\theta$ that
solves $\theta=\phi\left(r\right)$. Also, obviously $\theta$ and
$\phi\left(r\right)$ are defined in the range $0\rightarrow2\pi$
. In gauge theory this corresponds to the requirement that a single
physical configuration corresponds to a single gauge field configuration.
We shall now perform the trick which is central to Faddeev Popov gauge
fixing. Clearly we do not change the value of the integral if we were
to insert a $1$: \begin{equation}
I=\int\textrm{d}^{2}r\textrm{ }\times1\times e^{iS\left(r\right)}.\label{3.1.3}\end{equation}
Note that we can write $1=\int\textrm{d}\phi\left(r\right)\textrm{ }\delta\left(\theta-\phi\left(r\right)\right)$
where here one must regard the value of the function at some $r$,
$\phi\left(r\right)$, as a variable. Regard $r$ as an index for
the {}``variable'' $\phi$$\left(r\right)$. This means we can write,\begin{equation}
\begin{array}{rcl}
I & = & \int\textrm{d}^{2}r\left[\int\textrm{d}\phi\left(r\right)\delta\left(\theta-\phi\left(r\right)\right)\right]e^{iS\left(r\right)}\\
 & = & \int r\textrm{d}r\textrm{d}\theta d\phi\left(r\right)\delta\left(\theta-\phi\left(r\right)\right)e^{iS\left(r\right)}\\
 & = & \int r\textrm{d}r\int\textrm{d}\phi\left(r\right)\left[\int\textrm{d}\theta\delta\left(\theta-\phi\left(r\right)\right)e^{iS\left(r\right)}\right].\end{array}\label{3.1.4}\end{equation}
Note that though the integrand depends on $\phi\left(r\right)$ the
integral itself in the square brackets actually does not for no matter
what value of $r$ we take $\phi\left(r\right)$ will always be in
the range $0\rightarrow2\pi$ and so the integration over $\theta$
will always kill the $\phi\left(r\right)$; for every $r$ there is
one solution $\theta$ to $\theta=\phi\left(r\right)$. This means
we can actually shift the integral over $\phi\left(r\right)$ behind
the square bracket,\begin{equation}
I=\int r\textrm{d}r\int\textrm{d}\phi\left(r\right)\left[\int\textrm{d}\theta\delta\left(\theta-\phi\left(r\right)\right)e^{iS\left(r\right)}\right]=\int r\textrm{d}r\left[\int\textrm{d}\theta\delta\left(\theta-\phi\left(r\right)\right)e^{iS\left(r\right)}\right]\int\textrm{d}\phi\left(r\right).\label{3.1.6}\end{equation}
 Now for every value of $r$ the integral $\int\textrm{d}\phi\left(r\right)$
is exactly the same $\int\textrm{d}\phi\left(r\right)\equiv2\pi$
which gives us,\begin{equation}
I=2\pi\int r\textrm{d}r\int\textrm{d}\theta\delta\left(\theta-\phi\left(r\right)\right)e^{iS\left(r\right)}=2\pi\int\textrm{d}^{2}r\delta\left(\theta-\phi\left(r\right)\right)e^{iS\left(r\right)}=2\pi I_{\theta=\phi\left(r\right)}.\label{3.1.7}\end{equation}
 So we have managed to rewrite our integral as a product of a \emph{group}
\emph{volume} \emph{factor} $\left(2\pi\right)$ and a \emph{gauge}
\emph{fixed} version of our original integral which will only integrate
over \emph{gauge}/\emph{physically} \emph{inequivalent} configurations.

We shall now perform this example again but in a slightly more convoluted
way so as to make better contact with the gauge fixing formalism of
Faddeev and Popov. The complication is small, we ask the question
what happens if our gauge fixing condition is in an equivalent but
more complicated form \emph{e.g.} $f(r,\theta)=0$ where the gauge
fixing condition still has to obey the rules laid out before - one
gauge configuration (one $\theta$) for one physical configuration
(one $r$). So once again we insert a $1$, this time in the form,
$1=\int\textrm{d}\phi\left(r\right)\delta\left(f\left(r,\theta\right)\right)$,\begin{equation}
\begin{array}{rcl}
I & = & \int\textrm{d}^{2}r\times1\times e^{iS\left(r\right)}\\
 & = & \int\textrm{d}^{2}r\left[\int\textrm{d}\phi\left(r\right)\delta\left(f\left(r,\theta\right)\right)\right]e^{iS\left(r\right)}\\
 & = & \int r\textrm{d}r\textrm{d}\theta\textrm{d}\phi\left(r\right)\delta\left(f\left(r,\theta\right)\right)e^{iS\left(r\right)}\\
 & = & \int r\textrm{d}r\int\textrm{d}\phi\left(r\right)\left[\int\textrm{d}\theta\delta\left(f\left(r,\theta\right)\right)e^{iS\left(r\right)}\right]\end{array}\label{3.1.8}\end{equation}
The next thing to do is to make a change of variables. We know that
the constraint $f\left(r,\theta\right)=0$ can be written in the form/has
solutions $\theta=\phi\left(r\right)$. So the integral in square
brackets really has no $\phi\left(r\right)$ dependence just like
before when we had instead $\delta\left(\theta-\phi\left(r\right)\right)$.
Hence we can again shift the integral over $\int\textrm{d}\phi\left(r\right)$
to the other side of the square brackets where it becomes $2\pi$
like in the last example.\begin{equation}
I=2\pi\int r\textrm{d}r\textrm{d}\theta\delta\left(f\left(r,\theta\right)\right)e^{iS\left(r\right)}\label{3.1.9}\end{equation}
We must remember to be careful with our delta functions on account
of their functional form of their arguments,\begin{equation}
\begin{array}{rcl}
\textrm{d}\theta & = & \frac{\textrm{d}\theta}{\textrm{d}f\left(r,\theta\right)}df\left(r,\theta\right)\\
 & = & \frac{1}{\frac{\textrm{d}f\left(r,\theta\right)}{\textrm{d}\theta}}df\left(r,\theta\right)\end{array}\label{3.1.10}\end{equation}
 \begin{equation}
\begin{array}{crcl}
\Rightarrow & I & = & 2\pi\int r\textrm{d}r\textrm{d}\theta\delta\left(f\left(r,\theta\right)\right)e^{iS\left(r\right)}\\
 &  & = & 2\pi\int r\textrm{d}rdf\left(r,\theta\right)\left(\frac{df\left(r,\theta\right)}{\textrm{d}\theta}\right)^{-1}\delta\left(f\left(r,\theta\right)\right)e^{iS\left(r\right)}\end{array}.\label{3.1.11}\end{equation}
Like before we turned our integral into a group volume factor times
a \emph{gauge fixed} version of our original integral. The factor
$\left(\frac{df\left(r,\theta\right)}{\textrm{d}\theta}\right)^{-1}$now
appearing in the integrand typically appears as a Jacobian (determinant)
in systems with more degrees of freedom, in gauge theory it is known
as the Faddeev Popov determinant. In what follows we will see that
the Faddeev Popov determinant arising in the quantization of gauge
theories can be recast in the form of \emph{ghost fields}.

\bigskip{}
So let's see what we can do with QED.\begin{equation}
Z=\int D\bar{\psi}D\psi DA^{\mu}\textrm{ }e^{iS_{Q.E.D.}}\label{3.1.12}\end{equation}
 Now introduce a gauge fixing condition in the form of a {}``$1$''.
The gauge fixing condition takes the form of a \emph{functional} delta
function. Such a delta functional fixes the gauge differently at \emph{each}
\emph{point} \emph{in} \emph{space-time} hence we have a superscript
$\infty$ - we have a condition at each point in space-time:\begin{equation}
\Delta\left(A^{\mu}\right)=\int D\chi\textrm{ }\delta^{\infty}\left[f\left(A^{\mu}+\partial^{\mu}\chi\right)\right].\label{3.1.13}\end{equation}
Note that our object $\Delta$ is in fact gauge invariant itself:\begin{equation}
\begin{array}{rcl}
\Delta\left(A^{\mu}+\partial^{\mu}\phi\right) & = & \int D\chi\textrm{ }\delta^{\infty}\left[f\left(A^{\mu}+\partial^{\mu}\chi+\partial^{\mu}\phi\right)\right]\\
 & = & \int D\chi\textrm{ }\delta^{\infty}\left[f\left(A^{\mu}+\partial^{\mu}\left(\phi+\chi\right)\right)\right]\end{array}.\label{3.1.14}\end{equation}
 now shift the integration variables $\chi\left(x\right)$ (remember
for functionals $x$ is like the index, the indexed variables are
$\chi\left(x\right)$) at each point by a constant amount given by
$\phi\left(x\right)$:\begin{equation}
\begin{array}{rcl}
D\chi & = & \prod_{x}\int_{all\textrm{ }\chi\left(x\right)}d\chi\left(x\right)\\
 & = & \prod_{x}\int_{all\textrm{ }\chi\left(x\right)}d\left(\chi\left(x\right)+\phi\left(x\right)\right)\\
 & = & \prod_{x}\int_{all\textrm{ }\chi\left(x\right)+\phi\left(x\right)}d\left(\chi\left(x\right)+\phi\left(x\right)\right)\\
 & = & \int D\left(\phi+\chi\right)\end{array}.\label{3.1.15}\end{equation}
Hence we see that \begin{equation}
\begin{array}{rcl}
\Delta\left(A^{\mu}+\partial^{\mu}\phi\right) & = & \int D\left(\phi+\chi\right)\textrm{ }\delta^{\infty}\left[f\left(A^{\mu}+\partial^{\mu}\left(\chi+\phi\right)\right)\right]\\
 & = & \int D\chi\textrm{ }\delta^{\infty}\left[f\left(A^{\mu}+\partial^{\mu}\chi\right)\right]\\
 & = & \Delta\left(A^{\mu}\right)\end{array}\label{3.1.16}\end{equation}
 is as promised, gauge invariant. We now have a gauge invariant constraint
which we can insert into our integral. We need to be careful though
about inserting this condition, we must make sure we are in fact putting
{}``1'' into the integrand. $\Delta\left(A^{\mu}\right)$ in the
general form shown above will non-trivially affect the measure of
integration of the path integral. Just like in our previous example
with $r's$ and $\theta's$ we want to change the integration variable
from $\chi$ to $f\left(A^{\mu}+\partial^{\mu}\chi\right)$, in doing
so we will pick up a Jacobian which we will denote by $J$. \begin{equation}
\begin{array}{rcl}
\int D\chi & = & \prod\int_{x}d\chi\left(x\right)\\
 & = & \prod_{x}\int\textrm{d}f\left(x\right)\textrm{ Det}_{x,y}\left(\frac{\delta f\left(x\right)}{\delta\chi\left(y\right)}\right)^{-1}\\
 & = & \int Df\textrm{ }\frac{1}{J\left(f,\chi\right)}\end{array}\label{3.1.17}\end{equation}
\begin{equation}
\begin{array}{rcl}
\Rightarrow\Delta\left(A^{\mu}\right) & = & \int D\chi\textrm{ }\delta^{\infty}\left[f\left(A^{\mu}+\partial^{\mu}\chi\right)\right]\\
 & = & \int Df\textrm{ }\frac{1}{J\left(f,\chi\right)}\delta^{\infty}\left[f\left(A^{\mu}+\partial^{\mu}\chi\right)\right]\\
 & = & \left.\frac{1}{J\left(f,\chi\right)}\right|_{f=0}\int Df\textrm{ }\delta^{\infty}\left[f\left(A^{\mu}+\partial^{\mu}\chi\right)\right]\\
 & = & \left.\frac{1}{J\left(f,\chi\right)}\right|_{f=0}\times1\end{array}\label{3.1.18}\end{equation}
Hopefully it is clear that the $x,y$ subscript attached to the $\textrm{Det}$
above means that we are taking the determinant of $\left(...\right)$
with respect to the $x,y$ {}``indices''. We wish to insert the
gauge fixing condition, the {}``1'' in the form,\begin{equation}
\Rightarrow1=\left.J\left(f,\chi\right)\right|_{f=0}\Delta\left(A^{\mu}\right).\label{3.1.19}\end{equation}
The path integral becomes,\begin{equation}
\begin{array}{rl}
 & \int D\bar{\psi}D\psi DA^{\mu}\textrm{ }\left.J\left(f,\chi\right)\right|_{f=0}\Delta\left(A^{\mu}\right)\textrm{ }e^{iS_{Q.E.D.}}\\
= & \int D\bar{\psi}D\psi DA^{\mu}\textrm{ }\left.J\left(f,\chi\right)\right|_{f=0}\int D\chi\textrm{ }\delta^{\infty}\left[f\left(A^{\mu}+\partial^{\mu}\chi\right)\right]\textrm{ }e^{iS_{Q.E.D.}}.\end{array}\label{3.1.20}\end{equation}
The next bit of the trick involves changing variables $A^{\mu}\rightarrow A_{\left[\chi\right]}^{\mu}=A^{\mu}+\partial^{\mu}\chi$.
Note the action is invariant under such a transformation as it amounts
to a gauge transformation which would leave it invariant (one can
simultaneously transform the fermions in the usual way). The other
terms are also invariant, we showed above that $\Delta\left(A^{\mu}+\partial^{\mu}\phi\right)=\int D\chi\textrm{ }\delta^{\infty}\left[f\left(A^{\mu}+\partial^{\mu}\chi\right)\right]=\Delta\left(A^{\mu}\right)$
and we will therefore assume that $\Delta^{-1}\left(A^{\mu}\right)=\left.J\left(f,\chi\right)\right|_{f=0}$
is also invariant%
\footnote{\[
J\left(f,\chi\right)=\textrm{Det}_{x,y}\left(\frac{\delta f\left(A_{\left[\chi\right]}^{\mu}\left(x\right)\right)}{\delta\chi\left(y\right)}\right)\]
Now use the functional equivalent of taking the total derivative with
respect to a variable,\[
\Rightarrow J\left(f,\chi\right)=\textrm{Det}_{x,y}\left(\int\textrm{d}z\frac{\delta f\left(A_{\left[\chi\right]}\left(x\right)\right)}{\delta A_{\left[\chi\right]}^{\mu}\left(z\right)}\frac{\delta A_{\left[\chi\right]}^{\mu}\left(z\right)}{\delta\chi\left(y\right)}\right)=\textrm{Det}_{x,y}\left(\int\textrm{d}z\frac{\delta f\left(A_{\left[\chi\right]}\left(x\right)\right)}{\delta A_{\left[\chi\right]}^{\mu}\left(z\right)}\partial^{\mu}\delta\left(z-y\right)\right).\]
 This object depends on $A_{\left[\chi\right]}^{\mu}$ but is independent
of $\chi$.%
}. It then remains to prove that the measure $\int D\bar{\psi}D\psi DA^{\mu}$
is also invariant, consider for simplicity zero space-time dimensions
\emph{i.e.} all the fields exist at a single point $x_{0}$ in space-time.
Under the usual transformations $A^{\mu}\rightarrow A^{\mu}+\partial^{\mu}\chi$,
$\bar{\psi}\rightarrow e^{-i\chi}\bar{\psi}$, $\psi\rightarrow e^{i\chi}\psi$
we have \begin{equation}
\int D\bar{\psi}D\psi DA^{\mu}\rightarrow\int D\bar{\psi}D\psi DA^{\mu}\times\textrm{Det}\left(\begin{array}{rcl}
e^{-i\chi} & 0 & 0\\
0 & e^{i\chi} & 0\\
0 & 0 & 1\end{array}\right)=\int D\bar{\psi}D\psi DA^{\mu}.\label{3.1.21}\end{equation}
Generalizing to a space consisting of an infinity of points the matrix
above becomes an infinite dimensional Jacobian which is still diagonal
and still equal to one, due simply to the unitary nature of the transformation.
Therefore, \begin{equation}
Z=\int D\bar{\psi}D\psi DA_{\left[\chi\right]}^{\mu}D\chi\textrm{ }\left.J\left(f,\chi\right)\right|_{f=0}\textrm{ }\delta^{\infty}\left[f\left(A_{\left[\chi\right]}^{\mu}\right)\right]\textrm{ }e^{iS_{Q.E.D.}}.\label{3.1.22}\end{equation}
So now we have that the integrand does not depend on the gauge $\chi$,
this dependence has been absorbed into the gauge transformed field
$A_{\left[\chi\right]}^{\mu}$ instead! This means we can bring the
integral over $\chi$ to the front and absorb it in the normalization
of the path integral as an (infinite) group volume factor (\emph{i.e.}
forget about it)\begin{equation}
=\left(\int D\chi\right)\times\left(\int D\bar{\psi}D\psi DA_{\left[\chi\right]}^{\mu}\textrm{ }\left.J\left(f\right)\right|_{f=0}\textrm{ }\delta^{\infty}\left[f\left(A_{\left[\chi\right]}^{\mu}\left(x\right)\right)\right]\textrm{ }e^{iS_{Q.E.D.}}\right)\label{3.1.23}\end{equation}
leaving the Jacobian (aka Faddeev-Popov determinant) and gauge fixing
delta function as the remnants of gauge fixing in the path integral.
In calculating Green's functions the factor $\int D\chi$ cancels
in the numerator and denominator, recall Green's functions are of
the form, $G\left(x_{1},...,x_{n}\right)=\frac{1}{Z\left[0\right]}\left.\frac{\delta}{\delta J\left(x_{1}\right)}...\frac{\delta Z\left[J\right]}{\delta J\left(x_{n}\right)}\right|$.

Now we further simplify our situation by choosing our gauge fixing
condition such that it is of the form,\begin{equation}
f\left(A_{\left[\chi\right]}^{\mu}\right)=\omega\left(x\right)+\tilde{f}\left(A_{\left[\chi\right]}^{\mu}\right).\label{3.1.24}\end{equation}
The original path integral has had a {}``1'' inserted into it and
after some rewriting now looks like,\begin{equation}
Z=\left(\int D\chi\right)\times\left(\int D\bar{\psi}D\psi DA_{\left[\chi\right]}^{\mu}\textrm{ }\left.J\left(f\right)\right|_{f=0}\textrm{ }\delta^{\infty}\left[\omega\left(x\right)+\tilde{f}\left(A_{\left[\chi\right]}^{\mu}\right)\right]\textrm{ }e^{iS_{Q.E.D.}}\right).\label{3.1.25}\end{equation}
We should simplify this by noting that $A^{\mu}$ is a variable of
integration, a dummy variable, we should simply rename $A_{\left[\chi\right]}^{\mu}\rightarrow A^{\mu}$
\begin{equation}
Z=\left(\int D\chi\right)\times\left(\int D\bar{\psi}D\psi DA^{\mu}\textrm{ }\left.J\left(f\right)\right|_{f=0}\textrm{ }\delta^{\infty}\left[\omega\left(x\right)+\tilde{f}\left(A^{\mu}\right)\right]\textrm{ }e^{iS_{Q.E.D.}}\right).\label{3.1.26}\end{equation}
\emph{This relation is true independent of what $\omega\left(x\right)$
is} (we didn't specify an $\omega\left(x\right)$)\emph{, so it will
also be true for a linear combination of different $\omega\left(x\right)$'s}
- with some proper normalization! We will choose a linear combination
of \emph{all} \emph{possible} $\omega\left(x\right)$'s,\begin{equation}
N\left(\zeta\right)\int D\omega\textrm{ exp }-i\int\textrm{d}^{4}x\textrm{ }\frac{\omega\left(x\right)^{2}}{2\zeta}=1\label{3.1.27}\end{equation}
where $N\left(\zeta\right)$ is the normalization we just mentioned.
Therefore in our linear combination of $Z$'s each one is weighted
by a factor $N\left(\zeta\right)\textrm{ exp }-i\int\textrm{d}^{4}x\textrm{ }\frac{\omega\left(x\right)^{2}}{2\zeta}$
and the normalization $N\left(\zeta\right)$ is making sure that when
we add them all up we get just $Z$ again. Hence,\begin{equation}
\begin{array}{rcl}
Z & = & N\left(\zeta\right)\times\left(\int D\chi\right)\times\\
 &  & \int D\omega\textrm{ exp }-i\int\textrm{d}^{4}x\textrm{ }\frac{\omega\left(x\right)^{2}}{2\zeta}\left(\int D\bar{\psi}D\psi DA^{\mu}\textrm{ }\left.J\left(f\right)\right|_{f=0}\textrm{ }\delta^{\infty}\left[\omega\left(x\right)+\tilde{f}\left(A^{\mu}\right)\right]\textrm{ }e^{iS_{Q.E.D.}}\right)\\
 & = & N\left(\zeta\right)\times\left(\int D\chi\right)\times\\
 &  & \left(\int D\omega D\bar{\psi}D\psi DA^{\mu}\textrm{ }\left[\textrm{exp }-i\int\textrm{d}^{4}x\textrm{ }\frac{\omega\left(x\right)^{2}}{2\zeta}\right]\textrm{ }\left.J\left(f\right)\right|_{f=0}\textrm{ }\delta^{\infty}\left[\omega\left(x\right)+\tilde{f}\left(A^{\mu}\right)\right]\textrm{ }e^{iS_{Q.E.D.}}\right)\\
 & = & N\left(\zeta\right)\times\left(\int D\chi\right)\times\\
 &  & \left(\int D\bar{\psi}D\psi DA^{\mu}\textrm{ }\left.J\left(f\right)\right|_{f=0}\textrm{ }e^{iS_{Q.E.D.}}\textrm{exp }-i\int\textrm{d}^{4}x\textrm{ }\frac{\tilde{f}\left(x\right)^{2}}{2\zeta}\right)\end{array}\label{3.1.28}\end{equation}
In the last step we have integrated over $\omega\left(x\right)$ which
has removed the delta functional and turned $\textrm{exp }-i\int\textrm{d}^{4}x\textrm{ }\frac{\omega\left(x\right)^{2}}{2\xi}$
into $\textrm{exp }-i\int\textrm{d}^{4}x\textrm{ }\frac{\tilde{f}\left(x\right)^{2}}{2\xi}$.
It is worth taking a breath here and making sure you are happy with
the above which may look like smoke and mirrors, it is not. Remember
what we said above about the Physics (Green's functions) being independent
of the normalization of the path integral, consequently we will disregard
the normalization, instead we will work with $\bar{Z}=\frac{Z}{N\left(\xi\right)\times\left(\int D\chi\right)}$. 

To simplify further we need to actually specify a form for $\tilde{f}\left(A^{\mu}\right)$.
The form that is commonly used is the so-called covariant gauge (it
is clearly Lorentz invariant), \begin{equation}
\tilde{f}\left(A^{\mu}\right)=\partial^{\mu}A_{\mu}.\label{3.1.29}\end{equation}
The gauge condition above is such that if $F^{\mu\nu}$ is something
then $\tilde{F}^{\mu\nu}$ is something else, where $\tilde{F}^{\mu\nu}$
is the same as $F^{\mu\nu}$ but with the replacement, $A^{\mu}\rightarrow A^{\mu}+\partial^{\mu}\phi$.
So what does this make the Faddeev-Popov determinant? 

\begin{equation}
\textrm{Det}_{x,y}\left(\frac{\delta f\left(x\right)}{\delta\chi\left(y\right)}\right)^{-1}=\textrm{ }\frac{1}{J\left(f,\chi\right)}\label{3.1.30}\end{equation}
\begin{equation}
\begin{array}{rcl}
\Rightarrow\left.J\left(f,\chi\right)\right|_{f=0} & = & \left.\textrm{Det}_{x,y}\left(\frac{\delta f\left(x\right)}{\delta\chi\left(y\right)}\right)\right|_{f=0}\\
 & = & \left.\textrm{Det}_{x,y}\left(\frac{\delta\left(\partial_{\mu}A^{\mu}\left(x\right)+\partial_{\mu}\partial^{\mu}\chi\left(x\right)\right)}{\delta\chi\left(y\right)}\right)\right|_{f=0}\\
 & = & \left.\textrm{Det}_{x,y}\left(\partial_{\mu}\partial^{\mu}\delta\left(x-y\right)\right)\right|_{f=0}.\end{array}\label{3.1.31}\end{equation}
In the Faddeev-Popov gauge fixing formalism the determinant is rewritten
as a functional integral over an exponential of \emph{ghost fields}
with the operator above sandwiched between them. 

A Grassmann variable / number is an anticommuting number. For a set
of Grassmann variables $\theta_{i}$ this means, $\left\{ \theta_{i},\theta_{j}\right\} =0$,
this trivially means that the square (and therefore also higher powers
of a Grassmann variable) equals zero as anti-commutation requires
$\theta_{i}\theta_{i}+\theta_{i}\theta_{i}=0$. Consequently Taylor
expansions of functions of Grassmann variables terminate after a few
terms. These peculiar {}``numbers'' also generate ambiguity in the
definition of integration over them. In fact the most natural definition
of integration for Grassmann variables turns out to be such that it
is the same as differentiation. We attempt to bolster these statements
in the last section (5.3), for a dedicated discussion of Grassmann
numbers we refer the reader to {[}17{]}, for now we will merely state
the definitions. Taylor expanding a function $f\left(\theta_{i}\right)$
of a Grassmann variable $\theta_{i}$ we have, \begin{equation}
f\left(\theta_{i}\right)=a+b\theta_{i}.\label{3.1.32}\end{equation}
Integrations over $a$ and $b\theta_{i}$ are defined as ($a$ and
$b$ are {}``normal'' \emph{bosonic} numbers), \begin{equation}
\begin{array}{lclclcl}
\int\textrm{d}\theta_{i}\textrm{ }a & = & 0 & \& & \frac{d}{\textrm{d}\theta_{i}}\left(a\right) & = & 0\\
\int\textrm{d}\theta_{i}\textrm{ }b\theta_{i} & = & b & \& & \frac{d}{\textrm{d}\theta_{i}}\left(b\theta_{i}\right) & = & b\end{array}.\label{3.1.33}\end{equation}
\begin{equation}
\begin{array}{clclcl}
\Rightarrow & \int\textrm{d}\theta_{i}\textrm{ }f\left(\theta_{i}\right) & = & b\\
 & also & ...\\
 & \int\textrm{d}\left(c\theta_{i}\right)\textrm{ }a & = & \frac{d}{d\left(c\theta_{i}\right)}\left(a\right) & = & 0\\
 & \int\textrm{d}\left(c\theta_{i}\right)\textrm{ }b\theta_{i} & = & \frac{d}{d\left(c\theta_{i}\right)}\left(\frac{1}{c}bc\theta_{i}\right) & = & \frac{1}{c}b\end{array}\label{3.1.34}\end{equation}
An important point to take away from the last two equations above
is that for Grassmann variables the integration measure changes in
the opposite way that it does for normal variables \emph{i.e.}\begin{equation}
\int\textrm{d}\left(c\theta_{i}\right)\textrm{ }f\left(\theta_{i}\right)=\frac{1}{c}\int\textrm{d}\theta_{i}\textrm{ }f\left(\theta_{i}\right).\label{3.1.35}\end{equation}
Also, for complex $\theta_{i}$ we simply treat the two components
(real and imaginary) as two independent Grassmann variables. Finally,
as you may well  variables and therefore also the integrations, anticommute.

Now consider the following integral,\begin{equation}
\prod_{z}\int\textrm{d}\omega_{z}d\eta_{z}\textrm{ exp }i\Sigma_{x}\Sigma_{y}\eta_{x}\Gamma_{xy}\omega_{y}\label{3.1.36}\end{equation}
 where \emph{$x$ and $y$ are discrete indices} of complex Grassmann
variables $\eta_{x}$ and $\omega_{y}$. $\eta$ is understood to
be the complex conjugate of $\omega$ more commonly denoted in the
literature by a bar over $\omega$: $\eta=\bar{\omega}$) . $\Gamma_{xy}$
is some Hermitian matrix of ({}``normal'' / {}``bosonic'') numbers.
We would like to evaluate this integral. We don't know what to do
with it as it stands assuming $\Gamma_{xy}$ is not diagonal. First
we want to transform the integrand and variables of integration such
that the $\Gamma$ in the exponent is diagonal. To do this we introduce
a unitary transformation matrix $U$ as follows,\begin{equation}
\Sigma_{x}\Sigma_{y}\textrm{ }\eta_{x}\Gamma_{xy}\omega_{y}=\Sigma_{x}\Sigma_{y}\Sigma_{a}\Sigma_{b}\Sigma_{c}\Sigma_{d}\textrm{ }\eta_{x}U_{xa}^{\dagger}U_{ab}\Gamma_{bc}U_{cd}^{\dagger}U_{dy}\omega_{y}.\label{3.1.37}\end{equation}
We choose $U$ such that $\Sigma_{b}\Sigma_{c}U_{ab}\Gamma_{bc}U_{cd}^{\dagger}=\tilde{\Gamma}_{ad}$
is diagonal \emph{i.e.} $\tilde{\Gamma}_{ad}=\tilde{\Gamma}_{aa}\delta_{ad}$!
Note that the we show the summation over $x$ and $y$ explicitly
with $\Sigma$, a repeated index does not imply a sum over it. Now
we redefine our fields according to the unitary transformation such
that \begin{equation}
\Sigma_{x}\Sigma_{y}\textrm{ }\eta_{x}\Gamma_{xy}\omega_{y}=\Sigma_{a}\Sigma_{d}\textrm{ }\tilde{\eta}_{a}\tilde{\Gamma}_{aa}\tilde{\omega}_{a}\delta_{ad}=\Sigma_{a}\textrm{ }\tilde{\eta}_{a}\tilde{\Gamma}_{aa}\tilde{\omega}_{a}\label{3.1.38}\end{equation}
 \emph{i.e.},s\begin{equation}
\begin{array}{rcl}
\tilde{\eta}_{a} & = & \Sigma_{x}\eta_{x}U_{xa}^{\dagger}\\
\tilde{\omega}_{d} & = & \Sigma_{y}U_{dy}\omega_{y}.\end{array}\label{3.1.39}\end{equation}
Rewriting the integral \ref{3.1.35} in terms of the transformed variables
we have that,\begin{equation}
\prod_{z}\int\textrm{d}\omega_{z}d\eta_{z}\textrm{ exp }i\Sigma_{x}\Sigma_{y}\eta_{x}\Gamma_{xy}\omega_{y}=\prod_{z}\int\textrm{d}\left(\Sigma_{p}U_{zq}^{\dagger}\tilde{\omega}_{p}\right)d\left(\Sigma_{q}\tilde{\eta}_{q}U_{pz}\right)\textrm{ exp }i\Sigma_{a}\tilde{\eta}_{a}\tilde{\Gamma}_{aa}\tilde{\omega}_{a}.\label{3.1.40}\end{equation}
We need to get the integration measure in a more friendly form. As
we said before, the Grassmann integration measure transforms as the
inverse of the Jacobian of the transformation instead of just the
Jacobian as is the case for regular numbers. For a slightly better
proof of this consider the following integral over complex Grassmann
variables,\begin{equation}
\int\textrm{d}\theta_{N}...\int\textrm{d}\theta_{2}\int\textrm{d}\theta_{1}\textrm{ }\theta_{1}\theta_{2}...\theta_{N}=1\label{3.1.41}\end{equation}
as $\int\textrm{d}\theta_{i}\textrm{ }\theta_{i}=1$. Now we make
a change of variables $\theta_{i}\rightarrow\Sigma_{j}U_{ij}\theta_{j}$.
The change of variables cannot affect the value of the integral. How
does the integrand change? As the $\theta_{i}$'s are Grassmann we
can write, \begin{equation}
\theta_{i_{1}}\theta_{i_{2}}...\theta_{i_{N}}=\frac{1}{N!}\Sigma_{j_{1}}\Sigma_{j_{2}}...\Sigma_{j_{N}}\epsilon^{j_{1}j_{2}...j_{N}}\theta_{j_{1}}\theta_{j_{2}}...\theta_{j_{N}}\label{3.1.42}\end{equation}
\emph{i.e.} we sum over all $N!$ permutations of the $N$ variables
remembering to divide by $N!$ at the end and keeping track of the
minus signs with $\epsilon^{j_{1}j_{2}...j_{N}}$. Therefore we can
write the transformation of the integrand as,\begin{equation}
\theta_{1}\theta_{2}...\theta_{N}\rightarrow\tilde{\theta}_{1}\tilde{\theta}_{2}...\tilde{\theta}_{N}=\frac{1}{N!}\Sigma_{i_{1}}\Sigma_{i_{2}}...\Sigma_{i_{N}}\Sigma_{j_{1}}\Sigma_{j_{2}}...\Sigma_{j_{N}}\epsilon^{j_{1}j_{2}...j_{N}}\theta_{i_{1}}\theta_{i_{2}}...\theta_{i_{N}}\textrm{ }U_{i_{1}j_{1}}U_{i_{2}j_{2}}...U_{i_{N}j_{N}}\textrm{ }.\label{3.1.43}\end{equation}
We can use the anticommuting property again to simplify further and
write that for a given combination $\theta_{j_{1}}\theta_{j_{2}}...\theta_{j_{N}}$,\begin{equation}
\theta_{i_{1}}\theta_{i_{2}}...\theta_{i_{N}}=\epsilon^{i_{1}i_{2}...i_{N}}\theta_{1}\theta_{2}...\theta_{N}\label{3.1.44}\end{equation}
note that here we have no implied summation, there is no sum taking
place above and there are no indices on the thetas on the right!\begin{equation}
\begin{array}{rcl}
\theta_{1}\theta_{2}...\theta_{N} & \rightarrow & \tilde{\theta}_{1}\tilde{\theta}_{2}...\tilde{\theta}_{N}\\
 & = & \frac{1}{N!}\Sigma_{i_{1}}\Sigma_{i_{2}}...\Sigma_{i_{N}}\Sigma_{j_{1}}\Sigma_{j_{2}}...\Sigma_{j_{N}}\epsilon^{j_{1}j_{2}...j_{N}}\epsilon^{i_{1}i_{2}...i_{N}}\theta_{1}\theta_{2}...\theta_{N}\textrm{ }U_{i_{1}j_{1}}U_{i_{2}j_{2}}...U_{i_{N}j_{N}}\\
 & = & \left(\frac{1}{N!}\Sigma_{i_{1}}\Sigma_{i_{2}}...\Sigma_{i_{N}}\Sigma_{j_{1}}\Sigma_{j_{2}}...\Sigma_{j_{N}}\epsilon^{j_{1}j_{2}...j_{N}}\epsilon^{i_{1}i_{2}...i_{N}}\textrm{ }U_{i_{1}j_{1}}U_{i_{2}j_{2}}...U_{i_{N}j_{N}}\right)\theta_{1}\theta_{2}...\theta_{N}\end{array}\label{3.1.45}\end{equation}
Now the (definition of the) determinant of an $N\times N$ matrix
\textbf{$M$} is,\begin{equation}
\textrm{Det}\left(M_{ij}\right)=\frac{1}{N!}\Sigma_{\alpha_{1}}...\Sigma_{\alpha_{N}}\Sigma_{\beta_{1}}...\Sigma_{\beta_{N}}\epsilon^{\alpha_{1}...\alpha_{N}}\epsilon^{\beta_{1}...\beta_{N}}M_{\alpha_{1}\beta_{1}}...M_{\alpha_{N}\beta_{N}}\label{3.1.46}\end{equation}
 (feel free to check this) so,\begin{equation}
\theta_{1}\theta_{2}...\theta_{N}\rightarrow\tilde{\theta}_{1}\tilde{\theta}_{2}...\tilde{\theta}_{N}=\left(\textrm{Det}\left(U\right)\right)\theta_{1}\theta_{2}...\theta_{N}.\label{3.1.47}\end{equation}
Consequently if we need, \begin{equation}
\int\textrm{d}\theta_{N}...\int\textrm{d}\theta_{2}\int\textrm{d}\theta_{1}\textrm{ }\theta_{1}\theta_{2}...\theta_{N}=\int\textrm{d}\tilde{\theta}_{N}...\int\textrm{d}\tilde{\theta}_{2}\int\textrm{d}\tilde{\theta}_{1}\textrm{ }\tilde{\theta}_{1}\tilde{\theta}_{2}...\tilde{\theta}_{N}=1\label{3.1.48}\end{equation}
 then we must have that, \begin{equation}
\int\textrm{d}\tilde{\theta}_{N}...\int\textrm{d}\tilde{\theta}_{2}\int\textrm{d}\tilde{\theta}_{1}=\frac{1}{\textrm{Det}\left(U\right)}\int\textrm{d}\theta_{N}...\int\textrm{d}\theta_{2}\int\textrm{d}\theta_{1}.\label{3.1.49}\end{equation}
So invariance of the value of a Grassmann integral under a simple
change of variable means that the Grassmann integration measure transforms
as the \emph{inverse of the Jacobian} of the transformation as opposed
to just the Jacobian for regular numbers. 

In our case this means that, \begin{equation}
\prod_{x}\int\textrm{d}\omega_{x}\prod_{y}\int\textrm{d}\eta_{y}=\frac{1}{\textrm{Det}\left(U^{\dagger}\right)\textrm{Det}\left(U\right)}\prod_{x}\int\textrm{d}\tilde{\omega}_{x}\prod_{y}\int\textrm{d}\tilde{\eta}_{y}=\prod_{x}\int\textrm{d}\tilde{\omega}_{x}\prod_{y}\int\textrm{d}\tilde{\eta}_{y}\label{3.1.50}\end{equation}
as $U^{\dagger}U=I$. We can manipulate this equation above by moving
the $\eta$'s $\omega$'s and their corresponding $\tilde{\eta}$'s
and $\tilde{\omega}$'s in the same way on both sides of the equation
without worrying about minus signs. This is because every time we
reorder things on the left side we may generate a sign but the same
reordering on the right will naturally generate the same sign there
so we can forget about signs and say,\begin{equation}
\prod_{z}\int\textrm{d}\omega_{z}d\eta_{z}=\frac{1}{\textrm{Det}\left(U^{\dagger}\right)\textrm{Det}\left(U\right)}\prod_{z}\int\textrm{d}\tilde{\omega}_{z}d\tilde{\eta}_{z}=\prod_{z}\int\textrm{d}\tilde{\omega}_{z}d\tilde{\eta}_{z}\label{3.1.51}\end{equation}
The result is that the entire integral is invariant under the transformation
which diagonalizes $\Gamma_{xy}$,\begin{equation}
\prod_{z}\int\textrm{d}\omega_{z}d\eta_{z}\textrm{ exp }i\Sigma_{x}\Sigma_{y}\eta_{x}\Gamma_{xy}\omega_{y}=\prod_{z}\int\textrm{d}\tilde{\omega}_{z}d\tilde{\eta}_{z}\textrm{ exp }i\Sigma_{a}\tilde{\eta}_{a}\tilde{\Gamma}_{aa}\tilde{\omega}_{a}\label{3.1.52}\end{equation}
 Taylor expanding the integrand on the right gives,\begin{equation}
\prod_{z}\int\textrm{d}\omega_{z}d\eta_{z}\textrm{ exp }i\Sigma_{x}\Sigma_{y}\eta_{x}\Gamma_{xy}\omega_{y}=\prod_{z}\int\textrm{d}\tilde{\omega}_{z}d\tilde{\eta}_{z}\textrm{ }\left(1+\Sigma_{a}\tilde{\eta}_{a}\tilde{\Gamma}_{aa}\tilde{\omega}_{a}+\frac{1}{2!}\left(\Sigma_{a}\tilde{\eta}_{a}\tilde{\Gamma}_{aa}\tilde{\omega}_{a}\right)\left(\Sigma_{b}\tilde{\eta}_{b}\tilde{\Gamma}_{bb}\tilde{\omega}_{b}\right)+...\right)\label{3.1.53}\end{equation}
For a term in the  expansion to survive the integration it must have
one copy of each of $\tilde{\eta}_{x}$ and $\tilde{\omega}_{y}$
for all elements in the products of integrals. Take for example $\prod_{z}\int\textrm{d}\tilde{\omega}_{z}d\tilde{\eta}_{z}=\int\textrm{d}\tilde{\omega}_{z_{1}}d\tilde{\eta}_{z_{1}}\int\textrm{d}\tilde{\omega}_{z_{2}}d\tilde{\eta}_{z_{2}}$
\emph{i.e.} there are only two values of each index. The expansion
terminates after the second term (or more generally for $N$ values
of the indices, the $N$th term),\begin{equation}
\textrm{exp }i\Sigma_{a}\tilde{\eta}_{a}\tilde{\Gamma}_{aa}\tilde{\omega}_{a}=1+\Sigma_{a}\tilde{\eta}_{a}\tilde{\Gamma}_{aa}\tilde{\omega}_{a}+\frac{1}{2!}\left(\Sigma_{a}\tilde{\eta}_{a}\tilde{\Gamma}_{aa}\tilde{\omega}_{a}\right)\left(\Sigma_{b}\tilde{\eta}_{b}\tilde{\Gamma}_{bb}\tilde{\omega}_{b}\right)\label{3.1.54}\end{equation}
 as higher terms will have to involve the square of at least one of
the Grassmann variables. The first term $\left(1\right)$ vanishes
as $\int\textrm{d}\theta\textrm{ }1=0$ for any Grassmann variable
by definition and we are integrating over four different Grassmann
variables. The second term also vanishes for basically the same reason.
This term is simply, \begin{equation}
\tilde{\eta}_{x_{1}}\tilde{\Gamma}_{x_{1}x_{1}}\tilde{\omega}_{x_{1}}+\tilde{\eta}_{x_{2}}\tilde{\Gamma}_{x_{2}x_{2}}\tilde{\omega}_{x_{2}}\label{3.1.55}\end{equation}
 but the integral involves integrating each of these over four different
Grassmann variables so they both vanish. Only the last term makes
a contribution to the integral. It contains terms with one $\eta$
and $\omega$ for each $\eta$ and $\omega$ integration. The last
term is \begin{equation}
\tilde{\eta}_{x_{1}}\tilde{\Gamma}_{x_{1}x_{1}}\tilde{\omega}_{x_{1}}\tilde{\eta}_{x_{2}}\tilde{\Gamma}_{x_{2}x_{2}}\tilde{\omega}_{x_{2}}+\tilde{\eta}_{x_{2}}\tilde{\Gamma}_{x_{2}x_{2}}\tilde{\omega}_{x_{2}}\tilde{\eta}T_{x_{1}}\tilde{\Gamma}_{x_{1}x_{1}}\tilde{\omega}_{x_{1}}=\left(\tilde{\eta}_{x_{1}}\tilde{\omega}_{x_{1}}\tilde{\eta}_{x_{2}}\tilde{\omega}_{x_{2}}+\tilde{\eta}_{x_{2}}\tilde{\omega}_{x_{2}}\tilde{\eta}_{x_{1}}\tilde{\omega}_{x_{1}}\right)\tilde{\Gamma}_{x_{2}x_{2}}\tilde{\Gamma}_{x_{1}x_{1}}.\label{3.1.56}\end{equation}
where we have $\tilde{\eta}_{x_{1}}\tilde{\Gamma}_{x_{1}x_{1}}\tilde{\omega}_{x_{1}}\tilde{\eta}_{x_{1}}\tilde{\Gamma}_{x_{1}x_{1}}\tilde{\omega}_{x_{1}}=0$
as they contain squares of Grassmann variables \emph{i.e.} the last
term is a sum of the cross terms between the products of sums. If
we had $N$ values of the index it would be the $N$th term in the
expansion that would survive, for the same reasons. We are free to
split these terms into pairs of adjacent Grassmann variables and freely
move these pairs around without worrying about picking up minus signs,
such pairs commute with each other, \begin{equation}
\left[\theta_{1}\theta_{2},\theta_{3}\theta_{4}\right]=\theta_{1}\theta_{2}\theta_{3}\theta_{4}-\theta_{3}\theta_{4}\theta_{1}\theta_{2}=\theta_{1}\theta_{2}\theta_{3}\theta_{4}-\theta_{1}\theta_{2}\theta_{3}\theta_{4}=0\label{3.1.57}\end{equation}
 which basically means all of the aforementioned cross terms in the
integrand are the same. How many such cross terms are there? Easy,
we need a different $\eta\Gamma\omega$ from each sum, the $N$th
term is a product of $N$ such sums and each sum has $N$ different
$\eta\Gamma\omega$'s in it. So I can get the non-vanishing term $N!$
times corresponding to the number of permutations of the $N$ pairs.
\begin{equation}
\begin{array}{rcl}
\prod_{z}\int\textrm{d}\omega_{z}d\eta_{z}\textrm{ exp }i\Sigma_{x}\Sigma_{y}\eta_{x}\Gamma_{xy}\omega_{y} & = & \prod_{z}\int\textrm{d}\tilde{\omega}_{z}d\tilde{\eta}_{z}\textrm{ }\frac{1}{N!}\left(\Sigma_{\alpha}\tilde{\eta}_{a}\tilde{\Gamma}_{aa}\tilde{\omega}_{a}\right)...\left(\Sigma_{\beta}\tilde{\eta}_{\beta}\tilde{\Gamma}_{\beta\beta}\tilde{\omega}_{\beta}\right)\\
 & = & \prod_{z}\int\textrm{d}\tilde{\omega}_{z}d\tilde{\eta}_{z}\textrm{ }\frac{N!}{N!}\tilde{\eta}_{z_{1}}\tilde{\Gamma}_{z_{1}z_{1}}\tilde{\omega}_{z_{1}}...\tilde{\eta}_{z_{N}}\tilde{\Gamma}_{z_{N}z_{N}}\tilde{\omega}_{z_{N}}\\
 & = & \left(\prod_{z}\tilde{\Gamma}_{zz}\right)\left(\prod_{z}\int\textrm{d}\tilde{\omega}_{z}d\tilde{\eta}_{z}\textrm{ }\tilde{\eta}_{z_{N}}\tilde{\omega}_{z_{N}}...\tilde{\eta}_{z_{1}}\tilde{\omega}_{z_{1}}\right)\end{array}\label{3.1.58}\end{equation}
As $\tilde{\Gamma}$ is $\Gamma$ diagonalized this means that the
first product above is the product of the eigenvalues of $\Gamma$
which is the determinant of $\Gamma$:\begin{equation}
\begin{array}{rcl}
\textrm{Det}\left(\tilde{\Gamma}_{zz}\right) & = & \prod_{z}\tilde{\Gamma}_{zz}\\
 & = & \textrm{Det}\left(U\Gamma U^{\dagger}\right)\\
 & = & \textrm{Det}\left(U\right)\textrm{Det}\left(\Gamma\right)\textrm{Det}\left(U^{\dagger}\right)\\
 & = & \textrm{Det}\left(U^{\dagger}U\right)\textrm{Det}\left(\Gamma\right)\end{array}\label{3.1.59}\end{equation}
 as $\textrm{Det}\left(U^{\dagger}U\right)=\textrm{Det}\left(I\right)=1$.\begin{equation}
\begin{array}{rcl}
\prod_{z}\int\textrm{d}\omega_{z}d\eta_{z}\textrm{ exp }i\Sigma_{x}\Sigma_{y}\eta_{x}\Gamma_{xy}\omega_{y} & = & \left(\textrm{Det}\left(\Gamma\right)\right)\left(\prod_{z}\int\textrm{d}\tilde{\omega}_{z}d\tilde{\eta}_{z}\textrm{ }\tilde{\eta}_{z_{N}}\tilde{\omega}_{z_{N}}...\tilde{\eta}_{z_{1}}\tilde{\omega}_{z_{1}}\right)\\
 & = & \left(\textrm{Det}\left(\Gamma\right)\right)\left(\int\textrm{d}\tilde{\omega}_{z_{1}}d\tilde{\eta}_{z_{1}}...\int\textrm{d}\tilde{\omega}_{z_{N}}d\tilde{\eta}_{z_{N}}\textrm{ }\tilde{\eta}_{z_{N}}\tilde{\omega}_{z_{N}}...\tilde{\eta}_{z_{1}}\tilde{\omega}_{z_{1}}\right)\\
 & = & \left(\textrm{Det}\left(\Gamma\right)\right)\times1\end{array}\label{3.1.60}\end{equation}
Generalizing to the continuum limit we have,\begin{equation}
\int D\omega D\eta\textrm{ exp }i\int\textrm{d}^{4}x\int\textrm{d}^{4}y\textrm{ }\eta\left(x\right)\Gamma\left(x,y\right)\omega\left(y\right)=\textrm{Det}_{x,y}\left(\Gamma\left(x,y\right)\right)\label{3.1.61}\end{equation}
and finally we have a way of including the Faddeev-Popov determinant
in our path integral in terms of {}``Ghosts'' $\eta$ and $\omega$.
In the case of our covariant gauge we have, \begin{equation}
\begin{array}{rcl}
\textrm{Det}_{x,y}\left(\partial_{\mu}\partial^{\mu}\delta\left(x-y\right)\right) & = & \int D\omega D\eta\textrm{ exp }i\int\textrm{d}^{4}x\int\textrm{d}^{4}y\textrm{ }\eta\left(x\right)\partial_{\mu}\partial^{\mu}\delta\left(x-y\right)\omega\left(y\right)\\
 & = & \int D\omega D\eta\textrm{ exp }i\int\textrm{d}^{4}x\textrm{ }\eta\left(x\right)\partial_{\mu}\partial^{\mu}\omega\left(x\right)\end{array}\label{3.1.62}\end{equation}
Finally the gauge fixed path integral is,\begin{equation}
Z={\bkii{0}{0}}=\int DA^{\mu}D\bar{\psi}D\psi D\omega D\eta\textrm{ exp }i\int\textrm{d}^{4}x\textrm{ }\left\{ -\frac{1}{4}F_{\mu\nu}F^{\mu\nu}+\frac{1}{2\zeta}\left(\partial_{\mu}A^{\mu}\right)^{2}+\eta\partial^{\mu}\partial_{\mu}\omega+\bar{\psi}\left(i\not D-m\right)\psi\right\} \label{3.1.63}\end{equation}
with a kinetic term $\eta\partial^{\mu}\partial_{\mu}\omega$ appearing
in the original QED action due to gauge fixing. The ghosts are not
coupled to any of the physical fields and so they don't appear in
perturbation theory. In non-Abelian gauge theories fixing the gauge
results in ghosts as in QED but the resulting action will have a coupling
of the ghosts to the physical fields. This coupling results in the
ghosts appearing in Feynman diagrams, in perturbation theory. The
ghosts are so-called because they are unphysical for a number of reasons,
the most obvious one being that they are Grassmann fields yet they
are also (complex) scalar fields (pseudoscalar particles) they have
no spin! This unphysical spin-statistics relation means that ghosts
cannot appear as external particles / external legs on Feynman diagrams,
the perturbation expansion is such that (in non-Abelian theories)
the ghosts only appear in closed loops. 

Finally, as ghosts are an artifact of gauge fixing they manifest themselves
in the Lagrangian and hence in perturbation theory in different ways
according to the choice of gauge, in certain gauges the ghosts may
not even materialize in perturbation theory. However gauge fixing
affects the photon propagator (see next section) and in most instances
it turns out that calculations in perturbation theory are made much
easier by choosing a gauge which simplifies the propagator of some
physical particle at the price of including ghosts.

\newpage
\section{Feynman Rules in QED.}

We now have a fully gauge fixed path integral for QED ghosts \emph{etc}...(put
in fermion fields now). 

\begin{equation}
Z={\bkii{0}{0}}=\int DA^{\mu}D\bar{\psi}D\psi D\eta D\omega\textrm{ exp }i\int\textrm{d}^{4}x\textrm{ }\left\{ -\frac{1}{4}F_{\mu\nu}F^{\mu\nu}+\frac{1}{2\zeta}\left(\partial_{\mu}A^{\mu}\right)^{2}+\eta\partial^{\mu}\partial_{\mu}\omega+\bar{\psi}\left(i\not D-m\right)\psi\right\} .\label{3.2.1}\end{equation}
 Firstly we will try and derive the photon propagator, we therefore
want to calculate the following (we are normalizing everything by
$\frac{1}{Z}$),\begin{equation}
\begin{array}{rcl}
\frac{{\bkiii{0}{A^{\mu}\left(x_{1}\right)A^{\nu}\left(x_{2}\right)}{0}}}{{\bkii{0}{0}}} & = & \frac{\int DA^{\mu}...D\omega\textrm{ }A^{\mu}\left(x_{1}\right)A^{\nu}\left(x_{2}\right)\textrm{ exp }iS_{Q.E.D}}{Z}\\
 & = & \left.\frac{\delta^{2}}{\delta J_{\mu}\left(x_{1}\right)\delta J_{\nu}\left(x_{2}\right)}\ln\int DA^{\mu}...D\omega\textrm{ exp }i\left(S_{QED}-i\int\textrm{d}^{4}x\textrm{ }J_{\kappa}\left(x\right)A^{\kappa}\left(x\right)\right)\right|_{J_{\kappa}=0}\\
 & = & \left.\frac{\delta^{2}\ln Z\left[J_{\kappa}\right]}{\delta J_{\mu}\left(x_{1}\right)\delta J_{\nu}\left(x_{2}\right)}\right|_{J_{\kappa}=0}\end{array}.\label{3.2.2}\end{equation}
We have defined $Z\left[J^{\kappa}\right]=\int DA^{\mu}...D\omega\textrm{ exp }i\left(S_{QED}-i\int\textrm{d}^{4}x\textrm{ }J_{\kappa}\left(x\right)A^{\kappa}\left(x\right)\right)$
with $J^{\kappa}\left(x\right)$ a \emph{source} \emph{term}. The
source term allows us to rewrite our definition of the two point function
as above, it is of no real physical significance. Though $J^{\kappa}\left(x\right)$
is introduced ad hoc into the Lagrangian it is not affecting the Physics
as can be seen by the equivalence of everything in the working above.
The next thing we do is make a perturbative expansion in $e$. The
interaction part of the generating functional is expanded in powers
of the coupling constant. \begin{equation}
\textrm{exp }\int\textrm{d}^{4}x\textrm{ }e\bar{\psi}\not A\psi\textrm{ }=1+e\int\textrm{d}^{4}x\textrm{ }\bar{\psi}\not A\psi-{\cal {O}}\left(e^{2}\right)\label{3.2.3}\end{equation}
It is also necessary to do some rewriting of the free part of the
Lagrangian. Also up until now we have not had to worry about the (space-time)
variable that the fields depend on, we have safely been able to assume
everything depends on $x$ but things get a bit more complicated here.
Consider the bit for the photon (the same procedure works separately
for the fermions), \begin{equation}
\int\textrm{d}^{4}x\textrm{ }-\frac{1}{4}\left(\partial_{\mu}A\left(x\right)_{\nu}-\partial_{\nu}A\left(x\right)_{\mu}\right)\left(\partial^{\mu}A^{\nu}\left(x\right)-\partial^{\nu}A^{\mu}\left(x\right)\right)-\frac{1}{2\zeta}\left(\partial_{\mu}A^{\mu}\left(x\right)\right)^{2}-J_{\kappa}\left(x\right)A\left(x\right)^{\kappa}.\label{3.2.4}\end{equation}
 For reasons that will become apparent the next thing we want to do
is to put this in a form $\int\textrm{d}^{4}x\textrm{ }A_{\mu}Q^{\mu\nu}A_{\nu}$.
Forgetting about the source term we have\begin{equation}
\int\textrm{d}^{4}x\textrm{ }-\frac{1}{4}\left(2\partial_{\mu}A_{\nu}\left(x\right)\partial^{\mu}A^{\nu}\left(x\right)-2\partial_{\nu}A_{\mu}\left(x\right)\partial^{\mu}A^{\nu}\left(x\right)+\frac{2}{\zeta}\partial_{\mu}A^{\mu}\left(x\right)\partial_{\nu}A^{\nu}\left(x\right)\right).\label{3.2.5}\end{equation}
Integrating by parts gives\begin{equation}
\begin{array}{rl}
= & \int\textrm{d}^{4}x\textrm{ }-\frac{1}{2}\left(-A_{\nu}\left(x\right)\partial_{\mu}\partial^{\mu}A^{\nu}\left(x\right)+A_{\mu}\left(x\right)\partial^{\nu}\partial^{\mu}A_{\nu}\left(x\right)-\frac{1}{\zeta}A^{\mu}\left(x\right)\partial_{\mu}\partial_{\nu}A^{\nu}\left(x\right)\right)\\
= & \int\textrm{d}^{4}x\textrm{ }-\frac{1}{2}\left(A_{\lambda}\left(x\right)\left(-g^{\lambda\rho}\partial^{\mu}\partial_{\mu}+\left(1-\zeta^{-1}\right)\partial^{\lambda}\partial^{\rho}\right)A_{\rho}\left(x\right)\right)\end{array}.\label{3.2.6}\end{equation}
 We now need to sort out the source term which we so conveniently
forgot about, this is done by shifting our field $A^{\mu}\left(x\right)$
everywhere such that \[
A^{\mu}\left(x\right)\rightarrow A'^{\mu}\left(x\right)=A^{\mu}\left(x\right)+\int\textrm{d}^{4}y\textrm{ }\Delta^{\mu\kappa}\left(x-y\right)J_{\kappa}\left(y\right)\]
 where $Q^{\mu\nu}\Delta_{\nu\kappa}\left(x\right)=i\delta_{\kappa}^{\mu}\delta^{4}\left(x\right)$
\emph{i.e.} $Q^{\mu\nu}\int\textrm{d}^{4}y\textrm{ }\Delta_{\nu\kappa}\left(x-y\right)J^{\kappa}\left(y\right)=J^{\mu}\left(x\right)$,
this gives,\begin{equation}
\begin{array}{rcl}
-\frac{1}{2}A_{\mu}Q^{\mu\nu}A_{\nu}-J_{\kappa}A^{\kappa} & \rightarrow & -\frac{1}{2}\left(A'_{\mu}\left(x\right)-\int\textrm{d}^{4}y\textrm{ }\Delta_{\mu\kappa}\left(x-y\right)J^{\kappa}\left(y\right)\right)Q^{\mu\nu}\left(A'_{\nu}\left(x\right)-\int\textrm{d}^{4}z\textrm{ }\Delta_{\nu\lambda}\left(x-z\right)J^{\lambda}\left(z\right)\right)\\
 &  & -J_{\kappa}\left(x\right)A'^{\kappa}\left(x\right)+J_{\kappa}\left(x\right)\int\textrm{d}^{4}y\textrm{ }\Delta^{\kappa\nu}\left(x-y\right)J_{\nu}\left(y\right)\\
 & = & -\frac{1}{2}A'_{\mu}\left(x\right)Q^{\mu\nu}A'_{\nu}\left(x\right)+\frac{1}{2}A'_{\mu}\left(x\right)Q^{\mu\nu}\int\textrm{d}^{4}z\textrm{ }\Delta_{\nu\lambda}\left(x-z\right)J^{\lambda}\left(z\right)\\
 &  & +\frac{1}{2}\left(\int\textrm{d}^{4}y\textrm{ }\Delta_{\mu\kappa}\left(x-y\right)J^{\kappa}\left(y\right)\right)Q^{\mu\nu}A'_{\nu}\left(x\right)\\
 &  & -\frac{1}{2}\left(\int\textrm{d}^{4}y\textrm{ }\Delta_{\mu\kappa}\left(x-y\right)J^{\kappa}\left(y\right)\right)Q^{\mu\nu}\left(\int\textrm{d}^{4}y\textrm{ }\Delta_{\nu\gamma}\left(x-z\right)J^{\gamma}\left(z\right)\right)-J_{\kappa}\left(x\right)A'^{\mu}\left(x\right)\\
 &  & +J_{\kappa}\left(x\right)\int\textrm{d}^{4}y\textrm{ }\Delta^{\kappa\nu}\left(x-y\right)J_{\nu}\left(y\right)\\
 & = & -\frac{1}{2}A'_{\mu}\left(x\right)Q^{\mu\nu}A'_{\nu}\left(x\right)+\frac{1}{2}A'_{\mu}\left(x\right)J^{\mu}\left(x\right)+\frac{1}{2}\left(\int\textrm{d}^{4}y\textrm{ }\Delta_{\mu\kappa}\left(x-y\right)J^{\kappa}\left(y\right)\right)Q^{\mu\nu}A'_{\nu}\left(x\right)\\
 &  & -\frac{1}{2}\left(\int\textrm{d}^{4}y\textrm{ }\Delta_{\mu\kappa}\left(x-y\right)J^{\kappa}\left(y\right)\right)J^{\mu}\left(x\right)-J_{\mu}\left(x\right)A'^{\mu}\left(x\right)\\
 &  & +J_{\kappa}\left(x\right)\int\textrm{d}^{4}y\textrm{ }\Delta^{\kappa\nu}\left(x-y\right)J_{\nu}\left(y\right)\\
 & = & -\frac{1}{2}A'_{\mu}\left(x\right)Q^{\mu\nu}A'_{\nu}\left(x\right)-\frac{1}{2}A'_{\mu}\left(x\right)J^{\mu}\left(x\right)+\frac{1}{2}\left(\int\textrm{d}^{4}y\textrm{ }\Delta_{\mu\kappa}\left(x-y\right)J^{\kappa}\left(y\right)\right)Q^{\mu\nu}A'_{\nu}\left(x\right)\\
 &  & +\frac{1}{2}J_{\kappa}\left(x\right)\int\textrm{d}^{4}y\textrm{ }\Delta^{\kappa\nu}\left(x-y\right)J_{\nu}\left(y\right)\end{array}\label{3.2.7}\end{equation}
A long winded and tedious calculation%
\footnote{\[
\begin{array}{l}
\frac{1}{2}\left(\int\textrm{d}^{4}y\textrm{ }\Delta_{\mu\kappa}\left(x-y\right)J^{\kappa}\left(y\right)\right)Q^{\mu\nu}A'_{\nu}\left(x\right)\\
=\frac{1}{2}\left(\int\textrm{d}^{4}y\textrm{ }\Delta_{\mu\delta}\left(x-y\right)J^{\delta}\left(y\right)\right)\left(-g^{\mu\nu}\partial^{\kappa}\partial_{\kappa}+\left(1-\zeta^{-1}\right)\partial^{\mu}\partial^{\nu}\right)A'_{\nu}\left(x\right)\\
=\frac{1}{2}\partial^{\kappa}\left(\left(\int\textrm{d}^{4}y\textrm{ }\Delta_{\mu\delta}\left(x-y\right)J^{\delta}\left(y\right)\right)\left(-g^{\mu\nu}\partial_{\kappa}A'_{\nu}\left(x\right)\right)\right)-\left(\frac{1}{2}\partial^{\kappa}\int\textrm{d}^{4}y\textrm{ }\Delta_{\mu\delta}\left(x-y\right)J^{\delta}\left(y\right)\right)\left(-g^{\mu\nu}\partial_{\kappa}A'_{\nu}\left(x\right)\right)\\
+\frac{1}{2}\partial^{\mu}\left(\left(\int\textrm{d}^{4}y\textrm{ }\Delta_{\mu\delta}\left(x-y\right)J^{\delta}\left(y\right)\right)\left(\left(1-\zeta^{-1}\right)\partial^{\nu}A'_{\nu}\left(x\right)\right)\right)-\frac{1}{2}\left(\left(\partial^{\mu}\int\textrm{d}^{4}y\textrm{ }\Delta_{\mu\delta}\left(x-y\right)J^{\delta}\left(y\right)\right)\left(\left(1-\zeta^{-1}\right)\partial^{\nu}A'_{\nu}\left(x\right)\right)\right)\\
=\frac{1}{2}\partial_{\kappa}\partial^{\kappa}\left(\left(\int\textrm{d}^{4}y\textrm{ }\Delta_{\mu\delta}\left(x-y\right)J^{\delta}\left(y\right)\right)\left(-g^{\mu\nu}A'_{\nu}\left(x\right)\right)\right)-\frac{1}{2}\partial^{\kappa}\left(\left(\partial_{\kappa}\int\textrm{d}^{4}y\textrm{ }\Delta_{\mu\delta}\left(x-y\right)J^{\delta}\left(y\right)\right)\left(-g^{\mu\nu}A'_{\nu}\left(x\right)\right)\right)\\
-\partial_{\kappa}\left(\left(\frac{1}{2}\partial^{\kappa}\int\textrm{d}^{4}y\textrm{ }\Delta_{\mu\delta}\left(x-y\right)J^{\delta}\left(y\right)\right)\left(-g^{\mu\nu}A'_{\nu}\left(x\right)\right)\right)+\left(\frac{1}{2}\partial_{\kappa}\partial^{\kappa}\int\textrm{d}^{4}y\textrm{ }\Delta_{\mu\delta}\left(x-y\right)J^{\delta}\left(y\right)\right)\left(-g^{\mu\nu}A'_{\nu}\left(x\right)\right)\\
+\frac{1}{2}\partial^{\mu}\partial^{\nu}\left(\left(\int\textrm{d}^{4}y\textrm{ }\Delta_{\mu\delta}\left(x-y\right)J^{\delta}\left(y\right)\right)\left(\left(1-\zeta^{-1}\right)A'_{\nu}\left(x\right)\right)\right)-\frac{1}{2}\partial^{\mu}\left(\left(\partial^{\nu}\int\textrm{d}^{4}y\textrm{ }\Delta_{\mu\delta}\left(x-y\right)J^{\delta}\left(y\right)\right)\left(\left(1-\zeta^{-1}\right)A'_{\nu}\left(x\right)\right)\right)\\
-\frac{1}{2}\partial^{\nu}\left(\left(\partial^{\mu}\int\textrm{d}^{4}y\textrm{ }\Delta_{\mu\delta}\left(x-y\right)J^{\delta}\left(y\right)\right)\left(\left(1-\zeta^{-1}\right)A'_{\nu}\left(x\right)\right)\right)+\frac{1}{2}\left(\left(\partial^{\mu}\partial^{\nu}\int\textrm{d}^{4}y\textrm{ }\Delta_{\mu\delta}\left(x-y\right)J^{\delta}\left(y\right)\right)\left(\left(1-\zeta^{-1}\right)A'_{\nu}\left(x\right)\right)\right)\end{array}\]
Assuming that the field $A^{\mu}\left(x\right)$ and its first derivatives
vanish as $x\rightarrow\infty$ we can remove six of the above terms
on the grounds that they can be rewritten as surface integrals. \[
\begin{array}{rl}
= & \frac{1}{2}\left(-g^{\mu\nu}\partial_{\kappa}\partial^{\kappa}\int\textrm{d}^{4}y\textrm{ }\Delta_{\mu\delta}\left(x-y\right)J^{\delta}\left(y\right)\right)A'_{\nu}\left(x\right)+\frac{1}{2}\left(\left(1-\zeta^{-1}\right)\partial^{\mu}\partial^{\nu}\int\textrm{d}^{4}y\textrm{ }\Delta_{\mu\delta}\left(x-y\right)J^{\delta}\left(y\right)\right)A'_{\nu}\left(x\right)\\
= & \frac{1}{2}\left(Q^{\mu\nu}\int\textrm{d}^{4}y\textrm{ }\Delta_{\mu\delta}\left(x-y\right)J^{\delta}\left(y\right)\right)A'_{\nu}\left(x\right)\end{array}\]
} involving the use of the product rule and setting surface terms to
zero means that the 3rd term can be rewritten,\begin{equation}
\frac{1}{2}\left(\int\textrm{d}^{4}y\textrm{ }\Delta_{\mu\kappa}\left(x-y\right)J^{\kappa}\left(y\right)\right)Q^{\mu\nu}A'_{\nu}\left(x\right)=\frac{1}{2}\left(Q^{\mu\nu}\int\textrm{d}^{4}y\textrm{ }\Delta_{\mu\delta}\left(x-y\right)J^{\delta}\left(y\right)\right)A'_{\nu}\left(x\right)\label{3.2.8}\end{equation}
By definition of the function $\Delta_{\mu\delta}\left(x-y\right)$
from above, this equals\begin{equation}
\frac{1}{2}J^{\nu}\left(x\right)A'_{\nu}\left(x\right).\label{3.2.9}\end{equation}

So to first order in $e$ the generating functional can be written,\begin{equation}
Z\left[J\right]\approx\int DA'^{\mu}D\bar{\psi}D\psi D\eta D\omega\textrm{ }\left(1+e\int\textrm{d}^{4}y\textrm{ }\bar{\psi}\not A\psi\right)\textrm{ exp }iS_{Free\textrm{ }QED}\label{3.2.10}\end{equation}
where \begin{equation}
\begin{array}{rcl}
S_{Free\textrm{ }QED} & = & \int\textrm{d}^{4}x\textrm{ }\int\textrm{d}^{4}y\textrm{ }\frac{1}{2}J_{\kappa}\left(x\right)\Delta^{\kappa\nu}\left(x-y\right)J_{\nu}\left(y\right)\\
 &  & \int\textrm{d}^{4}x\textrm{ }-\frac{1}{2}A'_{\mu}\left(x\right)Q^{\mu\nu}A'_{\nu}\left(x\right)+\eta\left(x\right)\partial^{\mu}\partial_{\mu}\omega\left(x\right)+\bar{\psi}\left(x\right)\left(i\not\partial-m\right)\psi\left(x\right)\end{array}\label{eq:3.2.11}\end{equation}
\emph{i.e.} QED without the interaction term and the source and $A^{\mu}$
field terms rewritten. 

We now arrive at the common sense result that to get the photon propagator,
the photon two point function, we will be taking the zeroth order
of perturbation theory \emph{i.e.} ignore the interaction terms. If
we take higher orders the higher order terms in the  interacting part
of the Lagrangian will be playing a role in our correlator \emph{i.e.}
we would be having terms with $e\int\textrm{d}^{4}y\textrm{ }\bar{\psi}\not A\psi$
bits in them in our correlator. So to get the photon two point function
${\bkiii{0}{A^{\mu}\left(x_{1}\right)A^{\nu}\left(x_{2}\right)}{0}}$
to zeroth order in $e$ (normalized by $Z^{-1}$) we want to functionally
differentiate,\begin{equation}
Z\left[J\right]=\int DA^{\mu}D\bar{\psi}D\psi D\eta D\omega\textrm{ }\left(1+{\cal {O}}\left(e\right)+...\right)\textrm{exp }iS_{Free\textrm{ }Q.E.D.}\label{3.2.12}\end{equation}
with respect to $J_{\mu}\left(x_{1}\right)$ and $J_{\nu}\left(x_{2}\right)$,\begin{equation}
\begin{array}{rcl}
\frac{{\bkiii{0}{A^{\mu}\left(x_{1}\right)A^{\nu}\left(x_{2}\right)}{0}}}{{\bkii{0}{0}}} & = & \left.\frac{\delta^{2}\ln Z\left[J\right]}{\delta J_{\mu}\left(x_{1}\right)\delta J_{\nu}\left(x_{2}\right)}\right|_{J=0}\\
 & = & \int DA^{\mu}D\bar{\psi}D\psi D\eta D\omega\textrm{ }\Delta^{\mu\nu}\left(x_{1}-x_{2}\right)\textrm{ exp }iS_{Free\textrm{ }Q.E.D.}+{\cal {O}}\left(e\right)+...\\
 & = & \frac{1}{Z}\left({\bkiii{0}{\Delta^{\mu\nu}\left(x_{1}-x_{2}\right)}{0}}+{\cal {O}}\left(e\right)+...\right)\end{array}\label{3.2.13}\end{equation}
 So the propagator (to zeroth order in $e$) $\Delta^{\mu\nu}\left(x_{1}-x_{2}\right)$
is essentially just the inverse of the differential operator $Q^{\lambda\rho}$,\begin{equation}
Q^{\mu\nu}\Delta_{\nu\kappa}\left(x\right)=\left(-g^{\mu\nu}\partial^{\gamma}\partial_{\gamma}+\left(1-\zeta^{-1}\right)\partial^{\mu}\partial^{\nu}\right)\Delta_{\nu\kappa}\left(x\right)=i\delta_{\kappa}^{\mu}\delta^{4}\left(x\right)\label{3.2.14}\end{equation}
 Fourier transforming we have,\begin{equation}
\begin{array}{rcl}
\left(-g^{\mu\nu}\partial^{\gamma}\partial_{\gamma}+\left(1-\zeta^{-1}\right)\partial^{\mu}\partial^{\nu}\right)\int\textrm{d}^{4}k\textrm{ }\tilde{\Delta}_{\nu\kappa}\left(k\right)e^{ik.x} & = & i\delta_{\kappa}^{\mu}\delta^{4}\left(x\right)\\
\int\textrm{d}^{4}k\textrm{ }\tilde{\Delta}_{\nu\kappa}\left(k\right)\left(g^{\mu\nu}k^{2}-\left(1-\zeta^{-1}\right)k^{\mu}k^{\nu}\right)e^{ik.x} & = & i\delta_{\kappa}^{\mu}\delta^{4}\left(x\right)\end{array}.\label{3.2.15}\end{equation}
 So now the question is how do we solve $\tilde{\Delta}_{\nu\kappa}\left(k\right)\left(g^{\mu\nu}k^{2}-\left(1-\zeta^{-1}\right)k^{\mu}k^{\nu}\right)=i\delta_{\kappa}^{\mu}$
for $\tilde{\Delta}_{\nu\kappa}\left(k\right)$? To start with let's
rewrite $Q^{\lambda\rho}$:\begin{equation}
\begin{array}{rcl}
Q^{\lambda\rho} & = & \left(-g^{\lambda\rho}+\frac{k^{\lambda}k^{\rho}}{k^{2}}\right)k^{2}-\frac{1}{\zeta}.\frac{k^{\lambda}k^{\rho}}{k^{2}}.k^{2}\\
 & = & -k^{2}P_{T}^{\lambda\rho}-\frac{k^{2}}{\zeta}P_{L}^{\lambda\rho}.\end{array}\label{3.2.16}\end{equation}
 Where we have defined the projection operators, $P_{T}^{\lambda\rho}=g^{\lambda\rho}-\frac{k^{\lambda}k^{\rho}}{k^{2}}\textrm{ and }P_{L}^{\lambda\rho}=\frac{k^{\lambda}k^{\rho}}{k^{2}}$.
As one would expect $P_{T}P_{T}=P_{T}$,\begin{equation}
\begin{array}{rcl}
P_{T}^{\lambda\rho}g_{\rho\kappa}P_{T}^{\kappa\mu} & = & \left(g^{\lambda\rho}-\frac{k^{\lambda}k^{\rho}}{k^{2}}\right)g_{\rho\kappa}\left(g^{\kappa\mu}-\frac{k^{\kappa}k^{\mu}}{k^{2}}\right)\\
 & = & g^{\lambda\mu}-\frac{2k^{\lambda}k^{\mu}}{k^{2}}+\frac{k^{\lambda}k^{\mu}}{k^{2}}\\
 & = & g^{\lambda\mu}-\frac{k^{\lambda}k^{\mu}}{k^{2}}\\
 & = & P_{T}^{\lambda\mu}.\end{array}\label{3.2.17}\end{equation}
 The same is true of $P_{L}$, $P_{L}P_{L}=P_{L}.$ Also $P_{L}P_{T}=0=P_{T}P_{L}$,\begin{equation}
\begin{array}{rcl}
P_{L}^{\lambda\rho}g_{\rho\kappa}P_{T}^{\kappa\mu} & = & \frac{k^{\lambda}k^{\rho}}{k^{2}}g_{\rho\kappa}\left(g^{\kappa\mu}-\frac{k^{\kappa}k^{\mu}}{k^{2}}\right)\\
 & = & \frac{k^{\lambda}k^{\mu}}{k^{2}}-\frac{k^{\lambda}k^{\mu}}{k^{2}}\\
 & = & 0\end{array}.\label{3.2.18}\end{equation}
This last identity shows us that $P_{T}$ and $P_{L}$ project out
\emph{orthogonal} \emph{subspaces}, \emph{i.e.} the Hilbert space
of states is now split into two by $P_{T}$ and $P_{L}$, all the
states in one half being orthogonal to all the states in the other.

Returning to the derivation of the Feynman rules we recall that we
are essentially after the form of the object $Q^{-1}$. Given our
study of $P_{T}$ and $P_{L}$ and their relation to the space of
states we will assume that $Q^{-1}$ can be written as a linear combination
of these two operators: \begin{equation}
Q^{-1\textrm{ }\rho\mu}=AP_{T}^{\rho\mu}+BP_{L}^{\rho\mu}.\label{3.2.19}\end{equation}
 Naturally one $QQ^{-1}$ will give the identity or equivalently $P_{T}+P_{L}$:\begin{equation}
\begin{array}{rcl}
Q^{\textrm{ }\lambda\rho}g_{\rho\kappa}Q^{-1\textrm{ }\kappa\mu} & = & -k^{2}\left(P_{T}^{\lambda\rho}+\frac{1}{\zeta}P_{L}^{\lambda\rho}\right)g_{\rho\kappa}\left(AP_{T}^{\kappa\mu}+BP_{L}^{\kappa\mu}\right)\\
 & = & P_{L}^{\lambda\mu}+P_{T}^{\lambda\mu}\end{array}\label{3.2.20}\end{equation}
\begin{equation}
\begin{array}{crcl}
\Rightarrow & -k^{2}\left(AP_{T}^{\lambda\mu}+\frac{B}{\zeta}P_{L}^{\lambda\mu}\right) & = & P_{T}^{\lambda\mu}+P_{L}^{\lambda\mu}\\
\Rightarrow & A=-\frac{1}{k^{2}} & \textrm{and} & B=-\frac{\zeta}{k^{2}}\\
\Rightarrow & Q^{-1\textrm{ }\lambda\mu} & = & -\frac{1}{k^{2}}P_{T}^{\lambda\mu}-\frac{\zeta}{k^{2}}P_{L}^{\lambda\mu}\\
\Rightarrow & \textrm{Propagator} & = & -\frac{i}{k^{2}}\left(P_{T}^{\lambda\mu}+\zeta P_{L}^{\lambda\mu}\right).\end{array}\label{3.2.21}\end{equation}
Given the form of the photon propagator one might wonder why then
we don't see $\zeta's$ in our Feynman rules for QED. The answer is
simply that the photon has no longitudinal component. This is not
to say that if we were to put our photons inside some pathological
loop diagram we would not see the $\zeta's$, we would but they all
cancel in the final matrix element. The proof that all the $\zeta$'s
cancel is not obvious and we come back to it in the next lecture when
discussing the QED Ward identities. 

The same recipe can be used to give the following propagators for
the ghost fields and the fermion fields...

\begin{center}\includegraphics[%
  width=0.30\paperwidth,
  height=0.10\paperwidth]{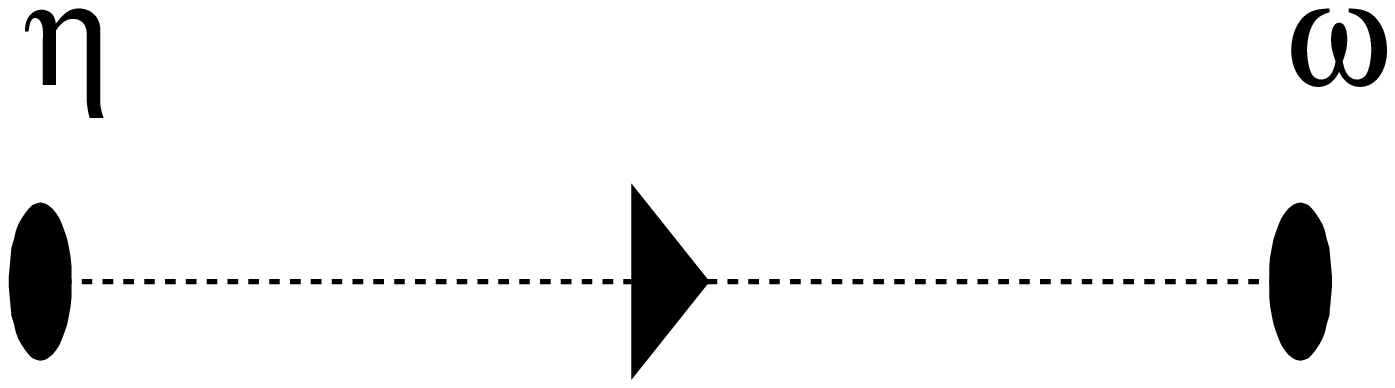}\end{center}

\begin{center}{\Large $\frac{i}{k^{2}+i\epsilon}$}\end{center}{\Large \par}
\vspace{0.3cm}

\begin{center}{\Large \includegraphics[%
  width=0.30\paperwidth,
  height=0.10\paperwidth]{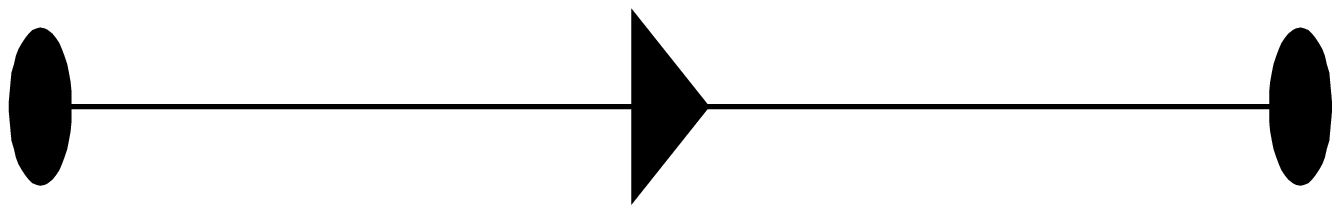}}\end{center}{\Large \par}

\begin{center}{\Large $\frac{i}{\not k-m+i\epsilon}$}\end{center}{\Large \par}

\noindent Note that the $\epsilon$'s in the above Feynman propagators
are put in by hand to regulate the poles in the Feynman propagators
(\emph{branch cuts}) where it is understood that in the calculation
of the amplitudes the limit $\epsilon\rightarrow0$ is taken at the
end.

\newpage
\section{BRS Symmetry.}

Continuing from lecture 7 - the generating functional for QED with
ghosts \emph{etc}...(put in fermion fields now). 

\begin{equation}
Z={\bkii{0}{0}}=\int DA^{\mu}D\bar{\psi}D\psi D\eta D\omega\textrm{ exp }i\int\textrm{d}^{4}x\textrm{ }\left\{ -\frac{1}{4}F_{\mu\nu}F^{\mu\nu}+\frac{1}{2\zeta}\left(\partial_{\mu}A^{\mu}\right)^{2}+\eta\partial^{\mu}\partial_{\mu}\omega+\bar{\psi}\left(i\not D-m\right)\psi\right\} .\label{3.3.1}\end{equation}
 This thing looks gauge invariant except for the $\partial^{\mu}A_{\mu}$
gauge fixing term. It turns out that despite the apparent breaking
of gauge invariance above there is in fact some residual symmetry
present, we shall come to this point shortly. This symmetry was the
result of work of Becchi, Rouet, Stora and (independently) Tyutin,
it is of crucial importance in the quantization of gauge (and other
more general) field theories. The BRS(T) symmetry is based on the
notion that we can have a further symmetry of the action if the ghost
fields transform non-trivially. BRS is a generalized gauge invariance.
In general the parameters of gauge transformations are arbitrary position
dependent functions. In the BRS arrangement this arbitrary function
is considered as being a product of two Grassmann quantities, one
a constant the other a function of space-time. Consequently \emph{anything
that is gauge invariant is BRS invariant} - \emph{i.e.} the original
action, prior to gauge fixing, is gauge and therefore BRS invariant.
It has been the nature of these notes to exhibit such obvious results
by brute force too.

Suppose that under these new BRS symmetry transformations we have:\begin{equation}
A_{\mu}\rightarrow A'_{\mu}=A_{\mu}+\epsilon\partial_{\mu}\omega\label{3.3.2}\end{equation}
where this $\epsilon$ is an \emph{infinitesimal} \emph{Grassmann}
\emph{parameter} and $\omega$ is our ghost field from before (also
Grassmann). Under this transformation the field strength tensor is
in fact invariant:\begin{equation}
\begin{array}{rcl}
F_{\mu\nu}\rightarrow F'_{\mu\nu} & = & \partial_{\mu}A'_{\nu}-\partial_{\nu}A'_{\mu}\\
 & = & \partial_{\mu}A_{\nu}-\partial_{\nu}A_{\mu}+\epsilon\partial_{\mu}\partial_{\nu}\omega-\epsilon\partial_{\nu}\partial_{\mu}\omega\\
 & = & \partial_{\mu}A_{\nu}-\partial_{\nu}A_{\mu}\\
 & = & F_{\mu\nu}\end{array}.\label{3.3.3}\end{equation}
What happens to the gauge fixing term under BRS? \begin{equation}
\begin{array}{rcl}
-\frac{1}{2\zeta}\left(\partial_{\mu}A^{\mu}\right)^{2} & \rightarrow & -\frac{1}{2\zeta}\left(\partial_{\mu}A^{\mu}+\epsilon\partial_{\mu}\partial^{\mu}\omega\right)^{2}\\
 & = & -\frac{1}{2\zeta}\left(\partial_{\mu}A^{\mu}\right)^{2}-\frac{1}{\zeta}\left(\partial_{\mu}A^{\mu}\right)\left(\epsilon\partial_{\nu}\partial^{\nu}\omega\right)\end{array}\label{3.3.4}\end{equation}
Well it changes, so how do we fix this? We must specify a way for
ghost field $\eta$ to transform so as to cancel the $-\frac{1}{\zeta}\left(\partial_{\mu}A^{\mu}\right)\left(\epsilon\partial_{\nu}\partial^{\nu}\omega\right)$
coming from the transformation of the gauge fixing term in the action.
We do this by making $\eta$ transform as $\eta\rightarrow\eta+\frac{1}{\zeta}\left(\partial_{\mu}A^{\mu}\right)\epsilon$
then the ghost term in the action $\left(\eta\partial_{\nu}\partial^{\nu}\omega\right)$
turns into itself plus a part which cancels the aforementioned term
from transforming the gauge fixing term. In addition to this we make
omega transform into itself $\omega\rightarrow\omega$ and the fermion
fields transform as $\psi\rightarrow e^{ie\epsilon\omega}\psi$, $\bar{\psi}\rightarrow e^{-ie\epsilon\omega}\bar{\psi}$.
It is worth Taylor expanding the exponential in the last pair of transformations.\begin{equation}
\begin{array}{rcl}
\psi\rightarrow\psi' & = & e^{ie\epsilon\omega}\psi\\
 & = & \left(1+ie\epsilon\omega+\frac{1}{2!}\left(ie\epsilon\omega\right)\left(ie\epsilon\omega\right)+...\right)\psi\end{array}\label{3.3.5}\end{equation}
Note that $\epsilon$ and $\omega$ are Grassmann numbers so $\epsilon^{2}=0$
\emph{etc} which means that the above Taylor series \ref{3.3.5} terminates
after the first two terms!\begin{equation}
\Rightarrow\psi\rightarrow\psi'=\left(1+ie\epsilon\omega\right)\psi\label{3.3.6}\end{equation}
Likewise we get for $\bar{\psi}$, \begin{equation}
\bar{\psi}\rightarrow\bar{\psi}'=\left(1-ie\epsilon\omega\right)\bar{\psi}.\label{3.3.7}\end{equation}
 So for the fermion part of the Lagrangian we have,\begin{equation}
\begin{array}{rcl}
\bar{\psi}\left(i\not D-m\right)\psi & \rightarrow & \bar{\psi}'\left(i\not D'-m\right)\psi'\\
 & = & \left(1-ie\epsilon\omega\right)\bar{\psi}\left(i\not\partial+e\not A+e\epsilon\left(\not\partial\omega\right)-m\right)\left(1+ie\epsilon\omega\right)\psi\\
 & = & \left(1-ie\epsilon\omega\right)\bar{\psi}\left(\left(i\not\partial+e\not A\right)\psi+e\epsilon\left(\not\partial\omega\right)\psi-m\psi-e\epsilon\not\partial\left(\omega\psi\right)+ie^{2}\epsilon\omega\not A\psi-ime\epsilon\omega\psi\right)\\
 & = & \bar{\psi}\left(\left(i\not\partial+e\not A\right)\psi+e\epsilon\left(\not\partial\omega\right)\psi-m\psi-e\epsilon\not\partial\left(\omega\psi\right)+ie^{2}\epsilon\omega\not A\psi-ime\epsilon\omega\psi\right)\\
 & - & ie\epsilon\omega\bar{\psi}\left(\left(i\not\partial+e\not A\right)\psi+e^{2}\epsilon\left(\not\partial\omega\right)\psi-m\psi-e\epsilon\not\partial\left(\omega\psi\right)+ie^{2}\epsilon\omega\not A\psi-ime\epsilon\omega\psi\right)\\
 & = & \bar{\psi}\left(\left(i\not D-m\right)\psi+e\epsilon\left(\not\partial\omega\right)\psi-e\epsilon\not\partial\left(\omega\psi\right)+ie^{2}\epsilon\not A\omega\psi-ime\epsilon\omega\psi\right)-ie\epsilon\omega\bar{\psi}\left(i\not D-m\right)\psi\\
 & = & \bar{\psi}\left(i\not D-m\right)\psi+e\bar{\psi}\epsilon\left(\not\partial\omega\right)\psi-e\bar{\psi}\epsilon\not\partial\left(\omega\psi\right)+ie^{2}\bar{\psi}\epsilon\not A\omega\psi-ime\bar{\psi}\epsilon\left(\omega\psi\right)\\
 & - & ie\epsilon\omega\bar{\psi}\left(i\not D-m\right)\psi\end{array}\label{3.3.8}\end{equation}
Plug in: $-e\bar{\psi}\epsilon\not\partial\left(\omega\psi\right)=-e\bar{\psi}\epsilon\left(\not\partial\omega\right)\psi-e\bar{\psi}\epsilon\omega\left(\not\partial\psi\right).$
\begin{equation}
\begin{array}{rl}
= & \bar{\psi}\left(i\not D-m\right)\psi-e\bar{\psi}\epsilon\omega\left(\not\partial\psi\right)+ie^{2}\bar{\psi}\epsilon\not A\left(\omega\psi\right)-ime\bar{\psi}\epsilon\left(\omega\psi\right)-ie\epsilon\omega\bar{\psi}\left(i\not D-m\right)\psi\\
= & \bar{\psi}\left(i\not D-m\right)\psi-e\bar{\psi}\epsilon\omega\left(\not\partial\psi\right)+ie^{2}\bar{\psi}\epsilon\not A\left(\omega\psi\right)-ime\bar{\psi}\epsilon\left(\omega\psi\right)-ie\epsilon\omega\bar{\psi}\left(i\not\partial+e\not A-m\right)\psi\\
= & \bar{\psi}\left(i\not D-m\right)\psi-e\bar{\psi}\epsilon\omega\left(\not\partial\psi\right)+ie^{2}\bar{\psi}\epsilon\not A\left(\omega\psi\right)-ime\bar{\psi}\epsilon\left(\omega\psi\right)+e\epsilon\omega\bar{\psi}\left(\not\partial\psi\right)-ie^{2}\epsilon\omega\bar{\psi}\not A\psi+ime\epsilon\omega\bar{\psi}\psi\end{array}\label{3.3.9}\end{equation}
Now if we shuffle the variables around in the 2nd 3rd and 4th terms
around taking into account the anticommuting nature of the Grassmann
variables $\psi,\bar{\psi},\omega\textrm{ and }\epsilon$ we have,\begin{equation}
\begin{array}{rcl}
-e\bar{\psi}\epsilon\omega\left(\not\partial\psi\right) & = & +e\epsilon\bar{\psi}\omega\left(\not\partial\psi\right)\\
 & = & -e\epsilon\omega\bar{\psi}\left(\not\partial\psi\right)\end{array}\label{3.3.10}\end{equation}
\begin{equation}
\begin{array}{rcl}
+ie^{2}\bar{\psi}\epsilon\not A\left(\omega\psi\right) & = & +ie^{2}\bar{\psi}\epsilon\omega\not A\psi\\
 & = & -ie^{2}\epsilon\bar{\psi}\omega\not A\psi\\
 & = & +ie^{2}\epsilon\omega\bar{\psi}\not A\psi\end{array}\label{3.3.11}\end{equation}
\begin{equation}
\begin{array}{rcl}
-ime\bar{\psi}\epsilon\left(\omega\psi\right) & = & +ime\epsilon\bar{\psi}\omega\psi\\
 & = & -ime\epsilon\omega\bar{\psi}\psi\end{array}\label{3.3.12}\end{equation}

\begin{equation}
\begin{array}{crcl}
\Rightarrow & \bar{\psi}\left(i\not D-m\right)\psi & \rightarrow & \bar{\psi}'\left(i\not D'-m\right)\psi'\\
 &  & = & \bar{\psi}\left(i\not D-m\right)\psi\end{array}\label{3.3.13}\end{equation}

We have a new symmetry of our \emph{gauge fixed} QED action. Is BRS
also a symmetry of the path integral measure? In other words is the
Jacobian between the variables $A^{\mu}\left(x\right),\bar{\psi}\left(x\right),\psi\left(x\right),\eta\left(x\right),\omega\left(x\right)$
and their transformed counterparts equal to one? Recall the BRS transformations,\begin{equation}
\begin{array}{rccclcl}
A^{\mu}\left(x\right) & \rightarrow & A'^{\mu}\left(x\right) & = & A^{\mu}\left(x\right) & + & \epsilon\partial^{\mu}\omega\left(x\right)\\
\bar{\psi}\left(x\right) & \rightarrow & \bar{\psi}'\left(x\right) & = & \bar{\psi}\left(x\right) & - & ie\epsilon\omega\bar{\psi}\left(x\right)\\
\psi\left(x\right) & \rightarrow & \psi'\left(x\right) & = & \psi\left(x\right) & + & ie\epsilon\omega\psi\left(x\right)\\
\eta\left(x\right) & \rightarrow & \eta'\left(x\right) & = & \eta\left(x\right) & + & \frac{1}{\zeta}\left(\partial_{\mu}A^{\mu}\left(x\right)\right)\epsilon\\
\omega\left(x\right) & \rightarrow & \omega'\left(x\right) & = & \omega\left(x\right)\end{array}.\label{3.3.14}\end{equation}
Imagine space-time consists of a single point $\left(x_{0}\right)$,
the Jacobian would be of the form:\begin{equation}
\begin{array}{rcl}
J & = & \textrm{Det}\left(\begin{array}{ccccc}
\frac{\partial A^{\mu}\left(x_{0}\right)}{\partial A^{\mu}\left(x_{0}\right)} & \frac{\partial\bar{\psi}\left(x_{0}\right)}{\partial A^{\mu}\left(x_{0}\right)} & \frac{\partial\psi\left(x_{0}\right)}{\partial A^{\mu}\left(x_{0}\right)} & \frac{\partial\eta\left(x_{0}\right)}{\partial A^{\mu}\left(x_{0}\right)} & \frac{\partial\omega\left(x_{0}\right)}{\partial A^{\mu}\left(x_{0}\right)}\\
\frac{\partial A^{\mu}\left(x_{0}\right)}{\partial\bar{\psi}\left(x_{0}\right)} & \frac{\partial\bar{\psi}\left(x_{0}\right)}{\partial\bar{\psi}\left(x_{0}\right)} & \frac{\partial\psi\left(x_{0}\right)}{\partial\bar{\psi}\left(x_{0}\right)} & \frac{\partial\eta\left(x_{0}\right)}{\partial\bar{\psi}\left(x_{0}\right)} & \frac{\partial\omega\left(x_{0}\right)}{\partial\bar{\psi}\left(x_{0}\right)}\\
\frac{\partial A^{\mu}\left(x_{0}\right)}{\partial\psi\left(x_{0}\right)} & \frac{\partial\bar{\psi}\left(x_{0}\right)}{\partial\psi\left(x_{0}\right)} & \frac{\partial\psi\left(x_{0}\right)}{\partial\psi\left(x_{0}\right)} & \frac{\partial\eta\left(x_{0}\right)}{\partial\psi\left(x_{0}\right)} & \frac{\partial\omega\left(x_{0}\right)}{\partial\psi\left(x_{0}\right)}\\
\frac{\partial A^{\mu}\left(x_{0}\right)}{\partial\eta\left(x_{0}\right)} & \frac{\partial\bar{\psi}\left(x_{0}\right)}{\partial\eta\left(x_{0}\right)} & \frac{\partial\psi\left(x_{0}\right)}{\partial\eta\left(x_{0}\right)} & \frac{\partial\eta\left(x_{0}\right)}{\partial\eta\left(x_{0}\right)} & \frac{\partial\omega\left(x_{0}\right)}{\partial\eta\left(x_{0}\right)}\\
\frac{\partial A^{\mu}\left(x_{0}\right)}{\partial\omega\left(x_{0}\right)} & \frac{\partial\bar{\psi}\left(x_{0}\right)}{\partial\omega\left(x_{0}\right)} & \frac{\partial\psi\left(x_{0}\right)}{\partial\omega\left(x_{0}\right)} & \frac{\partial\eta\left(x_{0}\right)}{\partial\omega\left(x_{0}\right)} & \frac{\partial\omega\left(x_{0}\right)}{\partial\omega\left(x_{0}\right)}\end{array}\right)\\
 & = & \textrm{Det}\left(\begin{array}{ccccc}
1 & 0 & 0 & {\cal {O}}\left(\epsilon\right) & 0\\
0 & 1-ie\epsilon\omega\left(x_{0}\right) & 0 & 0 & 0\\
0 & 0 & 1+ie\epsilon\omega\left(x_{0}\right) & 0 & 0\\
0 & 0 & 0 & 1 & 0\\
{\cal {O}}\left(\epsilon\right) & ie\epsilon\bar{\psi}\left(x_{0}\right) & ie\epsilon\psi\left(x_{0}\right) & 0 & 1\end{array}\right)\end{array}\label{eq:3.3.15}\end{equation}
Due to the nature of $\epsilon$ the off-diagonal elements do not
contribute so\begin{equation}
\begin{array}{rcl}
J & = & 1\times\left(1-ie\epsilon\omega\left(x_{0}\right)\right)\times\left(1+ie\epsilon\omega\left(x_{0}\right)\right)\times1\times1\\
 & = & 1-ie\epsilon\omega\left(x_{0}\right)+ie\epsilon\omega\left(x_{0}\right)+e^{2}\epsilon\omega\left(x_{0}\right)\epsilon\omega\left(x_{0}\right)\\
 & = & 1\end{array}.\label{3.3.16}\end{equation}
This is exactly what we want. In reality the Jacobian takes the form
of the determinant of an infinite dimensional matrix (because there
are an infinite number of space-time points in real life) but hopefully
it is still clear that it will still essentially be like the matrix
above repeated down the diagonal an infinite number of times (one
for every space-time point). The resulting determinant will appear
just like above with only terms linear in the $\omega$ (because all
other terms higher order in $\omega$ will have at least an $\epsilon^{2}$
attached to them and the linear terms cancel each other out at every
space-time point just as they did above. 

The partition function is totally invariant under BRS, so we can start
generating Ward identities in the usual fashion. Consider the general
case where we have some time ordered product of fields given by the
function $H\left(A^{\mu},\bar{\psi},\psi,\eta,\omega\right)$: \begin{equation}
\begin{array}{rcl}
{\bkiii{0}{\cal {T}}{H\left(A^{\mu},\bar{\psi},\psi,\eta,\omega\right)}{0}} & = & \int DA^{\mu}D\bar{\psi}D\psi D\eta D\omega\textrm{ }H\left(A^{\mu},\bar{\psi},\psi,\eta,\omega\right)\\
 &  & \textrm{exp }i\int\textrm{d}^{4}x\textrm{ }\left\{ \bar{\psi}\left(i\not D-m\right)\psi-\frac{1}{4}F_{\mu\nu}F^{\mu\nu}+\frac{1}{2\zeta}\left(\partial_{\mu}A^{\mu}\right)^{2}+\eta\partial^{\mu}\partial_{\mu}\omega\right\} \end{array}.\label{3.3.17}\end{equation}
In not specifying the form of $H$ we are essentially studying the
most general type of Green's function. Now we do the usual change
of variables trick on the above (BRS transformation):\begin{equation}
\begin{array}{rccclcl}
A^{\mu}\left(x\right) & \rightarrow & A'^{\mu}\left(x\right) & = & A^{\mu}\left(x\right) & + & \epsilon\partial^{\mu}\omega\left(x\right)\\
\bar{\psi}\left(x\right) & \rightarrow & \bar{\psi}'\left(x\right) & = & \bar{\psi}\left(x\right) & - & ie\epsilon\omega\bar{\psi}\left(x\right)\\
\psi\left(x\right) & \rightarrow & \psi'\left(x\right) & = & \psi\left(x\right) & + & ie\epsilon\omega\psi\left(x\right)\\
\eta\left(x\right) & \rightarrow & \eta'\left(x\right) & = & \eta\left(x\right) & + & \frac{1}{\zeta}\left(\partial_{\mu}A^{\mu}\left(x\right)\right)\epsilon\\
\omega\left(x\right) & \rightarrow & \omega'\left(x\right) & = & \omega\left(x\right)\end{array}.\label{3.3.18}\end{equation}
We know that $DA^{\mu}D\bar{\psi}D\psi D\eta D\omega$ and $\textrm{exp }i\int\textrm{d}^{4}x\textrm{ }\left\{ \bar{\psi}\left(i\not D-m\right)\psi-\frac{1}{4}F_{\mu\nu}F^{\mu\nu}+\frac{1}{2\zeta}\left(\partial_{\mu}A^{\mu}\right)^{2}+\eta\partial^{\mu}\partial_{\mu}\omega\right\} $
transform into themselves from our study above but this need not be
the case for $H$, it isn't necessarily BRS invariant. So what happens
to $H$? $H\rightarrow H'=H+\delta_{\epsilon}H.$\begin{equation}
\begin{array}{rcl}
\delta_{\epsilon}H & = & \frac{\partial H}{\partial A^{\mu}}\delta_{\epsilon}A^{\mu}+\frac{\partial H}{\partial\eta}\delta_{\epsilon}\eta+\frac{\partial H}{\partial\bar{\psi}}\delta_{\epsilon}\bar{\psi}+\frac{\partial H}{\partial\psi}\delta_{\epsilon}\psi+\frac{\partial H}{\partial\omega}\delta_{\epsilon}\omega\\
 & = & \frac{\partial H}{\partial A^{\mu}}\epsilon\partial^{\mu}\omega\left(x\right)+\frac{\partial H}{\partial\eta}\delta_{\epsilon}\eta-ie\frac{\partial H}{\partial\bar{\psi}}\epsilon\omega\bar{\psi}+ie\frac{\partial H}{\partial\psi}\epsilon\omega\bar{\psi}+\frac{\partial H}{\partial\omega}0\\
 & = & \frac{\partial H}{\partial A^{\mu}}\epsilon\partial^{\mu}\omega\left(x\right)+\frac{\partial H}{\partial\eta}\delta_{\epsilon}\eta-ie\frac{\partial H}{\partial\bar{\psi}}\epsilon\omega\bar{\psi}+ie\frac{\partial H}{\partial\psi}\epsilon\omega\bar{\psi}\end{array}\label{3.3.19}\end{equation}
Under the (BRS transformation) change of variables we have\begin{equation}
\begin{array}{rcl}
\int DA^{\mu}D\bar{\psi}D\psi D\eta D\omega\textrm{ }H\left(A^{\mu},\bar{\psi},\psi,\eta,\omega\right)\textrm{ }e^{iS}\textrm{ } & = & \int DA^{\mu}D\bar{\psi}D\psi D\eta D\omega\textrm{ }H\left(A^{\mu},\bar{\psi},\psi,\eta,\omega\right)\textrm{ }e^{iS}\\
 & + & \int DA^{\mu}D\bar{\psi}D\psi D\eta D\omega\textrm{ }\left(\delta_{\epsilon}H\left(A^{\mu},\bar{\psi},\psi,\eta,\omega\right)\right)\textrm{ }e^{iS}\end{array}\label{3.3.20}\end{equation}
Using Dirac notation and substituting in for $\delta_{\epsilon}H$
we have, \begin{equation}
\begin{array}{rcl}
{\bkiii{0}{{\cal T}\textrm{ }H\left(A^{\mu},\bar{\psi},\psi,\eta,\omega\right)}{0}} & = & {\bkiii{0}{{\cal T}\textrm{ }H\left(A^{\mu},\bar{\psi},\psi,\eta,\omega\right)}{0}}\\
 & + & {\bkiii{0}{{\cal T}\textrm{ }\delta_{\epsilon}H\left(A^{\mu},\bar{\psi},\psi,\eta,\omega\right)}{0}}\end{array}\label{3.3.21}\end{equation}
\emph{i.e.} we have the Ward identity \begin{equation}
{\bkiii{0}{{\cal T}\textrm{ }\frac{\partial H}{\partial A^{\mu}}\epsilon\partial^{\mu}\omega\left(x\right)+\frac{\partial H}{\partial\eta}\delta_{\epsilon}\eta-ie\frac{\partial H}{\partial\bar{\psi}}\epsilon\omega\bar{\psi}+ie\frac{\partial H}{\partial\psi}\epsilon\omega\bar{\psi}}{0}}=0\label{3.3.22}\end{equation}
Note that this Ward identity does not depend on $\epsilon$, we have
not required the parameter of the BRS transformation to be small!
The Grassmann nature of $\epsilon$ terminates any power expansions
after the linear term appears (the same reasoning applies to the case
of non-Abelian theories which is a simple extension of our formalism
above, instead of $\epsilon$ we have a vector of $\epsilon$'s, one
per gauge group generator). 
\vspace{0.3cm}\newpage

\section{BRS in QED.}

Recall the BRS transformations from the last lecture:\begin{equation}
\begin{array}{rccclcl}
A^{\mu}\left(x\right) & \rightarrow & A'^{\mu}\left(x\right) & = & A^{\mu}\left(x\right) & + & \epsilon\partial^{\mu}\omega\left(x\right)\\
\bar{\psi}\left(x\right) & \rightarrow & \bar{\psi}'\left(x\right) & = & \bar{\psi}\left(x\right) & - & ie\epsilon\omega\bar{\psi}\left(x\right)\\
\psi\left(x\right) & \rightarrow & \psi'\left(x\right) & = & \psi\left(x\right) & + & ie\epsilon\omega\psi\left(x\right)\\
\eta\left(x\right) & \rightarrow & \eta'\left(x\right) & = & \eta\left(x\right) & + & \frac{1}{\zeta}\left(\partial_{\mu}A^{\mu}\left(x\right)\right)\epsilon\\
\omega\left(x\right) & \rightarrow & \omega'\left(x\right) & = & \omega\left(x\right)\end{array}.\label{3.4.1}\end{equation}

\noindent We also discovered that the BRS symmetry gave us the following
general Ward identity for some string of fields $H$:\begin{equation}
{\bkiii{0}{{\cal T}\delta_{\epsilon}H}{0}}={\bkiii{0}{{\cal T}\textrm{ }\frac{\partial H}{\partial A^{\mu}}\epsilon\partial^{\mu}\omega\left(x\right)+\frac{\partial H}{\partial\eta}\delta_{\epsilon}\eta-ie\frac{\partial H}{\partial\bar{\psi}}\epsilon\omega\bar{\psi}+ie\frac{\partial H}{\partial\psi}\epsilon\omega\bar{\psi}}{0}}=0\label{3.4.2}\end{equation}
needless to say these BRS Ward identities can get arbitrarily complicated,
so let's consider an easy one first. Consider,\begin{equation}
H=\eta\left(x\right)A_{\mu}\left(y\right)\label{3.4.3}\end{equation}
\begin{equation}
\begin{array}{crcl}
\Rightarrow & \delta_{\epsilon}H & = & \eta\left(x\right)\epsilon\partial_{\mu}^{\left(y\right)}\omega\left(y\right)+\frac{1}{\zeta}\epsilon\partial_{\nu}^{\left(x\right)}A^{\nu}\left(x\right)A_{\mu}\left(y\right)\\
 &  & = & \epsilon\left(\frac{1}{\zeta}\partial_{\nu}^{\left(x\right)}A^{\nu}\left(x\right)A_{\mu}\left(y\right)-\eta\left(x\right)\partial_{\mu}^{\left(y\right)}\omega\left(y\right)\right)\end{array}\label{3.4.4}\end{equation}
Note $+\eta\left(x\right)\epsilon\partial_{\mu}\omega\left(y\right)$
has become $-\epsilon\eta\left(x\right)\partial_{\mu}\omega\left(y\right)$
on the grounds that $\epsilon$ and $\eta$ are Grassmann numbers.
Differentiation with respect to $x$ has been denoted $\partial_{\nu}^{\left(x\right)}$
and differentiation with respect to $y$ has been denoted $\partial_{\nu}^{\left(y\right)}$
\emph{etc}. So we get our first BRS Ward identity,\begin{equation}
\begin{array}{crcl}
 & {\bra{0}}{\cal T}\frac{1}{\zeta}\partial_{\nu}^{\left(x\right)}A^{\nu}\left(x\right)A_{\mu}\left(y\right){\ket{0}} & = & {\bra{0}}{\cal T}\eta\left(x\right)\partial_{\mu}^{\left(y\right)}\omega\left(y\right){\ket{0}}\\
\Rightarrow & \partial_{\nu}^{\left(x\right)}{\bra{0}}{\cal T}\frac{1}{\zeta}A^{\nu}\left(x\right)A_{\mu}\left(y\right){\ket{0}} & = & \partial_{\mu}^{\left(y\right)}{\bkiii{0}{{\cal T}\eta\left(x\right)\omega\left(y\right)}{0}}\end{array}\label{3.4.5}\end{equation}
 which relates the divergence of the photon 2-point function ${\bra{0}}{\cal T}\frac{1}{\zeta}A^{\nu}\left(x\right)A_{\mu}\left(y\right){\ket{0}}$
to that of the ghost two point function ${\bkiii{0}{{\cal T}\eta\left(x\right)\omega\left(y\right)}{0}}$.
Recall that in QED the ghosts do not couple to the other fields so
we already know what it is!

\begin{center}\includegraphics[%
  clip,
  width=0.30\paperwidth,
  height=0.1\paperwidth]{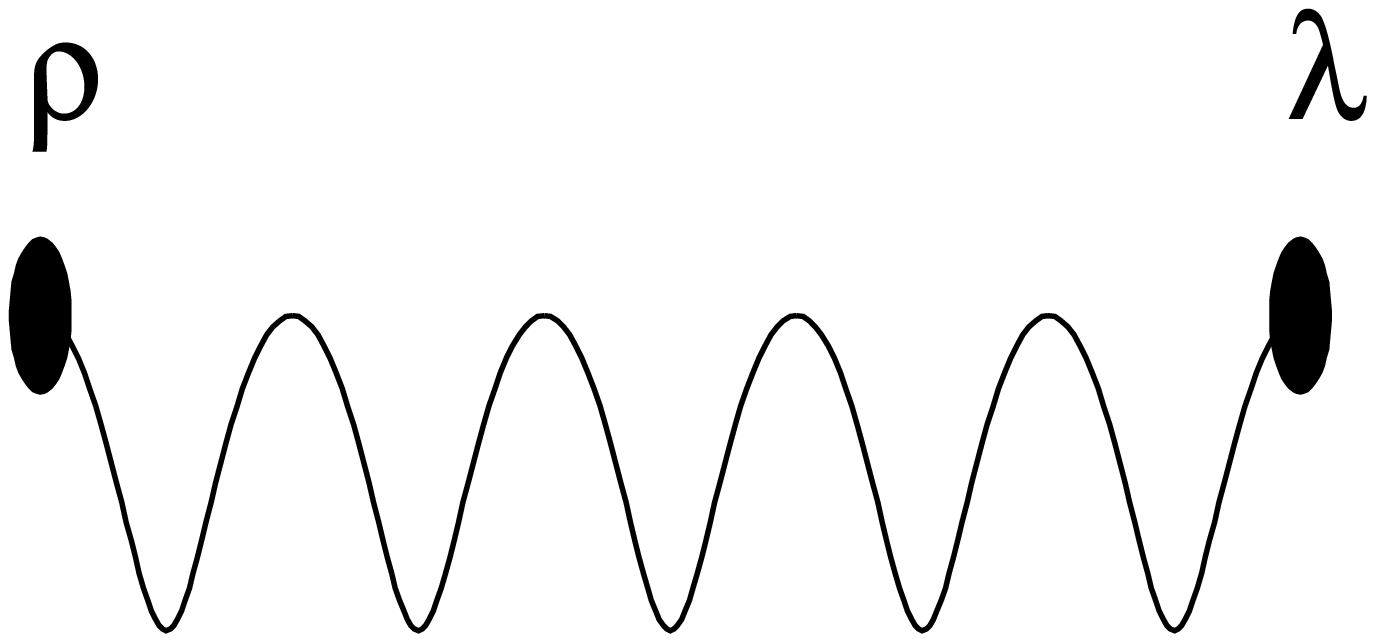}\begin{equation}
D_{\mu\nu}^{\left(0\right)}\left(k\right)=\frac{-i}{k^{2}+i\epsilon}\left(g_{\rho\lambda}-\frac{k_{\rho}k_{\lambda}}{k^{2}}\left(1-\zeta\right)\right)=\frac{-i}{k^{2}+i\epsilon}\left(P_{T\rho\lambda}+\zeta P_{L\rho\lambda}\right)\label{3.4.6}\end{equation}
\end{center}

\begin{center}{\Large \includegraphics[%
  width=0.30\paperwidth,
  height=0.1\paperwidth]{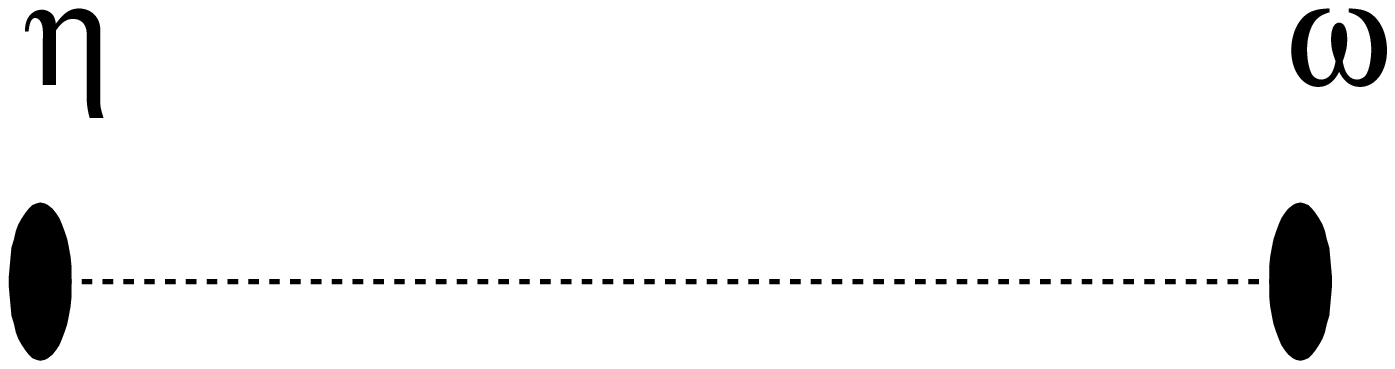}}\begin{equation}
\frac{i}{k^{2}+i\epsilon}\label{3.4.7}\end{equation}
\end{center}

\noindent Where $P_{T\rho\lambda}$ and $P_{L\rho\lambda}$ are the
transverse and longitudinal components of the photon two point function. 

We shall now attempt to simplify matters by Fourier transforming our
Ward identity. For the ghost part we have, \begin{equation}
\begin{array}{rcl}
\partial_{\mu}^{\left(y\right)}{\bkiii{0}{{\cal T}\eta\left(x\right)\omega\left(y\right)}{0}} & = & F.T.\left(\frac{i}{k^{2}+i\epsilon}\right)\\
 & = & \partial_{\mu}^{\left(y\right)}\int\frac{\textrm{d}^{4}k}{\left(2\pi\right)^{4}}\frac{i}{k^{2}+i\epsilon}e^{ik\left(x-y\right)}\\
 & = & \int\frac{\textrm{d}^{4}k}{\left(2\pi\right)^{4}}\frac{k_{\mu}}{k^{2}+i\epsilon}e^{ik\left(x-y\right)}.\end{array}\label{3.4.8}\end{equation}
For the time being we feign ignorance of the photon propagator in
the hope that being general may afford us some way of incorporating
modifications to to it by \emph{e.g.} fermion loops, hence we denote
it by $D_{\mu\nu}$ which means the photon part of the Ward identity
transforms as:\begin{equation}
\partial^{\left(x\right)\nu}\frac{1}{\zeta}\int\frac{\textrm{d}^{4}k}{\left(2\pi\right)^{4}}e^{ik\left(x-y\right)}D_{\mu\nu}\left(k\right)=\frac{1}{\zeta}\int\frac{\textrm{d}^{4}k}{\left(2\pi\right)^{4}}ik^{\nu}D_{\mu\nu}\left(k\right)e^{ik\left(x-y\right)}.\label{3.4.9}\end{equation}
Plugging this into our Ward identity we find,\begin{equation}
\int\frac{\textrm{d}^{4}k}{\left(2\pi\right)^{4}}e^{ik\left(x-y\right)}\left\{ \frac{i}{\zeta}k^{\nu}D_{\mu\nu}\left(k\right)-\frac{k_{\mu}}{k^{2}+i\epsilon}\right\} =0\label{3.4.10}\end{equation}
 There exists a theorem which states that if the Fourier transform
of something is zero then that something is also zero! Hence,\begin{equation}
\frac{i}{\zeta}k^{\nu}D_{\mu\nu}\left(k\right)=\frac{k_{\mu}}{k^{2}+i\epsilon}.\label{3.4.11}\end{equation}
 Now let's write the 2-point function as follows: 

\begin{center}\includegraphics[%
  width=0.60\textwidth,
  height=0.17\textwidth]{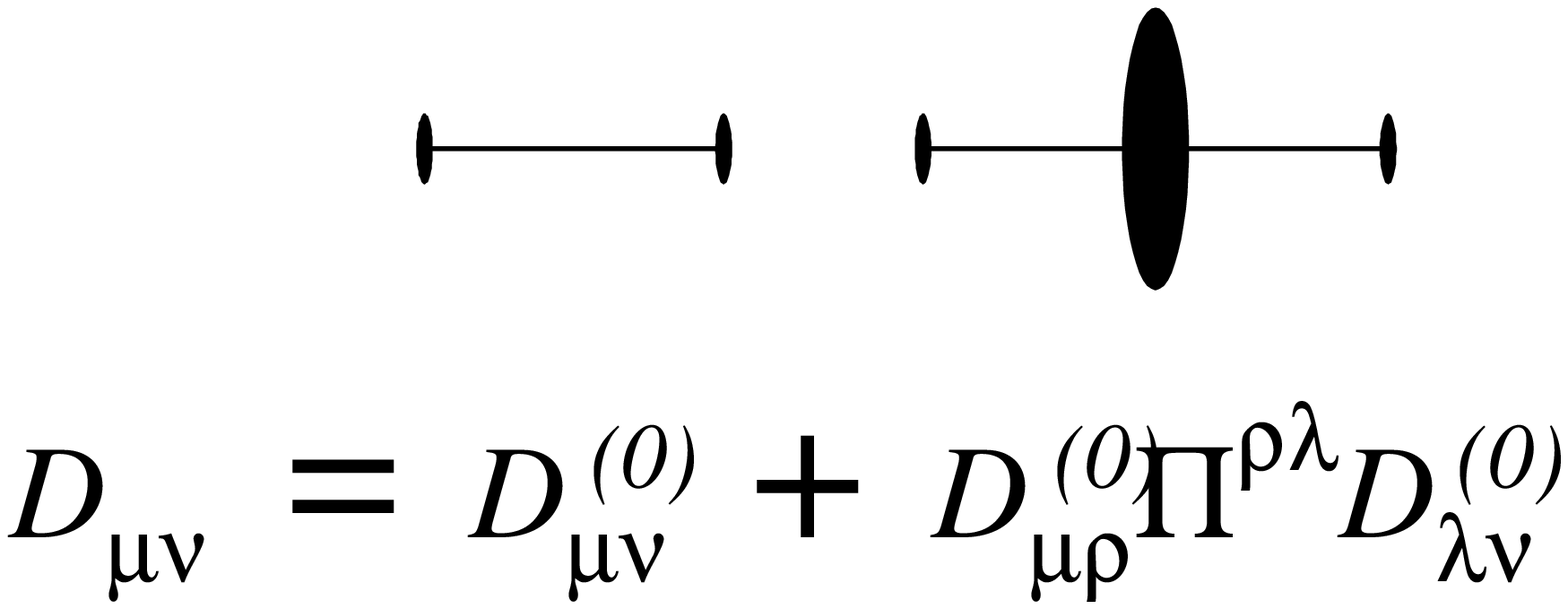} \end{center}

\noindent where $D_{\mu\nu}^{\left(0\right)}$ is the free photon
propagator given above. In our Ward identity we have $\frac{i}{\xi}k^{\nu}D_{\mu\nu}\left(k\right)$,
substituting in the above this gives\begin{equation}
\begin{array}{rcl}
\frac{i}{\zeta}k^{\nu}D_{\mu\nu}\left(k\right) & = & \frac{i}{\zeta}k^{\nu}\left\{ \frac{-i}{k^{2}+i\epsilon}\left(P_{T\mu\nu}+\zeta P_{L\mu\nu}\right)+D_{\mu\rho}^{\left(0\right)}\left(k\right)\Pi^{\rho\lambda}D_{\lambda\nu}^{\left(0\right)}\left(k\right)\right\} \\
 & = & \frac{k_{\mu}}{k^{2}+i\epsilon}.\end{array}\label{3.4.12}\end{equation}
Use the definitions of $P_{T\rho\lambda}$ and $P_{L\rho\lambda}$
(\emph{i.e.} in terms of $k$'s) to simplify the above first term
inside \{...\}\begin{equation}
\begin{array}{rcl}
\frac{i}{\zeta}k^{\nu}.\frac{-i}{k^{2}+i\epsilon}\left(\left(g_{\mu\nu}-\frac{k_{\mu}k_{\nu}}{k^{2}}\right)+\zeta\left(\frac{k_{\mu}k_{\nu}}{k^{2}}\right)\right) & = & \frac{1}{\zeta}.\frac{1}{k^{2}+i\epsilon}\left(\left(k_{\mu}-k_{\mu}\right)+\zeta k_{\mu}\right)\\
 & = & \frac{k_{\mu}}{k^{2}+i\epsilon}\end{array}.\label{3.4.13}\end{equation}
Substituting this into the Ward identity on the previous line gives
\begin{equation}
\begin{array}{rcl}
\frac{i}{\zeta}k^{\nu}D_{\mu\nu}\left(k\right) & = & \frac{k_{\mu}}{k^{2}+i\epsilon}+\frac{i}{\zeta}k^{\nu}D_{\mu\rho}^{\left(0\right)}\left(k\right)\Pi^{\rho\lambda}D_{\lambda\nu}^{\left(0\right)}\left(k\right)\\
 & = & \frac{k_{\mu}}{k^{2}+i\epsilon}\end{array}\label{3.4.14}\end{equation}
\begin{equation}
\Rightarrow\textrm{ }k^{\nu}D_{\mu\rho}^{\left(0\right)}\left(k\right)\Pi^{\rho\lambda}D_{\lambda\nu}^{\left(0\right)}\left(k\right)=0.\label{3.4.15}\end{equation}
This already looks pretty interesting as $D_{\mu\rho}^{\left(0\right)}\left(k\right)\Pi^{\rho\lambda}D_{\lambda\nu}^{\left(0\right)}\left(k\right)$,
which was a generic way of writing any and all radiative corrections
to the photon propagator, is looking heavily constrained if not totally
constrained as a result of our simple Ward identity. So let's take
a closer look. Let the $k^{\nu}$ hit the free photon propagator $D_{\lambda\nu}^{\left(0\right)}\left(k\right)$
($k^{\nu}$ is just a number we can move it through anything without
any worries):\begin{equation}
\begin{array}{rcl}
k^{\nu}D_{\lambda\nu}^{\left(0\right)}\left(k\right) & = & k^{\nu}\frac{-i}{k^{2}+i\epsilon}\left(g_{\lambda\nu}-\frac{k_{\lambda}k_{\nu}}{k^{2}}\left(1-\zeta\right)\right)\\
 & = & \frac{-i\zeta k_{\lambda}}{k^{2}+i\epsilon}\end{array}\label{3.4.16}\end{equation}
\begin{equation}
\begin{array}{crcl}
\Rightarrow & D_{\mu\rho}^{\left(0\right)}\left(k\right)\Pi^{\rho\lambda}\frac{-i\zeta k_{\lambda}}{k^{2}+i\epsilon} & = & 0\\
\Rightarrow & D_{\mu\rho}^{\left(0\right)}\left(k\right)\Pi^{\rho\lambda}k_{\lambda} & = & 0\end{array}.\label{3.4.17}\end{equation}
Now we note that the operator $D_{\mu\rho}^{\left(0\right)}$ is invertible
(if it wasn't invertible we wouldn't have been able to derive it in
the first place - see lecture 7). So if we apply the inverse of $D_{\mu\rho}^{\left(0\right)}$
(whatever that is - see lecture 7) to the above and we get,\begin{equation}
\Pi^{\rho\lambda}k_{\lambda}=0.\label{3.4.18}\end{equation}
This is telling us that $\Pi^{\rho\lambda}$ must be proportional
to the transverse projection operator $P_{T}^{\rho\lambda}$ because
$k_{\lambda}$ kills it: $k_{\lambda}P_{T}^{\rho\lambda}=k_{\lambda}\left(g^{\rho\lambda}-\frac{k^{\rho}k^{\lambda}}{k^{2}}\right)=k^{\rho}-k^{\rho}=0$.
It is this property which is making the radiative corrections to the
photon propagator cancel out $\left(k^{\nu}D_{\mu\rho}^{\left(0\right)}\left(k\right)\Pi^{\rho\lambda}D_{\lambda\nu}^{\left(0\right)}\left(k\right)=0\right).$
This is why the photon does not acquire a mass through radiative corrections!
For example in $\phi^{4}$ theory the self interaction radiative corrections
modify the free $\phi$ propagator such that it acquires a mass:

\vspace{0.3cm}
\begin{center}\includegraphics[%
  width=0.80\textwidth,
  height=0.15\textwidth]{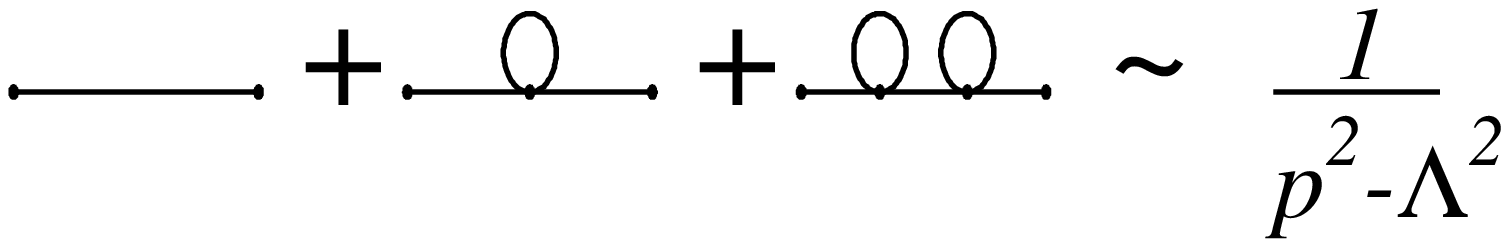}.\end{center}
\vspace{0.3cm}

\vspace{0.3cm}
Now for the next Ward identity. Consider now $H=\eta\left(x\right)\bar{\psi}\left(y\right)\psi\left(z\right)$,\begin{equation}
\begin{array}{rcl}
\delta_{\epsilon}H & = & \frac{1}{\zeta}\epsilon\left(\partial^{\mu}A_{\mu}\left(x\right)\right)\bar{\psi}\left(y\right)\psi\left(z\right)+\eta\left(x\right)\left(-ie\bar{\psi}\left(y\right)\epsilon\omega\left(y\right)\right)\psi\left(z\right)+\eta\left(x\right)\bar{\psi}\left(y\right)ie\epsilon\omega\left(z\right)\psi\left(z\right)\\
 & = & \epsilon\left\{ \frac{1}{\zeta}\left(\partial^{\mu}A_{\mu}\left(x\right)\right)\bar{\psi}\left(y\right)\psi\left(z\right)-ie\eta\left(x\right)\bar{\psi}\left(y\right)\omega\left(z\right)\psi\left(z\right)+ie\eta\left(x\right)\bar{\psi}\left(y\right)\omega\left(y\right)\psi\left(z\right)\right\} \end{array}.\label{3.4.19}\end{equation}
 Applying our Ward identity ${\bki{\delta_{\epsilon}H}}=0$ we get,\begin{equation}
\epsilon\frac{1}{\zeta}{\bki{{\cal T}\left(\partial^{\mu}A_{\mu}\left(x\right)\right)\bar{\psi}\left(y\right)\psi\left(z\right)}}-\epsilon{\bki{{\cal T}ie\eta\left(x\right)\bar{\psi}\left(y\right)\omega\left(z\right)\psi\left(z\right)}}+\epsilon{\bki{{\cal T}ie\eta\left(x\right)\bar{\psi}\left(y\right)\omega\left(y\right)\psi\left(z\right)}}=0.\label{3.4.20}\end{equation}
We can divide the $\epsilon$ away, it's just a (Grassmann) number.
Recall that in QED the ghosts are not coupled to any of the physical
fields in any gauge, this means that the last two vacuum expectation
values in the above simplify, \begin{equation}
\begin{array}{rcl}
{\bki{{\cal T}ie\eta\left(x\right)\bar{\psi}\left(y\right)\omega\left(z\right)\psi\left(z\right)}} & = & ie{\bki{{\cal T}\eta\left(x\right)\omega\left(z\right)}}{\bki{{\cal T}\bar{\psi}\left(y\right)\psi\left(z\right)}}\\
{\bki{{\cal T}ie\eta\left(x\right)\bar{\psi}\left(y\right)\omega\left(y\right)\psi\left(z\right)}} & = & ie{\bki{{\cal T}\eta\left(x\right)\omega\left(y\right)}}{\bki{{\cal T}\bar{\psi}\left(y\right)\psi\left(z\right)}}\end{array}\label{3.4.20}\end{equation}
and our Ward identity becomes,\begin{equation}
\Rightarrow\frac{1}{\zeta}{\bki{{\cal T}\left(\partial^{\mu}A_{\mu}\left(x\right)\right)\bar{\psi}\left(y\right)\psi\left(z\right)}}=ie{\bki{{\cal T}\eta\left(x\right)\omega\left(z\right)}}{\bki{{\cal T}\bar{\psi}\left(y\right)\psi\left(z\right)}}-ie{\bki{{\cal T}\eta\left(x\right)\omega\left(y\right)}}{\bki{{\cal T}\bar{\psi}\left(y\right)\psi\left(z\right)}}\label{3.4.21}\end{equation}
This Ward identity will give us a relationship between the fermion
2-point function and the 3-point function:

\begin{center}\includegraphics[%
  width=0.30\textwidth,
  height=0.15\textwidth]{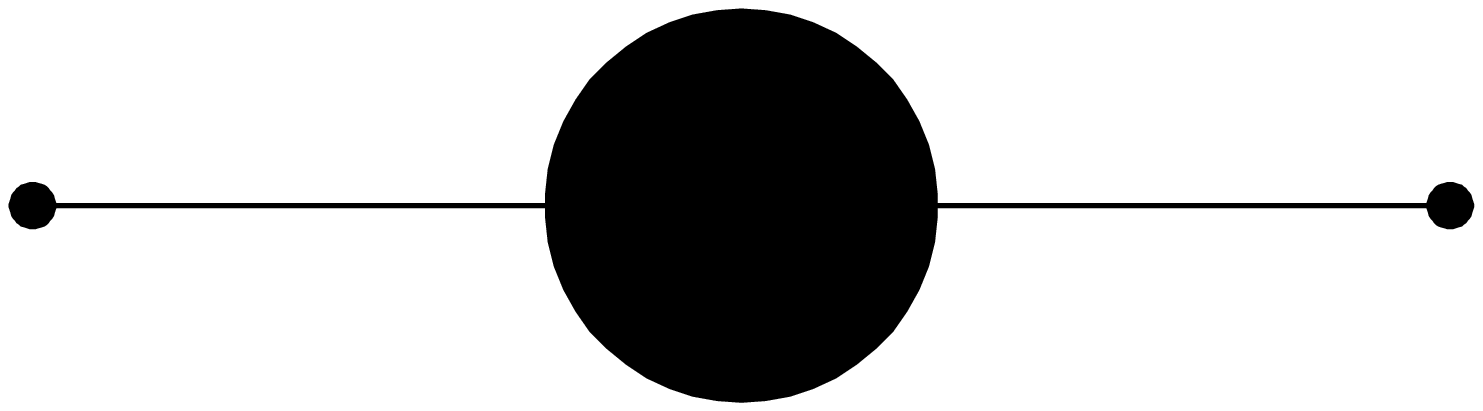} {\Huge \&} \includegraphics[%
  width=0.30\textwidth,
  height=0.15\textwidth]{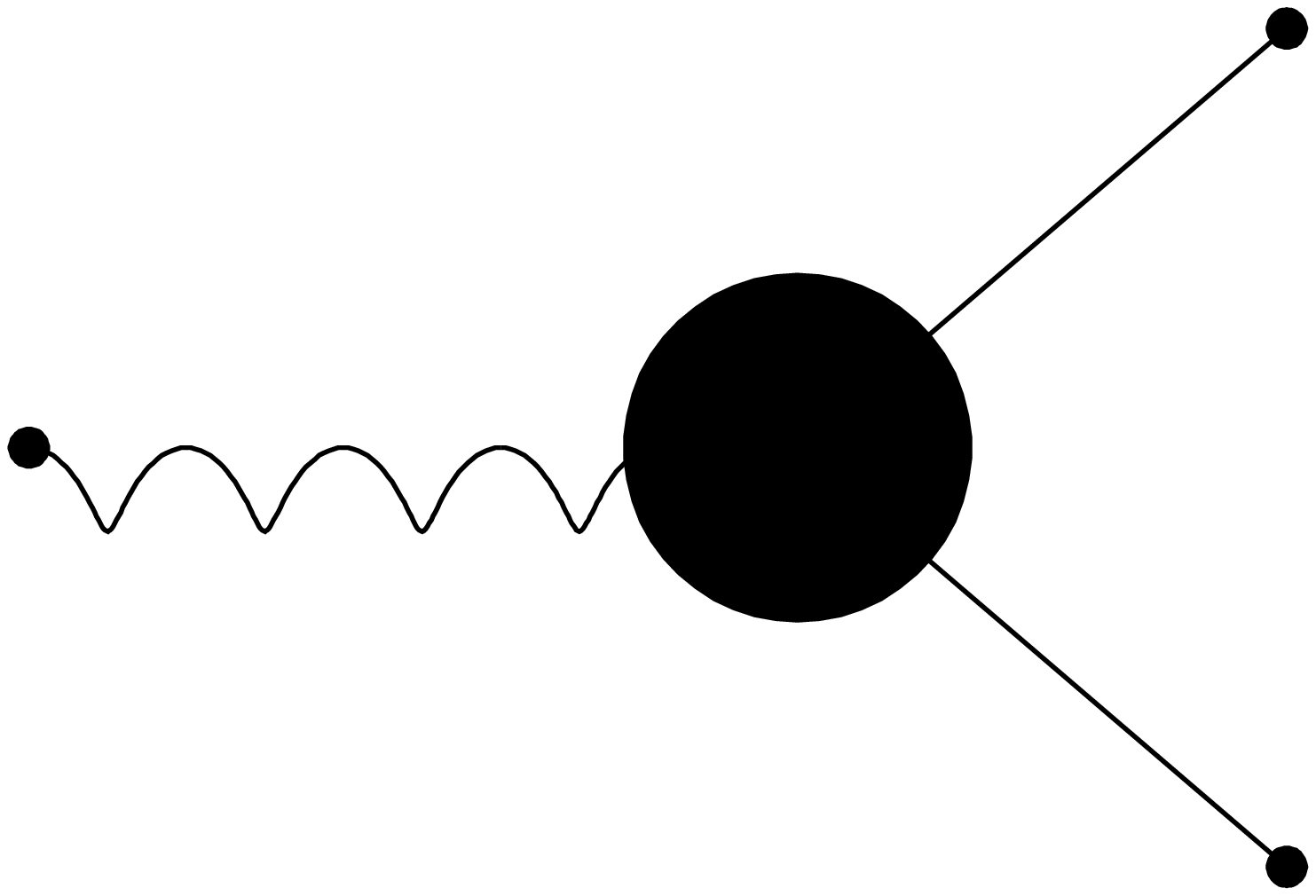}.\end{center}

As before let's start by Fourier transforming these vacuum expectation
values. We denote the full fermion 2-point function (the full fermion
propagator) $S_{F}\left(p\right)$ and the 3-point function $\Gamma^{\mu}\left(k;p,p'\right)$
(the vertex function). \begin{equation}
{\bki{\bar{\psi}\left(y\right)\psi\left(z\right)}}=\int\frac{\textrm{d}^{4}p}{\left(2\pi\right)^{4}}e^{ip.\left(y-z\right)}S_{F}\left(p\right)\label{3.4.22}\end{equation}
\begin{equation}
{\bki{A_{\mu}\left(x\right)\bar{\psi}\left(y\right)\psi\left(z\right)}}=\int\frac{\textrm{d}^{4}k}{\left(2\pi\right)^{4}}\frac{\textrm{d}^{4}p}{\left(2\pi\right)^{4}}\frac{\textrm{d}^{4}p}{\left(2\pi\right)^{4}}e^{i\left(k.x+p.y+p'.z\right)}\times\left(2\pi\right)^{4}\delta^{4}\left(k+p+p'\right)\times\Gamma^{\mu}\left(k;p,p'\right)\label{3.4.23}\end{equation}
If we now go ahead and insert these into the Ward identity \ref{3.4.21}
along with the expression for the ghost propagators, ${\bki{\eta\omega}}=\frac{i}{k^{2}+i\epsilon}$
we find, \begin{equation}
\begin{array}{rl}
 & \int\frac{\textrm{d}^{4}k}{\left(2\pi\right)^{4}}\frac{\textrm{d}^{4}p}{\left(2\pi\right)^{4}}\frac{\textrm{d}^{4}p}{\left(2\pi\right)^{4}}e^{i\left(k.x+p.y+p'.z\right)}\left(2\pi\right)^{4}\delta^{4}\left(k+p+p'\right)\frac{ik_{\mu}}{\zeta}\Gamma^{\mu}\left(k;p,p'\right)\\
+ & ie\int\frac{\textrm{d}^{4}k}{\left(2\pi\right)^{4}}e^{ik.\left(x-y\right)}\frac{i}{k^{2}+i\epsilon}\int\frac{\textrm{d}^{4}p}{\left(2\pi\right)^{4}}e^{ip.\left(y-z\right)}S_{F}\left(p\right)\\
- & ie\int\frac{\textrm{d}^{4}k}{\left(2\pi\right)^{4}}e^{ik.\left(x-z\right)}\frac{i}{k^{2}+i\epsilon}\int\frac{\textrm{d}^{4}p}{\left(2\pi\right)^{4}}e^{ip.\left(y-z\right)}S_{F}\left(p\right)\\
= & 0\end{array}\label{3.4.24}\end{equation}
Next we rearrange \begin{equation}
\begin{array}{rcl}
e^{ik.\left(x-y\right)}e^{ip.\left(y-z\right)} & = & e^{ik.\left(x-z\right)}e^{ik.\left(z-y\right)}e^{ip.\left(y-z\right)}\\
 & = & e^{ik.\left(x-z\right)}e^{i\left(p-k\right).\left(y-z\right)}\end{array}\label{3.4.25}\end{equation}
 and make a shift in the variables of integration to $p'=p-k$ to
yield\begin{equation}
\frac{ik_{\mu}}{\zeta}\Gamma^{\mu}\left(k;p,-k-p\right)+ie\frac{i}{k^{2}+i\epsilon}S_{F}\left(p+k\right)-ie\frac{i}{k^{2}+i\epsilon}S_{F}\left(p\right)=0.\label{3.4.26}\end{equation}
 
\vspace{0.3cm}

\vspace{0.3cm}
\begin{center}$\Gamma^{\mu}$=\includegraphics[%
  width=0.40\textwidth,
  height=0.20\textwidth]{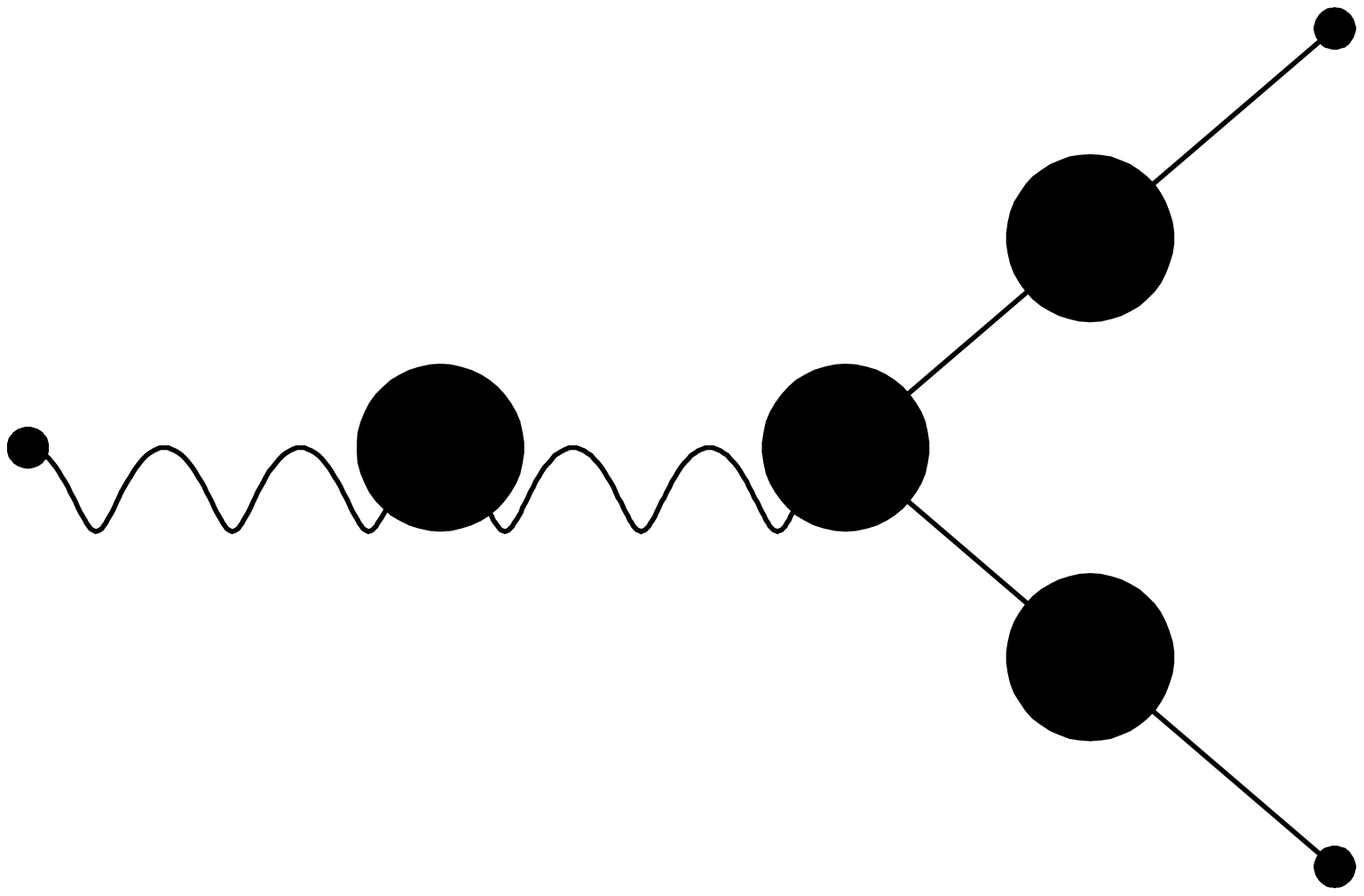}\begin{equation}
=D^{\mu\lambda}S_{F}\left(p\right)\tilde{\Gamma}_{\lambda}S_{F}\left(p+k\right).\label{3.4.27}\end{equation}
 \end{center}
\vspace{0.3cm}

\noindent Therefore the Ward identity is giving us the result that,\begin{equation}
\frac{i}{\zeta}k_{\mu}D^{\mu\lambda}\tilde{\Gamma}_{\lambda}+ie\frac{i}{k^{2}+i\epsilon}S_{F}^{-1}\left(p\right)-ie\frac{i}{k^{2}+i\epsilon}S_{F}^{-1}\left(p+k\right)=0\label{3.4.28}\end{equation}
From before we found that $\frac{i}{\zeta}k_{\mu}D^{\mu\lambda}=\frac{k^{\lambda}}{k^{2}+i\epsilon}$\begin{equation}
\Rightarrow k^{\lambda}\tilde{\Gamma}_{\lambda}-ieS_{F}^{-1}\left(p\right)+ieS_{F}^{-1}\left(p+k\right)=0.\label{3.4.29}\end{equation}
Let's see if this makes sense at by plugging in the appropriate tree
level values, we have $\tilde{\Gamma}_{\lambda}=ie\gamma_{\lambda}$,
$S_{F}^{-1}=\frac{\not p-m}{i}$ and so we must have, \begin{equation}
ie\not k-\frac{1}{i}e\left(\not p-m\right)+\frac{1}{i}e\left(\not p+\not k-m\right)=0\label{3.4.30}\end{equation}
 according to the Ward identity. It is trivial to add up the terms
on the left hand side and see that this is in fact the case. Beyond
tree level we have, $\tilde{\Gamma}_{\lambda}=ie\gamma_{\lambda}+\tilde{\Gamma}'_{\lambda}$
where $\tilde{\Gamma}'$ is representing the loop corrections to the
tree level vertex function and for the two point function, $S_{F}^{-1}\left(p\right)=\frac{\not p-m}{i}+\Sigma\left(p\right)$
where $\Sigma\left(p\right)$ represents the loop corrections to the
fermion propagator (the self energy). Plugging these into the Ward
identity above gives the following relation:\begin{equation}
k^{\lambda}\tilde{\Gamma}'_{\lambda}+e\Sigma\left(p+k\right)-e\Sigma\left(p\right)=0.\label{3.4.31}\end{equation}

\chapter{Renormalization and QED. }

\section{One Loop Correction to the Photon Propagator.}

We now discuss loop corrections and renormalization in QED. Firstly
we discuss the photon two-point function:

\vspace{0.375cm}
\begin{center}\includegraphics[%
  width=0.30\paperwidth,
  height=0.15\paperwidth]{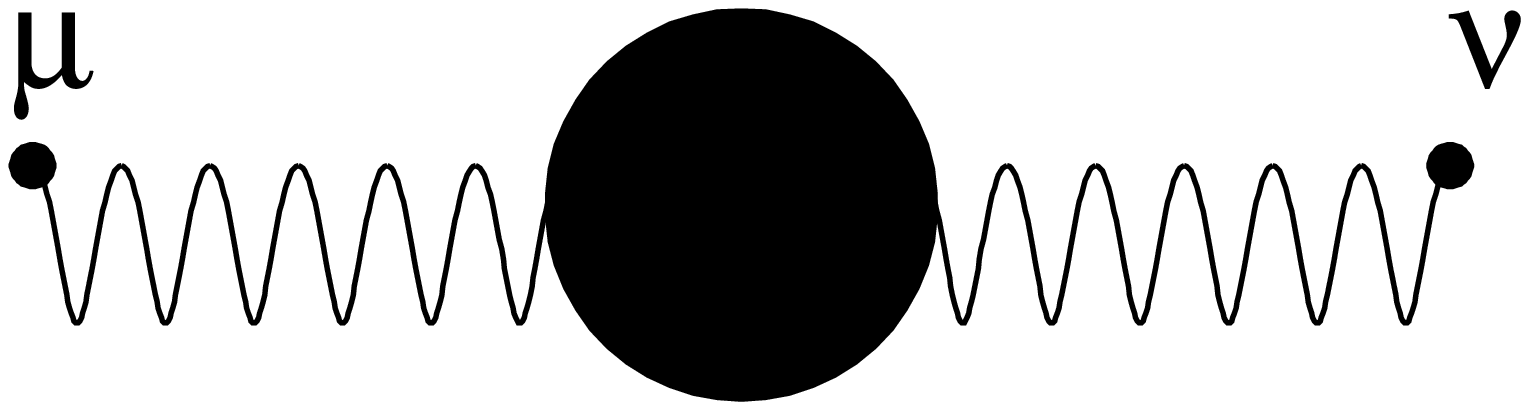}\end{center}
\vspace{0.375cm}

\noindent which is superficially divergent%
\footnote{The superficial degree is an exercise in counting the number of powers
of the loop momentum $\left(k\right)$ appearing in the amplitude
for a given process. In QED we have a $\textrm{d}^{4}k$ integration
measure and a $\frac{1}{k}$ for every internal fermion and $\frac{1}{k^{2}}$
for every internal boson. So in QED we have that a diagram has superficial
degree of divergence $D$ if $D=4-Internal\textrm{ }Loop\textrm{ }Fermions-Internal\textrm{ }Loop\textrm{ }Bosons\ge0$ %
} $\left(4-Internal\textrm{ }Loop\textrm{ }Fermions-2\times Internal\textrm{ }Loop\textrm{ }Bosons\right)=2\ge0$.
From last lecture we had that the Ward identity gave us that the self
energy $\Pi^{\mu\nu}\propto P_{T}^{\mu\nu}\Pi\left(k^{2}\right)$.
It is worth noting that the BRS symmetry which we used to derive this
(and other Ward identities) is independent of the number of space-time
dimensions. This implies that one can use dimensional regularization
techniques to treat ultraviolet divergences. Before proceeding with
the calculation of the full two point function in dimensional regularization
let us first set up some conventions. Our gauge fixed QED action in
$n$ dimensions is, \begin{equation}
S=\int\textrm{d}^{n}x\textrm{ }-\frac{1}{4}F_{\mu\nu}F^{\mu\nu}+i\bar{\psi}\not\partial\psi+e\bar{\psi}\not A\psi-m\bar{\psi}\psi+\frac{1}{2\zeta}\left(\partial_{\mu}A^{\mu}\right)^{2}+\eta\partial^{\mu}\partial_{\mu}\omega\label{4.1.1}\end{equation}
 We must have that the action is dimensionless, this means the quantities
inside the action must have the following dimensions (we use $\left[X\right]$
to denote the dimensions of $X$):\begin{equation}
\begin{array}{lclcl}
\left[d^{n}x\right] & = & n\\
\left[A\right] & = & \frac{n-2}{2}\\
\left[\psi\right] & = & \frac{n-1}{2}\\
\left[\bar{\psi}\not A\psi\right] & = & \frac{n-2}{2}+n-1 & = & \frac{3n}{2}-2\\
\left[e\right] & = & n-\left(\frac{3n}{2}-2\right) & = & 2-\frac{n}{2}\end{array}\label{4.1.2}\end{equation}
We will ultimately want to look at the limit $n\rightarrow4$ so (as
usual in dimensional regularization) we write, $n=4-2\epsilon$, where
$\epsilon$ is some small positive quantity. Note that $e$ is dimensionless,
a dimensionless coupling constant is a prerequisite of a renormalizable
theory. In defining a renormalized theory it is necessary to introduce
a mass scale $\mu$, known as the t'Hooft scaling parameter, which
will let us keep $e$ dimensionless through dimensional regularization,
we write $e'=e\mu^{2-\frac{n}{2}}=e\mu^{\epsilon}$. The $\mu$ essentially
compensates for the changing of the dimensions of quantities we will
calculate due to the changing of dimensions of the integration measure,
$d^{n}x$. Introducing the t'Hooft scale parameter changes our Feynman
rules slightly.

\begin{itemize}
\item Propagators $\rightarrow$ No change, we had calculated free propagators
which never depended on $e$ in the first place.
\item \includegraphics[%
  clip,
  width=0.1\paperwidth,
  height=0.1\paperwidth]{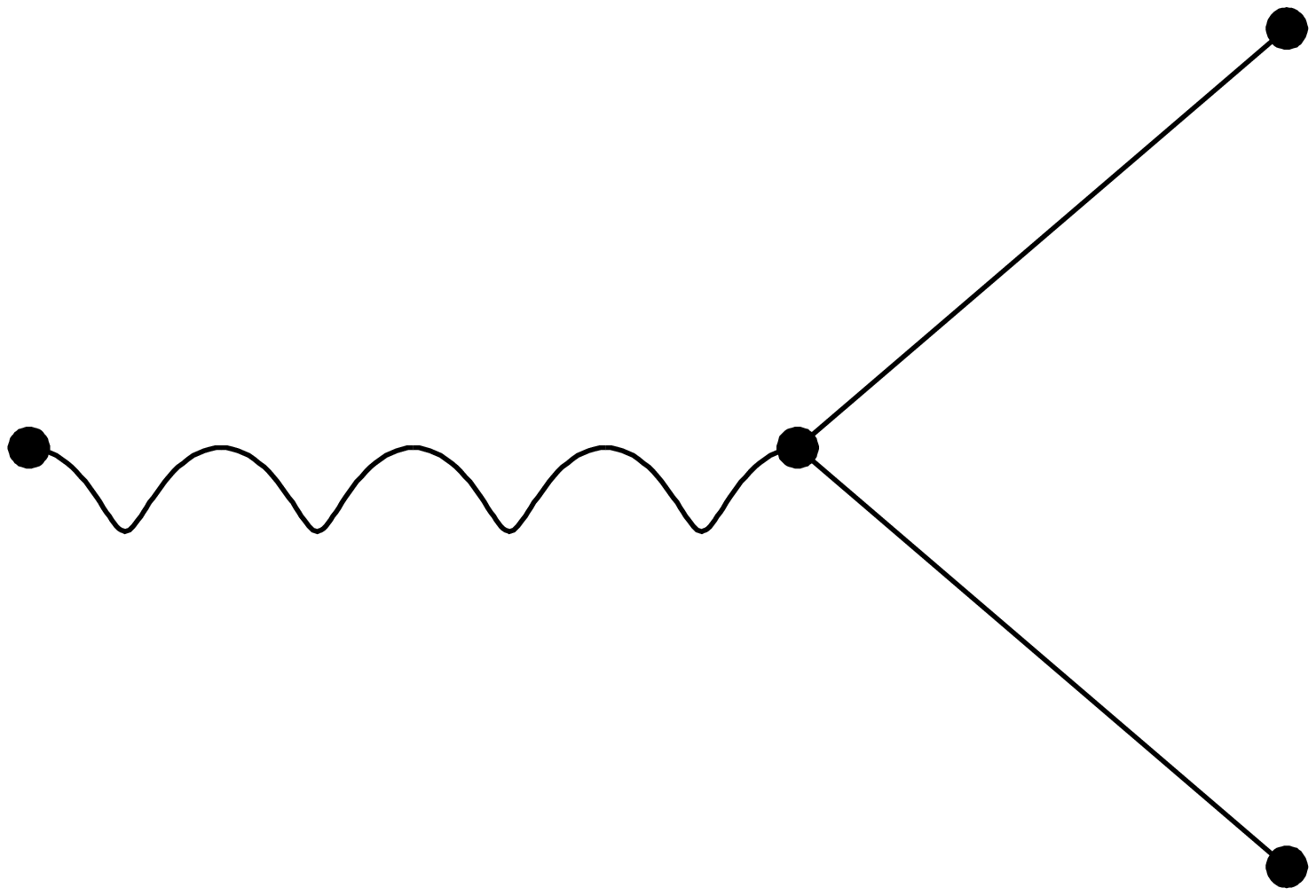}Vertices, $ie\gamma^{\mu}\rightarrow ie\mu^{\epsilon}\gamma^{\mu}$.
\end{itemize}
The following rules also depended on the dimensionality of space-time
and change as follows.

\begin{itemize}
\item $\int\textrm{d}^{4}x\rightarrow\int\textrm{d}^{n}x$, $\int\frac{\textrm{d}^{4}p}{\left(2\pi\right)^{4}}\rightarrow\int\frac{d^{n}p}{\left(2\pi\right)^{n}}$
.
\item $g^{\mu\nu}g_{\nu\mu}=n$
\item $Tr\left[I\right]=4\textrm{ }\rightarrow\textrm{ }Tr\left[I\right]=n$
\item $\gamma^{\mu}\gamma^{\nu}\gamma_{\mu}=-2\gamma^{\nu}\textrm{ }\rightarrow\textrm{ }\gamma^{\mu}\gamma^{\nu}\gamma_{\mu}=-2\left(1-\epsilon\right)\gamma^{\nu}$
\item $\gamma^{\mu}\gamma^{\rho}\gamma^{\sigma}\gamma_{\mu}=4g^{\rho\sigma}\textrm{ }\rightarrow\textrm{ }\gamma^{\mu}\gamma^{\rho}\gamma^{\sigma}\gamma_{\mu}=4g^{\rho\sigma}-2\epsilon\gamma^{\rho}\gamma^{\sigma}$
\item $\gamma^{\mu}\gamma^{\rho}\gamma^{\sigma}\gamma^{\tau}\gamma_{\mu}=-2\gamma^{\tau}\gamma^{\sigma}\gamma^{\rho}\textrm{ }\rightarrow\textrm{ }\gamma^{\mu}\gamma^{\rho}\gamma^{\sigma}\gamma^{\tau}\gamma_{\mu}=-2\gamma^{\tau}\gamma^{\sigma}\gamma^{\rho}+2\epsilon\gamma^{\rho}\gamma^{\sigma}\gamma^{\tau}$
\item Traces of $\gamma$ matrices $\rightarrow$ Not changed unless the
trace contains a $\gamma^{5}$ which depends on the space-time dimensionality. 
\end{itemize}
Now let's get on with our analysis of the photon two point function.
Using our (modified) Feynman rules we can write down the amplitude,

\vspace{0.375cm}
\begin{center}\includegraphics[%
  width=0.30\paperwidth,
  height=0.15\paperwidth]{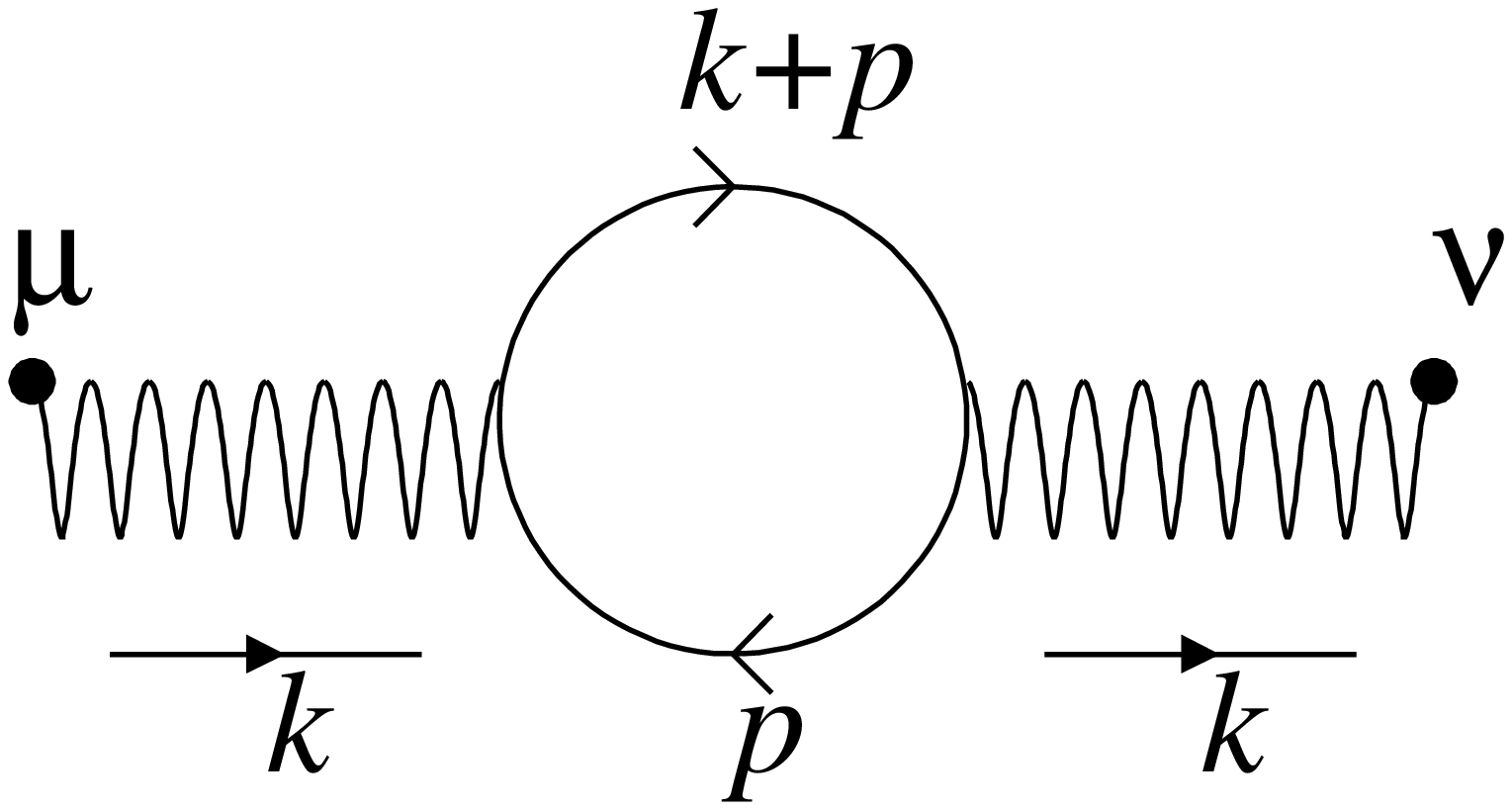}\end{center}

\noindent \begin{equation}
i\Pi^{\mu\nu}\left(k\right)=-e^{2}\mu^{2\epsilon}\int\frac{d^{n}p}{\left(2\pi\right)^{n}}\frac{Tr\left[\gamma^{\mu}\left(\not k+\not p+m\right)\gamma^{\nu}\left(\not p+m\right)\right]}{\left(\left(k+p\right)^{2}-m^{2}\right)\left(p^{2}-m^{2}\right)}.\label{4.1.3}\end{equation}
Counting the powers of the loop momentum $p$, in the numerator we
get $n+2$ ($n$ from the integration measure and two inside the trace),
the denominator clearly has four so the amplitude looks badly divergent
in $n=4$ but appears to exist/be well defined for $n<2$. It turns
out however that this divergence is an illusion, the Ward identity,\begin{equation}
\Pi^{\mu\nu}\left(k\right)=\left(k^{2}g_{\mu\nu}-k_{\mu}k_{\nu}\right)\Pi\left(k^{2}\right)\label{4.1.4}\end{equation}
 kills the divergence above. We just discussed above that $\Pi^{\mu\nu}$
has dimensions of momentum $n-2$, so in $n=4$ we have $\left[\Pi^{\mu\nu}\right]=2$
in which case the above equation tells us that $\left[\Pi\right]=0$. 
\vspace{0.375cm}

First we  use the fact that the trace of an odd number of gamma matrices
vanishes to halve the number of terms. Finally we use the identities
$Tr\left[\gamma^{\mu}\gamma^{\nu}\gamma^{\rho}\gamma^{\sigma}\right]=4\left(g^{\mu\nu}g^{\rho\sigma}-g^{\mu\rho}g^{\nu\sigma}+g^{\mu\sigma}g^{\nu\rho}\right)$
and $Tr\left[\gamma^{\mu}\gamma^{\nu}\right]=4g^{\mu\nu}$ to give
us a sum products of n-vectors. \begin{equation}
\begin{array}{rcl}
Tr\left[\gamma^{\mu}\left(\not k+\not p+m\right)\gamma^{\nu}\left(\not p+m\right)\right] & = & Tr\left[\gamma^{\mu}\not k\gamma^{\nu}\not p+\gamma^{\mu}\not p\gamma^{\nu}\not p+m\gamma^{\mu}\gamma^{\nu}\not p\right.\\
 & = & \left.+m\gamma^{\mu}\not k\gamma^{\nu}+m\gamma^{\mu}\not p\gamma^{\nu}+m^{2}\gamma^{\mu}\gamma^{\nu}\right]\\
 & = & Tr\left[\gamma^{\mu}\not k\gamma^{\nu}\not p+\gamma^{\mu}\not p\gamma^{\nu}\not p+m^{2}\gamma^{\mu}\gamma^{\nu}\right]\\
 & = & 4\left(k^{\mu}p^{\nu}-g^{\mu\nu}k.p+k^{\nu}p^{\mu}\right)+4\left(p^{\mu}p^{\nu}-g^{\mu\nu}p^{2}+p^{\mu}p^{\nu}\right)+4m^{2}g^{\mu\nu}\\
 & = & 4k^{\mu}p^{\nu}+4k^{\nu}p^{\mu}-4g^{\mu\nu}k.p+8p^{\mu}p^{\nu}-4\left(p^{2}-m^{2}\right)g^{\mu\nu}\end{array}\label{4.1.5}\end{equation}
The next thing to do is rewrite the denominator in terms of integrals
over Feynman parameters, this will eventually let us integrate out
our loop momentum. To do this we need the identity\begin{equation}
\frac{1}{A_{1}A_{2}...A_{n}}=\int_{0}^{1}\textrm{d}x_{1}\textrm{d}x_{2}...\textrm{d}x_{n}\textrm{ }\delta\left(\Sigma x_{i}-1\right)\frac{\left(n-1\right)!}{\left[x_{1}A_{1}+x_{2}A_{2}+...+x_{n}A_{n}\right]^{n}}.\label{4.1.6}\end{equation}
In our case we have $A_{1}=\left(\left(k+p\right)^{2}-m^{2}\right)$
and $A_{2}=\left(p^{2}-m^{2}\right)$:\begin{equation}
\begin{array}{rcl}
\frac{1}{\left(\left(k+p\right)^{2}-m^{2}\right)\left(p^{2}-m^{2}\right)} & = & \int_{0}^{1}\textrm{d}xdy\textrm{ }\delta\left(x+y-1\right)\textrm{ }\frac{1}{\left(x\left(\left(k+p\right)^{2}-m^{2}\right)+y\left(p^{2}-m^{2}\right)\right)^{2}}\\
 & = & \int_{0}^{1}\textrm{d}x\textrm{ }\frac{1}{\left(x\left(\left(k+p\right)^{2}-m^{2}\right)+\left(1-x\right)\left(p^{2}-m^{2}\right)\right)^{2}}\end{array}\label{4.1.7}\end{equation}
Now we try to complete the square in $p$ in the denominator,\begin{equation}
\begin{array}{rcl}
x\left(\left(k+p\right)^{2}-m^{2}\right)+\left(1-x\right)\left(p^{2}-m^{2}\right) & = & xk^{2}+xp^{2}+2xk.p-xm^{2}+p^{2}-m^{2}-xp^{2}+xm^{2}\\
 & = & p^{2}+2xk.p+xk^{2}-m^{2}\\
 & = & \left(p+xk\right)^{2}-x^{2}k^{2}+xk^{2}-m^{2}\\
 & = & \left(p+xk\right)^{2}+xk^{2}\left(1-x\right)-m^{2}\\
 & = & \tilde{p}^{2}-\left(xk^{2}\left(x-1\right)+m^{2}\right)\end{array}\label{4.1.8}\end{equation}
 In the last line we have defined $\tilde{p}=p+xk$, this amounts
to a simple constant shift in the integration variable. Seeing as
the shift is constant we have that the integration measure is invariant
$\int\textrm{d}^{n}p=\int\textrm{d}^{n}\tilde{p}$ and so we can equivalently
integrate over $\tilde{p}$, which makes life easier. The photon self
energy can therefore be written, \begin{equation}
\begin{array}{rcl}
\Pi^{\mu\nu}\left(k\right) & = & ie^{2}\mu^{2\epsilon}\int\frac{d^{n}p}{\left(2\pi\right)^{n}}\frac{Tr\left[\gamma^{\mu}\left(\not k+\not p+m\right)\gamma^{\nu}\left(\not p+m\right)\right]}{\left(\left(k+p\right)^{2}-m^{2}\right)\left(p^{2}-m^{2}\right)}\\
 & = & ie^{2}\mu^{2\epsilon}\int_{0}^{1}\textrm{d}x\int\frac{d^{n}\tilde{p}}{\left(2\pi\right)^{n}}\frac{4k^{\mu}\tilde{p}^{\nu}-4xk^{\mu}k^{\nu}+4k^{\nu}\tilde{p}^{\mu}-4xk^{\nu}k^{\mu}-4g^{\mu\nu}k.\tilde{p}+4xg^{\mu\nu}k^{2}+8\left(\tilde{p}-xk\right)^{\mu}\left(\tilde{p}-xk\right)^{\nu}-4\left(\left(\tilde{p}-xk\right)^{2}-m^{2}\right)g^{\mu\nu}}{\left(\tilde{p}^{2}-\left(xk^{2}\left(x-1\right)+m^{2}\right)\right)^{2}}\end{array}\label{4.1.9}\end{equation}
 As the integral is symmetric in $\tilde{p}$ we can drop terms in
the numerator which are odd powers of $\tilde{p}$. \begin{equation}
\begin{array}{rcl}
\Pi^{\mu\nu}\left(k\right) & = & ie^{2}\mu^{2\epsilon}\int_{0}^{1}\textrm{d}x\int\frac{d^{n}\tilde{p}}{\left(2\pi\right)^{n}}\frac{-4xk^{\mu}k^{\nu}-4xk^{\nu}k^{\mu}+4xg^{\mu\nu}k^{2}+8\tilde{p}^{\mu}\tilde{p}^{\nu}+8x^{2}k^{\mu}k^{\nu}-4\left(\tilde{p}^{2}+x^{2}k^{2}-m^{2}\right)g^{\mu\nu}}{\left(\tilde{p}^{2}-\left(xk^{2}\left(x-1\right)+m^{2}\right)\right)^{2}}\\
 & = & ie^{2}\mu^{2\epsilon}\int_{0}^{1}\textrm{d}x\int\frac{d^{n}\tilde{p}}{\left(2\pi\right)^{n}}\frac{-8xk^{\mu}k^{\nu}+4xg^{\mu\nu}k^{2}+8\tilde{p}^{\mu}\tilde{p}^{\nu}+8x^{2}k^{\mu}k^{\nu}-4\tilde{p}^{2}g^{\mu\nu}-4x^{2}k^{2}g^{\mu\nu}+4m^{2}g^{\mu\nu}}{\left(\tilde{p}^{2}-\left(xk^{2}\left(x-1\right)+m^{2}\right)\right)^{2}}\\
 & = & 4ie^{2}\mu^{2\epsilon}\int_{0}^{1}\textrm{d}x\int\frac{d^{n}\tilde{p}}{\left(2\pi\right)^{n}}\frac{-g^{\mu\nu}\tilde{p}^{2}-2x\left(1-x\right)k^{\nu}k^{\mu}+2\tilde{p}^{\mu}\tilde{p}^{\nu}+g^{\mu\nu}\left(xk^{2}\left(1-x\right)+m^{2}\right)}{\left(\tilde{p}^{2}-\left(xk^{2}\left(x-1\right)+m^{2}\right)\right)^{2}}\end{array}\label{4.1.10}\end{equation}
 The symmetry of the integral also then allows one to replace $\tilde{p}^{\mu}\tilde{p}^{\nu}\rightarrow\frac{1}{n}\tilde{p}^{2}g^{\mu\nu}$
in the numerator (terms off diagonal in $\mu$ and $\nu$ look linear
in their components with respect to whatever component of $p$ is
being integrated over). Defining $\Delta=xk^{2}\left(x-1\right)+m^{2}$
we have,\begin{equation}
\Pi^{\mu\nu}\left(k\right)=4ie^{2}\mu^{2\epsilon}\int_{0}^{1}\textrm{d}x\int\frac{d^{n}\tilde{p}}{\left(2\pi\right)^{n}}\frac{\left(\frac{2}{n}-1\right)\tilde{p}^{2}g^{\mu\nu}}{\left(\tilde{p}^{2}-\Delta\right)^{2}}-\frac{2x\left(1-x\right)k^{\mu}k^{\nu}-g^{\mu\nu}\left(m^{2}+xk^{2}\left(1-x\right)\right)}{\left(\tilde{p}^{2}-\Delta\right)^{2}}.\label{4.1.11}\end{equation}

These $n$-dimensional integrals are now in a form that is readily
found in tables of integrals,%
\footnote{Note these integrals are in Minkowski space.%
} \begin{equation}
\begin{array}{rcl}
\int\frac{d^{n}l}{\left(2\pi\right)^{n}}\frac{1}{\left(l^{2}-\Delta\right)^{m}} & = & \frac{\left(-1\right)^{m}i}{\left(4\pi\right)^{n/2}}\frac{\Gamma\left(m-\frac{n}{2}\right)}{\Gamma\left(m\right)}\left(\frac{1}{\Delta}\right)^{m-\frac{n}{2}}\\
\int\frac{d^{n}l}{\left(2\pi\right)^{n}}\frac{l^{2}}{\left(l^{2}-\Delta\right)^{m}} & = & \frac{\left(-1\right)^{m-1}i}{\left(4\pi\right)^{n/2}}\frac{n}{2}\frac{\Gamma\left(m-\frac{n}{2}-1\right)}{\Gamma\left(m\right)}\left(\frac{1}{\Delta}\right)^{m-\frac{n}{2}-1}\end{array}\label{4.1.12}\end{equation}
Applying these identities to the two terms in our  energy we get (remember
in our convention $n=4-2\epsilon$, $\epsilon=2-\frac{n}{2}$), \begin{equation}
\Rightarrow\begin{array}{rcl}
\int\frac{d^{n}\tilde{p}}{\left(2\pi\right)^{n}}\frac{2x\left(1-x\right)k^{\mu}k^{\nu}-g^{\mu\nu}\left(m^{2}+xk^{2}\left(1-x\right)\right)}{\left(\tilde{p}^{2}-\Delta\right)^{2}} & = & \frac{i\left(2x\left(1-x\right)k^{\mu}k^{\nu}-g^{\mu\nu}\left(m^{2}+xk^{2}\left(1-x\right)\right)\right)}{\left(4\pi\right)^{2-\epsilon}}\frac{\Gamma\left(\epsilon\right)}{\Gamma\left(2\right)}\left(\frac{1}{\Delta}\right)^{\epsilon}\\
\int\frac{d^{n}\tilde{p}}{\left(2\pi\right)^{n}}\frac{\left(\frac{2}{n}-1\right)g^{\mu\nu}\tilde{p}^{2}}{\left(\tilde{p}^{2}-\Delta\right)^{2}} & = & \left(\frac{2}{n}-1\right)g^{\mu\nu}\frac{-i}{\left(4\pi\right)^{n/2}}\frac{n}{2}\frac{\Gamma\left(1-\frac{n}{2}\right)}{\Gamma\left(2\right)}\left(\frac{1}{\Delta}\right)^{1-\frac{n}{2}}\\
 & = & \frac{-ig^{\mu\nu}\left(1-\frac{n}{2}\right)}{\left(4\pi\right)^{n/2}}\frac{\Gamma\left(1-\frac{n}{2}\right)}{\Gamma\left(2\right)}\left(\frac{1}{\Delta}\right)^{1-\frac{n}{2}}\end{array}\label{4.1.13}\end{equation}
A special property of the Euler gamma function is that $z\Gamma\left(z\right)=\Gamma\left(z+1\right)$
\emph{i.e.} $\left(1-\frac{n}{2}\right)\Gamma\left(1-\frac{n}{2}\right)=\Gamma\left(2-\frac{n}{2}\right)$.
If we use this and also the fact that $\Gamma\left(2\right)=1$ second
integral above becomes,\begin{equation}
\begin{array}{rcl}
\int\frac{d^{n}\tilde{p}}{\left(2\pi\right)^{n}}\frac{\left(\frac{2}{n}-1\right)g^{\mu\nu}\tilde{p}^{2}}{\left(\tilde{p}_{E}^{2}-\Delta\right)^{2}} & = & \frac{-ig^{\mu\nu}}{\left(4\pi\right)^{n/2}}\Gamma\left(2-\frac{n}{2}\right)\left(\frac{1}{\Delta}\right)^{2-\frac{n}{2}}\Delta\\
 & = & \frac{-ig^{\mu\nu}}{\left(4\pi\right)^{2-\epsilon}}\Gamma\left(\epsilon\right)\left(\frac{1}{\Delta}\right)^{\epsilon}\Delta\end{array}\label{4.1.14}\end{equation}
Inserting these integrals into $\Pi^{\mu\nu}$ we have,\begin{equation}
\Pi^{\mu\nu}\left(k\right)=-4e^{2}\mu^{2\epsilon}\left(4\pi\right)^{\epsilon-2}\int_{0}^{1}\textrm{d}x\textrm{ }\left(-g^{\mu\nu}\Delta-2x\left(1-x\right)k^{\mu}k^{\nu}+g^{\mu\nu}\left(m^{2}+xk^{2}\left(1-x\right)\right)\right)\Gamma\left(\epsilon\right)\left(\frac{1}{\Delta}\right)^{\epsilon}.\label{4.1.15}\end{equation}
Substituting in for $\Delta$, \begin{equation}
\begin{array}{rcl}
\Pi^{\mu\nu}\left(k\right) & = & \frac{-4e^{2}\mu^{2\epsilon}}{\left(4\pi\right)^{2-\epsilon}}\int_{0}^{1}\textrm{d}x\textrm{ }\left(-g^{\mu\nu}\left(xk^{2}\left(x-1\right)+m^{2}\right)-2x\left(1-x\right)k^{\mu}k^{\nu}+g^{\mu\nu}\left(m^{2}+xk^{2}\left(1-x\right)\right)\right)\Gamma\left(\epsilon\right)\left(\frac{1}{\Delta}\right)^{\epsilon}\\
 & = & \frac{-e^{2}}{4\pi^{2}}\int_{0}^{1}\textrm{d}x\textrm{ }2x\left(1-x\right)\left(g^{\mu\nu}k^{2}-k^{\mu}k^{\nu}\right)\Gamma\left(\epsilon\right)\left(\frac{4\pi\mu^{2}}{\Delta}\right)^{\epsilon}\\
 & = & -k^{2}\left(g^{\mu\nu}-\frac{k^{\mu}k^{\nu}}{k^{2}}\right)\left(\frac{e^{2}}{4\pi^{2}}\int_{0}^{1}\textrm{d}x\textrm{ }2x\left(1-x\right)\Gamma\left(\epsilon\right)\left(\frac{4\pi\mu^{2}}{\Delta}\right)^{\epsilon}\right)\\
 & = & -k^{2}P_{T}^{\mu\nu}\Pi\left(k^{2}\right)\end{array}\label{4.1.16}\end{equation}

The photon's self energy is proportional to the transverse projection
operator just like the Ward identity from the last lecture said it
was! Had we believed the magical Ward identity from lecture 8 we could
have accelerated the calculation dropping all terms not proportional
to $P_{T}^{\mu\nu}=g^{\mu\nu}-\frac{k^{\mu}k^{\nu}}{k^{2}}$. The
above $\Pi^{\mu\nu}\left(k\right)$ contains divergences as $n\rightarrow4$
\emph{i.e.} as $\epsilon\rightarrow0$, we shall now take a closer
look at these divergences. For small $\epsilon$, $\Gamma\left(\epsilon\right)$
is approximately given by, $\Gamma\left(\epsilon\right)\approx\frac{1}{\epsilon}-\gamma+O\left(\epsilon\right)$
where $\gamma$ is Euler's number $\left(\approx0.577\right)$. We
will also be using the following $A^{\epsilon}=\textrm{exp}\left(\epsilon\log A\right)\approx1+\epsilon\log A$
and hence we also approximate $\Gamma\left(\epsilon\right)A^{\epsilon}$
by $\Gamma\left(\epsilon\right)A^{\epsilon}\approx\frac{1}{\epsilon}+\log A-\gamma$.
Taking the limit $n\rightarrow4$ \emph{i.e.} using these $\Pi$ we
have, 

\begin{equation}
\begin{array}{rcl}
\Pi\left(k^{2}\right) & = & \lim_{\epsilon\rightarrow0}\frac{e^{2}}{2\pi^{2}}\int_{0}^{1}\textrm{d}x\textrm{ }x\left(1-x\right)\Gamma\left(\epsilon\right)\left(\frac{4\pi\mu^{2}}{\Delta}\right)^{\epsilon}\\
 & = & \lim_{\epsilon\rightarrow0}\frac{e^{2}}{2\pi^{2}}\int_{0}^{1}\textrm{d}x\textrm{ }x\left(1-x\right)\left(\frac{1}{\epsilon}+\log\left(\frac{\mu^{2}}{\Delta}\right)+\log\left(4\pi\right)-\gamma\right)\\
 & = & \lim_{\epsilon\rightarrow0}\frac{2\alpha}{\pi}\int_{0}^{1}\textrm{d}x\textrm{ }x\left(1-x\right)\left(\frac{1}{\epsilon}+\log\left(\frac{\mu^{2}}{m^{2}+xk^{2}\left(x-1\right)}\right)+\log\left(4\pi\right)-\gamma\right)\\
 & = & \lim_{\epsilon\rightarrow0}\frac{2\alpha}{\pi}\int_{0}^{1}\textrm{d}x\textrm{ }x\left(1-x\right)\log\left(\frac{\mu^{2}}{m^{2}+xk^{2}\left(x-1\right)}\right)+\frac{2\alpha}{\pi}\left(\frac{1}{\epsilon}+\log\left(4\pi\right)-\gamma\right)\left[\frac{1}{2}x^{2}-\frac{1}{3}x^{3}\right]_{0}^{1}\\
 & = & \lim_{\epsilon\rightarrow0}\frac{2\alpha}{\pi}\left(\frac{1}{6}\left(\frac{1}{\epsilon}+\log\left(4\pi\right)-\gamma\right)+\int_{0}^{1}\textrm{d}x\textrm{ }x\left(1-x\right)\log\left(\frac{\mu^{2}}{m^{2}+xk^{2}\left(x-1\right)}\right)\right)\end{array}\label{4.1.17}\end{equation}

\noindent where we have introduced the fine structure constant $\alpha$
which, in natural units, is equal to $\frac{e^{2}}{4\pi}$. 

Clearly the above is logarithmically divergent in the limit $n\rightarrow4$
dimensions as is indicated by the $\frac{1}{\epsilon}$ outside the
first term. This is bad at a superficial level clearly as it implies
that amplitude for our self energy Feynman diagram is infinite which
would predict infinity for various measurable quantities \emph{e.g.}
the cross section for $e^{+}e^{-}\rightarrow\mu^{+}\mu^{-}$ . To
obtain a sensible result we have to renormalize some parameters in
the theory \emph{i.e.} we will the \emph{Ultra}-\emph{Violet} infinity
(\emph{i.e.} those infinities that occur due to the integrand of a
momentum integral becoming infinite as the loop momentum becomes infinite)
by renormalizing the parameters of the theory.

\newpage
\section{Resummation of loops\label{sec:Resummation.}.}

Imagine the QED Lagrangian was rewritten,\begin{equation}
\begin{array}{rcl}
S & = & \int\textrm{d}^{n}x\textrm{ }-\frac{1}{4}F_{\mu\nu}F^{\mu\nu}+i\bar{\psi}\not\partial\psi+eZ_{1}\bar{\psi}\not A\psi-m\bar{\psi}\psi+\frac{1}{2\zeta}\left(\partial_{\mu}A^{\mu}\right)^{2}+\eta\partial^{\mu}\partial_{\mu}\omega\\
 & = & \int\textrm{d}^{n}x\textrm{ }-\frac{1}{4}Z_{3}F_{R,\mu\nu}F_{R}^{\mu\nu}+iZ_{2}\bar{\psi_{R}}\not\partial\psi_{R}+e_{R}Z_{1}\bar{\psi_{R}}\not A_{R}\psi_{R}\\
 &  & -m_{R}Z_{0}\bar{\psi_{R}}\psi_{R}+\frac{Z_{3}}{2\zeta_{R}Z_{\zeta}}\left(\partial_{\mu}A_{R}^{\mu}\right)^{2}+Z_{G}\eta_{R}\partial^{\mu}\partial_{\mu}\omega_{R}\end{array}\label{eq:4.2.1}\end{equation}
where $Z_{3},Z_{2},Z_{1},Z_{0},Z_{\zeta}$,$Z_{G}$ are real constants
and the subscript $R$ denotes a renormalized field or parameter.
This is not the best parametrization perhaps, it would be better to
have rewritten the action rescaling explicitly the fields and the
parameters by constants but we shall see that the above is essentially
the same. Consider the part relating to the fermions coupled to the
$A^{\mu}$ field. \begin{equation}
Z_{2}\left(i\bar{\psi_{R}}\not\partial\psi_{R}+e_{R}\frac{Z_{1}}{Z_{2}}\bar{\psi}_{R}\not A_{R}\psi_{R}-m_{R}\frac{Z_{0}}{Z_{2}}\bar{\psi}_{R}\psi_{R}\right)\label{4.2.2}\end{equation}
We could interpret $Z_{2}$ as a renormalization of the fermion field
$\psi=\sqrt{Z_{2}}\psi_{R}$, which would mean that $\frac{Z_{0}}{Z_{2}}$
represents a renormalization of the mass parameter, $m=\frac{Z_{0}}{Z_{2}}m_{R}$.
The renormalization of the charge $e$ and the $A^{\mu}$ field are
then somehow tied up in $\frac{Z_{1}}{Z_{2}}$ as indicated by the
second term above. We can untangle the renormalization of $e$ and
$A^{\mu}$ by looking at the first term in the action,\begin{equation}
\begin{array}{rcl}
-\frac{1}{4}Z_{3}F_{R,\mu\nu}F_{R}^{\mu\nu} & = & -\frac{1}{4}Z_{3}\left(\partial_{\mu}A_{R,\nu}-\partial_{\nu}A_{R,\mu}\right)\left(\partial^{\mu}A_{R}^{\nu}-\partial^{\nu}A_{R}^{\mu}\right)\\
 & = & -\frac{1}{4}Z_{3}\left(\left(\partial_{\mu}A_{R,\nu}\right)\left(\partial^{\mu}A_{R}^{\nu}\right)-\left(\partial_{\mu}A_{R\nu}\right)\left(\partial^{\nu}A_{R}^{\mu}\right)-\left(\partial_{\nu}A_{R,\mu}\right)\left(\partial^{\mu}A_{R}^{\nu}\right)+\left(\partial_{\nu}A_{R,\mu}\right)\left(\partial^{\nu}A_{R}^{\mu}\right)\right)\\
 & = & -\frac{1}{4}\left(\left(\partial_{\mu}A_{\nu}\right)\left(\partial^{\mu}A^{\nu}\right)-\left(\partial_{\mu}A_{\nu}\right)\left(\partial^{\nu}A^{\mu}\right)-\left(\partial_{\nu}A_{\mu}\right)\left(\partial^{\mu}A^{\nu}\right)+\left(\partial_{\nu}A_{\mu}\right)\left(\partial^{\nu}A^{\mu}\right)\right)\\
 & = & -\frac{1}{4}F_{\mu\nu}F^{\mu\nu}\end{array}\label{4.2.3}\end{equation}
where we have interpreted the re-parametrization as a renormalization
of the $A^{\mu}$ field as $A^{\mu}=\sqrt{Z_{3}}A_{R}^{\mu}$. Consequently,
from the term coupling the photon and the fermions $eZ_{1}\bar{\psi}\not A\psi$
we can deduce that the electric charge is being renormalized too.
In terms of the renormalized quantities this term will equal $e_{R}\bar{\psi}_{R}\not A_{R}\psi_{R}$.
Hence,\begin{equation}
\begin{array}{rcl}
e\bar{\psi}\not A\psi & = & e_{R}Z_{1}\bar{\psi_{R}}\not A_{R}\psi_{R}\\
 & = & e_{R}\frac{Z_{1}}{Z_{2}\sqrt{Z_{3}}}\bar{\psi}_{R}\not A_{R}\psi_{R}\\
\Rightarrow e & = & e_{R}\frac{Z_{1}}{Z_{2}\sqrt{Z_{3}}}\end{array}.\label{4.2.4}\end{equation}

\noindent Finally for the gauge fixing term we have \begin{equation}
\begin{array}{rcl}
\frac{1}{\zeta}\left(\partial^{\mu}A_{\mu}\right)^{2} & = & \frac{Z_{3}}{\zeta_{R}Z_{\zeta}}\left(\partial^{\mu}A_{R,\mu}\right)^{2}\\
 & = & \frac{1}{\zeta_{R}Z_{\zeta}}\left(\partial^{\mu}A_{\mu}\right)^{2}\\
\Rightarrow\zeta & = & Z_{\zeta}\zeta_{R}\end{array}.\label{4.2.5}\end{equation}

What have we done? We have multiplied all the terms in the Lagrangian
by six real numbers and reinterpreted this in terms of a \emph{renormalization}
of the fermion fields, the photon field, the electric charge and the
mass. These $Z$'s are where we hide the infinities that come up in
the loop diagrams. We interpret $S$ above as being the \emph{bare
action} with \emph{bare fields and parameters} (those quantities with
\emph{no} subscript $R$). This raises some questions. Does multiplying
all these constants into the Lagrangian \emph{ad hoc} break the original
gauge invariance (neglecting the gauge fixing term of course)? At
face value gauge invariance is completely broken, however we have
not specified the values of these numbers. We shall later essentially
demand gauge invariance of this Lagrangian by imposing the Ward identities
which will result in relations between the $Z's$ such that the gauge
invariance of the Lagrangian is restored. What about the gauge fixing
term? We do not need to worry about the gauge fixing term, though
it is influenced by quantum corrections, as all of our S-Matrix elements
are gauge independent anyway \emph{i.e.} we can effectively set $Z_{\zeta}=1$.
What about the ghosts? Again we don't have to worry about what happens
to the ghosts as trivially in QED the ghosts are decoupled from the
rest of the theory and go around as non-interacting (unphysical) complex
scalars i.e. there is no renormalization of ghosts as they have no
interactions $Z_{G}=1$. Finally we might worry that we broke BRS
invariance? The answer to this is analogous to the answer about gauge
invariance, \emph{i.e.} we will impose the Ward identities on the
one loop divergent diagrams and this is tantamount to demanding BRS
invariance as that is where the Ward identities came from. Hopefully
the rest of this section will give weight to these answers.

Let's try and work out what our one loop calculation would give using
the renormalized Lagrangian. The renormalization should affect the
form of the Feynman rules. We derived the photon propagator at the
start of lecture 7. To do that we rewrote the terms relating to the
free photon in the form $A^{\mu}Q_{\mu\nu}A^{\nu}$, then the propagator
was found to be the inverse of the differential operator $Q_{\mu\nu}$
in Fourier space. We want to know how our renormalization coefficients
affect the form of the propagator. Previously our starting point was
therefore,

\begin{equation}
\int\textrm{d}^{4}x\textrm{ }-\frac{1}{4}\left(\partial_{\mu}A\left(x\right)_{\nu}-\partial_{\nu}A\left(x\right)_{\mu}\right)\left(\partial^{\mu}A^{\nu}\left(x\right)-\partial^{\nu}A^{\mu}\left(x\right)\right)-\frac{1}{2\zeta}\left(\partial_{\mu}A^{\mu}\left(x\right)\right)^{2}\label{4.2.6}\end{equation}
which has now become,\begin{equation}
\int\textrm{d}^{4}x\textrm{ }-\frac{Z_{3}}{4}\left(\partial_{\mu}A_{R}\left(x\right)_{\nu}-\partial_{\nu}A_{R}\left(x\right)_{\mu}\right)\left(\partial^{\mu}A_{R}^{\nu}\left(x\right)-\partial^{\nu}A_{R}^{\mu}\left(x\right)\right)-\frac{Z_{3}}{2\zeta_{R}Z_{\zeta}}\left(\partial_{\mu}A_{R}^{\mu}\left(x\right)\right)^{2}.\label{4.2.7}\end{equation}
We perform the same steps as before to get the propagator. First we
put the free photon part of the Lagrangian in the form $A^{\mu}Q_{\mu\nu}A^{\nu}$\begin{equation}
\int\textrm{d}^{4}x\textrm{ }-\frac{Z_{3}}{4}\left(2\partial_{\mu}A_{R,\nu}\left(x\right)\partial^{\mu}A_{R}^{\nu}\left(x\right)-2\partial_{\nu}A_{R,\mu}\left(x\right)\partial^{\mu}A_{R}^{\nu}\left(x\right)+\frac{2}{\zeta_{R}Z_{\zeta}}\partial_{\mu}A_{R}^{\mu}\left(x\right)\partial_{\nu}A_{R}^{\nu}\left(x\right)\right),\label{4.2.8}\end{equation}
and integrate by parts\begin{equation}
\begin{array}{rl}
= & \int\textrm{d}^{4}x\textrm{ }-\frac{Z_{3}}{2}\left(-A_{R,\nu}\left(x\right)\partial_{\mu}\partial^{\mu}A_{R}^{\nu}\left(x\right)+A_{R,\mu}\left(x\right)\partial^{\nu}\partial^{\mu}A_{R,\nu}\left(x\right)-\frac{1}{\zeta_{R}Z_{\zeta}}A_{R}^{\mu}\left(x\right)\partial_{\mu}\partial_{\nu}A_{R}^{\nu}\left(x\right)\right)\\
= & \int\textrm{d}^{4}x\textrm{ }-\frac{Z_{3}}{2}\left(A_{R,\lambda}\left(x\right)\left(-g^{\lambda\rho}\partial^{\mu}\partial_{\mu}+\left(1-\frac{1}{\zeta_{R}Z_{\zeta}}\right)\partial^{\lambda}\partial^{\rho}\right)A_{R,\rho}\left(x\right)\right)\end{array}.\label{4.2.9}\end{equation}

\noindent Thus once again we find we need to invert the differential
operator between the two photon we fields. We proceed exactly as before,
we need to solve $\tilde{\Delta}_{\nu\kappa}\left(k\right)Z_{3}\left(g^{\mu\nu}k^{2}-\left(1-\frac{1}{\zeta_{R}Z_{\zeta}}\right)k^{\mu}k^{\nu}\right)=i\delta_{\kappa}^{\mu}$
for $\tilde{\Delta}_{\nu\kappa}\left(k\right)$. \begin{equation}
Q^{\lambda\rho}=Z_{3}\left(-g^{\lambda\rho}+\frac{k^{\lambda}k^{\rho}}{k^{2}}\right)k^{2}-\frac{Z_{3}}{\zeta_{R}Z_{\zeta}}.\frac{k^{\lambda}k^{\rho}}{k^{2}}.k^{2}=-k^{2}Z_{3}P_{T}^{\lambda\rho}-\frac{k^{2}Z_{3}}{\zeta_{R}Z_{\zeta}}P_{L}^{\lambda\rho}.\label{4.2.10}\end{equation}

\noindent Where we have defined the projection operators, $P_{T}^{\lambda\rho}=g^{\lambda\rho}-\frac{k^{\lambda}k^{\rho}}{k^{2}}\textrm{ and }P_{L}^{\lambda\rho}=\frac{k^{\lambda}k^{\rho}}{k^{2}}$,
just as before. Consequently, as before we have $P_{T}P_{T}=P_{T}$,
$P_{L}P_{L}=P_{L}$ and $P_{L}P_{T}=0=P_{T}P_{L}$ so $P_{T}$ and
$P_{L}$ again project out orthogonal subspaces. We assume that $Q^{-1}$
can be written as a linear combination of these two operators: \begin{equation}
Q^{-1\textrm{ }\rho\mu}=AP_{T}^{\rho\mu}+BP_{L}^{\rho\mu}.\label{4.2.11}\end{equation}
 Naturally one expects $QQ^{-1}$ will give the identity or equivalently
$P_{T}+P_{L}$:\begin{equation}
\begin{array}{rcl}
Q^{\textrm{ }\lambda\rho}g_{\rho\kappa}Q^{-1\textrm{ }\kappa\mu} & = & -Z_{3}k^{2}\left(P_{T}^{\lambda\rho}+\frac{1}{\zeta_{R}Z_{\zeta}}P_{L}^{\lambda\rho}\right)g_{\rho\kappa}\left(AP_{T}^{\kappa\mu}+BP_{L}^{\kappa\mu}\right)\\
 & = & P_{L}^{\lambda\mu}+P_{T}^{\lambda\mu}\end{array}\label{4.2.12}\end{equation}
\begin{equation}
\begin{array}{crcl}
\Rightarrow & -Z_{3}k^{2}\left(AP_{T}^{\lambda\mu}+\frac{B}{\zeta_{R}Z_{\zeta}}P_{L}^{\lambda\mu}\right) & = & P_{T}^{\lambda\mu}+P_{L}^{\lambda\mu}\\
\Rightarrow & A=-\frac{1}{k^{2}Z_{3}} & \textrm{and} & B=-\frac{\zeta_{R}Z_{\zeta}}{k^{2}Z_{3}}\\
\Rightarrow & Q^{-1\textrm{ }\lambda\mu} & = & -\frac{1}{k^{2}Z_{3}}\left(P_{T}^{\lambda\mu}+\zeta_{R}Z_{\zeta}P_{L}^{\lambda\mu}\right)\end{array}\label{4.2.13}\end{equation}
\begin{equation}
\begin{array}{crcl}
\Rightarrow & Free\textrm{ }Propagator^{\lambda\mu} & = & D_{0}^{\lambda\mu}\left(k^{2}\right)\\
 &  & = & -\frac{i}{k^{2}Z_{3}}\left(P_{T}^{\lambda\mu}+\zeta_{R}Z_{\zeta}P_{L}^{\lambda\mu}\right)\\
 &  & = & -\frac{i}{k^{2}Z_{3}}\left(g^{\lambda\mu}+\left(\zeta_{R}Z_{\zeta}-1\right)\frac{k^{\lambda}k^{\mu}}{k^{2}}\right).\end{array}\label{4.2.14}\end{equation}
If we define the full propagator as the sum of all these one-loop
corrections, \emph{i.e.} the free propagator plus the free propagator
with a loop plus the free propagator with two separate loops plus...
ad infinitum we have the \emph{Dyson} \emph{re-summed} propagator:
\begin{equation}
D_{\lambda\rho}\left(k^{2}\right)=D_{0\lambda\rho}\left(k^{2}\right)+D_{0\lambda\mu}\left(k^{2}\right)i\Pi^{\mu\nu}\left(k^{2}\right)D_{0\nu\rho}\left(k^{2}\right)+D_{0\lambda\mu}\left(k^{2}\right)i\Pi^{\mu\nu}\left(k^{2}\right)D_{0\nu\kappa}\left(k^{2}\right)i\Pi^{\kappa\alpha}\left(k^{2}\right)D_{0\alpha\rho}\left(k^{2}\right)+...\label{4.2.15}\end{equation}
This can be expressed in a much more compact form by noting it is
of the form\[
A\left(1-B\right)^{-1}C=A\left(1+B+B^{2}+B^{3}...\right)C,\]
this is known as resummation. To do this we first simplify the terms
of the form $D_{0}^{\kappa\nu}\left(k^{2}\right)i\Pi_{\nu\alpha}\left(k^{2}\right)$.
Substituting in we have,\begin{equation}
\begin{array}{rcl}
D_{0}^{\kappa\nu}\left(k^{2}\right)i\Pi_{\nu\alpha}\left(k^{2}\right) & = & \frac{-1}{k^{2}Z_{3}}\left(P_{T}^{\kappa\nu}+\zeta_{R}Z_{\zeta}P_{L}^{\kappa\nu}\right)P_{T\nu\alpha}k^{2}\Pi\left(k^{2}\right)\\
 & = & \frac{-\Pi\left(k^{2}\right)}{Z_{3}}P_{T\alpha}^{\kappa}\end{array}\label{4.2.16}\end{equation}
So the sum for the full photon propagator is,\begin{equation}
\begin{array}{rcl}
D_{\lambda\rho}\left(k^{2}\right) & = & D_{0\lambda\rho}\left(k^{2}\right)+\frac{-\Pi\left(k^{2}\right)}{Z_{3}}P_{T\lambda}^{\alpha}D_{0\alpha\rho}\left(k^{2}\right)+\frac{-\Pi\left(k^{2}\right)}{Z_{3}}P_{T\lambda}^{\alpha}\frac{-\Pi\left(k^{2}\right)}{Z_{3}}P_{T\alpha}^{\beta}D_{0\beta\rho}\left(k^{2}\right)+..\\
 & = & \left(\delta_{\lambda}^{\alpha}-P_{T\lambda}^{\alpha}\right)D_{0\alpha\rho}\left(k^{2}\right)+\left(P_{T\lambda}^{\alpha}D_{0\alpha\rho}\left(k^{2}\right)+\frac{-\Pi\left(k^{2}\right)}{Z_{3}}P_{T\lambda}^{\alpha}D_{0\alpha\rho}\left(k^{2}\right)+\right.\\
 &  & \left.\frac{-\Pi\left(k^{2}\right)}{Z_{3}}P_{T\lambda}^{\alpha}\frac{-\Pi\left(k^{2}\right)}{Z_{3}}P_{T\alpha}^{\beta}D_{0\beta\rho}\left(k^{2}\right)+...\right)\\
 & = & \left(\delta_{\lambda}^{\alpha}-P_{T\lambda}^{\alpha}\right)D_{0\alpha\rho}\left(k^{2}\right)+\left(1+\frac{-\Pi\left(k^{2}\right)}{Z_{3}}+\left(\frac{-\Pi\left(k^{2}\right)}{Z_{3}}\right)^{2}+...\right)P_{T\lambda}^{\alpha}D_{0\alpha\rho}\left(k^{2}\right)\\
 & = & \left(\delta_{\lambda}^{\alpha}-P_{T\lambda}^{\alpha}\right)D_{0\alpha\rho}\left(k^{2}\right)+\frac{P_{T\lambda}^{\alpha}}{1-\frac{-\Pi\left(k^{2}\right)}{Z_{3}}}D_{0\alpha\rho}\left(k^{2}\right)\\
 & = & \left(\delta_{\lambda}^{\alpha}-P_{T\lambda}^{\alpha}\right)\left(-\frac{i}{k^{2}Z_{3}}\left(P_{T\alpha\rho}+\zeta_{R}Z_{\zeta}P_{L\alpha\rho}\right)\right)+\frac{P_{T\lambda}^{\alpha}}{1+\frac{\Pi\left(k^{2}\right)}{Z_{3}}}\left(-\frac{i}{k^{2}Z_{3}}\left(P_{T\alpha\rho}+\zeta_{R}Z_{\zeta}P_{L\alpha\rho}\right)\right)\textrm{ }\\
 & = & \left(-\frac{i\delta_{\lambda}^{\alpha}}{k^{2}Z_{3}}\left(P_{T\alpha\rho}+\zeta_{R}Z_{\zeta}P_{L\alpha\rho}\right)+\frac{iP_{T\lambda}^{\alpha}}{k^{2}Z_{3}}\left(P_{T\alpha\rho}+\zeta_{R}Z_{\zeta}P_{L\alpha\rho}\right)\right)-\frac{iP_{T\lambda\rho}}{k^{2}\left(Z_{3}+\Pi\left(k^{2}\right)\right)}\\
 & = & -\frac{iP_{T\lambda\rho}}{k^{2}\left(Z_{3}+\Pi\left(k^{2}\right)\right)}-\frac{i\zeta_{R}Z_{\zeta}P_{L\lambda\rho}}{k^{2}Z_{3}}\\
 & = & \frac{-i}{k^{2}\left(Z_{3}+\Pi\left(k^{2}\right)\right)}\left(P_{T\lambda\rho}+\frac{\zeta_{R}Z_{\zeta}\left(Z_{3}+\Pi\left(k^{2}\right)\right)}{Z_{3}}P_{L\lambda\rho}\right)\end{array}\label{4.2.17}\end{equation}
where,\begin{equation}
Z_{3}+\Pi\left(k^{2}\right)=Z_{3}+\lim_{\epsilon\rightarrow0}\frac{2\alpha}{\pi}\left(\frac{1}{6}\left(\frac{1}{\epsilon}+\log\left(4\pi\right)-\gamma\right)+\int_{0}^{1}\textrm{d}x\textrm{ }x\left(1-x\right)\log\left(\frac{\mu^{2}}{m^{2}+xk^{2}\left(x-1\right)}\right)\right).\label{4.2.18}\end{equation}

\noindent We can choose $Z_{3}$ so as to cancel the divergences in
$\Pi\left(k^{2}\right)$ in many different ways, these basically constitute
different renormalization schemes. We firstly have the condition that
this be finite, but how should we fix the finite parts of $Z_{3}$?
We could insist that the residue of the propagator was one at its
pole $k^{2}=0$. Note that fortunately the radiative correction has
not shifted the position of the pole - the photon is still massless.
Such a procedure is called \emph{mass shell renormalization}. It is
a special case of \emph{momentum subtraction renormalization}. In
the momentum subtraction renormalization scheme we demand that there
be no corrections to the propagator at $k^{2}=p^{2}$ (choosing $p^{2}=m^{2}$
gives one). Another approach is to fix $Z_{3}$ just so it cancels
$\frac{1}{\epsilon}$, this is called \emph{minimal subtraction} (MS),
or we can choose $Z_{3}$ such that it cancels the $\frac{1}{\epsilon}-\gamma+\log4\pi$
which is known as the modified momentum subtraction or $\overline{MS}$
 scheme. In these renormalization schemes the scale $\mu$ is left
over, it's numerical value may be chosen at will - in relationships
between physical measurements / processes it drops out. In $\overline{MS}$,
\begin{equation}
\begin{array}{rcl}
Z_{3} & = & -\frac{\alpha}{3\pi}\left(\frac{1}{\epsilon}-\gamma+\log4\pi\right)\\
Z_{3}+\Pi\left(k^{2}\right) & = & \frac{2\alpha}{\pi}\int_{0}^{1}\textrm{d}x\textrm{ }x\left(1-x\right)\log\left(\frac{\mu^{2}}{m^{2}+xk^{2}\left(x-1\right)}\right)\end{array}\label{4.2.19}\end{equation}
 We should also choose $Z_{\zeta}=Z_{3}$ to render the longitudinal
part of the full propagator finite (see last line of \ref{4.2.17})!
Then, letting \begin{equation}
\tilde{\Pi}\left(k^{2}\right)=\frac{2\alpha}{\pi}\int_{0}^{1}\textrm{d}x\textrm{ }x\left(1-x\right)\log\left(\frac{\mu^{2}}{m^{2}+xk^{2}\left(x-1\right)}\right)\label{4.2.20}\end{equation}
we have for the one loop corrected photon propagator: \begin{equation}
D_{\lambda\rho}\left(k^{2}\right)=\frac{-i}{k^{2}\tilde{\Pi}\left(k^{2}\right)}\left(P_{T\lambda\rho}+\zeta_{R}\tilde{\Pi}\left(k^{2}\right)P_{L\lambda\rho}\right).\label{4.2.21}\end{equation}
 Note that the pole has not shifted so the photon has not acquired
a mass through these radiative corrections but instead the gauge has
shifted! The fact that it is attached to the longitudinal part of
the propagator is telling us that we should not expect contributions
to matrix elements from the longitudinal part of the resummed propagator.
Also for $k^{2}>4m^{2}$, $\widetilde{\Pi}$$\left(k^{2}\right)$
has an imaginary piece, this is associated with the fact that for
$k^{2}>4m^{2}$ there is enough energy to create real fermion-antifermion
pairs in the loop. 
\newpage

\section{The Electron Self Energy and the Vertex Function\label{sec:The-Electron-Self}. }

\noindent In this section we will return to the Lagrangian and Feynman
rules of the bare Lagrangian \emph{i.e.} everything is bare, there
are no $Z$'s. At the end of the section we will shift back to the
renormalized parametrization of 4.2, look out for that. We could proceed
to calculate the one loop corrections to the fermion self energy:

\vspace{0.375cm}
\begin{center}\includegraphics[%
  width=0.30\paperwidth,
  height=0.15\paperwidth]{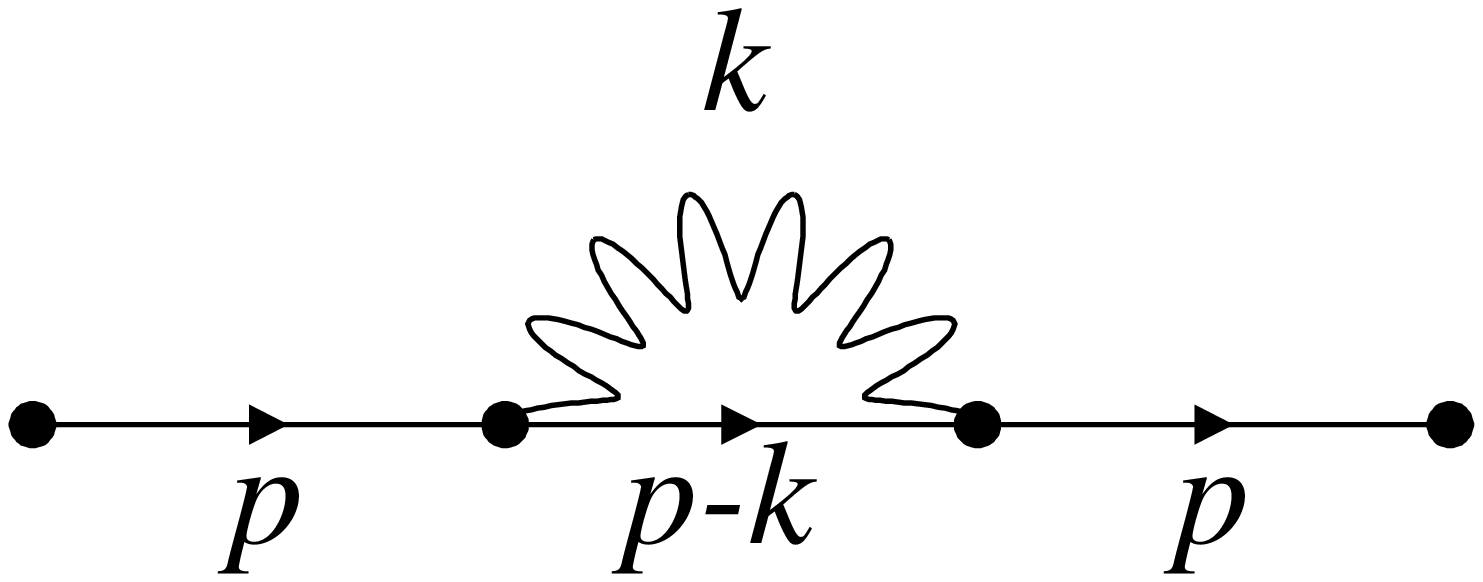}\end{center}
\vspace{0.375cm}

\begin{center}{\LARGE $=i\Sigma\left(p\right)$}\end{center}{\LARGE \par}

\noindent and the vertex function,

\vspace{0.375cm}
\begin{center}\includegraphics[%
  width=0.30\paperwidth,
  height=0.15\paperwidth]{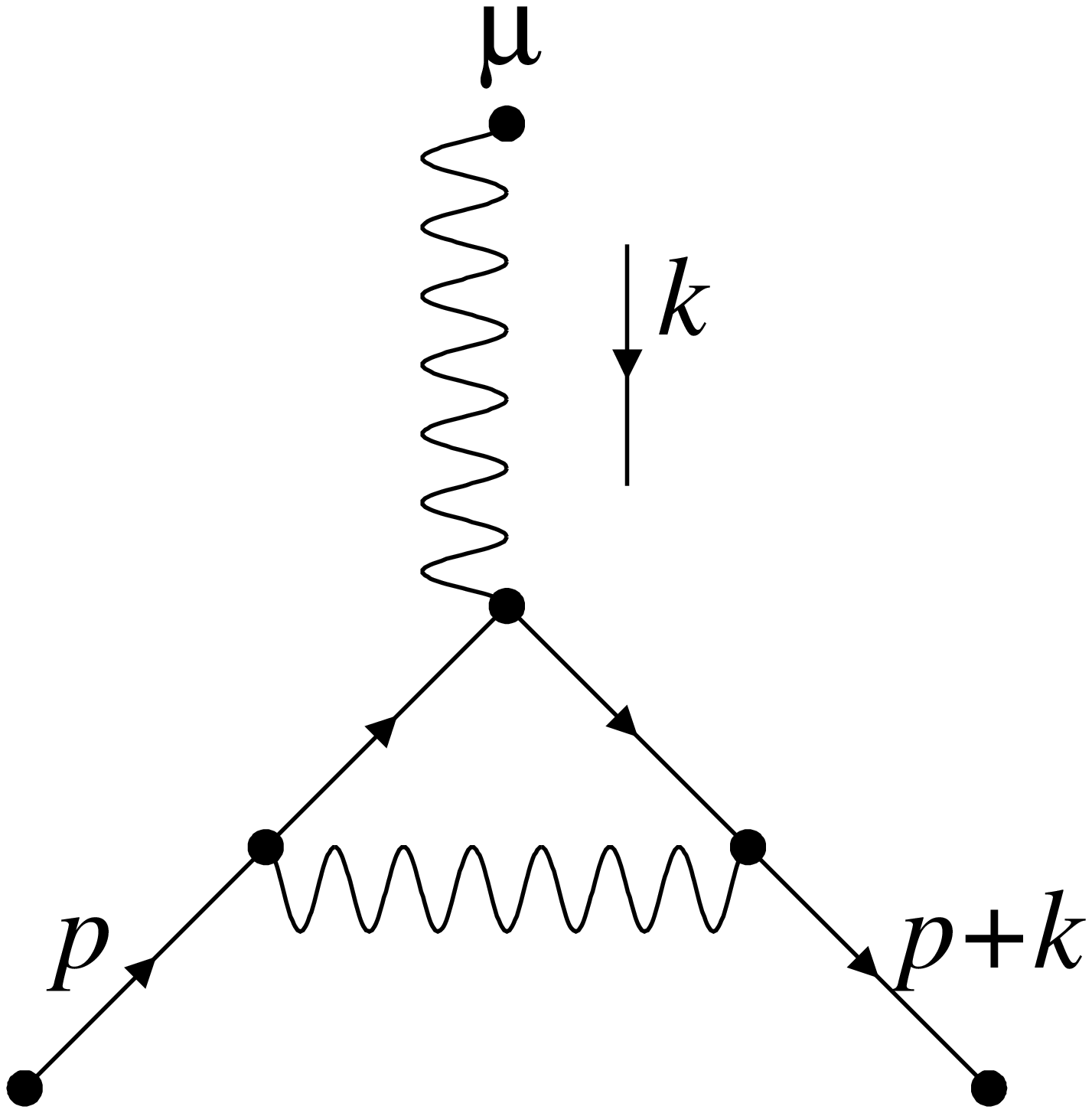}\end{center}
\vspace{0.375cm}

\begin{center}{\LARGE $=\widetilde{\Gamma_{c}^{\mu}}\left(k,p\right)$}\end{center}{\LARGE \par}

\noindent In terms of the self energy the full fermion propagator
satisfies:\begin{equation}
\begin{array}{rlcl}
 & S_{F} & = & S_{F}^{0}+S_{F}^{0}i\Sigma\left(p\right)S_{F}\\
\Rightarrow & S_{F}^{0-1} & = & S_{F}^{-1}+i\Sigma\left(p\right)\\
\Rightarrow & S_{F}^{-1} & = & -i\left(\not p-m+\Sigma\left(p\right)\right)\end{array}\label{4.3.1}\end{equation}
 If we use the Ward identity\begin{equation}
k_{\mu}\widetilde{\Gamma^{\mu}}\left(k,p\right)=-eS_{F}\left(p+k\right)^{-1}+eS_{F}\left(p\right)^{-1}\label{4.3.2}\end{equation}
 we find that \begin{equation}
k_{\mu}\widetilde{\Gamma_{c}^{\mu}}\left(k,p\right)=ie\left(\Sigma\left(p+k\right)-\Sigma\left(p\right)\right)\label{4.3.3}\end{equation}
 and remember that this is true \emph{in any dimension} \emph{\underbar{}}\emph{i.e.}
for arbitrary $\epsilon$. Bearing this in mind let us look a little
more closely at the structure of the divergent diagrams.

\vspace{0.375cm}
\begin{center}{\Large \includegraphics[%
  width=0.30\textwidth,
  height=0.18\textwidth]{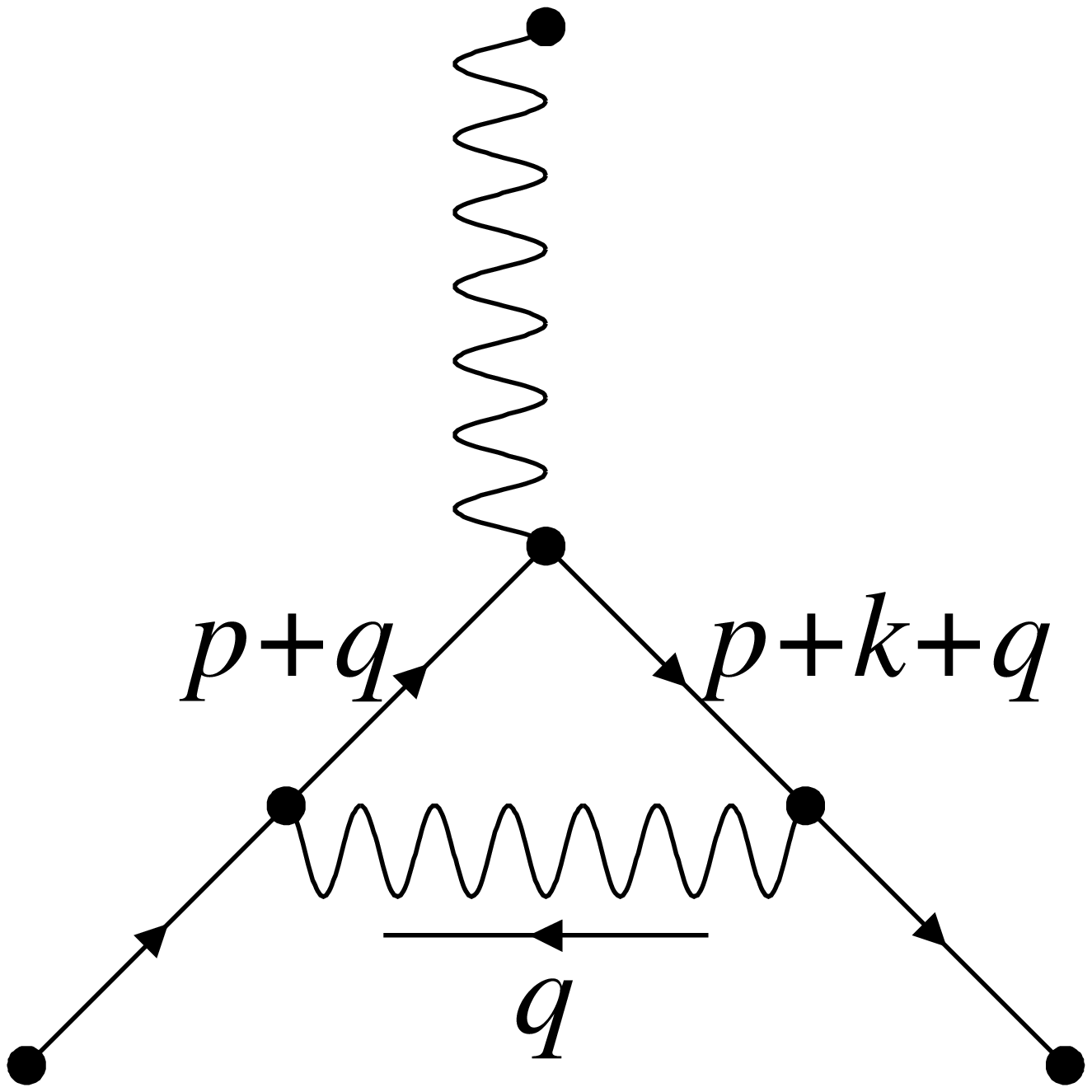}}\end{center}{\Large \par}
\vspace{0.375cm}

\noindent \begin{equation}
\sim\int\textrm{d}^{4}q\textrm{ }\gamma^{\nu}\frac{1}{\not p+\not q-m}\gamma^{\mu}\frac{1}{\not p+\not k+\not q-m}\gamma^{\lambda}\frac{P_{\nu\lambda}}{q^{2}}\label{4.3.4}\end{equation}
where $P_{\nu\lambda}$ is the photon tensor structure (for whichever
gauge) and is \emph{dimensionless}. Rewriting the integral,\begin{equation}
\int\textrm{d}^{4}q\frac{\gamma^{\nu}\left(\not p+\not q-m\right)\gamma^{\mu}\left(\not p+\not k+\not q-m\right)\gamma^{\lambda}P_{\nu\lambda}}{\left(\left(p+q\right)^{2}-m^{2}\right)\left(\left(p+k+q\right)^{2}-m^{2}\right)q^{2}}\label{4.3.5}\end{equation}
 we see that, only when both $\not q$ factors are taken in the numerator
is it divergent. Doing the integral we know that $\not q...\not q=q_{\sigma}...q_{\tau}\gamma^{\sigma}...\gamma^{\tau}$.
Note that $\gamma^{\lambda}\gamma^{\mu}\gamma^{\lambda}=\left(2-n\right)\gamma^{\mu}\textrm{ }$
$\Rightarrow$ $\not q...\not q=\frac{q^{2}g_{\sigma\tau}}{n}\gamma^{\sigma}...\gamma^{\tau}$.
So the structure of the divergent term must be simply $\gamma^{\mu}$!
Other pieces proportional to $p^{\mu},k^{\mu},m\gamma^{\mu}$ \emph{etc}
\emph{must} \underbar{}\emph{be} \underbar{}\emph{finite}! \begin{equation}
\widetilde{\Gamma_{c}^{\mu}}=ie\gamma^{\mu}\left(\frac{c}{\epsilon}+h\left(p^{2}\right)\right)+other\textrm{ }finite\textrm{ }terms.\label{4.3.6}\end{equation}
 Now let us turn our attention to the fermion self energy:\begin{equation}
\int\textrm{d}^{4}q\textrm{ }\gamma^{\mu}\frac{1}{\not p+\not q-m}\gamma^{\nu}\frac{P_{\nu\mu}}{q^{2}}\sim\int\textrm{d}^{n}q\frac{\gamma^{\mu}\left(\not p+\not q-m\right)\gamma^{\nu}P_{\nu\mu}}{q^{2}\left(\left(p+q\right)^{2}-m^{2}\right)}\label{4.3.7}\end{equation}
Now this integral is apparently linearly divergent; however we know
that this must actually integrate to zero leaving us with logarithmic
divergences. We will find that,\begin{equation}
\Sigma=\not p\left(\frac{A}{\epsilon}+g\left(p^{2}\right)\right)+m\left(\frac{B}{\epsilon}+f\left(p^{2}\right)\right)\label{4.3.8}\end{equation}
where $g$ and $f$ are regular functions. Returning to the Ward identity
and substituting in we find that \begin{equation}
\begin{array}{rcl}
iek_{\mu}\left(\gamma^{\mu}\left(\frac{C}{\epsilon}+h\left(p^{2}\right)\right)+finite\textrm{ }terms\right) & = & +ie\left\{ \left(\not p+\not k\right)\left(\frac{A}{\epsilon}+g\left(\left(p+k\right)^{2}\right)\right)+m\left(\frac{B}{\epsilon}+f\left(\left(p+k\right)^{2}\right)\right)\right.\\
 &  & -\left.\not p\left(\frac{A}{\epsilon}+g\left(p^{2}\right)\right)-m\left(\frac{B}{\epsilon}+f\left(p^{2}\right)\right)\right\} \\
\Rightarrow ie\not k\frac{C}{\epsilon}+finite\textrm{ }terms & = & +ie\not k\frac{A}{\epsilon}+finite\textrm{ }terms\end{array}\label{4.3.9}\end{equation}
Since this is true for arbitrary $\epsilon$ we conclude that $C=+A$.
That is to say, the divergent term in the proper vertex is related
to the divergent term in the wave function renormalization of the
fermion self energy. 

\noindent Now we flip back to the renormalized parametrization \ref{eq:4.2.1}
multiply the (bare) fermion fields by $Z_{f}^{\frac{1}{2}}$ (in notation
of section 4.2 $Z_{2}=Z_{f}$),\begin{equation}
\psi=Z_{f}^{\frac{1}{2}}\psi_{R}\label{4.3.10}\end{equation}
 and write,\begin{equation}
m=\frac{Z_{0}}{Z_{2}}m_{R}=m_{R}+\delta m,\label{4.3.11}\end{equation}
 then \begin{equation}
S_{F}^{-1}=-i\left(\left(\not p-m_{R}-\delta m\right)Z_{f}+\not p\frac{A}{\epsilon}+m_{R}\frac{B}{\epsilon}+...\right)\label{4.3.12}\end{equation}
 must be finite. In $\bar{MS}$ \begin{equation}
Z_{f}=1-\frac{A}{\epsilon}\label{4.3.13}\end{equation}
 then $Z_{f}m_{R}+\delta m=m_{R}\left(\frac{B}{\epsilon}+1\right)$
\emph{i.e.}\begin{equation}
-\delta m=\left(1-\frac{A}{\epsilon}\right)m_{R}-\left(1+\frac{B}{\epsilon}\right)m_{R}=-m_{R}\frac{A+B}{\epsilon}.\label{4.3.14}\end{equation}
 Now for the interaction term we have,\begin{equation}
e\bar{\psi_{R}}\not A_{R}\psi_{R}Z_{f}\sqrt{Z_{3}}\label{4.3.15}\end{equation}
 finally we multiply $e$ by $Z_{e}$ ($=\frac{Z_{1}}{Z_{2}\sqrt{Z_{3}}}$
in notation of section 4.2) to give,\begin{equation}
e\bar{_{R}\psi_{R}}\not A_{R}\psi_{R}Z_{f}Z_{e}\sqrt{Z_{3}}\label{4.3.16}\end{equation}
so that $ie\gamma^{\mu}\left(Z_{f}Z_{e}\sqrt{Z_{3}}+\frac{C}{\epsilon}+finite\right)$
is finite. However, $C=A=\epsilon-\epsilon Z_{f}$ so to cancel the
infinities here we require, \begin{equation}
\begin{array}{lrcccl}
\Rightarrow & Z_{f}Z_{e}\sqrt{Z_{3}} & = & -\frac{C}{\epsilon} & = & Z_{f}-1\\
\Rightarrow & \left(Z_{e}\sqrt{Z_{3}}-1\right) & = & \frac{1}{Z_{f}} & = & 0\end{array}\label{4.3.17}\end{equation}
\begin{equation}
\Rightarrow\textrm{ }Z_{3}^{\frac{1}{2}}Z_{e}=1\label{4.3.18}\end{equation}
In the notation of the last section $\left(Z_{e}=\frac{Z_{1}}{Z_{2}Z_{3}^{\frac{1}{2}}}\right)$
this result is,\begin{equation}
Z_{1}=Z_{2}.\label{4.3.19}\end{equation}
 To summarize we see that the Ward identities mean that:

\begin{itemize}
\item the terms multiplying $\bar{\psi}\not\partial\psi$ and $e\bar{\psi}\not A\psi$
in ${\cal {L}}$ are the same \ref{4.3.19}.
\item the charge renormalization $\left(Z_{e}\right)$ depends only on the
photon wave function renormalization $\left(\sqrt{Z_{3}}\right)$.
\end{itemize}
From the discussions of sections 4.2 and \ref{sec:The-Electron-Self}
we have trivially $Z_{G}=1$ and not trivially $Z_{1}=Z_{2}$, $Z_{\zeta}=Z_{3}$.
Substituting these expressions into the bare action we have,\begin{equation}
S=\int\textrm{d}^{n}x\textrm{ }-\frac{1}{4}Z_{3}F_{R,\mu\nu}F_{R}^{\mu\nu}+Z_{1}\bar{\psi_{R}}\left(i\not\partial+e\not_{R}A\right)\not\psi_{R}-m_{R}Z_{0}\bar{\psi_{R}}\psi_{R}+\frac{1}{2\zeta}\left(\partial_{\mu}A_{R}^{\mu}\right)^{2}+\eta\partial^{\mu}\partial_{\mu}\omega\label{eq:4.3.20}\end{equation}

If we now write the action in terms of a renormalized action plus
a counter-term Lagrangian \emph{viz} $S=S_{R}+S_{CT}$, we have,\begin{equation}
\begin{array}{lcl}
S_{R} & = & \int\textrm{d}^{n}x\textrm{ }-\frac{1}{4}F_{R,\mu\nu}F_{R}^{\mu\nu}+\bar{\psi_{R}}\left(i\not\partial+e\not A\right)\not\psi_{R}-m_{R}\bar{\psi_{R}}\psi_{R}+\frac{1}{2\zeta}\left(\partial_{\mu}A_{R}^{\mu}\right)^{2}+\eta\partial^{\mu}\partial_{\mu}\omega\\
S_{CT} & = & \int\textrm{d}^{n}x\textrm{ }-\frac{1}{4}\delta_{3}F_{R,\mu\nu}F_{R}^{\mu\nu}+\delta_{1}\bar{\psi_{R}}\left(i\not\partial+e\not A\right)\not\psi_{R}-\delta m\bar{\psi_{R}}\psi_{R}\end{array}\label{eq:4.3.21}\end{equation}
 \begin{equation}
\begin{array}{lclclcl}
\delta_{1} & = & Z_{1}-1 &  & \delta m & = & Z_{0}-1\\
\delta_{3} & = & Z_{3}-1\end{array}\label{eq:4.3.22}\end{equation}
Note the ghost plus gauge fixing sector here is the same as in chapter
\ref{cha:Principles-of-Gauge}. In that chapter we saw that the basic
part of the Lagrangian (kinetic and mass terms for physical particles
and their interaction) is BRS invariant because gauge transformations
are contained within BRS transformations (albeit with a fancy function
comprised of Grassmann number times Grassmann function) and that the
gauge fixing term plus ghost sector are \emph{together} also BRS invariant.
The gauge fixing and ghost sector is therefore BRS invariant, as before.
The photon kinetic term was gauge invariant (and so BRS invariant)
and so is the renormalized version in $S_{R}$ but now under (renormalized)
transformations of the same form as the original ones but with unrenormalized
quantities replaced by renormalized ones. The same goes for $\bar{\psi_{R}}\left(i\not\partial+e_{R}\not A\right)\psi_{R}$
and $m_{R}\bar{\psi_{R}}\psi_{R}$, they continue to be separately
gauge invariant under the gauge transformations of the same form as
before but with unrenormalized quantities replaced by renormalized
ones. The renormalized action is invariant under the original gauge
transformations with unrenormalized quantities replaced by renormalized
ones, the counter-term Lagrangian is invariant under renormalized
gauge transformations.

We now recap what has gone on and add some words of warning. In the
case of a theory with a \emph{global} symmetry it is easy to show
that the counter-term Lagrangian is also invariant under the symmetry
(see \emph{e.g.} \cite{key-5,key-8,key-14}). One might therefore
expect that the counter-term Lagrangian of gauge theories would behave
likewise. This isn't true though, gauge fixing broke gauge symmetry
long before any talk of renormalization and showed us that the true
symmetry of gauge theory is in fact the BRS invariance related to
the gauge group of the invariance of the un-fixed Lagrangian. So is
the counter-term Lagrangian supposed to be BRS invariant? In the case
of a general gauge theory \emph{i.e.} non-Abelian this is not the
case either, the obvious symmetries that we can see in the counter-term
and renormalized actions are unique due to the Abelian nature of QED.
The coefficients of gauge variant counter-terms vanish in QED. In
the general case the gauge variant counter-terms can arise, they are
such that the \emph{renormalized} Lagrangian is invariant under \emph{renormalized}
BRS transformations. For more on the general case see \cite{key-32,key-30}.
We would also like to emphasize that what has gone before is \emph{not}
a proof of the renormalizability of QED. Here by renormalization we
mean not just in the power counting sense of dimensionless couplings
but that renormalization and gauge invariance are compatible, that
the renormalized action is gauge invariant under some representation
of the gauge group which the bare action is invariant under. Our treatment
has been merely exploratory, a tour of renormalizability in the sense
just described. We arrive at a gauge invariant renormalized Lagrangian
by inferring $Z_{1}=Z_{2}$ which was in turn derived from \emph{demanding}
that the action including the renormalization constants $Z_{i}$ obey
the Ward identities, which was equivalent to demanding the renormalized
action be gauge invariant. A nicer (longer) thing to do would have
been to calculate explicitly, with a gauge invariant regularization,
the graphs above and show the infinities cancel naturally rather than
demanding it occur. Formally the proof of renormalizability is an
inductive one. It is possible to expand the (effective) bare action
in powers of $\hbar$ which is equivalent to expanding the action
in loops, the power of $\hbar$ in the expansion corresponds the number
of loops associated with that order. The induction proof requires
that one show that the loop expansion at $n$ loops and at $n+1$
loops obeys the Slavnov-Taylor identities%
\footnote{Slavnov-Taylor identities are what Ward identities are called in the
context of non-Abelian gauge theories, they are derived by analogy
to what we have been doing so far. Generally {}``Ward-Slavnov-Taylor
identities'', {}``Slavnov-Taylor identities'' and {}``Ward identities''
are somewhat interchangeable names.%
}. In other words the proof of {}``renormalizability'' amounts to
showing that the (BRS) symmetry, expressed by the Slavnov-Taylor identities,
exists at each order of the loop expansion. In fact the BRS transformations
are not necessary, it is possible to obtain Ward / Slavnov-Taylor
identities without it but the BRS machinery greatly simplifies them
which in turn makes the problem of proof of renormalizability of gauge
theory a tractable one. We recommend the reader to explore this technology
further in the literature of some of its founding fathers J.C.Taylor,
B.W.Lee, J.Zinn-Justin and G.'tHooft \cite{key-32,key-33,key-14,key-34}.

\chapter{Anomalies.}

\section{Chiral Symmetry%
\footnote{Unless otherwise stated the terms 'chiral symmetry' and `axial symmetry'
are interchangeable in these notes.%
} .}

We just observed that there is no radiative correction to fermion
masses if the mass parameter $m\rightarrow0$. Thus a massless fermion
will remain massless to all orders in perturbation theory, behind
this lies chiral symmetry. Chiral symmetry is an \emph{internal} \emph{symmetry}.
We shall see that a massless free theory (in even dimensions) possesses
$U\left(1\right)$ chiral symmetry. To begin with we will study chiral
symmetry in a slightly modified version of QED, \emph{axial} \emph{electrodynamics}:\begin{equation}
\begin{array}{rcl}
S & = & \int\textrm{d}^{4}x\textrm{ }{\cal {L}}\\
 & = & \int\textrm{d}^{4}x\textrm{ }-\frac{1}{4}F_{\mu\nu}F^{\mu\nu}-\frac{1}{4}G_{\mu\nu}G^{\mu\nu}+\bar{\psi}\left(i\not\partial+q\not V+g\not A\gamma^{5}-m\right)\psi+Ghosts+Gauge\textrm{ }Fixing\end{array}\label{5.1.1}\end{equation}
 where $F_{\mu\nu}=\partial_{\mu}V_{\nu}-\partial_{\nu}V_{\mu}$ and
$G_{\mu\nu}=\partial_{\mu}A_{\nu}-\partial_{\nu}A_{\mu}$. First consider
local vector gauge transformations $U_{V}\left(1\right)$ \begin{equation}
\begin{array}{crcl}
 & \psi & \rightarrow & e^{+iq\alpha\left(x\right)}\psi\\
\Rightarrow & \bar{\psi} & \rightarrow & \bar{\psi}e^{-iq\alpha\left(x\right)}\\
 & V_{\mu} & \rightarrow & V_{\mu}+\partial_{\mu}\alpha\left(x\right)\end{array}\label{5.1.2}\end{equation}
These are essentially the regular $U\left(1\right)$ gauge transformations
and they only act on what is basically the QED Lagrangian \emph{i.e.}
we know that this will leave things invariant by inspection barring
the term $\bar{\psi}i\gamma^{\mu}igA_{\mu}\gamma^{5}\psi$ which is
also trivially invariant:\begin{equation}
\begin{array}{rcl}
\bar{\psi}g\gamma^{\mu}A_{\mu}\gamma^{5}\psi & \rightarrow & \bar{\psi}e^{-iq\alpha\left(x\right)}g\gamma^{\mu}A_{\mu}\gamma^{5}e^{+iq\alpha\left(x\right)}\psi\\
 & = & \bar{\psi}g\gamma^{\mu}A_{\mu}\gamma^{5}\psi\end{array}\label{5.1.3}\end{equation}
 We also know that by using the classical equations of motion the
symmetry transformations of the fermions (\emph{i.e.} only the first
two transformations above) result in the conserved vector current
$j_{V}^{\mu}=\bar{\psi}\gamma^{\mu}\psi$ so-called because it transforms
as a 4-vector under Lorentz transformations. So we have local $U_{V}\left(1\right)$
gauge invariance, does the Lagrangian also have $U_{A}\left(1\right)$
gauge invariance \emph{i.e.} is the Lagrangian invariant under,\begin{equation}
\begin{array}{crcl}
 & \psi & \rightarrow & e^{+Pig\beta\left(x\right)\gamma^{5}}\psi\\
\Rightarrow & \bar{\psi} & \rightarrow & \bar{\psi}e^{+Pig\beta\left(x\right)\gamma^{5}}\\
 & A_{\mu} & \rightarrow & A_{\mu}+Q\partial_{\mu}\beta\left(x\right)\end{array}\label{5.1.4}\end{equation}
? We have generalized the transformations with real numbers $P$ and
$Q$ for reasons we will come to. Let us work through the terms in
the Lagrangian. Clearly the kinetic term $-\frac{1}{4}F_{\mu\nu}F^{\mu\nu}$
of the vector field $V_{\mu}$ is unaffected so we need not consider
it. The term $-\frac{1}{4}G_{\mu\nu}G^{\mu\nu}$ has basically the
same composition and transformation as $-\frac{1}{4}F_{\mu\nu}F^{\mu\nu}$
which is itself invariant so we can also add it to our list of invariants
and we need only concern ourselves with the $\bar{\psi}\left(...\right)\psi$
part. To do this we will need the following trivial identities,\begin{equation}
\gamma_{\mu}e^{if\left(x\right)\gamma^{5}}=\gamma_{\mu}\left(1+if\left(x\right)\gamma^{5}+\frac{1}{2}\left(if\left(x\right)\gamma^{5}\right)^{2}+...\right)=\left(1-if\left(x\right)\gamma^{5}+\frac{1}{2}\left(if\left(x\right)\gamma^{5}\right)^{2}+...\right)\gamma_{\mu}=e^{-if\left(x\right)\gamma^{5}}\gamma_{\mu}\label{5.1.5}\end{equation}
\begin{equation}
\gamma^{5}e^{if\left(x\right)\gamma^{5}}=e^{if\left(x\right)\gamma^{5}}\gamma^{5}\label{5.1.6}\end{equation}
\begin{equation}
\partial_{\mu}e^{if\left(x\right)\gamma^{5}}=i\gamma^{5}\left(\partial_{\mu}f\left(x\right)\right)e^{if\left(x\right)\gamma^{5}}.\label{5.1.7}\end{equation}
Now we use these identities in considering the rest of the transformation,

\begin{equation}
\begin{array}{l}
\bar{\psi}\left(i\not\partial+q\not V+g\not A\gamma^{5}-m\right)\psi\textrm{ }\rightarrow\textrm{ }\bar{\psi}e^{Pig\beta\left(x\right)\gamma^{5}}\left(i\not\partial+q\not V+g\not A\gamma^{5}+g\left(Q\not\partial\beta\left(x\right)\right)\gamma^{5}-m\right)e^{Pig\beta\left(x\right)\gamma^{5}}\psi\\
=\bar{\psi}e^{Pig\beta\left(x\right)\gamma^{5}}\left(i\left(-i\gamma^{5}\left(gP\not\partial\beta\left(x\right)\right)e^{Pig\beta\left(x\right)\gamma^{5}}\psi+e^{-Pig\beta\left(x\right)\gamma^{5}}\not\partial\psi\right)\right.\\
+\left.\left(q\not V+g\not A\gamma^{5}+gQ\not\partial\beta\left(x\right)\gamma^{5}-m\right)e^{Pig\beta\left(x\right)\gamma^{5}}\psi\right)\\
=\bar{\psi}e^{Pig\beta\left(x\right)\gamma^{5}}\left(e^{-Pig\beta\left(x\right)\gamma^{5}}\gamma^{5}\left(gP\not\partial\beta\left(x\right)\right)\psi+ie^{-Pig\beta\left(x\right)\gamma^{5}}\not\partial\psi\right.\\
+\left.e^{-Pig\beta\left(x\right)\gamma^{5}}\left(q\not V+g\not A\gamma^{5}+gQ\not\partial\beta\left(x\right)\gamma^{5}\right)\psi-me^{Pig\beta\left(x\right)\gamma^{5}}\psi\right)\\
=\bar{\psi}\left(\gamma^{5}\left(gP\not\partial\beta\left(x\right)\right)\psi+i\not\partial\psi+q\not V\psi+g\not A\gamma^{5}\psi+\left(gQ\not\partial\beta\left(x\right)\right)\gamma^{5}\psi-me^{2Pig\beta\left(x\right)\gamma^{5}}\psi\right)\\
=\bar{\psi}\left(i\not\partial+q\not V+g\not A\gamma^{5}-m\right)\psi+m\bar{\psi}\left(1-e^{2Pig\beta\left(x\right)\gamma^{5}}\right)\psi+\left(Q-P\right)g\left(\partial_{\mu}\beta\left(x\right)\right)\bar{\psi}\gamma^{\mu}\gamma^{5}\psi\\
=\bar{\psi}\left(i\not\partial+q\not V+g\not A\gamma^{5}-m\right)\psi+m\bar{\psi}\left(1-e^{2Pig\beta\left(x\right)\gamma^{5}}\right)\psi-g\left(Q-P\right)\beta\left(x\right)\left(\partial_{\mu}j_{A}^{\mu}\right)+g\left(Q-P\right)\partial_{\mu}\left(\beta\left(x\right)j_{A}^{\mu}\right)\end{array}\label{5.1.8}\end{equation}
We have defined the \emph{axial} \emph{current} $j_{A}^{\mu}=\bar{\psi}\gamma^{\mu}\gamma^{5}\psi$
so-called because it transforms as an axial vector under Lorentz transformations
\emph{i.e.} when $x\rightarrow-x$, $j_{A}^{\mu}\rightarrow j_{A}^{\mu}$.
In the action, the last term in the above is a total divergence which
we can rewrite as a surface term, the surface being at infinity. Assuming
that the fields and their first derivatives are vanishing at infinity
we set $\int\textrm{d}^{4}x\textrm{ }g\left(Q-P\right)\partial_{\mu}\left(\beta\left(x\right)j_{A}^{\mu}\right)$
to zero - we did this all the time at the start of the notes. \emph{To
first order in $\beta\left(x\right)$ we have}\emph{\underbar{,}}\begin{equation}
S\rightarrow S+\int\textrm{d}^{4}x\textrm{ }g\beta\left(x\right)\left(\left(P-Q\right)\partial_{\mu}j_{A}^{\mu}-2imP\bar{\psi}\gamma^{5}\psi\right).\label{5.1.9}\end{equation}
If we take the standard gauge transformations $P=Q=1$ we have that
the action is invariant up to $\int\textrm{d}^{4}x\textrm{ }2im\bar{\psi}\gamma^{5}\psi$,
so in the limit of massless fermions the action is invariant under
local chiral gauge transformations. Massless fermions are a general
feature of chiral invariant actions. On the other hand had we just
looked at local gauge transformations of the fermion fields \emph{i.e.}
$P=1,\textrm{ }Q=0$ such that $A_{\mu}\rightarrow A_{\mu}$ we have
from demanding invariance of the action that the axial current must
obey the following conservation law,\begin{equation}
\partial_{\mu}j_{A}^{\mu}=2im\bar{\psi}\gamma^{5}\psi\label{5.1.10}\end{equation}
\emph{as $\beta\left(x\right)$ is an arbitrary function}. $\bar{\psi}\gamma^{5}\psi$
is a pseudoscalar under parity operations $x\rightarrow-x$ it transforms
as $\bar{\psi}\gamma^{5}\psi\rightarrow-\bar{\psi}\gamma^{5}\psi$,
the four divergence of the axial current is proportional to the mass
density of the pseudoscalar. At the classical level we can show that
this current relation holds by considering what happens with the classical
equations of motion for $\psi$ and $\bar{\psi}$. Hopefully getting
the equation of motion for $\psi$ is trivial, to get the equation
of motion for $\bar{\psi}$ one can conjugate that for $\psi$ using
the fact that $\gamma^{0},\gamma^{0}\gamma^{\mu}\textrm{ and }\gamma^{5}$
are Hermitian (see below) or rewrite the Lagrangian using the product
rule and vanishing surface terms so that the derivative in the kinetic
term acts instead on $\bar{\psi}$.%
\footnote{$\overleftarrow{\partial_{\mu}}$ indicates that $\partial_{\mu}$
is acting on all the stuff to the left instead of the right.%
}\begin{equation}
\begin{array}{rrcl}
 & \left(i\not\partial+\not V+\not A\gamma^{5}-m\right)\psi & = & 0\\
\Rightarrow & \gamma^{0}\left(\gamma^{0}\gamma^{\mu}\right)\left(i\partial+V+A\gamma^{5}-m\right)\gamma^{0}\left(\gamma^{0}\psi\right) & = & 0\\
\Rightarrow & \bar{\psi}\gamma^{0}\left(-i\overleftarrow{\partial_{\mu}}+V+A\gamma^{5}-m\right)\left(\gamma^{0}\gamma^{\mu}\right)\gamma^{0} & = & 0\\
\Rightarrow & \bar{\psi}\left(-i\overleftarrow{\partial_{\mu}}+V-A\gamma^{5}-m\right)\gamma^{0}\left(\gamma^{0}\gamma^{\mu}\right)\gamma^{0} & = & 0\\
\Rightarrow & \bar{\psi}\left(i\overleftarrow{\not\partial}-\not V-\not A\gamma^{5}+m\right) & = & 0\end{array}\label{5.1.11}\end{equation}
To derive the classical current equation we use these equations to
substitute into the current equation $\bar{\psi}\overleftarrow{\not\partial}$
and $\not\partial\psi$:\begin{equation}
\begin{array}{rclcl}
\partial_{\mu}\left(\bar{\psi}\gamma^{\mu}\gamma^{5}\psi\right) & = & \bar{\psi}\overleftarrow{\not\partial}\gamma^{5}\psi & - & \bar{\psi}\gamma^{5}\not\partial\psi\\
 & = & -i\bar{\psi}\left(\not V+\not A\gamma^{5}-m\right)\gamma^{5}\psi & + & i\bar{\psi}\gamma^{5}\left(-\not V-\not A\gamma^{5}+m\right)\psi\\
 & = & -i\bar{\psi}\left(\not V+\not A\gamma^{5}-m\right)\gamma^{5}\psi & + & i\bar{\psi}\left(\not V+\not A\gamma^{5}+m\right)\gamma^{5}\psi\\
 & = & 2im\bar{\psi}\gamma^{5}\psi\end{array}\label{5.1.12}\end{equation}
 At the quantum level what we have done is effectively redefine our
variables of integration (the fields) by $\bar{\psi}\rightarrow\bar{\psi}e^{i\beta\left(x\right)\gamma^{5}}$
and $\psi\rightarrow e^{i\beta\left(x\right)\gamma^{5}}\psi$ so that,
(omitting spectator fields for brevity),\begin{equation}
\begin{array}{rcl}
\int D\bar{\psi}D\psi\textrm{ exp }i\int\textrm{d}^{4}x\textrm{ }S\left[\bar{\psi},\psi\right] & = & \int D\bar{\psi}'D\psi'\textrm{ exp }i\int\textrm{d}^{4}x\textrm{ }S\left[\bar{\psi}',\psi'\right]\\
 & = & \int D\bar{\psi}D\psi\textrm{ exp }i\int\textrm{d}^{4}x\textrm{ }S\left[\bar{\psi},\psi\right]+g\beta\left(x\right)\left(\partial_{\mu}j_{A}^{\mu}-2im\bar{\psi}\gamma^{5}\psi\right)+{\cal {O}}\left(\beta^{2}\right)\end{array}\label{5.1.13}\end{equation}
 We have assumed (naively) that because,\begin{equation}
\textrm{Det}\left(e^{i\beta\left(x\right)\gamma^{5}}\right)=\textrm{Det}\left(\begin{array}{cccc}
e^{-i\beta\left(x\right)} & 0 & 0 & 0\\
0 & e^{-i\beta\left(x\right)} & 0 & 0\\
0 & 0 & e^{i\beta\left(x\right)} & 0\\
0 & 0 & 0 & e^{i\beta\left(x\right)}\end{array}\right)=1\label{5.1.14}\end{equation}
 the Jacobian associated with the path integral measure under the
transformation is also one hence,\begin{equation}
\Rightarrow{\bki{\partial_{\mu}j_{A}^{\mu}}}=2im{\bki{\bar{\psi}\gamma^{5}\psi}}\label{5.1.15}\end{equation}
 This gives us the (\emph{naive}) axial vector current Ward identity.
It is naive because it is only true at the classical level, in our
abbreviated path integral description above we neglect the possibility
that the path integral measure may change non-trivially under the
transformation of the fields, it may contribute something in the form
of a Jacobian. When the Jacobian is not one we have an \emph{anomaly}.
Anomalies are of fundamental importance in field theory.

\newpage
\section{\noindent The ABJ Anomaly. }

\noindent Before plunging into a calculation of the change in the
path integral measure we will detour to some phenomenology which preceded
it. As we said before the anomaly presents itself by making the path
integral measure over the fermions transform so as to produce a Jacobian
not equal to one. It also presents itself in the calculation of certain
S-matrix elements, the axial anomaly was first discovered in the calculation
of the VVA (vector, vector, axial) and VVP (vector, vector, pseudoscalar)
triangle diagrams.

\noindent \begin{center}\includegraphics[%
  width=0.20\textwidth,
  height=0.17\textwidth]{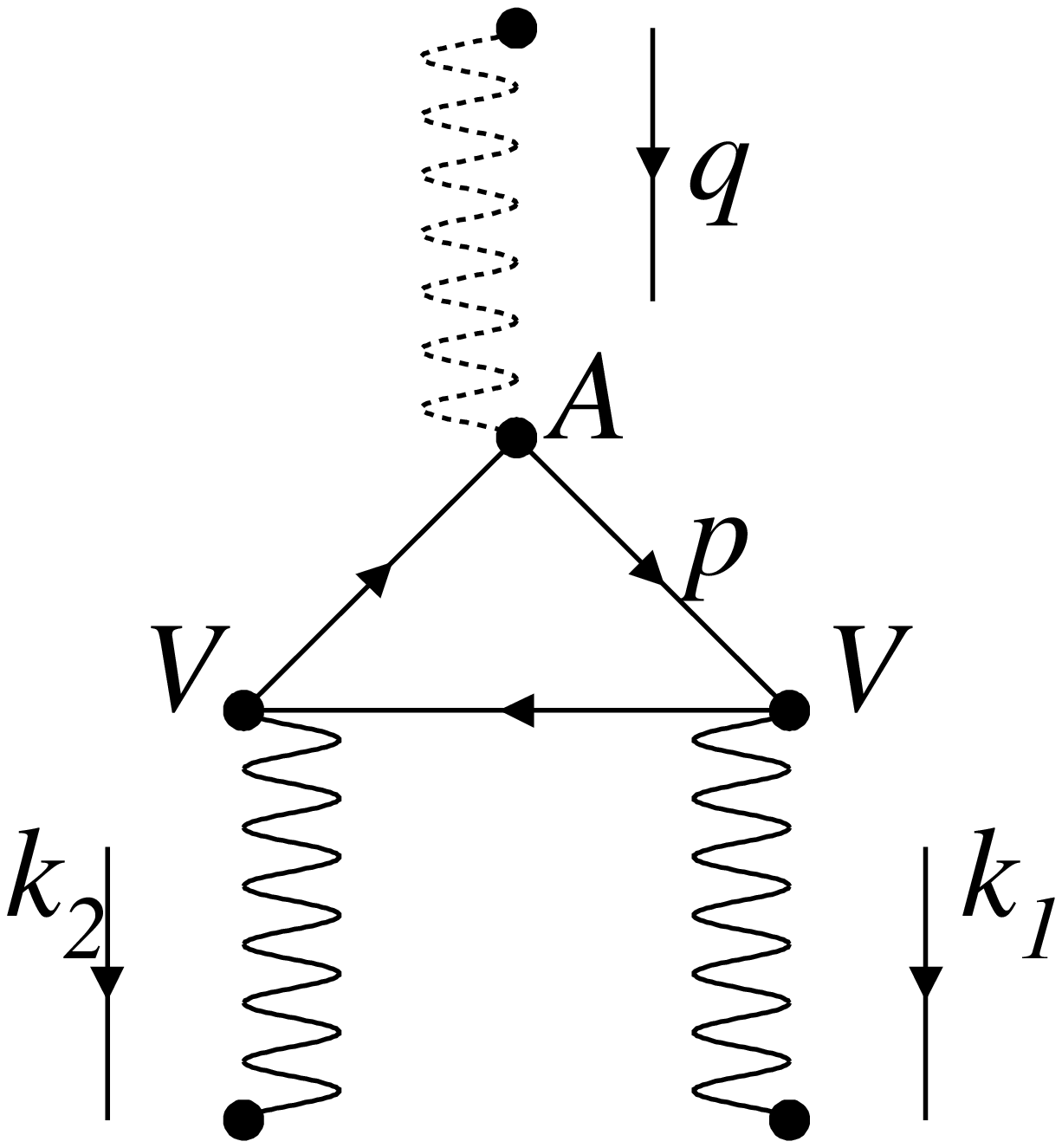} \includegraphics[%
  width=0.20\textwidth,
  height=0.17\textwidth]{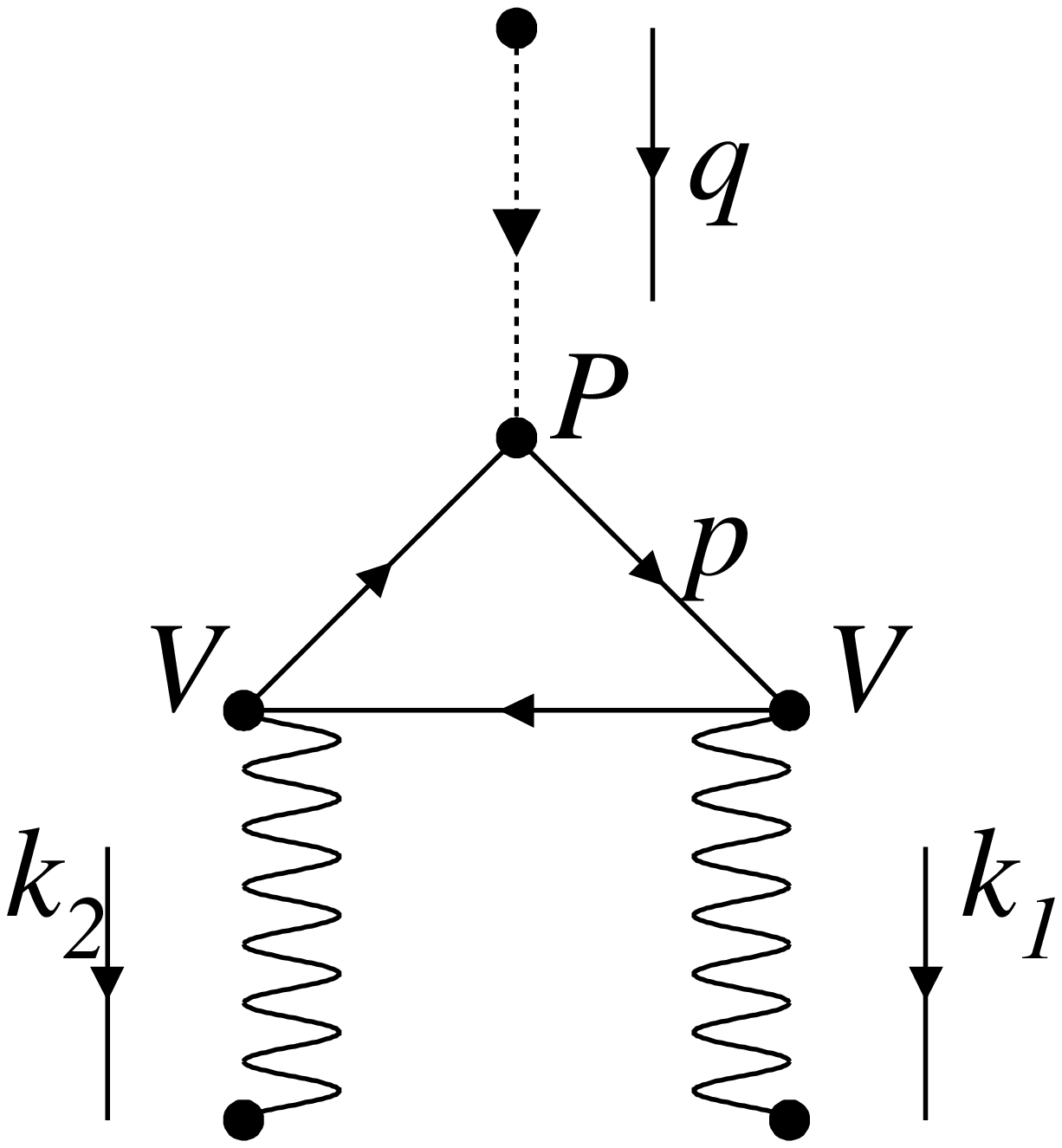}\end{center}

\noindent These correspond to the following 3-point functions,\begin{equation}
{\bkiii{0}{{\cal {T}}j_{\mu}\left(x\right)j_{\nu}\left(y\right)j_{A,\lambda}\left(z\right)}{0}}\label{5.2.1}\end{equation}
\begin{equation}
{\bkiii{0}{{\cal {T}}j_{\mu}\left(x\right)j_{\nu}\left(y\right)P\left(z\right)}{0}}\label{5.2.2}\end{equation}
In momentum space these are,\begin{equation}
T_{\mu\nu\lambda}\left(k_{1},k_{2},q\right)=i\int\textrm{d}^{4}x\textrm{d}^{4}y\textrm{d}^{4}z\textrm{ }e^{ik_{1}x+ik_{2}y-iqz}\textrm{ }{\bkiii{0}{{\cal {T}}j_{\mu}\left(x\right)j_{\nu}\left(y\right)j_{A,\lambda}\left(z\right)}{0}}\label{5.2.3}\end{equation}
\begin{equation}
T_{\mu\nu}\left(k_{1},k_{2},q\right)=i\int\textrm{d}^{4}x\textrm{d}^{4}y\textrm{d}^{4}z\textrm{ }e^{ik_{1}x+ik_{2}y-iqz}\textrm{ }{\bkiii{0}{{\cal {T}}j_{\mu}\left(x\right)j_{\nu}\left(y\right)P\left(z\right)}{0}}\label{5.2.4}\end{equation}
with $P\left(z\right)$ the pseudoscalar $\bar{\psi}\left(z\right)\gamma^{5}\psi\left(z\right)$
(probability) density. We are trying to test our naive chiral ward
identity. The first amplitude contains an axial current and the second
contains a pseudoscalar, the Ward identity relates the divergence
of the axial vector to the pseudoscalar density, so should provide
a relation between these two amplitudes. To do this we will clearly
want to be differentiating the first amplitude so as to get the divergence
of the axial vector current. This will involve differentiating the
\underbar{time ordered} product of operators $j_{\mu}\left(x\right)j_{\nu}\left(y\right)j_{A,\lambda}\left(z\right)$
which is not easy. \begin{equation}
\begin{array}{l}
\frac{\partial}{\partial z^{\mu}}\left[{\cal {T}}j^{\mu}\left(x\right)j^{\nu}\left(y\right)j_{A}^{\lambda}\left(z\right)\right]\\
=\frac{\partial}{\partial z^{\mu}}\left\{ \begin{array}{rccl}
 & \theta\left(x_{0}-y_{0}\right)\theta\left(y_{0}-z_{0}\right)j^{\mu}\left(x\right)j^{\nu}\left(y\right)j_{A}^{\lambda}\left(z\right) & + & \theta\left(x_{0}-z_{0}\right)\theta\left(z_{0}-y_{0}\right)j^{\mu}\left(x\right)j_{A}^{\lambda}\left(z\right)j^{\nu}\left(y\right)\\
+ & \theta\left(y_{0}-x_{0}\right)\theta\left(x_{0}-z_{0}\right)j^{\nu}\left(y\right)j^{\mu}\left(x\right)j_{A}^{\lambda}\left(z\right) & + & \theta\left(y_{0}-z_{0}\right)\theta\left(z_{0}-x_{0}\right)j^{\nu}\left(y\right)j_{A}^{\lambda}\left(z\right)j^{\mu}\left(x\right)\\
+ & \theta\left(z_{0}-x_{0}\right)\theta\left(x_{0}-y_{0}\right)j_{A}^{\lambda}\left(z\right)j^{\mu}\left(x\right)j^{\nu}\left(y\right) & + & \theta\left(z_{0}-y_{0}\right)\theta\left(y_{0}-x_{0}\right)j_{A}^{\lambda}\left(z\right)j^{\nu}\left(y\right)j^{\mu}\left(x\right)\end{array}\right\} \\
=\left[{\cal {T}}j^{\mu}\left(x\right)j^{\nu}\left(y\right)\left(\partial_{\lambda}j_{A}^{\lambda}\left(z\right)\right)\right]\\
\begin{array}{rccl}
- & \delta\left(z_{0}-y_{0}\right)\theta\left(x_{0}-y_{0}\right)j^{\mu}\left(x\right)j^{\nu}\left(y\right)j_{A}^{0}\left(z\right) & - & \delta\left(z_{0}-x_{0}\right)\theta\left(z_{0}-y_{0}\right)j^{\mu}\left(x\right)j_{A}^{0}\left(z\right)j^{\nu}\left(y\right)\\
- & \delta\left(z_{0}-x_{0}\right)\theta\left(y_{0}-x_{0}\right)j^{\nu}\left(y\right)j^{\mu}\left(x\right)j_{A}^{0}\left(z\right) & - & \delta\left(z_{0}-y_{0}\right)\theta\left(z_{0}-x_{0}\right)j^{\nu}\left(y\right)j_{A}^{0}\left(z\right)j^{\mu}\left(x\right)\\
+ & \delta\left(z_{0}-x_{0}\right)\theta\left(x_{0}-y_{0}\right)j_{A}^{0}\left(z\right)j^{\mu}\left(x\right)j^{\nu}\left(y\right) & + & \delta\left(z_{0}-y_{0}\right)\theta\left(y_{0}-x_{0}\right)j_{A}^{0}\left(z\right)j^{\nu}\left(y\right)j^{\mu}\left(x\right)\\
+ & \delta\left(z_{0}-y_{0}\right)\theta\left(x_{0}-z_{0}\right)j^{\mu}\left(x\right)j_{A}^{0}\left(z\right)j^{\nu}\left(y\right) & + & \delta\left(z_{0}-x_{0}\right)\theta\left(y_{0}-z_{0}\right)j^{\nu}\left(y\right)j_{A}^{0}\left(z\right)j^{\mu}\left(x\right)\end{array}\\
=\left[{\cal {T}}j^{\mu}\left(x\right)j^{\nu}\left(y\right)\left(\partial_{\lambda}j_{A}^{\lambda}\left(z\right)\right)\right]\\
\begin{array}{cc}
+ & \delta\left(z_{0}-y_{0}\right)\left(\theta\left(y_{0}-x_{0}\right)j_{A}^{0}\left(z\right)j^{\nu}\left(y\right)j^{\mu}\left(x\right)-\theta\left(x_{0}-y_{0}\right)j^{\mu}\left(x\right)j^{\nu}\left(y\right)j_{A}^{0}\left(z\right)\right)\\
+ & \delta\left(z_{0}-x_{0}\right)\left(\theta\left(y_{0}-z_{0}\right)j^{\nu}\left(y\right)j_{A}^{0}\left(z\right)j^{\mu}\left(x\right)-\theta\left(z_{0}-y_{0}\right)j^{\mu}\left(x\right)j_{A}^{0}\left(z\right)j^{\nu}\left(y\right)\right)\\
+ & \delta\left(z_{0}-x_{0}\right)\left(\theta\left(x_{0}-y_{0}\right)j_{A}^{0}\left(z\right)j^{\mu}\left(x\right)j^{\nu}\left(y\right)-\theta\left(y_{0}-x_{0}\right)j^{\nu}\left(y\right)j^{\mu}\left(x\right)j_{A}^{0}\left(z\right)\right)\\
+ & \delta\left(z_{0}-y_{0}\right)\left(\theta\left(x_{0}-z_{0}\right)j^{\mu}\left(x\right)j_{A}^{0}\left(z\right)j^{\nu}\left(y\right)-\theta\left(z_{0}-x_{0}\right)j^{\nu}\left(y\right)j_{A}^{0}\left(z\right)j^{\mu}\left(x\right)\right)\end{array}\end{array}\label{5.2.5}\end{equation}

\noindent In actual fact the thing that we are differentiating is
inside an integral (it's inside $T_{\mu\nu\lambda}\left(k_{1},k_{2},q\right)$
at the top of the page) so when we do the integration the delta functions
will set some of the $x_{0}s$, $y_{0}s$ and $z_{0}s$ equal to each
other. We can therefore safely rewrite the $\theta$ functions as
if this has already occurred.\begin{equation}
\begin{array}{l}
=\left[{\cal {T}}\left(\partial_{x,\mu}j^{\mu}\left(x\right)\right)j^{\nu}\left(y\right)j_{A}^{\lambda}\left(z\right)\right]\\
\begin{array}{cc}
+ & \delta\left(z_{0}-y_{0}\right)\left(\theta\left(y_{0}-x_{0}\right)j_{A}^{0}\left(z\right)j^{\nu}\left(y\right)j^{\mu}\left(x\right)-\theta\left(x_{0}-y_{0}\right)j^{\mu}\left(x\right)j^{\nu}\left(y\right)j_{A}^{0}\left(z\right)\right)\\
+ & \delta\left(z_{0}-x_{0}\right)\left(\theta\left(y_{0}-x_{0}\right)j^{\nu}\left(y\right)j_{A}^{0}\left(z\right)j^{\mu}\left(x\right)-\theta\left(x_{0}-y_{0}\right)j^{\mu}\left(x\right)j_{A}^{0}\left(z\right)j^{\nu}\left(y\right)\right)\\
+ & \delta\left(z_{0}-x_{0}\right)\left(\theta\left(x_{0}-y_{0}\right)j_{A}^{0}\left(z\right)j^{\mu}\left(x\right)j^{\nu}\left(y\right)-\theta\left(y_{0}-x_{0}\right)j^{\nu}\left(y\right)j^{\mu}\left(x\right)j_{A}^{0}\left(z\right)\right)\\
+ & \delta\left(z_{0}-y_{0}\right)\left(\theta\left(x_{0}-y_{0}\right)j^{\mu}\left(x\right)j_{A}^{0}\left(z\right)j^{\nu}\left(y\right)-\theta\left(y_{0}-x_{0}\right)j^{\nu}\left(y\right)j_{A}^{0}\left(z\right)j^{\mu}\left(x\right)\right)\end{array}\\
=\left[{\cal {T}}\left(\partial_{x,\mu}j^{\mu}\left(x\right)\right)j^{\nu}\left(y\right)j_{A}^{\lambda}\left(z\right)\right]\\
\begin{array}{cc}
+ & \left(\theta\left(y_{0}-x_{0}\right)\left(\delta\left(z_{0}-y_{0}\right)\left[j_{A}^{0}\left(z\right),j^{\nu}\left(y\right)\right]\right)j^{\mu}\left(x\right)+\theta\left(x_{0}-y_{0}\right)j^{\mu}\left(x\right)\left(\delta\left(z_{0}-y_{0}\right)\left[j_{A}^{0}\left(z\right),j^{\nu}\left(y\right)\right]\right)\right)\\
+ & \left(\theta\left(y_{0}-x_{0}\right)j^{\nu}\left(y\right)\left(\delta\left(z_{0}-x_{0}\right)\left[j_{A}^{0}\left(z\right),j^{\mu}\left(x\right)\right]\right)+\theta\left(x_{0}-y_{0}\right)\left(\delta\left(z_{0}-x_{0}\right)\left[j_{A}^{0}\left(z\right),j^{\mu}\left(x\right)\right]\right)j^{\nu}\left(y\right)\right)\end{array}\end{array}\label{5.2.6}\end{equation}
Now consider the commutators, they are all accompanied by a delta
function which gets integrated over in $T_{\mu\nu\lambda}\left(k_{1},k_{2},q\right)$
effectively making them equal time commutators. We know the equal
time commutation relations (ETCRs) for fermion field operators such
as those above from canonical quantization:\begin{equation}
\begin{array}{rcl}
\left\{ \psi_{\alpha}\left(x\right),\psi_{\beta}\left(y\right)\right\}  & = & 0\\
\left\{ \psi_{\alpha}^{\dagger}\left(x\right),\psi_{\beta}^{\dagger}\left(y\right)\right\}  & = & 0\\
\left\{ \psi_{\alpha}^{\dagger}\left(x\right),\psi_{\beta}\left(y\right)\right\}  & = & \delta_{\alpha\beta}\delta^{3}\left(\mathbf{x}-\mathbf{y}\right)\end{array}.\label{5.2.7}\end{equation}
The $\alpha$ and $\beta$ are Dirac indices. Given these ETCRs consider
a commutator of the form,\begin{equation}
\left[\psi_{\alpha}^{\dagger}\left(x\right)\Gamma_{\alpha\beta}^{\mu}\psi_{\beta}\left(x\right),\psi_{\gamma}^{\dagger}\left(y\right)\Lambda_{\gamma\delta}^{\nu}\psi_{\delta}\left(y\right)\right]\label{5.2.8}\end{equation}
 Where $\Gamma^{\mu}$ and $\Lambda^{\nu}$ can be $\gamma^{\mu}$
or $\gamma^{\mu}\gamma^{5}$. Written in terms of Dirac indices the
$\Gamma^{\mu}$ and $\Lambda^{\nu}$ are just numbers (as opposed
to operators - $\psi^{\dagger}s$ and $\psi s$) and can be pulled
through as such.\begin{equation}
\begin{array}{rl}
= & \left(\psi_{\alpha}^{\dagger}\left(x\right)\psi_{\beta}\left(x\right)\psi_{\gamma}^{\dagger}\left(y\right)\psi_{\delta}\left(y\right)-\psi_{\gamma}^{\dagger}\left(y\right)\psi_{\delta}\left(y\right)\psi_{\alpha}^{\dagger}\left(x\right)\psi_{\beta}\left(x\right)\right)\Gamma_{\alpha\beta}^{\mu}\Lambda_{\gamma\delta}^{\nu}\\
 & \textrm{Pull the }\psi_{\alpha}^{\dagger}\left(x\right)\psi_{\beta}\left(x\right)\textrm{ through to the other side of the first term}.\\
= & \left(\delta^{3}\left(\mathbf{x}-\mathbf{y}\right)\delta_{\beta\gamma}\psi_{\alpha}^{\dagger}\left(x\right)\psi_{\delta}\left(y\right)-\delta^{3}\left(\mathbf{x}-\mathbf{y}\right)\delta_{\alpha\delta}\psi_{\gamma}^{\dagger}\left(y\right)\psi_{\beta}\left(x\right)\right.\\
+ & \left.\psi_{\gamma}^{\dagger}\left(y\right)\psi_{\delta}\left(y\right)\psi_{\alpha}^{\dagger}\left(x\right)\psi_{\beta}\left(x\right)-\psi_{\gamma}^{\dagger}\left(y\right)\psi_{\delta}\left(y\right)\psi_{\alpha}^{\dagger}\left(x\right)\psi_{\beta}\left(x\right)\right)\Gamma_{\alpha\beta}^{\mu}\Lambda_{\gamma\delta}^{\nu}\\
= & \delta^{3}\left(\mathbf{x}-\mathbf{y}\right)\left(\psi_{\alpha}^{\dagger}\left(x\right)\Gamma_{\alpha\beta}^{\mu}\Lambda_{\beta\delta}^{\nu}\psi_{\delta}\left(y\right)-\psi_{\alpha}^{\dagger}\left(y\right)\Lambda_{\alpha\beta}^{\nu}\Gamma_{\beta\delta}^{\mu}\psi_{\delta}\left(x\right)\right)\\
+ & \left(\psi_{\gamma}^{\dagger}\left(y\right)\psi_{\delta}\left(y\right)\psi_{\alpha}^{\dagger}\left(x\right)\psi_{\beta}\left(x\right)-\psi_{\gamma}^{\dagger}\left(y\right)\psi_{\delta}\left(y\right)\psi_{\alpha}^{\dagger}\left(x\right)\psi_{\beta}\left(x\right)\right)\Gamma_{\alpha\beta}^{\mu}\Lambda_{\gamma\delta}^{\nu}\\
= & \delta^{3}\left(\mathbf{x}-\mathbf{y}\right)\left(\psi_{\alpha}^{\dagger}\left(x\right)\Gamma_{\alpha\beta}^{\mu}\Lambda_{\beta\delta}^{\nu}\psi_{\delta}\left(y\right)-\psi_{\alpha}^{\dagger}\left(y\right)\Lambda_{\alpha\beta}^{\nu}\Gamma_{\beta\delta}^{\mu}\psi_{\delta}\left(x\right)\right)\end{array}\label{5.2.9}\end{equation}

\noindent Given that the above is multiplied by $\delta\left(x_{0}-y_{0}\right)\delta^{3}\left(\mathbf{x}-\mathbf{y}\right)$
and we integrate over $x$ and $y$, after one of the integrations
this becomes,\begin{equation}
\psi^{\dagger}\left(x\right)\left[\Gamma^{\mu},\Lambda^{\nu}\right]\psi\left(x\right).\label{5.2.10}\end{equation}
 In our particular case we always have a $j^{0}\left(x\right)$ ($=\psi^{\dagger}\left(x\right)\psi$$\left(x\right)$)
in the commutator \emph{i.e.} in terms of the above we are always
dealing with $\left[\psi_{\alpha}^{\dagger}\left(x\right)\Gamma_{\alpha\beta}^{0}\psi_{\beta}\left(x\right),\psi_{\gamma}^{\dagger}\left(y\right)\Lambda_{\gamma\delta}^{\nu}\psi_{\delta}\left(y\right)\right]=\psi^{\dagger}\left(x\right)\left[\Gamma^{0},\Lambda^{\nu}\right]\psi\left(x\right)$
with $\Gamma^{0}=I$ the identity, everything commutes with the identity
so all our commutators vanish giving,\begin{equation}
\begin{array}{rcl}
\partial_{z}^{\lambda}T_{\mu\nu\lambda}\left(k_{1},k_{2},q\right) & = & i\int\textrm{d}^{4}x\textrm{d}^{4}y\textrm{d}^{4}z\textrm{ }e^{ik_{1}x+ik_{2}y-iqz}\textrm{ }\left(\begin{array}{rl}
- & iq^{\lambda}{\bkiii{0}{{\cal {T}}j_{\mu}\left(x\right)j_{\nu}\left(y\right)j_{A,\lambda}\left(z\right)}{0}}\\
+ & \partial_{z}^{\lambda}{\bkiii{0}{{\cal {T}}j_{\mu}\left(x\right)j_{\nu}\left(y\right)j_{A,\lambda}\left(z\right)}{0}}\end{array}\right)\\
 & = & i\int\textrm{d}^{4}x\textrm{d}^{4}y\textrm{d}^{4}z\textrm{ }e^{ik_{1}x+ik_{2}y-iqz}\left(\begin{array}{rl}
- & iq^{\lambda}{\bkiii{0}{{\cal {T}}j_{\mu}\left(x\right)j_{\nu}\left(y\right)j_{A,\lambda}\left(z\right)}{0}}\\
+ & {\bkiii{0}{{\cal {T}}j_{\mu}\left(x\right)j_{\nu}\left(y\right)\left(\partial_{z}^{\lambda}j_{A,\lambda}\left(z\right)\right)}{0}}\end{array}\right)\textrm{ }\end{array}.\label{5.2.11}\end{equation}
 We can now use the naive Ward identity to rewrite the 4-divergence
of the axial current as $\partial_{z}^{\lambda}j_{A,\lambda}\left(z\right)=2imP\left(z\right)$
where $P\left(z\right)=\bar{\psi}\left(z\right)\gamma^{5}\psi\left(z\right)$
is the pseudoscalar (probability) density.\begin{equation}
\begin{array}{rcl}
\partial_{z}^{\lambda}T_{\mu\nu\lambda}\left(k_{1},k_{2},q\right) & = & i\int\textrm{d}^{4}x\textrm{d}^{4}y\textrm{d}^{4}z\textrm{ }e^{ik_{1}x+ik_{2}y-iqz}\left(\begin{array}{rl}
- & iq^{\lambda}{\bkiii{0}{{\cal {T}}j_{\mu}\left(x\right)j_{\nu}\left(y\right)j_{A,\lambda}\left(z\right)}{0}}\\
+ & 2im{\bkiii{0}{{\cal {T}}j_{\mu}\left(x\right)j_{\nu}\left(y\right)P\left(z\right)}{0}}\end{array}\right)\end{array}\label{5.2.12}\end{equation}
 In fact $\partial_{z}^{\lambda}T_{\mu\nu\lambda}\left(k_{1},k_{2},q\right)$
is an integration over a 4-divergence which we can take to vanish
using the same trick were we write it as an integral over a surface
at infinity and assume all the fields and their derivatives vanish
at infinity. Hence,\begin{equation}
\int\textrm{d}^{4}x\textrm{d}^{4}y\textrm{d}^{4}z\textrm{ }2m{\bkiii{0}{{\cal {T}}j_{\mu}\left(x\right)j_{\nu}\left(y\right)P\left(z\right)}{0}}-q^{\lambda}{\bkiii{0}{{\cal {T}}j_{\mu}\left(x\right)j_{\nu}\left(y\right)j_{A,\lambda}\left(z\right)}{0}}\label{5.2.13}\end{equation}
Consequently we get the naive Ward identity for our two three point
functions,\begin{equation}
q^{\lambda}T_{\mu\nu\lambda}\left(k_{1},k_{2},q\right)=2mT_{\mu\nu}\left(k_{1},k_{2},q\right).\label{5.2.14}\end{equation}
It is also worth pointing out that there are also vector Ward identities
associated with the VVA amplitude which are derived in exactly the
same way as the axial Ward identity. In this case we differentiate
with respect to one of the other coordinates in the time ordered product,
$x$ or $y$. This gives a time ordered product of the divergence
of one of the vector currents with the other two currents and a bunch
of equal time commutators of the form $\left[j^{0},j^{\mu}\right]$
all of which vanish just as they did in calculating the axial Ward
identity. Applying the classical conservation of the vector current
to the time ordered product gives the vector Ward identities for the
VVA diagram, \begin{equation}
\begin{array}{rcl}
k_{1}^{\mu}T_{\mu\nu\lambda}\left(k_{1},k_{2},q\right) & = & 0\\
k_{2}^{\nu}T_{\mu\nu\lambda}\left(k_{1},k_{2},q\right) & = & 0\end{array}.\label{5.2.15}\end{equation}
Hopefully it's clear that for VVV diagrams we would get three of these
and for VAA we would get one vector and two axial Ward identities. 

Thus we have a Ward identity relating the VVP and VVA amplitudes which
was derived assuming that the classical chiral Ward identity $\partial_{\mu}j_{A}^{\mu}=2imP$
remained true at the quantum level. We can now test this hypothesis
by simply evaluating the two diagrams and see if the relation above
is obeyed. It turns out that the divergence of the axial current does
not obey the classical relation. The classical relation is modified
by an \emph{anomaly} and the Ward identity above relating our three
point functions is missing a term ${\cal {A}}_{\mu\nu}=-\frac{1}{2\pi^{2}e}\epsilon_{\mu\nu\alpha\beta}k_{1}^{\alpha}k_{2}^{\beta}$
on the right, the so-called \emph{ABJ} \emph{anomaly} (Adler $\left[1969\right]$
Bell and Jackiw $\left[1969\right]$). 

Hopefully it is obvious that the Feynman rule for the axial current
vertex is just $e\gamma^{\mu}\gamma^{5}$ as opposed to $e\gamma^{\mu}$
for the usual vector current vertex. With the usual Feynman rules
we then get the following  the amplitudes,\begin{equation}
\begin{array}{rcl}
T_{\mu\nu\lambda}\left(k_{1},k_{2},q\right) & = & \int\frac{\textrm{d}^{4}p}{\left(2\pi\right)^{4}}\times-1\times Tr\left[\begin{array}{rl}
 & \frac{i}{\not p-m}\gamma_{\lambda}\gamma^{5}\frac{i}{\not p-\not q-m}\gamma_{\nu}\frac{i}{\not p-\not k_{1}-m}\gamma_{\mu}\\
+ & \frac{i}{\not p-m}\gamma_{\lambda}\gamma^{5}\frac{i}{\not p-\not q-m}\gamma_{\mu}\frac{i}{\not p-\not k_{2}-m}\gamma_{\nu}\end{array}\right]\\
T_{\mu\nu}\left(k_{1},k_{2},q\right) & = & \int\frac{\textrm{d}^{4}p}{\left(2\pi\right)^{4}}\times-1\times Tr\left[\begin{array}{rl}
 & \frac{i}{\not p-m}\gamma^{5}\frac{i}{\not p-\not q-m}\gamma_{\nu}\frac{i}{\not p-\not k_{1}-m}\gamma_{\mu}\\
+ & \frac{i}{\not p-m}\gamma^{5}\frac{i}{\not p-\not q-m}\gamma_{\mu}\frac{i}{\not p-\not k_{2}-m}\gamma_{\nu}\end{array}\right]\end{array}\label{5.2.16}\end{equation}
 To check the Ward identities we can use the following,\begin{equation}
\left\{ \not q,\gamma^{5}\right\} =0=\left\{ \not p,\gamma^{5}\right\} -2m\gamma^{5}+2m\gamma^{5}\label{5.2.17}\end{equation}
\begin{equation}
\Rightarrow\not q\gamma^{5}=\gamma^{5}\left(\not p-\not q-m\right)+\left(\not p-m\right)\gamma^{5}+2m\gamma^{5}\label{5.2.18}\end{equation}
gives, \begin{equation}
\begin{array}{rcl}
q^{\lambda}T_{\mu\nu\lambda}\left(k_{1},k_{2},q\right) & = & \int\frac{\textrm{d}^{4}p}{\left(2\pi\right)^{4}}\times-1\times Tr\left[\begin{array}{rl}
 & \frac{i}{\not p-m}\not q\gamma^{5}\frac{i}{\not p-\not q-m}\gamma_{\nu}\frac{i}{\not p-\not k_{1}-m}\gamma_{\mu}\\
+ & \frac{i}{\not p-m}\not q\gamma^{5}\frac{i}{\not p-\not q-m}\gamma_{\mu}\frac{i}{\not p-\not k_{2}-m}\gamma_{\nu}\end{array}\right]\\
 & = & -\int\frac{\textrm{d}^{4}p}{\left(2\pi\right)^{4}}Tr\left[\begin{array}{rl}
 & i\frac{i}{\not p-m}\gamma^{5}\gamma_{\nu}\frac{i}{\not p-\not k_{1}-m}\gamma_{\mu}\\
+ & i\frac{i}{\not p-m}\gamma^{5}\gamma_{\mu}\frac{i}{\not p-\not k_{2}-m}\gamma_{\nu}\end{array}\right]+Tr\left[\begin{array}{rl}
 & i\gamma^{5}\frac{i}{\not p-\not q-m}\gamma_{\nu}\frac{i}{\not p-\not k_{1}-m}\gamma_{\mu}\\
+ & i\gamma^{5}\frac{i}{\not p-\not q-m}\gamma_{\mu}\frac{i}{\not p-\not k_{2}-m}\gamma_{\nu}\end{array}\right]\\
 & + & 2m\int\frac{\textrm{d}^{4}p}{\left(2\pi\right)^{4}}\times-1\times Tr\left[\begin{array}{rl}
 & \frac{i}{\not p-m}\gamma^{5}\frac{i}{\not p-\not q-m}\gamma_{\nu}\frac{i}{\not p-\not k_{1}-m}\gamma_{\mu}\\
+ & \frac{i}{\not p-m}\gamma^{5}\frac{i}{\not p-\not q-m}\gamma_{\mu}\frac{i}{\not p-\not k_{2}-m}\gamma_{\nu}\end{array}\right]\\
 & = & R_{\mu\nu}^{(A1)}+R_{\mu\nu}^{(A2)}+2mT_{\mu\nu}\left(k_{1},k_{2},q\right)\end{array}\label{5.2.19}\end{equation}
$R_{\mu\nu}^{(A1)}$ and $R_{\mu\nu}^{(A2)}$ are the so-called \emph{rest}
\emph{terms}, \begin{equation}
\begin{array}{rcl}
R_{\mu\nu}^{\left(A1\right)} & = & i\int\frac{\textrm{d}^{4}p}{\left(2\pi\right)^{4}}Tr\left[\begin{array}{rl}
 & \frac{1}{\not p-m}\gamma^{5}\gamma_{\nu}\frac{1}{\not p-\not k_{1}-m}\gamma_{\mu}\\
+ & \frac{1}{\not p-m}\gamma^{5}\gamma_{\mu}\frac{1}{\not p-\not k_{2}-m}\gamma_{\nu}\end{array}\right]\\
R_{\mu\nu}^{\left(A2\right)} & = & i\int\frac{\textrm{d}^{4}p}{\left(2\pi\right)^{4}}Tr\left[\begin{array}{rl}
 & \gamma^{5}\frac{1}{\not p-\not q-m}\gamma_{\nu}\frac{1}{\not p-\not k_{1}-m}\gamma_{\mu}\\
+ & \gamma^{5}\frac{1}{\not p-\not q-m}\gamma_{\mu}\frac{1}{\not p-\not k_{2}-m}\gamma_{\nu}\end{array}\right]\end{array}\label{5.2.20}\end{equation}
We can rewrite $R_{\mu\nu}^{\left(A1\right)}$ with $\left\{ \gamma^{5},\gamma^{\mu}\right\} $
and the cyclic property of the trace so as to closer resemble $R_{\mu\nu}^{\left(A2\right)}$,\begin{equation}
\begin{array}{rcl}
R_{\mu\nu}^{\left(A1\right)} & = & -i\int\frac{\textrm{d}^{4}p}{\left(2\pi\right)^{4}}Tr\left[\begin{array}{rl}
 & \frac{1}{\not p-m}\gamma_{\nu}\gamma^{5}\frac{1}{\not p-\not k_{1}-m}\gamma_{\mu}\\
+ & \frac{1}{\not p-m}\gamma_{\mu}\gamma^{5}\frac{1}{\not p-\not k_{2}-m}\gamma_{\nu}\end{array}\right]\\
 & = & -i\int\frac{\textrm{d}^{4}p}{\left(2\pi\right)^{4}}Tr\left[\begin{array}{rl}
 & \gamma^{5}\frac{1}{\not p-\not k_{1}-m}\gamma_{\mu}\frac{1}{\not p-m}\gamma_{\nu}\\
+ & \gamma^{5}\frac{1}{\not p-\not k_{2}-m}\gamma_{\nu}\frac{1}{\not p-m}\gamma_{\mu}\end{array}\right]\end{array}.\label{5.2.21}\end{equation}
So we have the axial current Ward identity if the rest terms,\begin{equation}
\begin{array}{rcl}
R_{\mu\nu}^{\left(A1\right)} & = & -i\int\frac{\textrm{d}^{4}p}{\left(2\pi\right)^{4}}Tr\left[\begin{array}{rl}
 & \gamma^{5}\frac{1}{\not p-\not k_{2}-m}\gamma_{\nu}\frac{1}{\not p-m}\gamma_{\mu}\\
+ & \gamma^{5}\frac{1}{\not p-\not k_{1}-m}\gamma_{\mu}\frac{1}{\not p-m}\gamma_{\nu}\end{array}\right]\\
R_{\mu\nu}^{\left(A2\right)} & = & +i\int\frac{\textrm{d}^{4}p}{\left(2\pi\right)^{4}}Tr\left[\begin{array}{rl}
 & \gamma^{5}\frac{1}{\not p-\not k_{1}-\not k_{2}-m}\gamma_{\nu}\frac{1}{\not p-\not k_{1}-m}\gamma_{\mu}\\
+ & \gamma^{5}\frac{1}{\not p-\not k_{1}-\not k_{2}-m}\gamma_{\mu}\frac{1}{\not p-\not k_{2}-m}\gamma_{\nu}\end{array}\right]\end{array}\label{5.2.22}\end{equation}
 cancel each other (they're clearly not both zero). As written above
it looks like simply shifting the integration variable $p$ by a constant
to $p+k_{1}$ in the first term in $R_{\mu\nu}^{\left(A1\right)}$
and to $p+k_{2}$ in the second term in $R_{\mu\nu}^{\left(A2\right)}$
we would have $R_{\mu\nu}^{\left(A2\right)}=-R_{\mu\nu}^{\left(A1\right)}$,
a cancellation of the rest terms and agreement with the axial Ward
identity! This is not true however as these integrals are divergent,
shifting the variable of integration in divergent integrals is not
generally trivial and there may be a finite difference between the
shifted and unshifted values. It is equivalent to saying that the
amplitude $T_{\mu\nu\lambda}\left(k_{1},k_{2},q\right)$ depends on
whether we use $p$ for our loop momenta or $p+constant$, naturally
one expects this not to matter, $p$ is a dummy variable and the limits
of integration are at infinity, however in this way the amplitude
is ambiguous. We have a similar situation for the vector Ward identities
$k_{1}^{\mu}T_{\mu\nu\lambda}\left(k_{1},k_{2},q\right)=k_{2}^{\nu}T_{\mu\nu\lambda}\left(k_{1},k_{2},q\right)=0.$
To see this we use the identities\begin{equation}
\not k_{1}\frac{1}{\not p-m}=1-\left(-\not k_{1}+\not p-m\right)\frac{1}{\not p-m}\textrm{ and }\frac{1}{\not p-\not q-m}\not k_{1}=\frac{1}{\not p-\not q-m}\left(\not p-\not k_{2}-m\right)-1\label{5.2.23}\end{equation}
\begin{equation}
\begin{array}{rl}
\Rightarrow & k_{1}^{\mu}T_{\mu\nu\lambda}\left(k_{1},k_{2},q\right)\\
= & \int\frac{\textrm{d}^{4}p}{\left(2\pi\right)^{4}}\times-1\times Tr\left[\begin{array}{rl}
 & \frac{i}{\not p-m}\gamma_{\lambda}\gamma^{5}\frac{i}{\not p-\not q-m}\gamma_{\nu}\frac{i}{\not p-\not k_{1}-m}\not k_{1}\\
+ & \frac{i}{\not p-m}\gamma_{\lambda}\gamma^{5}\frac{i}{\not p-\not q-m}\not k_{1}\frac{i}{\not p-\not k_{2}-m}\gamma_{\nu}\end{array}\right]\\
= & -\int\frac{\textrm{d}^{4}p}{\left(2\pi\right)^{4}}Tr\left[\begin{array}{rl}
 & i\gamma_{\lambda}\gamma^{5}\frac{i}{\not p-\not q-m}\gamma_{\nu}\frac{i}{\not p-\not k_{1}-m}\\
- & i\frac{i}{\not p-m}\gamma_{\lambda}\gamma^{5}\frac{i}{\not p-\not k_{2}-m}\gamma_{\nu}\end{array}\right]+Tr\left[\begin{array}{rl}
- & i\left(-\not k_{1}+\not p-m\right)\frac{1}{\not p-m}\textrm{ }\gamma_{\lambda}\gamma^{5}\frac{i}{\not p-\not q-m}\gamma_{\nu}\frac{i}{\not p-\not k_{1}-m}\\
+ & i\frac{i}{\not p-m}\gamma_{\lambda}\gamma^{5}\frac{1}{\not p-\not q-m}\left(\not p-\not k_{2}-m\right)\frac{i}{\not p-\not k_{2}-m}\gamma_{\nu}\end{array}\right]\\
= & -\int\frac{\textrm{d}^{4}p}{\left(2\pi\right)^{4}}Tr\left[\begin{array}{rl}
 & i\gamma_{\lambda}\gamma^{5}\frac{i}{\not p-\not k_{1}-\not k_{2}-m}\gamma_{\nu}\frac{i}{\not p-\not k_{1}-m}\\
- & i\gamma_{\lambda}\gamma^{5}\frac{i}{\not p-\not k_{2}-m}\gamma_{\nu}\frac{i}{\not p-m}\end{array}\right]+Tr\left[\begin{array}{rl}
 & i\frac{1}{\not p-m}\textrm{ }\gamma_{\lambda}\gamma^{5}\frac{1}{\not p-\not q-m}\gamma_{\nu}\\
- & i\frac{1}{\not p-m}\gamma_{\lambda}\gamma^{5}\frac{1}{\not p-\not q-m}\gamma_{\nu}\end{array}\right]\\
= & -\int\frac{\textrm{d}^{4}p}{\left(2\pi\right)^{4}}Tr\left[\begin{array}{rl}
 & i\gamma_{\lambda}\gamma^{5}\frac{i}{\not p-\not k_{1}-\not k_{2}-m}\gamma_{\nu}\frac{i}{\not p-\not k_{1}-m}\\
- & i\gamma_{\lambda}\gamma^{5}\frac{i}{\not p-\not k_{2}-m}\gamma_{\nu}\frac{i}{\not p-m}\end{array}\right]=R_{\lambda\nu}^{(V1)}+R_{\lambda\nu}^{(V2)}\end{array}\label{5.2.24}\end{equation}
An analogous calculation gives a similar result for $k_{2}^{\nu}T_{\mu\nu\lambda}\left(k_{1},k_{2},q\right)$,
in both cases we see that shifting the variable of integration in
one of the terms by a constant would give a cancellation and the classical
vector current Ward identity. Like the axial case we cannot trivially
do this as the integrals are linearly divergent. We will now calculate
the rest terms and treat properly the shift of the loop momentum.

We will first attempt to calculate the rest terms in the axial case
starting with the trace. The trace of the first term in $R_{\mu\nu}^{(A2)}$
is,\begin{equation}
\begin{array}{rl}
 & Tr\left[\gamma^{5}\left(\not p-\not k_{1}-\not k_{2}+m\right)\gamma_{\nu}\left(\not p-\not k_{1}+m\right)\gamma_{\mu}\right]\\
= & Tr\left[\gamma^{5}\left(\not p-\not k_{1}-\not k_{2}\right)\gamma_{\nu}\left(\not p-\not k_{1}\right)\gamma_{\mu}+m^{2}\gamma^{5}\gamma_{\nu}\gamma_{\mu}\right]\\
 & Tr\left[\gamma^{5}\gamma_{\nu}\gamma_{\mu}\right]=0\textrm{ and }Tr\left[\left(\not p-q\right)\gamma_{\nu}\left(\not p-\not k_{1}\right)\gamma_{\mu}\gamma^{5}\right]=-4i\epsilon_{\alpha\nu\beta\mu}\left(p-q\right)^{\alpha}\left(p-k_{1}\right)^{\beta}\\
= & 4i\epsilon_{\alpha\beta\nu\mu}\left(p-q\right)^{\alpha}\left(p-k_{1}\right)^{\beta}\end{array}\label{5.2.25}\end{equation}
 \begin{equation}
\Rightarrow\begin{array}{rcl}
R_{\mu\nu}^{\left(A1\right)} & = & -4\epsilon_{\alpha\beta\mu\nu}\int\frac{\textrm{d}^{4}p}{\left(2\pi\right)^{4}}\frac{1}{p^{2}-m^{2}}\left(\begin{array}{rl}
 & \frac{\left(p-k_{2}\right)^{\alpha}p^{\beta}}{\left(p-k_{2}\right)^{2}-m^{2}}\\
- & \frac{\left(p-k_{1}\right)^{\alpha}p^{\beta}}{\left(p-k_{1}\right)^{2}-m^{2}}\end{array}\right)\\
R_{\mu\nu}^{\left(A2\right)} & = & +4\epsilon_{\alpha\beta\mu\nu}\int\frac{\textrm{d}^{4}p}{\left(2\pi\right)^{4}}\frac{1}{\left(p-k_{1}-k_{2}\right)^{2}-m^{2}}\left(\begin{array}{cc}
 & \frac{\left(p-k_{1}-k_{2}\right)^{\alpha}\left(p-k_{1}\right)^{\beta}}{\left(p-k_{1}\right)^{2}-m^{2}}\\
- & \frac{\left(p-k_{1}-k_{2}\right)^{\alpha}\left(p-k_{2}\right)^{\beta}}{\left(p-k_{2}\right)^{2}-m^{2}}\end{array}\right)\end{array}\label{5.2.26}\end{equation}
It is easy to check that terms of the form $\epsilon_{\alpha\beta\mu\nu}p^{\alpha}p^{\beta}$
are identically equal to zero this is what kills the apparent quadratic
divergence making the integrals linearly divergent instead. To get
a cancellation of the rest terms above we need to shift $p\rightarrow p-k_{1}$
in the first term of $R_{\mu\nu}^{\left(A1\right)}$ and $p\rightarrow p-k_{2}$
in the second term of $R_{\mu\nu}^{\left(A1\right)}$. We can write
the rest terms as functions of the loop momentum, we parametrize the
first case as having no shift \emph{i.e.}\begin{equation}
\begin{array}{rlcl}
 & R_{\mu\nu}^{\left(A1\right)} & = & -\int\frac{\textrm{d}^{4}p}{\left(2\pi\right)^{4}}\left(\Delta_{\mu\nu}\left(p\right)-k_{1}\leftrightarrow k_{2}\right)\\
\Rightarrow & R_{\mu\nu}^{\left(A2\right)} & = & +\int\frac{\textrm{d}^{4}p}{\left(2\pi\right)^{4}}\left(\Delta_{\mu\nu}\left(p-k_{1}\right)-k_{1}\leftrightarrow k_{2}\right)\end{array}.\label{5.2.27}\end{equation}
 where the $\Delta$s correspond to the first and second terms in
$R_{\mu\nu}^{\left(A1\right)}$ and $R_{\mu\nu}^{\left(A2\right)}$.
So if the integrals were convergent such shifts would have no effects,
we would find $\int\frac{\textrm{d}^{4}p}{\left(2\pi\right)^{4}}\left(\Delta_{\mu\nu}\left(p\right)-k_{1}\leftrightarrow k_{2}\right)=\int\frac{\textrm{d}^{4}p}{\left(2\pi\right)^{4}}\left(\Delta_{\mu\nu}\left(p-k_{1}\right)-k_{1}\leftrightarrow k_{2}\right)$
and the rest terms would cancel but the integrals are divergent so
this is not the case. To find the anomaly we need to know the extent
to which $R_{\mu\nu}^{\left(A1\right)}$ and $R_{\mu\nu}^{\left(A2\right)}$
don't cancel \emph{i.e.} we must determine how the rest terms are
affected by shifts in the loop momentum.\begin{equation}
\begin{array}{rcl}
\int\frac{\textrm{d}^{4}p}{\left(2\pi\right)^{4}}\left(\Delta_{\mu\nu}\left(p+a\right)-\Delta_{\mu\nu}\left(p\right)\right) & = & 4\epsilon_{\alpha\beta\mu\nu}\int\frac{\textrm{d}^{4}p}{\left(2\pi\right)^{4}}\frac{\left(p+a-k_{2}\right)^{\alpha}\left(p+a\right)^{\beta}}{\left(\left(p+a\right)^{2}-m^{2}\right)\left(\left(p+a-k_{2}\right)^{2}-m^{2}\right)}\\
 & - & 4\epsilon_{\alpha\beta\mu\nu}\int\frac{\textrm{d}^{4}p}{\left(2\pi\right)^{4}}\frac{\left(p-k_{2}\right)^{\alpha}p^{\beta}}{\left(p^{2}-m^{2}\right)\left(\left(p-k_{2}\right)^{2}-m^{2}\right)}\end{array}\label{5.2.28}\end{equation}
We can Taylor expand $\Delta_{\mu\nu}\left(p+a\right)$ about $\Delta_{\mu\nu}\left(p\right)$
to give,\begin{equation}
\Delta_{\mu\nu}\left(p+a\right)-\Delta_{\mu\nu}\left(p\right)=a^{\kappa}\partial_{\kappa}\Delta_{\mu\nu}\left(p\right)+a^{\kappa}a^{\lambda}\partial_{\kappa}\partial_{\lambda}\Delta_{\mu\nu}\left(p\right)+...\label{5.2.29}\end{equation}
hence,\begin{equation}
\begin{array}{rcl}
\int\frac{\textrm{d}^{4}p}{\left(2\pi\right)^{4}}\left(\Delta_{\mu\nu}\left(p+a\right)-\Delta_{\mu\nu}\left(p\right)\right) & = & \int\frac{\textrm{d}^{4}p}{\left(2\pi\right)^{4}}a^{\kappa}\partial_{\kappa}\left(\Delta_{\mu\nu}\left(p\right)+a^{\lambda}\partial_{\lambda}\Delta_{\mu\nu}\left(p\right)+...\right)\\
 & = & \frac{1}{\left(2\pi\right)^{4}}\int_{A}dS^{\tau}a_{\tau}\left(\Delta_{\mu\nu}\left(p\right)+a^{\lambda}\partial_{\lambda}\Delta_{\mu\nu}\left(p\right)+...\right)\end{array}\label{5.2.30}\end{equation}

\noindent where in the last line we have used Gauss' law to rewrite
the volume integral $\int\textrm{d}^{4}p\textrm{ }\partial_{\kappa}\left(a^{\kappa}f_{\mu\nu}\left(p\right)\right)$
as a surface integral $\int_{A}dS^{\tau}a_{\tau}f_{\mu\nu}\left(p\right)$.
The surface $A$ is a 4d (hyper)sphere of infinite radius (the volume
being integrated over was symmetric and infinite). The integration
measure $dS^{\tau}$ is an infinitesimal 4-vector defined on $A$
and perpendicular to it. Now to evaluate the surface integral (we
replace $\left(p-k_{2}\right)^{\alpha}p^{\beta}$ with $-k_{2}^{\alpha}p^{\beta}$
in the numerators of the integrand as $\epsilon_{\alpha\beta\mu\nu}p^{\alpha}p^{\beta}\equiv0$).
\begin{equation}
\frac{1}{\left(2\pi\right)^{4}}\int_{A}dS^{\tau}a_{\tau}\left(\Delta_{\mu\nu}\left(p\right)+a^{\lambda}\partial_{\lambda}\Delta_{\mu\nu}\left(p\right)+...\right)=\frac{-4\epsilon_{\alpha\beta\mu\nu}k_{2}^{\alpha}}{\left(2\pi\right)^{4}}\int_{A}dS^{\tau}a_{\tau}\textrm{ }\left(\begin{array}{rr}
 & \frac{p^{\beta}}{\left(p^{2}-m^{2}\right)\left(\left(p-k_{2}\right)^{2}-m^{2}\right)}\\
+ & a^{\lambda}\partial_{\lambda}\frac{p^{\beta}}{\left(p^{2}-m^{2}\right)\left(\left(p-k_{2}\right)^{2}-m^{2}\right)}\\
+ & .........\end{array}\right)\label{5.2.31}\end{equation}
 As the surface of the sphere we are integrating over is at infinity
we can greatly simplify the denominators above by taking the limit
$p^{2}\rightarrow\infty$.\begin{equation}
\frac{1}{\left(2\pi\right)^{4}}\int_{A}dS^{\tau}a_{\tau}\left(\Delta_{\mu\nu}\left(p\right)+a^{\lambda}\partial_{\lambda}\Delta_{\mu\nu}\left(p\right)+...\right)=\frac{-4\epsilon_{\alpha\beta\mu\nu}k_{2}^{\alpha}}{\left(2\pi\right)^{4}}\int_{A}dS^{\tau}a_{\tau}\textrm{ }\left(\frac{p^{\beta}}{p^{4}}+a^{\lambda}\partial_{\lambda}\left(\frac{p^{\beta}}{p^{4}}\right)+...\right)\label{5.2.32}\end{equation}
In three dimensions the integration measure is easily found to be
$dS^{i}=p^{i}\left|\mathbf{p}\right|sin\theta\textrm{ d}\theta\textrm{d}\phi$,
where $p^{i}\equiv\mathbf{p}$ is a vector from the centre to some
point on the surface of the sphere (\emph{i.e.} it is radial) and
$\theta,\phi$ are the usual spherical polar and azimuthal angles.
Analogous to this in four dimensions the integration measure is proportional
to $p^{\tau}\left|p\right|^{2}$ consequently the second term above
does not contribute as it has fewer and fewer powers of $p$ which
makes that integrand tend to zero all over the surface of integration
$\left(\left|p\right|\rightarrow\infty\right)$. The same is true
for the other higher derivative terms in the Taylor  expansion. However
we can see that the first term contains an equal number of powers
of $p$ in the numerator and denominator in the integrand which making
it non-vanishing, we will evaluate this below. Had the integral been
quadratically divergent the second term would have contributed, cubic
divergences would have made the third term contribute \emph{etc}.
Following this reasoning we find that only in the case of integrals
which are convergent or at most logarithmically divergent does a shift
in the integration variable not contribute any such terms and the
final integration is unaffected. We can also generalize our 3-d spherical
surface element to a 4-d one, by analogy it will be of the form $dS^{\tau}=p^{\tau}p^{2}f\left(\theta,\phi\right)\textrm{d}\theta\textrm{d}\phi\textrm{d}\psi$
where $\theta,\textrm{ }\phi$ and $\psi$ are the spherical polar
angles of Minkowski space. Also the 4-vector $a^{\tau}$ is a constant
and so it can be brought outside the integral. \begin{equation}
\begin{array}{rccl}
\Rightarrow & \frac{1}{\left(2\pi\right)^{4}}\int_{A}dS^{\tau}a_{\tau}\left(\Delta_{\mu\nu}\left(p\right)+a^{\lambda}\partial_{\lambda}\Delta_{\mu\nu}\left(p\right)+...\right) & = & \frac{-4\epsilon_{\alpha\beta\mu\nu}k_{2}^{\alpha}}{\left(2\pi\right)^{4}}\int_{A}p^{\tau}\left|p\right|^{2}f\left(\theta,\phi\right)a_{\tau}\frac{p^{\beta}}{p^{4}}\textrm{ d}\theta\textrm{d}\phi\textrm{d}\psi\\
 &  & = & \frac{-4\epsilon_{\alpha\beta\mu\nu}k_{2}^{\alpha}a_{\tau}}{\left(2\pi\right)^{4}}\int_{A}\textrm{ }\frac{p^{\tau}p^{\beta}}{p^{2}}\textrm{ }f\left(\theta,\phi\right)\textrm{ d}\theta\textrm{d}\phi\textrm{d}\psi\end{array}\label{5.2.33}\end{equation}
Needless to say an honest evaluation of this integral is highly tedious.
Instead of showing the 16 integrations we sketch how they are done.
First to properly determine the integration measure $f\left(\theta,\psi\right)\textrm{ d}\theta\textrm{d}\phi\textrm{d}\psi$
one has to evaluate $f\left(\theta,\phi\right)$, this is the infinitesimal
surface area element. It is equivalent to the determinant of the metric
(of the sphere). In 3-d the metric of a sphere of radius $R$ is,\begin{equation}
\left(\begin{array}{cc}
g_{\theta\theta} & g_{\theta\phi}\\
g_{\phi\theta} & g_{\phi\phi}\end{array}\right)=\left(\begin{array}{cc}
R^{2} & 0\\
0 & R^{2}sin^{2}\theta\end{array}\right).\label{5.2.34}\end{equation}
To get the metric all you have to do is take a point on the sphere
parametrized in spherical polars and make infinitesimal rotations
in all the angles then compute $ds^{2}$, the infinitesimal distance
between the new and old points. The root of the determinant of the
metric above is $\sqrt{R^{2}\times R^{2}sin^{2}\theta}$ which gives
the well known $R^{2}sin\theta\textrm{ d}\theta\textrm{d}\phi$ as
the surface are element. Pictorially this corresponds to the area
of an infinitesimal square on the sphere made by moving a point on
the surface by two orthogonal rotations (the metric is diagonal) $\theta\rightarrow\theta+\textrm{d}\theta$
and $\phi\rightarrow\phi+\textrm{d}\phi$. This method generalizes
to higher dimensions, in Minkowski space a sphere has the form,\begin{equation}
t^{2}-x^{2}-y^{2}-z^{2}=R^{2}\label{5.2.35}\end{equation}
\begin{equation}
\begin{array}{rcl}
x & = & iR\textrm{ }sin\psi\textrm{ }sin\phi\textrm{ }sin\theta\\
y & = & iR\textrm{ }cos\psi\textrm{ }sin\phi\textrm{ }sin\theta\\
z & = & iR\textrm{ }cos\phi\textrm{ }sin\theta\\
t & = & R\textrm{ }cos\theta\end{array}\label{5.2.35b}\end{equation}
The metric when calculated using the method just described is,\begin{equation}
\left(\begin{array}{rcl}
g_{\theta\theta} & g_{\theta\phi} & g_{\theta\psi}\\
g_{\phi\theta} & g_{\phi\phi} & g_{\phi\psi}\\
g_{\psi\theta} & g_{\psi\phi} & g_{\psi\psi}\end{array}\right)=\left(\begin{array}{rcl}
R^{2} & 0 & 0\\
0 & R^{2}sin^{2}\theta & 0\\
0 & 0 & R^{2}sin^{2}\theta\textrm{ }sin^{2}\phi\end{array}\right)\label{5.2.36}\end{equation}
giving $R^{3}sin^{2}\theta\textrm{ }sin\phi\textrm{ d}\theta\textrm{d}\phi\textrm{d}\psi$
as the surface area element. In terms of our integral over momenta
we now have, \begin{equation}
\Rightarrow\frac{1}{\left(2\pi\right)^{4}}\int_{A}dS^{\tau}a_{\tau}\left(\Delta_{\mu\nu}\left(p\right)+a^{\lambda}\partial_{\lambda}\Delta_{\mu\nu}\left(p\right)+...\right)=\frac{-4\epsilon_{\alpha\beta\mu\nu}k_{2}^{\alpha}a_{\tau}}{\left(2\pi\right)^{4}}\int_{0}^{2\pi}\textrm{d}\psi\int_{0}^{\pi}\textrm{d}\phi\int_{0}^{\pi}\textrm{d}\theta\textrm{ }\frac{p^{\tau}p^{\beta}}{p^{2}}sin^{2}\theta\textrm{ }sin\phi.\label{5.2.37}\end{equation}
 It is now a matter of grinding out the integrals by substituting
in $p^{0},p^{1},p^{2},p^{3}$ completely analogous to $x,y,z,t$ above.
The integrals with $\tau\ne\beta$ are quick to do, they all vanish.
For $\tau=\beta=1,2,3$ the integrals give $-\frac{\pi^{2}}{2}$ and
for $\tau=\beta=0$ the integral is $\frac{\pi^{2}}{2}$ so the integral
is in fact equal to $\frac{\pi^{2}}{2}g^{\tau\beta}$ giving, \begin{equation}
\int\frac{\textrm{d}^{4}p}{\left(2\pi\right)^{4}}\left(\Delta_{\mu\nu}\left(p+a\right)-\Delta_{\mu\nu}\left(p\right)\right)=\frac{1}{\left(2\pi\right)^{4}}\int_{A}dS^{\tau}a_{\tau}\left(\Delta_{\mu\nu}\left(p\right)+a^{\lambda}\partial_{\lambda}\Delta_{\mu\nu}\left(p\right)+...\right)=\frac{1}{8\pi^{2}}\epsilon_{\alpha\beta\mu\nu}a^{\alpha}k_{2}^{\beta}.\label{5.2.38}\end{equation}
The actual rest terms were, from before, \begin{equation}
R_{\mu\nu}^{\left(A1\right)}+R_{\mu\nu}^{\left(A2\right)}=\int\frac{\textrm{d}^{4}p}{\left(2\pi\right)^{4}}\left(\left(\Delta_{\mu\nu}\left(p-k_{1}\right)-k_{1}\leftrightarrow k_{2}\right)-\left(\Delta_{\mu\nu}\left(p\right)-k_{1}\leftrightarrow k_{2}\right)\right)\label{5.2.39}\end{equation}
 so the first term is the third term with a shift $p\rightarrow p-k_{1}$
in the integrand giving $-\frac{1}{8\pi^{2}}\epsilon_{\alpha\beta\mu\nu}k_{1}^{\alpha}k_{2}^{\beta}$
and the second term is the same as the fourth with a shift $p\rightarrow p-k_{2}$
giving $\frac{1}{8\pi^{2}}\epsilon_{\alpha\beta\mu\nu}k_{2}^{\alpha}k_{1}^{\beta}$,
where in the latter case we have swapped $k_{1}\leftrightarrow k_{2}$
in our formula for the shift. By asymmetry of $\epsilon_{\alpha\beta\mu\nu}$
this gives, $R_{\mu\nu}^{\left(A1\right)}+R_{\mu\nu}^{\left(A2\right)}=-\frac{1}{4\pi^{2}}\epsilon_{\alpha\beta\mu\nu}k_{1}^{\alpha}k_{2}^{\beta}$
and consequently the anomalous/quantum axial Ward identity,\begin{equation}
q^{\lambda}T_{\mu\nu\lambda}\left(k_{1},k_{2},q\right)=2mT_{\mu\nu}\left(k_{1},k_{2},q\right)-\frac{1}{4\pi^{2}}\epsilon_{\alpha\beta\mu\nu}k_{1}^{\alpha}k_{2}^{\beta}\label{5.2.40}\end{equation}
 with $-\frac{1}{4\pi^{2}}\epsilon_{\alpha\beta\mu\nu}k_{1}^{\alpha}k_{2}^{\beta}$
being the \emph{anomaly}.

What about the rest terms in the vector Ward identity, do they cancel
each other out? From before we had,\begin{equation}
\begin{array}{rl}
\Rightarrow & k_{1}^{\mu}T_{\mu\nu\lambda}\left(k_{1},k_{2},q\right)=i\int\frac{\textrm{d}^{4}p}{\left(2\pi\right)^{4}}Tr\left[\begin{array}{rl}
 & \gamma_{\lambda}\gamma^{5}\frac{1}{\not p-\not k_{1}-\not k_{2}-m}\gamma_{\nu}\frac{1}{\not p-\not k_{1}-m}\\
- & \gamma_{\lambda}\gamma^{5}\frac{1}{\not p-\not k_{2}-m}\gamma_{\nu}\frac{1}{\not p-m}\end{array}\right]=R_{\lambda\nu}^{(V1)}-R_{\lambda\nu}^{(V2)}\end{array}\label{5.2.41}\end{equation}
 The trace evaluates to,\begin{equation}
\begin{array}{rcl}
Tr\left[\gamma_{\lambda}\gamma^{5}\left(\not p-\not k_{1}-\not k_{2}+m\right)\gamma_{\nu}\left(\not p-\not k_{1}+m\right)\right] & = & Tr\left[\gamma_{\lambda}\gamma^{5}\left(\not p-\not k_{1}-\not k_{2}\right)\gamma_{\nu}\left(\not p-\not k_{1}\right)\right]+Tr\left[m^{2}\gamma_{\lambda}\gamma^{5}\gamma_{\nu}\right]\\
 &  & \textrm{Use }Tr\left[\gamma_{\mu}\gamma_{\nu}\gamma^{5}\right]=0\\
 & = & -Tr\left[\gamma^{5}\gamma_{\lambda}\left(\not p-\not k_{1}-\not k_{2}\right)\gamma_{\nu}\left(\not p-\not k_{1}\right)\right]\\
 &  & \textrm{Use }Tr\left[\gamma^{5}\gamma_{\mu}\gamma_{\nu}\gamma_{\alpha}\gamma_{\beta}\right]=4i\epsilon_{\mu\nu\alpha\beta}\\
 & = & -4i\epsilon_{\alpha\beta\lambda\nu}k_{2}^{\alpha}\left(p-k_{1}\right)^{\beta}\end{array}\label{5.2.42}\end{equation}
\begin{equation}
\Rightarrow k_{1}^{\mu}T_{\mu\nu\lambda}\left(k_{1},k_{2},q\right)=\frac{4\epsilon_{\alpha\beta\lambda\nu}k_{2}^{\alpha}}{\left(2\pi\right)^{4}}\int\textrm{d}^{4}p\left(\frac{\left(p-k_{1}\right)^{\beta}}{\left(p-k_{1}-k_{2}\right)^{2}-m^{2}}-\frac{p^{\beta}}{\left(p-k_{2}\right)^{2}-m^{2}}\right).\label{5.2.43}\end{equation}
As before we can write this integral as a surface integral as we did
in the case of the axial Ward identity. Using the same techniques
as for the axial case we have,\begin{equation}
\begin{array}{rcl}
\Rightarrow k_{1}^{\mu}T_{\mu\nu\lambda}\left(k_{1},k_{2},q\right) & = & \frac{4\epsilon_{\alpha\beta\lambda\nu}k_{2}^{\alpha}}{\left(2\pi\right)^{4}}\int_{A}dS^{\tau}k_{1,\tau}\frac{p^{\beta}}{\left(p-k_{2}\right)^{2}-m^{2}}\\
 & = & \frac{4\epsilon_{\alpha\beta\lambda\nu}k_{2}^{\alpha}k_{1,\tau}}{\left(2\pi\right)^{4}}\int_{0}^{2\pi}\textrm{d}\psi\int_{0}^{\pi}\textrm{d}\phi\int_{0}^{\pi}\textrm{d}\theta\textrm{ }\frac{p^{\tau}p^{\beta}}{p^{2}}sin^{2}\theta\textrm{ }sin\phi\\
 & = & \frac{1}{8\pi^{2}}\epsilon_{\alpha\beta\nu\lambda}k_{1}^{\alpha}k_{2}^{\beta}.\end{array}\label{5.2.44}\end{equation}
The vector Ward identity is also anomalous. 

As I said earlier the linear divergences essentially generate an ambiguity
of the VVA amplitude, that the amplitude depends on how one initially
defines the loop momentum. We can show this explicitly. Imagine that
instead of defining the loop momentum as $p$ in the initial diagram
we defined it as $p+a$ where a is some constant. For momentum conservation
at each of the vertices this will mean that $a$ is some linear combination
of $k_{1}$ and $k_{2}$. The amplitude for $\tilde{T}_{\mu\nu\lambda}\left(k_{1},k_{2},q\right)$
is now,\begin{equation}
\tilde{T}_{\mu\nu\lambda}\left(k_{1},k_{2},q\right)=i\int\frac{\textrm{d}^{4}p}{\left(2\pi\right)^{4}}Tr\left[\begin{array}{rl}
 & \frac{1}{\not p+\not a-m}\gamma_{\lambda}\gamma^{5}\frac{1}{\not p+\not a-\not q-m}\gamma_{\nu}\frac{1}{\not p+\not a-\not k_{1}-m}\gamma_{\mu}\\
+ & \frac{1}{\not p+\not a-m}\gamma_{\lambda}\gamma^{5}\frac{1}{\not p+\not a-\not q-m}\gamma_{\mu}\frac{1}{\not p+\not a-\not k_{2}-m}\gamma_{\nu}\end{array}\right].\label{5.2.45}\end{equation}
Previously the amplitude was,\begin{equation}
T_{\mu\nu\lambda}\left(k_{1},k_{2},q\right)=i\int\frac{\textrm{d}^{4}p}{\left(2\pi\right)^{4}}Tr\left[\begin{array}{rl}
 & \frac{1}{\not p-m}\gamma_{\lambda}\gamma^{5}\frac{1}{\not p-\not q-m}\gamma_{\nu}\frac{1}{\not p-\not k_{1}-m}\gamma_{\mu}\\
+ & \frac{1}{\not p-m}\gamma_{\lambda}\gamma^{5}\frac{1}{\not p-\not q-m}\gamma_{\mu}\frac{1}{\not p-\not k_{2}-m}\gamma_{\nu}\end{array}\right].\label{5.2.46}\end{equation}
If I can simply shift the loop momentum by the constant $a$, $p\rightarrow p+a$
in the original amplitude then there is clearly no difference in the
two amplitudes. We know that this is not the case as shifts of the
variable of integration in greater than logarithmically divergent
integrals are not trivial. Defining,\begin{equation}
\begin{array}{rcl}
\Delta_{\mu\nu\lambda}\left(p\right) & = & Tr\left[\frac{1}{\not p-m}\gamma_{\lambda}\gamma^{5}\frac{1}{\not p-\not q-m}\gamma_{\nu}\frac{1}{\not p-\not k_{1}-m}\gamma_{\mu}\right]\\
 & = & \frac{Tr\left[\left(\not p+m\right)\gamma_{\lambda}\gamma^{5}\left(\not p-\not q+m\right)\gamma_{\nu}\left(\not p-\not k_{1}+m\right)\gamma_{\mu}\right]}{\left(p^{2}-m^{2}\right)\left(\left(p-q\right)^{2}-m^{2}\right)\left(\left(p-k_{1}\right)^{2}-m^{2}\right)}\end{array}\label{5.2.47}\end{equation}
using our surface integral technology we can write, \begin{equation}
\begin{array}{rcl}
\tilde{T}_{\mu\nu\lambda}\left(k_{1},k_{2},q\right)-T_{\mu\nu\lambda}\left(k_{1},k_{2},q\right) & = & \frac{i}{\left(2\pi\right)^{4}}\int\textrm{d}S^{\tau}a_{\tau}\left(\Delta_{\mu\nu\lambda}\left(p\right)+\textrm{derivatives of }\Delta_{\mu\nu\lambda}\left(p\right)...+\begin{array}{c}
k_{1}\leftrightarrow k_{2}\\
\mu\leftrightarrow\nu\end{array}\right)\\
 & = & \frac{i}{\left(2\pi\right)^{4}}\int_{0}^{2\pi}\textrm{d}\psi\int_{0}^{\pi}\textrm{d}\phi\int_{0}^{\pi}\textrm{d}\theta\textrm{ }sin^{2}\theta sin\phi\textrm{ }p^{\tau}p^{2}a_{\tau}\left(\Delta_{\mu\nu\lambda}\left(p\right)+\right.\\
 &  & \left.\textrm{derivatives of }\Delta_{\mu\nu\lambda}\left(p\right)...\right)+\begin{array}{c}
k_{1}\leftrightarrow k_{2}\\
\mu\leftrightarrow\nu\end{array}\end{array}\label{5.2.48}\end{equation}
looking at the trace the largest power of $p$ possible in it is ${\cal {O}}\left(p^{3}\right)$,
we get another three powers of $p$ from the surface element $p^{\tau}p^{2}$.
Our surface of integration is a sphere of infinite radius $\left|p\right|=\infty$.
Taking the limit of $p^{2}\rightarrow\infty$ in the denominator it
becomes $p^{6}$, consequently the integrand above is ${\cal {O}}\left(p^{0}\right)$
and the higher derivative terms above are ${\cal {O}}\left(p^{-1}\right)$
and below. This being the case we need only concern ourselves with
the first term in the Taylor expansion of $\Delta_{\mu\nu\lambda}\left(p+a\right)-\Delta_{\mu\nu\lambda}\left(p\right)$,\begin{equation}
\tilde{T}_{\mu\nu\lambda}\left(k_{1},k_{2},q\right)-T_{\mu\nu\lambda}\left(k_{1},k_{2},q\right)=\frac{i}{\left(2\pi\right)^{4}}\int_{0}^{2\pi}\textrm{d}\psi\int_{0}^{\pi}\textrm{d}\phi\int_{0}^{\pi}\textrm{d}\theta\textrm{ }sin^{2}\theta sin\phi\textrm{ }p^{\tau}p^{2}a_{\tau}\left(\Delta_{\mu\nu\lambda}\left(p\right)+\begin{array}{c}
k_{1}\leftrightarrow k_{2}\\
\mu\leftrightarrow\nu\end{array}\right).\label{5.2.49}\end{equation}
 In addition, considering again the trace inside $\Delta_{\mu\nu\lambda}\left(p\right)$
we need terms ${\cal {O}}\left(p^{3}\right)$ to combine with $p^{\tau}p^{2}$
of the surface term and hence counter the $p^{6}$ denominator so
as to have the integrand non zero at $\left|p\right|=\infty$. We
can therefore save ourselves several pages of trace algebra by only
considering the ${\cal {O}}\left(p^{3}\right)$ terms in the trace,
for $\Delta_{\mu\nu\lambda}\left(p\right)$ this means, \begin{equation}
Tr\left[\left(\not p+m\right)\gamma_{\lambda}\gamma^{5}\left(\not p-\not q+m\right)\gamma_{\nu}\left(\not p-\not k_{1}+m\right)\gamma_{\mu}\right]\rightarrow Tr\left[\not p\gamma_{\lambda}\gamma^{5}\not p\gamma_{\nu}\not p\gamma_{\mu}\right]\label{5.2.50}\end{equation}

The trace is complicated and requires the use of the gamma matrix
identity $\gamma_{\alpha}\gamma_{\beta}\gamma_{\gamma}=g_{\alpha\beta}\gamma_{\gamma}+g_{\beta\gamma}\gamma_{\alpha}-g_{\alpha\gamma}\gamma_{\beta}+i\epsilon_{\mu\alpha\beta\gamma}\gamma^{\mu}\gamma^{5}$,\begin{equation}
\begin{array}{llll}
 & Tr\left[\gamma^{5}\not p\gamma_{\nu}\not p\gamma_{\mu}\not p\gamma_{\lambda}\right]\\
 & \textrm{Use identity above}.\\
= & p_{\mu}Tr\left[\gamma^{5}\not p\gamma_{\nu}\not p\gamma_{\lambda}\right] & + & p_{\lambda}Tr\left[\gamma^{5}\not p\gamma_{\nu}\not p\gamma_{\mu}\right]\\
- & g_{\mu\lambda}Tr\left[\gamma^{5}\not p\gamma_{\nu}\not p\not p\right] & + & i\epsilon_{\alpha\mu\gamma\lambda}p^{\gamma}Tr\left[\gamma^{5}\not p\gamma_{\nu}\not p\gamma^{\alpha}\gamma^{5}\right]\\
 & \gamma^{5}\gamma^{5}=I\textrm{ in last term}.\\
= & p_{\mu}Tr\left[\gamma^{5}\not p\gamma_{\nu}\not p\gamma_{\lambda}\right] & + & p_{\lambda}Tr\left[\gamma^{5}\not p\gamma_{\nu}\not p\gamma_{\mu}\right]\\
- & g_{\mu\lambda}Tr\left[\gamma^{5}\not p\gamma_{\nu}\not p\not p\right] & + & i\epsilon_{\alpha\mu\gamma\lambda}p^{\gamma}Tr\left[\not p\gamma_{\nu}\not p\gamma^{\alpha}\right]\\
 & \textrm{Use standard traces}.\\
= & 4i\epsilon_{\alpha\nu\beta\lambda}p^{\alpha}p^{\beta}p_{\mu} & - & 4i\epsilon_{\alpha\beta\nu\mu}p^{\alpha}p^{\beta}p_{\lambda}\\
+ & 4ig_{\mu\lambda}\epsilon_{\alpha\beta\nu\gamma}p^{\alpha}p^{\beta}p^{\gamma} & - & 4i\epsilon_{\alpha\beta\mu\lambda}p^{\beta}p_{\nu}p^{\alpha}\\
- & 4i\epsilon_{\alpha\beta\mu\lambda}p^{\beta}p_{\nu}p^{\alpha} & - & 4i\epsilon_{\nu\mu\gamma\lambda}p^{\gamma}p^{2}\\
 & \epsilon_{\mu\nu\alpha\beta}p^{\alpha}p^{\beta}\textrm{ type terms}=0\\
= & -4i\epsilon_{\nu\mu\gamma\lambda}p^{\gamma}p^{2}\end{array}\label{5.2.50}\end{equation}
\begin{equation}
\begin{array}{rl}
\Rightarrow & \frac{i}{\left(2\pi\right)^{4}}\int_{0}^{2\pi}\textrm{d}\psi\int_{0}^{\pi}\textrm{d}\phi\int_{0}^{\pi}\textrm{d}\theta\textrm{ }sin^{2}\theta sin\phi\textrm{ }p^{\tau}p^{2}a_{\tau}\Delta_{\mu\nu\lambda}\left(p\right)\\
= & \frac{i}{\left(2\pi\right)^{4}}\int_{0}^{2\pi}\textrm{d}\psi\int_{0}^{\pi}\textrm{d}\phi\int_{0}^{\pi}\textrm{d}\theta\textrm{ }sin^{2}\theta sin\phi\textrm{ }p^{\tau}p^{2}a_{\tau}\frac{-4i\epsilon_{\nu\mu\gamma\lambda}p^{\gamma}p^{2}}{p^{6}}\\
= & \frac{4a_{\tau}\epsilon_{\nu\mu\gamma\lambda}}{\left(2\pi\right)^{4}}\int_{0}^{2\pi}\textrm{d}\psi\int_{0}^{\pi}\textrm{d}\phi\int_{0}^{\pi}\textrm{d}\theta\textrm{ }sin^{2}\theta sin\phi\textrm{ }\frac{p^{\tau}p^{\gamma}}{p^{2}}\end{array}\label{5.2.51}\end{equation}
We previously evaluated this integral in calculating the rest terms
for the axial Ward identity, it was $\frac{1}{2}\pi^{2}g^{\tau\gamma}$
\begin{equation}
\Rightarrow\tilde{T}_{\mu\nu\lambda}\left(k_{1},k_{2},q\right)-T_{\mu\nu\lambda}\left(k_{1},k_{2},q\right)=\frac{\epsilon_{\nu\mu\gamma\lambda}a^{\gamma}}{8\pi^{2}}+\begin{array}{c}
k_{1}\leftrightarrow k_{2}\\
\mu\leftrightarrow\nu\end{array}\label{5.2.52}\end{equation}
Remember we said that to conserve momenta at the vertices $a^{\gamma}$
must be a linear combination of $k_{1}$ and $k_{2}$ (it is not also
a linear combination of $q$ as $q=k_{1}+k_{2}$ from simple momentum
conservation), writing $a=xk_{1}+\left(x-y\right)k_{2}$ we have,\begin{equation}
\Rightarrow\tilde{T}_{\mu\nu\lambda}\left(k_{1},k_{2},q\right)-T_{\mu\nu\lambda}\left(k_{1},k_{2},q\right)=\frac{y}{8\pi^{2}}\epsilon_{\mu\nu\lambda\gamma}\left(k_{1}-k_{2}\right)^{\gamma}.\label{5.2.53}\end{equation}
does the same thing happen to the amplitude for the VVP diagram, how
does it behave if the loop momentum is defined differently? By analogy
with the VVA diagram the corresponding (surface) integrand in the
case of the VVP diagram is the same as that of the VVA diagram but
with the axial vector coupling $\gamma_{\lambda}\gamma^{5}$ replaced
by the pseudoscalar coupling $\gamma^{5}$, \begin{equation}
\Delta_{\mu\nu}\left(p\right)=\frac{Tr\left[\left(\not p+m\right)\gamma^{5}\left(\not p-\not q+m\right)\gamma_{\nu}\left(\not p-\not k_{1}+m\right)\gamma_{\mu}\right]}{\left(p^{2}-m^{2}\right)\left(\left(p-q\right)^{2}-m^{2}\right)\left(\left(p-k_{1}\right)^{2}-m^{2}\right)}.\label{5.2.54}\end{equation}
Just as in the case of the VVA diagram the surface integral will let
us take the limit $\left|p\right|\rightarrow\infty$ in the denominator
making it $p^{6}$ and the integration measure will contribute $p^{\tau}p^{2}$
to the numerator so like the VVA case we need only consider terms
${\cal {O}}$$\left(p^{3}\right)$ in the trace. In other words the
trace simplifies,\begin{equation}
Tr\left[\left(\not p+m\right)\gamma^{5}\left(\not p-\not q+m\right)\gamma_{\nu}\left(\not p-\not k_{1}+m\right)\gamma_{\mu}\right]\rightarrow Tr\left[\not p\gamma^{5}\not p\gamma_{\nu}\not p\gamma_{\mu}\right]=Tr\left[\textrm{odd number of }\gamma\textrm{s}\right]=0.\label{5.2.55}\end{equation}
 Crucially the same ambiguity is therefore not true of the VVP diagram
it does not change under constant shifts of the loop momentum $p\rightarrow p+constant$,
unlike the VVA diagram, the VVP amplitude is \emph{unambiguous}.

Hence had we written down our VVA diagram with $p+a$ where previously
there had been just $p$ we would have had the extra term above dotted
with the different 4-vectors appearing in our Ward identities! In
the case of the axial Ward identity we would have,\begin{equation}
\begin{array}{rcl}
q^{\lambda}\tilde{T}_{\mu\nu\lambda}\left(k_{1},k_{2},q\right) & = & q^{\lambda}T_{\mu\nu\lambda}\left(k_{1},k_{2},q\right)+\frac{y}{8\pi^{2}}\epsilon_{\mu\nu\lambda\gamma}q^{\lambda}\left(k_{1}-k_{2}\right)^{\gamma}\\
 & = & q^{\lambda}T_{\mu\nu\lambda}\left(k_{1},k_{2},q\right)+\frac{y}{8\pi^{2}}\epsilon_{\mu\nu\lambda\gamma}\left(k_{2}^{\lambda}k_{1}^{\gamma}-k_{1}^{\lambda}k_{2}^{\gamma}\right)\end{array}.\label{5.2.56}\end{equation}
Recall we calculated for the axial Ward identity,\begin{equation}
\begin{array}{crcl}
 & q^{\lambda}T_{\mu\nu\lambda}\left(k_{1},k_{2},q\right) & = & 2mT_{\mu\nu}\left(k_{1},k_{2},q\right)-\frac{1}{4\pi^{2}}\epsilon_{\alpha\beta\mu\nu}k_{1}^{\alpha}k_{2}^{\beta}\\
\Rightarrow & \Rightarrow q^{\lambda}\tilde{T}_{\mu\nu\lambda}\left(k_{1},k_{2},q\right) & = & 2mT_{\mu\nu}\left(k_{1},k_{2},q\right)-\frac{1}{4\pi^{2}}\epsilon_{\alpha\beta\mu\nu}\left(1+y\right)k_{1}^{\alpha}k_{2}^{\beta}\end{array}.\label{5.2.57}\end{equation}
For the vector Ward identity we would have,\begin{equation}
\begin{array}{rcl}
k_{1}^{\mu}\tilde{T}_{\mu\nu\lambda}\left(k_{1},k_{2},q\right) & = & k_{1}^{\mu}T_{\mu\nu\lambda}\left(k_{1},k_{2},q\right)+\frac{y}{8\pi^{2}}\epsilon_{\mu\nu\lambda\gamma}k_{1}^{\mu}\left(k_{1}-k_{2}\right)^{\gamma}\\
 & = & k_{1}^{\mu}T_{\mu\nu\lambda}\left(k_{1},k_{2},q\right)-\frac{y}{8\pi^{2}}\epsilon_{\mu\nu\lambda\gamma}k_{1}^{\mu}k_{2}^{\gamma}\end{array}.\label{5.2.58}\end{equation}
Again, in explicit calculation of the quantum vector Ward identity
we found,\begin{equation}
k_{1}^{\mu}T_{\mu\nu\lambda}\left(k_{1},k_{2},q\right)=\frac{1}{8\pi^{2}}\epsilon_{\alpha\beta\nu\lambda}k_{1}^{\alpha}k_{2}^{\beta}.\label{5.2.59}\end{equation}
 Substituting this in we find,\begin{equation}
k_{1}^{\mu}\tilde{T}_{\mu\nu\lambda}\left(k_{1},k_{2},q\right)=\frac{1}{8\pi^{2}}\epsilon_{\alpha\beta\nu\lambda}\left(1-y\right)k_{1}^{\alpha}k_{2}^{\beta}.\label{5.2.60}\end{equation}
Hopefully it's clear that the same result is found for the other vector
Ward identity,\begin{equation}
k_{2}^{\nu}\tilde{T}_{\mu\nu\lambda}\left(k_{1},k_{2},q\right)=\frac{1}{8\pi^{2}}\epsilon_{\alpha\beta\mu\lambda}\left(1-y\right)k_{1}^{\alpha}k_{2}^{\beta}.\label{5.2.61}\end{equation}
Therefore by a judicious choice of loop momentum in the VVA diagram
(remember nothing happens to the VVP diagram under a constant shift
of the loop momentum) we can recover the classical Ward identities
for the axial vector and vector currents. For $y=-1$ we regain the
classical equation of motion for the the axial current (conservation
of the axial vector current when the loop fermions are massless) but
the classical vector Ward identity is violated by the quantum anomaly
above. For $y=1$ the opposite is true and for all other values of
$y$ both the classical axial vector and vector current conservation
equations contain extra quantum anomalies. It is a matter of convention
in doing perturbation theory that one defines the loop momenta of
such ambiguous graphs such that the vector current is conserved and
the axial vector current contains the anomaly.

\newpage
\section{Fujikawa's Method for determining the Chiral Anomaly. }

In the last lecture we saw that our naive chiral Ward identity did
not survive quantization. This can be attributed to the variance of
the functional integration measure under local transformations of
(only the fermion) fields giving an extra contribution to the path
integral under the transformation which we called the anomaly function.
We will now attempt an alternative calculation of the (Abelian) anomaly
first following the method of K.Fujikawa \cite{key-23,key-24,key-25},
this is a delicate calculation.

In our initial derivation of the axial vector Ward identity we found
that our action of \emph{axial} \emph{electrodynamics},\begin{equation}
S=\int\textrm{d}^{4}x\textrm{ }-\frac{1}{4}F_{\mu\nu}F^{\mu\nu}-\frac{1}{4}G_{\mu\nu}G^{\mu\nu}+\bar{\psi}\left(i\not\partial+q\not V+g\not A\gamma^{5}-m\right)\psi+Ghosts+Gauge\textrm{ }Fixing\label{5.3.1}\end{equation}
 transformed under local $U_{A}\left(1\right)$ transformations $\psi\left(x\right)\rightarrow e^{+iq\beta\left(x\right)\gamma^{5}}\psi\left(x\right)$,
$\bar{\psi}\left(x\right)\rightarrow\bar{\psi}\left(x\right)e^{-iq\beta\left(x\right)\gamma^{5}}$,
of the fermion fields as, \begin{equation}
S\rightarrow S+\int\textrm{d}^{4}x\textrm{ }g\beta\left(x\right)\left(\partial_{\mu}j_{A}^{\mu}-2im\bar{\psi}\gamma^{5}\psi\right).\label{5.3.2}\end{equation}
We worried that the Jacobian of the path integral measure would not
remain invariant under such redefinitions of the fields and this is
indeed the case. The essence of Fujikawa's anomaly calculation is
to properly work out the change in the path integral measure under
the transformations of the fields above. For reasons which will become
apparent Fujikawa's anomaly calculation is only well defined in Euclidean
space, we can transform the anomaly back to Minkowski space at the
end. To first get into Euclidean space we have to make a Wick rotation
to Euclidean space \emph{i.e.} we rotate the upper index $0$th component
of all vectors as $ia^{0}=a^{4}$ and replace $g^{\mu\nu}=-\delta^{\mu\nu}$.
So in Euclidean space we have the following identifications, $ix^{0}=x^{4}$,
$i\gamma^{0}=\gamma^{4}$, $\partial_{0}=i\frac{\partial}{\partial x^{4}}=i\partial_{4}$,
$A_{0}=iA_{4}$. Hence $\not D=\gamma^{\mu}D_{\mu}=g^{\mu\nu}\gamma_{\nu}D_{\mu}=-\gamma_{\mu}D_{\mu}=-\gamma^{1}D_{1}-\gamma^{2}D_{2}-\gamma^{3}D_{3}-\gamma^{4}D_{4}$.
As $\gamma^{4}=i\gamma^{0}$ in Euclidean space all gamma matrices
are anti-Hermitian (in Minkowski space only $\gamma^{0}$ was Hermitian
$\gamma^{1},\gamma^{2}$ and $\gamma^{3}$ were anti-Hermitian). The
$\gamma^{5}$ matrix is still Hermitian in Euclidean space though,\begin{equation}
\gamma^{5}=i\gamma^{0}\gamma^{1}\gamma^{2}\gamma^{3}=\gamma^{4}\gamma^{1}\gamma^{2}\gamma^{3}=-\gamma^{1}\gamma^{2}\gamma^{3}\gamma^{4}\label{5.3.3}\end{equation}
and we still have $\left\{ \gamma^{5},\gamma^{\mu}\right\} =0$. Therefore
in Euclidean space the Dirac operator above $i\not D=i\not\partial+\not V+\gamma^{5}\not A$
is Hermitian (see the end of section 1.3 for a discussion of the Hermiticity
of the gauge fields). 

The next thing to do is decompose the fermion fields into eigenstates
of the full Dirac operator, \emph{i.e.} eigenstates of $i\not\partial+q\not V+g\not A\gamma^{5}-m$
which we will henceforth denote $i\not D$. \begin{equation}
\begin{array}{c}
i\not D\phi_{i}\left(x\right)=\lambda_{i}\phi_{i}\left(x\right)\\
\phi_{i}^{\dagger}\left(x\right)i\not D=\lambda_{i}\phi_{i}^{\dagger}\left(x\right)\end{array}\label{5.3.4}\end{equation}
Seeing as $i\not D$ is a Hermitian operator the eigenvalues $\lambda_{i}$
are real! For no background fields $V_{\mu},\textrm{ }A_{\mu}$ these
eigenstates are the usual Dirac eigenfunctions of definite momentum
$p^{\mu}$ as $\not D=\not\partial$ and $\lambda_{i}^{2}=p^{2}=m^{2}$.
When the background fields are non-zero but well behaved and fixed,
the eigenfunctions above are still the eigenfunctions of $i\not D$
but only in the asymptotic limit of large momentum, in which case
we have $p\gg V,A$ \begin{equation}
\begin{array}{crcl}
\Rightarrow & i\not D\phi\left(x\right)\sim\left(i\not\partial+\not V+\not A\gamma^{5}\right)\Psi e^{-ip.x} & = & \left(\not p+q\not V+g\not A\gamma^{5}\right)\Psi e^{-ip.x}\\
 &  & \approx & \not p\Psi e^{-ip.x}\end{array}.\label{5.3.5}\end{equation}
This is an important point but it is more important to note that although
we will be appealing to this limit later we will at no point be assuming
that we are in a plane wave basis, we are strictly dealing with the
eigenfunctions of the full Dirac operator including the gauge field(s)
$i\not D$. Note that in Euclidean space we can choose our Dirac matrices
so that $\phi^{\dagger}=\bar{\phi}$. These eigenfunctions are orthogonal
and form a complete set just like in the case of the free Dirac equation,
\begin{equation}
\begin{array}{rcl}
\psi\left(x\right) & = & \Sigma_{i}a_{i}\phi_{i}\left(x\right)\\
\bar{\psi}\left(x\right) & = & \Sigma_{i}\phi_{i}^{\dagger}\left(x\right)\bar{b}_{i}\end{array}\label{5.3.6}\end{equation}
\begin{equation}
\begin{array}{rcl}
\int\textrm{d}^{4}x\textrm{ }\phi_{i,\alpha}^{\dagger}\left(x\right)\phi_{j,\beta}\left(x\right) & = & \delta_{ij}\delta_{\alpha\beta}\\
\Sigma_{n}\phi_{n,\alpha}^{\dagger}\left(y\right)\phi_{n,\beta}\left(x\right) & = & \delta\left(x-y\right)\delta_{\alpha\beta}\end{array}\label{5.3.7}\end{equation}
with $a_{i}$ and $\bar{b}_{j}$ \emph{Grassmann} variables $\left(\left\{ a_{i},a_{j}\right\} =0,\textrm{ }\left\{ \bar{b}_{i},\bar{b}_{j}\right\} =0,\textrm{ }\left\{ a_{i},\bar{b}_{j}\right\} =0\right)$,
$\alpha,\beta$ spinor indices and $\phi_{n}\left(x\right)$ eigenfunctions
(\emph{normal} numbers). The fields are now completely specified by
these two (Grassmann) variables, consequently an integral over all
possible field configurations $\psi$ and $\bar{\psi}$ is equivalent
to an integral over all possible combinations of values of $a_{i}$
and $\bar{b}_{i}$. This being the case we can now redefine the path
integral measure of the fields as \begin{equation}
\int D\bar{\psi}D\psi\rightarrow\prod_{n}\int\textrm{d}a_{n}\prod_{m}\int\textrm{d}\bar{b}_{m}.\label{5.3.8}\end{equation}
So how do the Grassmann variables change under the axial transformation?
\begin{equation}
\begin{array}{rl}
 & \psi\left(x\right)\rightarrow e^{+i\beta\left(x\right)\gamma^{5}}\psi\left(x\right)\\
\Rightarrow & \Sigma_{j}a_{j}\phi_{j}\left(x\right)\rightarrow\left(1+i\beta\left(x\right)\gamma^{5}\right)\Sigma_{j}a_{j}\phi_{j}\left(x\right)\\
\Rightarrow & \int\textrm{d}^{4}x\textrm{ }\phi_{i}^{\dagger}\left(x\right)\Sigma_{j}a_{j}\phi_{j}\left(x\right)\rightarrow\int\textrm{d}^{4}x\textrm{ }\phi_{i}^{\dagger}\left(x\right)\Sigma_{j}\left(1+i\beta\left(x\right)\gamma^{5}\right)a_{j}\phi_{j}\left(x\right)\\
\Rightarrow & a_{i}\rightarrow\Sigma_{j}\left(\delta_{ij}+\int\textrm{d}^{4}x\textrm{ }i\beta\left(x\right){\bkii{i}{x}}\gamma^{5}{\bkii{x}{j}}\right)a_{j}\\
\Rightarrow & a_{i}\rightarrow a_{i}+\Sigma_{j}\left(i\int\textrm{d}^{4}x\textrm{ }\beta\left(x\right){\bkii{i}{x}}\gamma^{5}{\bkii{x}{j}}\right)a_{j}.\end{array}\label{5.3.9}\end{equation}
We will abbreviate the matrix $\int\textrm{d}^{4}x\textrm{ }\beta\left(x\right){\bkii{i}{x}}\gamma^{5}{\bkii{x}{j}}$
by $M_{ij}$ so the change in $a_{i}$ is given by,\begin{equation}
a_{i}\rightarrow\left(\delta_{ij}+M_{ij}\right)a_{j}\label{5.3.10}\end{equation}
where the repeated index implies a summation. Likewise we have\begin{equation}
\begin{array}{rl}
 & \bar{\psi}\left(x\right)\rightarrow\bar{\psi}\left(x\right)e^{+i\beta\left(x\right)\gamma^{5}}\\
\Rightarrow & \Sigma_{j}\phi_{j}^{\dagger}\left(x\right)\bar{b}_{j}\rightarrow\left(\Sigma_{j}\phi_{j}^{\dagger}\left(x\right)\bar{b}_{j}\right)\left(1+i\beta\left(x\right)\gamma^{5}\right)\\
\Rightarrow & \int\textrm{d}^{4}x\textrm{ }\left(\Sigma_{j}\phi_{j}^{\dagger}\left(x\right)\bar{b}_{j}\right)\phi_{i}\left(x\right)\rightarrow\int\textrm{d}^{4}x\textrm{ }\left(\Sigma_{j}\phi_{j}^{\dagger}\left(x\right)\bar{b}_{j}\left(1+i\beta\left(x\right)\gamma^{5}\right)\right)\phi_{i}\left(x\right)\\
\Rightarrow & \bar{b}_{i}\rightarrow\Sigma_{j}\left(\delta_{ij}+\int\textrm{d}^{4}x\textrm{ }i\beta\left(x\right){\bkii{j}{x}}\gamma^{5}{\bkii{x}{i}}\right)\bar{b}_{j}\\
\Rightarrow & \bar{b}_{i}\rightarrow\bar{b}_{i}+i\Sigma_{j}\left(\int\textrm{d}^{4}x\textrm{ }\beta\left(x\right){\bkii{j}{x}}\gamma^{5}{\bkii{x}{i}}\right)\bar{b}_{j}.\end{array}\label{5.3.11}\end{equation}
 So $\bar{b}_{i}$ transforms as, \begin{equation}
\bar{b}_{i}\rightarrow\left(\delta_{ij}+M_{ji}\right)\bar{b}_{j}\label{5.3.12}\end{equation}
a little different to $a_{i}$. Hence the total path integral measure
changes as,\begin{equation}
\prod_{n}\int\textrm{d}a_{n}\prod_{m}\int\textrm{d}\bar{b}_{m}\rightarrow\prod_{n}\int\textrm{d}\left(\Sigma_{j}\left(\delta_{nj}+M_{nj}\right)a_{j}\right)\prod_{m}\int\textrm{d}\left(\Sigma_{j}\left(\delta_{mj}+M_{jm}\right)\bar{b}_{j}\right).\label{5.3.13}\end{equation}
We would like to simplify this so that the connection with the old
measure is clearer \emph{i.e.} we want the Jacobian whatever that
is. To do this it will be good to know a few basic things about integrating
over Grassmann variables. 

We shall now revisit our earlier discussion of Grassmann variables.
Consider a function $f\left(a_{i}\right)$ of our Grassmann variable
$a_{i}$, we can Taylor expand it, as with functions of regular numbers,
but because $a_{i}$ is Grassmann $a_{i}a_{i}=-a_{i}a_{i}=0$ so all
terms containing powers of $a_{i}$ greater than one vanish. Hence
the Taylor  expansion of any function of a single Grassmann variable
$a_{i}$ terminates after two terms,\begin{equation}
f\left(a_{i}\right)=P+a_{i}Q\label{5.3.14}\end{equation}
 where $P$ and $Q$ are the coefficients of the expansion.

Integration over Grassmann variables is, at least mathematically,
somewhat ambiguous, consider the double derivative of $f\left(a_{i}\right)$,
$\partial_{a_{i}}^{2}f\left(a_{i}\right)=0$ for any function $f$,
so in the world of Grassmann numbers there would appear to be no inverse
operation to left or right differentiation (for Grassmann variables
differentiating from the left is different to differentiating from
the right as $\partial_{\alpha_{i}}$ anticommutes with Grassmann
numbers). Instead of using the formal definition of integration being
the inverse operation of differentiation we make do with an alternative
definition which makes the best possible analogy with that of bosonic
integration. It is convention in Physics to insist that Grassmann
integration is analogous to bosonic integration when shifting the
variable of integration by a constant as this is something we often
find ourselves doing in Physics. We want to define our integration
such that under $a_{i}\rightarrow a_{i}+c_{i}$ with $c_{i}$ a constant
number, the integral of the function over all possible values of the
Grassmann variable $a_{i}$ is unchanged:\begin{equation}
\begin{array}{rcl}
\int\textrm{d}a_{i}\textrm{ }f\left(a_{i}\right) & = & \int\textrm{d}\left(a_{i}+c_{i}\right)\textrm{ }f\left(a_{i}+c_{i}\right)\\
 & = & \int\textrm{d}a_{i}\textrm{ }f\left(a_{i}+c_{i}\right)\\
 & = & \int\textrm{d}a_{i}\textrm{ }P+\left(a_{i}+c_{i}\right)Q\\
 & = & \int\textrm{d}a_{i}\textrm{ }f\left(a_{i}\right)+\int\textrm{d}a_{i}\textrm{ }c_{i}Q\end{array}\label{5.3.15}\end{equation}
 To get the desired invariance under the shifts of variable above
we must clearly have Grassmann integration such that $\int\textrm{d}a_{i}\textrm{ }Qc_{i}=0$.
To do this we \emph{define} for integration of any Grassmann variable
over the domain of all possible Grassmann numbers, \begin{equation}
\begin{array}{lcl}
\int\textrm{d}a_{i}\textrm{ } & = & 0\\
\int\textrm{d}a_{i}\textrm{ }a_{i} & = & 1.\end{array}\label{5.3.16}\end{equation}
The first of these definitions arises because we want $\int\textrm{d}a_{i}\textrm{ }c_{i}Q=0$
for any function $f\left(a_{i}\right)$ and any shift $c_{i}$. The
second definition is merely a matter of convention, $\int\textrm{d}a_{i}\textrm{ }a_{i}$
is some number, it is convention to take it as one as a matter of
normalization. These are known as the \emph{Berezin} \emph{Integration}
\emph{Rules}. Another motivation for them is that with these definitions
the integration $\int\textrm{d}a_{i}$ is equivalent to differentiating
from the left (\emph{i.e.} differentiation in the usual sense). Recall
we are interested in the Jacobian involved in the transformation of
a multiple integration over Grassmann variables. With the definitions
above we have, for $c_{i}$ a real bosonic number,\begin{equation}
\begin{array}{rcl}
\int\textrm{d}\left(c_{i}a_{i}\right)\textrm{ }c_{i}a_{i} & = & \frac{\partial}{\partial\left(c_{i}a_{i}\right)}\left(c_{i}a_{i}\right)\\
 & = & \frac{\partial}{\partial a_{i}}a_{i}\\
 & = & \int\textrm{d}a_{i}\textrm{ }a_{i}\end{array}\label{5.3.17}\end{equation}
\begin{equation}
\Rightarrow\textrm{ }d\left(c_{i}a_{i}\right)=\frac{1}{c_{i}}da_{i}\label{5.3.18}\end{equation}
the Jacobian is the inverse of what it normally is! This persists
to multiple integrals over Grassmann variables. Note that the ordering
of integration of Grassmann variables is important, defining integration
of Grassmann variables the same as differentiation means that the
integrals anticommute like the derivatives. This being the case we
can get some useful expressions for our products of integrals, for
instance we can rewrite,\begin{equation}
\prod_{n}\int\textrm{d}a_{n}=\Sigma_{i_{1}}\Sigma_{i_{2}}...\Sigma_{i_{N}}\frac{1}{N!}\epsilon^{i_{1}i_{2}...i_{N}}\int\textrm{d}a_{i_{1}}\int\textrm{d}a_{i_{2}}...\int\textrm{d}a_{i_{N}}\label{5.3.19}\end{equation}
 where the  which are all the same up to a minus sign, which the $\epsilon^{i_{1}i_{2}...i_{N}}$
tensor takes care of. This gives $N!$ terms which are all the same
as $\prod_{n}\int\textrm{d}a_{n}$ taking into account the anticommuting
property and the whole lot gets divided by $N!$ giving the equivalence
with $\prod_{n}\int\textrm{d}a_{n}$.

In our case we have, \begin{equation}
\begin{array}{rl}
 & \prod_{n}\int\textrm{d}a_{n}\\
= & \Sigma_{i_{1}}...\Sigma_{i_{N}}\frac{1}{N!}\epsilon^{i_{1}...i_{N}}\int\textrm{d}a_{i_{1}}...\int\textrm{d}a_{i_{N}}\\
\rightarrow & \Sigma_{i_{1}}...\Sigma_{i_{N}}\frac{1}{N!}\epsilon^{i_{1}...i_{N}}\int\textrm{d}\left(\Sigma_{j_{1}}\left(\delta_{i_{1}j_{1}}+M_{i_{1}j_{1}}\right)a_{j_{1}}\right)...\int\textrm{d}\left(\Sigma_{j_{2}}\left(\delta_{i_{N}j_{N}}+M_{i_{N}j_{N}}\right)a_{j_{N}}\right)\\
= & \Sigma_{i_{1}}...\Sigma_{i_{N}}\Sigma_{j_{1}}...\Sigma_{j_{N}}\frac{1}{N!}\epsilon^{i_{1}...i_{N}}\left(\delta_{i_{1}j_{1}}+M_{i_{1}j_{1}}\right)^{-1}...\left(\delta_{i_{N}j_{N}}+M_{i_{N}j_{N}}\right)^{-1}\int\textrm{d}a_{j_{1}}...\int\textrm{d}a_{j_{N}},\end{array}\label{5.3.20}\end{equation}
where in the last line we have taken the summation signs outside the
integration measures they were in and used the rule for Grassmann
integration (described above) $\int\textrm{d}\left(ca\right)=\frac{1}{c}\int\textrm{d}a$.
Because the integrals $\int\textrm{d}a_{j_{1}}...\int\textrm{d}a_{j_{N}}$
all anticommute with each other we can make the replacement,\begin{equation}
\int\textrm{d}a_{j_{1}}...\int\textrm{d}a_{j_{N}}=\epsilon^{j_{1}...j_{N}}\left(\prod_{n}\int\textrm{d}a_{n}\right)\label{5.3.21}\end{equation}
hence,\begin{equation}
\begin{array}{rcl}
\prod_{n}\int\textrm{d}a_{n} & \rightarrow & \Sigma_{i_{1}}...\Sigma_{i_{N}}\Sigma_{j_{1}}...\Sigma_{j_{N}}\frac{1}{N!}\epsilon^{i_{1}...i_{N}}\left(\delta_{i_{1}j_{1}}+M_{i_{1}j_{1}}\right)^{-1}...\left(\delta_{i_{N}j_{N}}+M_{i_{N}j_{N}}\right)^{-1}\epsilon^{j_{1}...j_{N}}\left(\prod_{n}\int\textrm{d}a_{n}\right)\\
 & = & \left(\prod_{n}\int\textrm{d}a_{n}\right)\Sigma_{i_{1}}...\Sigma_{i_{N}}\Sigma_{j_{1}}...\Sigma_{j_{N}}\frac{1}{N!}\epsilon^{j_{1}...j_{N}}\epsilon^{i_{1}...i_{N}}\left(\delta_{i_{1}j_{1}}+M_{i_{1}j_{1}}\right)^{-1}...\left(\delta_{i_{N}j_{N}}+M_{i_{N}j_{N}}\right)^{-1}\end{array}.\label{5.3.22}\end{equation}
Hopefully it is clear that the analysis for $\prod_{n}\int\textrm{d}\bar{b}_{n}$
is exactly the same with the replacements $a\rightarrow\bar{b}$ and
$M_{ij}\rightarrow M_{ji}$ or equivalently $M\rightarrow M^{T}$
everywhere. We can also use the fact that the determinant of an $N\times N$
matrix \textbf{$T_{ij}$} is\begin{equation}
\textrm{Det}\left(T_{ij}\right)=\frac{1}{N!}\Sigma_{\alpha_{1}}...\Sigma_{\alpha_{N}}\Sigma_{\beta_{1}}...\Sigma_{\beta_{N}}\epsilon^{\alpha_{1}...\alpha_{N}}\epsilon^{\beta_{1}...\beta_{N}}T_{\alpha_{1}\beta_{1}}...T_{\alpha_{N}\beta_{N}}.\label{5.3.23}\end{equation}
Recall that $\beta\left(x\right)$ is infinitesimal (we are working
with an infinitesimal chiral transformation) $M_{ii}\ll\delta_{ii}$
so we can expand $\left(\delta_{ii}+M_{ii}\right)^{-1}\approx\delta_{ii}-M_{ii}$.
Using this definition in the working above we finally have, \begin{equation}
\begin{array}{rcl}
\prod_{n}\int\textrm{d}a_{n} & \rightarrow & \textrm{Det}\left(\tilde{M}_{ij}\right)\prod_{n}\int\textrm{d}a_{n}\\
\prod_{n}\int\textrm{d}\bar{b}_{n} & \rightarrow & \textrm{Det}\left(\tilde{M}_{ji}\right)\prod_{n}\int\textrm{d}\bar{b}_{n}\end{array}\label{5.3.25}\end{equation}
where \[
\tilde{M}_{ij}\approx\left\{ \begin{array}{ll}
\delta_{ii}-M_{ii} & i=j\\
\frac{1}{M_{ij}} & i\ne j\end{array}\right..\]
 We can write this determinant using the identity $\textrm{ln Det}\left(\tilde{M}_{ij}\right)=\textrm{ln Tr}\left[\tilde{M}_{ij}\right]$
which is true for any matrix which can be diagonalized by transformations
of the form $\mathbf{UMU}^{-1}$. Given $\beta\left(x\right)$ is
small (we are considering an infinitesimal chiral transformation)
so for $i=j$,\begin{equation}
\textrm{ln}\left[\delta_{ij}-i\int\textrm{d}^{4}x\textrm{ }\beta\left(x\right){\bkii{i}{x}}\gamma^{5}{\bkii{x}{j}}\right]\approx-i\int\textrm{d}^{4}x\textrm{ }\beta\left(x\right){\bkii{i}{x}}\gamma^{5}{\bkii{x}{j}}\textrm{ }\forall\textrm{ }i=j.\label{5.3.27A}\end{equation}
 As we are interested in the trace of the above we need not concern
ourselves with the case $i\ne j$. \begin{equation}
\begin{array}{rccl}
\Rightarrow & \textrm{Det}\left(\tilde{M}_{ij}\right) & = & \textrm{exp }-Tr\textrm{ }\left[i\int\textrm{d}^{4}x\textrm{ }\beta\left(x\right){\bkii{i}{x}}\gamma^{5}{\bkii{x}{j}}\textrm{ }\right]\\
 &  & = & \textrm{exp }-i\Sigma_{n}\textrm{ }\int\textrm{d}^{4}x\textrm{ }\beta\left(x\right){\bkii{n}{x}}\gamma^{5}{\bkii{x}{n}}\textrm{ }.\end{array}\label{5.3.28}\end{equation}
Clearly we get exactly the same  for $\prod_{n}\int\textrm{d}\bar{b}_{n}$.
So the total transformation of the measure is,\begin{equation}
\prod_{n}\int\textrm{d}a_{n}\int\textrm{d}\bar{b}_{n}\rightarrow\left(\textrm{exp }-2i\Sigma_{n}\textrm{ }\int\textrm{d}^{4}x\textrm{ }\beta\left(x\right){\bkii{n}{x}}\gamma^{5}{\bkii{x}{n}}\right)\prod_{n}\int\textrm{d}a_{n}\int\textrm{d}\bar{b}_{n}.\label{5.3.29}\end{equation}
 If we show the Dirac indices $\left(\rho,\sigma\right)$ in the exponent
we have, \begin{equation}
\begin{array}{rccl}
\Rightarrow & \textrm{Det}\left(\delta_{ij}-i\int\textrm{d}^{4}x\textrm{ }\beta\left(x\right){\bkii{j}{x}}\gamma^{5}{\bkii{x}{i}}\right)\times i\leftrightarrow j & = & \textrm{exp }-2i\Sigma_{n}\textrm{ }\int\textrm{d}^{4}x\textrm{ }\beta\left(x\right)\phi_{n,\rho}^{\dagger}\left(x\right)\gamma_{\rho\sigma}^{5}\phi_{n,\sigma}\left(x\right)\\
 &  & = & \textrm{exp }-2i\int\textrm{d}^{4}x\textrm{ }\beta\left(x\right)\gamma_{\rho\sigma}^{5}\left(\Sigma_{n}\phi_{n,\rho}^{\dagger}\left(x\right)\phi_{n,\sigma}\left(x\right)\right)\textrm{ }\\
 &  &  & \textrm{Use completeness}:\textrm{ }\Sigma_{n}\phi_{n,\alpha}^{\dagger}\left(x\right)\phi_{n,\beta}\left(y\right)=\delta_{\alpha\beta}\delta\left(x-y\right)\\
 &  & = & \textrm{exp }-2i\textrm{ }\int\textrm{d}^{4}x\textrm{ }\beta\left(x\right)\textrm{ }\left(Tr\textrm{ }\left[\gamma^{5}\right].\delta\left(0\right)\right)\end{array}\label{5.3.30}\end{equation}
 The sum in the integrand is badly defined, $Tr\left[\gamma^{5}\right]=0$,
$\delta\left(0\right)=\infty$ and we need to regulate it. There are
various methods of regularization open to us at this point, the crucial
point is that they will all give us the same anomaly. We will follow
Fujikawa's regularization which was to insert a Gaussian cut-off into
the sum so as to damp the contributions from the large eigenvalues.
For a detailed discussion of all other regularization methods the
reader is referred to $\left[4\right]$. \begin{equation}
\Sigma_{n}\phi_{n}^{\dagger}\left(x\right)\gamma^{5}\phi_{n}\left(x\right)=\textrm{lim}_{M\rightarrow\infty}\Sigma_{n}\phi_{n}^{\dagger}\left(x\right)\gamma^{5}\textrm{ exp}\left[-\frac{\lambda_{n}^{2}}{M^{2}}\right]\phi_{n}\left(x\right)\label{5.3.31}\end{equation}
with $\lambda_{n}$ the eigenvalues of the Dirac operator $i\not D$,\begin{equation}
=\textrm{lim}_{M\rightarrow\infty}\Sigma_{n}{\bkiii{n}{\gamma^{5}\textrm{ exp}\left[\frac{\not D^{2}}{M^{2}}\right]}{n}}.\label{5.3.32}\end{equation}
This regularization is $U_{V}\left(1\right)$ and $U_{A}\left(1\right)$
gauge invariant (this is easy to see if one considers the first form
of the regulator \emph{i.e.} the regulator in terms of eigenvalues
instead of $\not D$).

Now consider an eigenstate of $i\not D$ $\left|i\right>$ with eigenvalue
$\lambda_{i}$ this means that another of the eigenstates $\left|\tilde{i}\right>=\gamma^{5}\left|i\right>$
has the opposite eigenvalue:\begin{equation}
\begin{array}{rcl}
i\not D\left|\tilde{i}\right> & = & i\not D\gamma^{5}\left|i\right>\\
 & = & -\gamma^{5}i\not D\left|i\right>\\
 & = & -\lambda_{i}\left|i\right>\end{array}.\label{5.3.33}\end{equation}
Consequently the element of the sum vanishes as all the eigenstates
in the complete set with different eigenvalues are orthogonal. \begin{equation}
{\bkiii{n}{\gamma^{5}\textrm{ exp}\left[\frac{\not D^{2}}{M^{2}}\right]}{n}}=\textrm{exp}\left[-\frac{\lambda_{n}^{2}}{M^{2}}\right]{\bkii{\tilde{n}}{n}}=0\label{5.3.34}\end{equation}
The only exception occurs when the eigenvalue of the state in question
is zero in which case the eigenstates need not be orthogonal. These
so-called zero modes $i\not D\left|i,0\right>=0$ provide the sole
contribution to the sum \begin{equation}
{\bkiii{n,0}{\gamma^{5}\textrm{ exp}\left[\frac{\not D^{2}}{M^{2}}\right]}{n,0}}=\textrm{exp}\left[\frac{0^{2}}{M^{2}}\right]{\bkii{n,0}\gamma^{5}{n,0}}={\bkiii{n,0}{\gamma^{5}}{n,0}}\label{5.3.35}\end{equation}
and hence the anomaly. We can classify the zero modes according to
their chirality \begin{equation}
\begin{array}{rcl}
\gamma^{5}\left|n_{+},0\right> & = & +\left|n_{+},0\right>\\
\gamma^{5}\left|n_{-},0\right> & = & -\left|n_{-},0\right>.\end{array}\label{5.3.36}\end{equation}
Therefore our sum over all states is equal to,\begin{equation}
\begin{array}{rcl}
\textrm{lim}_{M\rightarrow\infty}\Sigma_{n}{\bkiii{n}{\gamma^{5}\textrm{ exp}\left[\frac{\not D^{2}}{M^{2}}\right]}{n}} & = & \Sigma_{n\textrm{ }}{\bkiii{n_{+},0}{\gamma^{5}}{n_{+},0}}+{\bkiii{n_{-},0}{\gamma^{5}}{n_{-},0}}\\
 & = & N_{+}-N_{-}\end{array}\label{5.3.37}\end{equation}
where $N_{+}$ and $N_{-}$ are the number of positive and negative
chirality zero modes respectively. The above is a statement of a topological
theorem known as the the Atiyah-Singer index theorem $\left[5\right]$.
The Atiyah-Singer index theorem relates the index of the Dirac operator
to the topological charge $q=N_{+}-N_{-}$, this also goes under the
name Pontryagin index. In the case of an $SU\left(2\right)$ gauge
theory it is the instanton number and for $U\left(1\right)$ gauge
theory in three dimensions it is known as the (magnetic) monopole
charge. For those more familiar with topological jargon the anomaly
/ topological charge is the winding number of the map of the surface
at the boundary of space to the gauge group of the theory in question.
Consequently for trivial homotopy groups $\pi_{n}\left(G\right)=0$
we have no anomaly. For a detailed topological analysis of the anomaly
the reader is encouraged to read more in $\left[4\right]$.

Note that these zero modes live in a subspace of the Hilbert space
where effectively $\left[i\not D,\gamma^{5}\right]=0$. On the other
hand all the non-zero modes in the rest of Hilbert space are not invariant
under chiral transformations recall that $\gamma^{5}\left|n\right>$
and $\left|n\right>$ have opposite eigenvalues of $\not D$ so for
the space of non-zero modes $\left[i\not D,\gamma^{5}\right]=2i\not D\gamma^{5}$.
It is the case that in our basis of eigenstates of the full Dirac
operator $i\not D$, the chiral asymmetry (anomaly) is contained in
the sub-space of zero modes. If we change to a different basis by
some unitary transformation the chiral asymmetry (anomaly) will move
into the other space of non-zero modes. We can show this for plane
waves, let's Fourier transform our eigenfunctions so we are in a plane
wave basis,\begin{equation}
\phi_{n}\left(x\right)=\int\frac{\textrm{d}^{4}k}{\left(2\pi\right)^{2}}e^{ik.x}\tilde{\phi}_{n}\left(k\right)\label{5.3.38}\end{equation}
\begin{equation}
\begin{array}{rcl}
\Sigma_{n}\phi_{n}^{\dagger}\left(x\right)\gamma^{5}\phi_{n}\left(x\right) & = & \textrm{lim}_{M\rightarrow\infty}\Sigma_{n}\phi_{n}^{\dagger}\left(x\right)\gamma^{5}\textrm{ exp}\left[\frac{\not D^{2}}{M^{2}}\right]\phi_{n}\left(x\right)\\
 & = & \textrm{lim}_{M\rightarrow\infty}\int\frac{\textrm{d}^{4}k_{1}}{\left(2\pi\right)^{2}}\frac{\textrm{d}^{4}k_{2}}{\left(2\pi\right)^{2}}\Sigma_{n}\tilde{\phi}_{n}^{\dagger}\left(k_{1}\right)e^{-ik_{2}.x}\gamma^{5}\textrm{ exp}\left[\frac{\not D^{2}}{M^{2}}\right]\tilde{\phi}_{n}\left(k_{2}\right)e^{ik_{1}.x}\\
 & = & \textrm{lim}_{M\rightarrow\infty}\int\frac{\textrm{d}^{4}k_{1}}{\left(2\pi\right)^{2}}\frac{\textrm{d}^{4}k_{2}}{\left(2\pi\right)^{2}}Tr\left[e^{-ik_{2}.x}\gamma^{5}\textrm{ exp}\left[\frac{\not D^{2}}{M^{2}}\right]e^{ik_{1}.x}\right]\delta\left(k_{1}-k_{2}\right).\end{array}\label{5.3.39}\end{equation}
Where in the last line we have used the completeness of $\tilde{\phi}_{n}$
, performed the sum and used the cyclic property of the trace in the
same way as we did on the last page where we showed the Jacobian was
ill-defined and needed to be regulated. We can rewrite the exponent
using some Dirac matrix gymnastics as, \begin{equation}
\begin{array}{rcl}
\not D^{2} & = & \gamma^{\mu}\gamma^{\nu}D_{\mu}D_{\nu}\\
 & = & \frac{1}{2}\left\{ \gamma^{\mu},\gamma^{\nu}\right\} D_{\mu}D_{\nu}+\frac{1}{2}\left[\gamma^{\mu},\gamma^{\nu}\right]D_{\mu}D_{\nu}\\
 & = & D^{\mu}D_{\mu}+\frac{1}{4}\left[\gamma^{\mu},\gamma^{\nu}\right]D_{\mu}D_{\nu}-\frac{1}{4}\left[\gamma^{\mu},\gamma^{\nu}\right]D_{\nu}D_{\mu}\\
 & = & D^{\mu}D_{\mu}+\frac{1}{4}\left[\gamma^{\mu},\gamma^{\nu}\right]\left[D_{\mu},D_{\nu}\right]\\
 & = & D^{\mu}D_{\mu}+\frac{1}{4}\left[\gamma^{\mu},\gamma^{\nu}\right]F_{\mu\nu}.\end{array}\label{5.3.40}\end{equation}
Where we have used the result from 1.3 for the field strength tensor
$F_{\mu\nu}=\left[D_{\mu},D_{\nu}\right]$. We can also integrate
over $k_{2}$ which sets $k_{1}=k_{2}$ on account of the delta function
that we picked up from the completeness relation. \begin{equation}
\Sigma_{n}\phi_{n}^{\dagger}\left(x\right)\gamma^{5}\phi_{n}\left(x\right)=\textrm{lim}_{M\rightarrow\infty}\int\frac{\textrm{d}^{4}k_{1}}{\left(2\pi\right)^{4}}Tr\left[e^{-ik_{1}.x}\gamma^{5}\textrm{ exp}\left[\frac{D^{\mu}D_{\mu}}{M^{2}}+\frac{\frac{1}{4}\left[\gamma^{\mu},\gamma^{\nu}\right]F_{\mu\nu}}{M^{2}}\right]e^{ik_{1}.x}\right]\label{5.3.41}\end{equation}
The next thing to do is move the factor of $e^{ik_{1}.x}$ through
to meet the other one which gives a delta function of $k_{1}$ and
$k_{2}$. This is easy but we need to look out for the differential
operator in $D_{\mu}D^{\mu}$ acting on the $e^{ik_{1}.x}$ factor.
In the limit $M\rightarrow\infty$ we can \begin{equation}
e^{-ik_{1}.x}\gamma^{5}\textrm{ exp}\left[\frac{D^{\mu}D_{\mu}}{M^{2}}\right]e^{ik_{1}.x}=\gamma^{5}e^{-ik_{1}.x}\left(1+\frac{D^{\mu}D_{\mu}}{M^{2}}+\frac{1}{2}\left(\frac{D^{\mu}D_{\mu}}{M^{2}}\right)^{2}...\right)e^{ik_{1}.x}.\label{5.3.42}\end{equation}
It is now necessary to simplify the $D_{\mu}D^{\mu}$ term, all the
while we have to consider that this is an operator equation, \emph{i.e.}
that $D_{\mu}D^{\mu}$ is acting on something else on the left besides
$e^{ik_{1}.x}$. \begin{equation}
\begin{array}{rcl}
e^{-ik_{1}.x}D_{\mu}D^{\mu}e^{ik_{1}.x} & = & e^{-ik_{1}.x}\left(i\partial_{\mu}+A_{\mu}\right)\left(i\partial^{\mu}+A^{\mu}\right)e^{ik_{1}.x}\\
 & = & e^{-ik_{1}.x}\left(i\partial_{\mu}+A_{\mu}\right)\left(-e^{ik_{1}.x}k_{1}^{\mu}+e^{ik_{1}.x}i\partial^{\mu}+e^{ik_{1}.x}A^{\mu}\right)\\
 & = & e^{-ik_{1}.x}\left(i\partial_{\mu}+A_{\mu}\right)e^{ik_{1}.x}\left(-k_{1}^{\mu}+i\partial^{\mu}+A^{\mu}\right)\end{array}\label{5.3.43}\end{equation}
Now we move $e^{ik_{1}.x}$ left through the next $D_{\mu}$, this
has the same effect as above and in addition the two exponentials
cancel, \begin{equation}
\begin{array}{crcl}
\Rightarrow & e^{-ik_{1}.x}D_{\mu}D^{\mu}e^{ik_{1}.x} & = & \left(-k_{1\mu}+i\partial_{\mu}+A_{\mu}\right)\left(ik_{1}^{\mu}+\partial^{\mu}+A^{\mu}\right)\\
 &  & = & \left(D_{\mu}-k_{1\mu}\right)\left(D^{\mu}-k_{1}^{\mu}\right).\end{array}\label{5.3.44}\end{equation}
 Repeating this for the following terms in the series $e^{-ik_{1}.x}\frac{1}{2}\left(\frac{D_{\mu}D^{\mu}}{M^{2}}\right)^{2}e^{ik_{1}.x}$
gives a completely analogous result \emph{i.e.} simply replace $D\rightarrow D-k_{1}$
and remove the exponentials. This being the case we have, \begin{equation}
\begin{array}{rcl}
\Sigma_{n}\phi_{n}^{\dagger}\left(x\right)\gamma^{5}\phi_{n}\left(x\right) & = & \textrm{lim}_{M\rightarrow\infty}\int\frac{\textrm{d}^{4}k_{1}}{\left(2\pi\right)^{4}}Tr\left[\gamma^{5}\textrm{ exp}\left[\frac{\left(D^{\mu}-k_{1}^{\mu}\right)\left(D_{\mu}-k_{1\mu}\right)}{M^{2}}+\frac{\frac{1}{4}\left[\gamma^{\mu},\gamma^{\nu}\right]F_{\mu\nu}}{M^{2}}\right]\right]\\
 & = & \textrm{lim}_{M\rightarrow\infty}\int\frac{\textrm{d}^{4}k_{1}}{\left(2\pi\right)^{4}}e^{\frac{+k_{1}^{\mu}k_{1\mu}}{M^{2}}}Tr\left[\gamma^{5}\textrm{ exp}\left[-\frac{2k_{\mu}D^{\mu}}{M^{2}}+\frac{D_{\mu}D^{\mu}}{M^{2}}+\frac{\frac{1}{4}\left[\gamma^{\mu},\gamma^{\nu}\right]F_{\mu\nu}}{M^{2}}\right]\right]\\
 & = & \textrm{lim}_{M\rightarrow\infty}\int\frac{\textrm{d}^{4}k_{1}}{\left(2\pi\right)^{4}}M^{4}e^{+k_{1}^{\mu}k_{1\mu}}Tr\left[\gamma^{5}\textrm{ exp}\left[-\frac{2k_{\mu}D^{\mu}}{M}+\frac{D_{\mu}D^{\mu}}{M^{2}}+\frac{\gamma^{\mu}\gamma^{\nu}F_{\mu\nu}}{2M^{2}}\right]\right],\end{array}\label{5.3.45}\end{equation}
 where in the last step we simply rescaled the momentum integral as
$k_{1}\rightarrow Mk_{1}$ and used the fact that $F_{\mu\nu}$ is
$\mu\leftrightarrow\nu$ symmetric. 

In Taylor expanding the exponential the first term that will be non-zero
is the one quadratic in $F_{\mu\nu}$. This is because of the $\gamma^{5}$
in the trace (the trace was over the eigenfunctions of $\not D$ in
position space (\emph{i.e.} $x$) but also naturally over the Dirac
indices). The first non vanishing trace including a $\gamma^{5}$
is the trace of $\gamma^{5}$ with four other $\gamma$ matrices,
$Tr\left[\gamma^{5}\gamma^{\mu}\gamma^{\nu}\gamma^{\alpha}\gamma^{\beta}\right]=-4\epsilon^{\mu\nu\alpha\beta}$
(in Euclidean space). The first term in the expansion to have this
property which is least suppressed by powers of $M$ is clearly, \begin{equation}
\frac{1}{2}\left(\frac{\gamma^{\mu}\gamma^{\nu}F_{\mu\nu}}{2M^{2}}\right)^{2}.\label{5.3.46}\end{equation}
This is also the only non vanishing term in the $M\rightarrow\infty$,
it is the only such term with four powers of $M$ in the numerator
and the denominator (we picked up a factor $M^{4}$ on rescaling the
momentum), all other terms with non-vanishing traces will have at
least one power of $M$ in the denominator and so vanish as $M\rightarrow\infty$.
Therefore,\begin{equation}
\begin{array}{rcl}
\Sigma_{n}\phi_{n}^{\dagger}\left(x\right)\gamma^{5}\phi_{n}\left(x\right) & = & \textrm{lim}_{M\rightarrow\infty}\int\frac{\textrm{d}^{4}k_{1}}{\left(2\pi\right)^{4}}M^{4}e^{+k_{1}^{\mu}k_{1\mu}}Tr\left[\gamma^{5}\textrm{ }\frac{1}{2!}\left(\frac{\gamma^{\mu}\gamma^{\nu}F_{\mu\nu}}{2M^{2}}\right)^{2}\right]\\
 & = & \int\frac{\textrm{d}^{4}k_{1}}{\left(2\pi\right)^{4}}e^{+k_{1}^{\mu}k_{1\mu}}Tr\left[\gamma^{5}\textrm{ }\frac{1}{8}\left(\gamma^{\mu}\gamma^{\nu}F_{\mu\nu}\right)^{2}\right]\\
 & = & \int\frac{\textrm{d}^{4}k_{1}}{\left(2\pi\right)^{4}}e^{+k_{1}^{\mu}k_{1\mu}}\frac{-1}{2}\epsilon^{\mu\nu\alpha\beta}F_{\mu\nu}F_{\alpha\beta}\end{array}\label{5.3.47}\end{equation}
(the minus sign appears from the trace over Dirac indices in Euclidean
space $Tr\left[\gamma^{5}\gamma^{\mu}\gamma^{\nu}\gamma^{\alpha}\gamma^{\beta}\right]=-4\epsilon^{\mu\nu\alpha\beta}$).
Now we recall that in going from Minkowski space to Euclidean space
we have $k_{1}^{\mu}k_{1\mu}=-k_{1\mu}k_{1\mu}$ where the $k_{1\mu}k_{1\mu}$
vectors are Euclidean. We can therefore perform the integral over
$k_{1}$ ($F_{\mu\nu}$ does not depend on $k_{1}$) as it is now
just a product for four regular Gaussian integrals (one for each dimension).\begin{equation}
\int\textrm{d}^{4}k_{1}\textrm{ }e^{-k_{1\mu}k_{1\mu}}=\pi^{2}\label{5.3.48}\end{equation}
\begin{equation}
\Rightarrow\Sigma_{n}{\bkiii{n}{\gamma^{5}}{n}}=-\frac{1}{32\pi^{2}}\epsilon^{\mu\nu\alpha\beta}F_{\mu\nu}F_{\alpha\beta}\label{5.3.49}\end{equation}
 Going back a few pages we found that the path integral measure changed
as,\begin{equation}
\prod_{n}\int\textrm{d}a_{n}\int\textrm{d}\bar{b}_{n}\rightarrow\left(\textrm{exp }-2i\Sigma_{n}\textrm{ }\int\textrm{d}^{4}x\textrm{ }\beta\left(x\right){\bkii{n}{x}}\gamma^{5}{\bkii{x}{n}}\right)\prod_{n}\int\textrm{d}a_{n}\int\textrm{d}\bar{b}_{n}\label{5.3.50}\end{equation}
substituting in the result above we have,\begin{equation}
\begin{array}{rcl}
\prod_{n}\int\textrm{d}a_{n}\int\textrm{d}\bar{b}_{n} & \rightarrow & \left(\textrm{exp }\frac{i}{16\pi^{2}}\int\textrm{d}^{4}x\textrm{ }\beta\left(x\right)\epsilon^{\mu\nu\alpha\beta}F_{\mu\nu}F_{\alpha\beta}\right)\prod_{n}\int\textrm{d}a_{n}\int\textrm{d}\bar{b}_{n}\\
 & = & \left(\textrm{exp }-\int\textrm{d}^{4}x\textrm{ }\beta\left(x\right){\cal {A}}\left[A_{\mu}\right]\right)\prod_{n}\int\textrm{d}a_{n}\int\textrm{d}\bar{b}_{n}\end{array}\label{5.3.51}\end{equation}
where, \begin{equation}
{\cal {A}}\left[A_{\mu}\right]=\frac{-i}{16\pi^{2}}\epsilon^{\mu\nu\alpha\beta}F_{\mu\nu}F_{\alpha\beta}\label{5.3.52}\end{equation}
 is the so-called singlet / axial / chiral anomaly. To get the anomaly
in Minkowski space we need to make one change which is that taking
the trace of the $\gamma$ matrices above in Minkowski space gives
$-4i\epsilon^{\mu\nu\alpha\beta}F_{\mu\nu}F_{\alpha\beta}$, which
means it is different by a factor $-i$,\begin{equation}
{\cal {A}}\left[A_{\mu}\right]=\frac{1}{16\pi^{2}}\epsilon^{\mu\nu\alpha\beta}F_{\mu\nu}F_{\alpha\beta}.\label{5.3.53}\end{equation}
 So in performing the usual transformation of variables technique
to derive the axial current conservation equation we have the same
terms as before but now we also have the anomaly term,\begin{equation}
\Rightarrow{\bki{\partial_{\mu}j_{A}^{\mu}}}=2im{\bki{\bar{\psi}\gamma^{5}\psi}}+{\bki{{\cal {A}}\left[A_{\mu}\right]}}.\label{5.3.54}\end{equation}
It turns out that this expression is the same as that which we calculated
in perturbation theory from the triangle diagrams, to do this you
consider the field strength tensor $F_{\mu\nu}$ associated with the
two photons in the triangle diagram with momenta $k_{1}$, $k_{2}$
and polarization vectors $\epsilon_{1}$ and $\epsilon_{2}$. This
is one of a long list of peculiarities of the anomaly, that the lowest
order calculation of the anomaly (\emph{i.e.} with the triangle diagrams)
receives no higher order corrections, it agrees with the non-perturbative
calculation in this section. 

To finish off we list briefly some of the broader issues and spin
offs associated with this calculation, unfortunately time and space
does not permit more detail. 

The first thing is related to an old problem associated with the electromagnetic
decay of the pion $\pi^{0}\rightarrow\gamma\gamma$, this decay is
a triangle diagram of the form calculated in the previous section.
Pions are pseudoscalar particles and in the limit of soft interactions
$\left(q\approx0\right)$ we can effectively consider them as a point
like particle with a pseudoscalar coupling (\emph{i.e.} only a $\gamma^{5}$
at the vertex) thus their decay basically looks like the VVP diagram
in section 5.2. However the naive Ward identity we derived between
the VVA and VVP diagrams told us that,\begin{equation}
q^{\lambda}T_{\mu\nu\lambda}\left(k_{1},k_{2},q\right)=2mT_{\mu\nu}\left(k_{1},k_{2},q\right)\label{5.3.55}\end{equation}
\emph{i.e.} that the amplitude of the VVP diagram is effectively zero
in the limit $q\rightarrow0$, or equivalently, the $\pi^{0}$ doesn't
decay! This was called the \emph{Sutherland}-\emph{Veltman} \emph{Paradox}.
It is one of the great successes of the discovery of the anomaly that
the $\pi^{0}$ lifetime agrees very well with experiment when one
takes into account the anomaly in the triangle diagrams of low energy
effective theory of the $\pi^{0}$. Such a calculation of the $\pi^{0}$
lifetime can be found in $\left[3\right]$.

Anomalies are more well known these days for their role in constraining
physical theories. The constraints arise because innocuous as it may
seem the fact that anomalies violate our classical Ward identities
is another way of saying that the gauge invariance associated with
the anomaly does not exist at the quantum level. The existence of
the quantum Ward identities is crucial to the renormalizability of
a theory, we saw how they killed off the nasty quadratic divergences
in QED for example. It also enabled us to derive relations between
divergent diagrams which in turn gave a cancellation of those divergences
(perhaps a better way of saying this is that it gave us relations
between the renormalization coefficients). The fact that these identities
are not true ruins the Ward identities and so we lose the ability
to have such cancellations, this was very clear in the case of trying
to do perturbation theory with the VVA diagram which showed an ambiguous
amplitude, the naive Ward identity between the VVA and VVP diagrams
was untrue. It is also the case that anomalies give more nonsense
in so far as the S-Matrix acquires a gauge dependence, see $\left[9\right]$.
A sensible unambiguous (and renormalizable) theory is necessarily
anomaly free, this is the aforementioned constraint which aids model
building. Had we considered a non-Abelian gauge theory in the treatment
of the last two sections we would have had various factors involving
the gauge group generators appearing, around the vertices of the triangle
diagrams in 5.2 and in the transformations of the fermion fields in
5.3. The modification to the calculations is trivial and the calculations
proceed as they do in 5.2 and 5.3 (but with more mess). The upshot
is that the anomaly contains a factor proportional to the gauge group
generators,\begin{equation}
A^{abc}=tr\left[t^{a}\left\{ t^{b}t^{c}+t^{c}t^{b}\right\} \right]\label{5.3.56}\end{equation}
where the trace is over the indices of the internal symmetry space.
Substituting in the generators of Electroweak theory the anomaly from
the VVA diagrams can yield a term such that,\begin{equation}
tr\left[t^{a}\left\{ t^{b}t^{c}+t^{c}t^{b}\right\} \right]=tr\left[t^{a}\left\{ t^{b}Y+Yt^{b}\right\} \right]\propto\delta^{ab}tr\textrm{ }Q.\label{5.3.57}\end{equation}
Thus anomaly cancellation in the Standard Model requires that the
sum of the electric charges of all the fermions should vanish%
\footnote{Remember a colour factor $N_{C}=3$ when adding up the quarks charges.%
}.

\chapter*{Acknowledgments}

\addcontentsline{toc}{chapter}{\numberline{}{Acknowledgments}} 

KMH is grateful for the financial support of PPARC. The author would
also like to express his gratitude for support and patience from his
family, Katy, G.G.Ross, P.Renton and Subir. Finally the author would
like to thank, in advance, any readers for emailing corrections and
criticisms (\emph{email: k.hamilton1@physics.ox.ac.uk}).

\end{document}